\pdfoutput=1
\documentclass[phd, icsa, logo, fullspacing, twoside, 12pt, leftchapter, sansheadings, notimes, romanprepages]{infthesis}

\usepackage[style=numeric,backend=bibtex8,minnames=20,maxnames=20,sorting=ydnt,giveninits=true,sortcites,backref=true]{biblatex}

\usepackage{comment}
\usepackage{appendix}
\usepackage{graphicx}
\usepackage{listings}
\usepackage{blindtext}
\usepackage{lipsum}
\usepackage[normalem]{ulem}
\usepackage[table,dvipsnames]{xcolor}
\definecolor[named]{ACMBlue}{cmyk}{1,0.1,0,0.1}
\definecolor[named]{ACMYellow}{cmyk}{0,0.16,1,0}
\definecolor[named]{ACMOrange}{cmyk}{0,0.42,1,0.01}
\definecolor[named]{ACMRed}{cmyk}{0,0.90,0.86,0}
\definecolor[named]{ACMLightBlue}{cmyk}{0.49,0.01,0,0}
\definecolor[named]{ACMGreen}{cmyk}{0.20,0,1,0.19}
\definecolor[named]{ACMPurple}{cmyk}{0.55,1,0,0.15}
\definecolor[named]{ACMDarkBlue}{cmyk}{1,0.58,0,0.21}
\PassOptionsToPackage{hyphens}{url}\usepackage[bookmarks=true,breaklinks=true,colorlinks,linkcolor=ACMPurple,citecolor=ACMPurple,urlcolor=ACMDarkBlue,filecolor=ACMDarkBlue]{hyperref}
\usepackage{microtype}
\usepackage[caption=false]{subfig}
\usepackage[style=base]{caption}
\usepackage{floatrow}
\usepackage[export]{adjustbox}%
\usepackage{numprint}
\npthousandsep{,}
\npdecimalsign{.}
\usepackage{multirow}
\usepackage{array,booktabs}
\usepackage{tabularx}
\usepackage{ragged2e}
\usepackage{hhline}
\usepackage{makecell}
\newcolumntype{L}[1]{>{\raggedright}m{#1}}
\newcolumntype{C}[1]{>{\centering}m{#1}}
\newcolumntype{R}[1]{>{\raggedleft}m{#1}}
\newcommand{\tn}{\tabularnewline}
\usepackage{amsfonts}

\hypersetup{
    bookmarksopen = true,
    bookmarksdepth= 3,%
    pdfauthor     = {Priyank Faldu},
    pdftitle      = {Addressing Variability in Reuse Prediction for Last-Level Caches by Priyank Faldu},
    pdfsubject    = {Addressing Variability in Reuse Prediction for Last-Level Caches by Priyank Faldu},
    pdfkeywords   = {Priyank Faldu, PhD, Thesis, University of Edinburgh, School of Informatics}
}

\usepackage{libertine}
\usepackage{libertinust1math}
\usepackage[utf8]{inputenc}
\usepackage[T1]{fontenc}

\emergencystretch=\maxdimen
\usepackage{xspace}
\usepackage{floatrow}
\usepackage[caption=false]{subfig}
\usepackage[style=base]{caption}
\usepackage{transparent}
\newcommand\stitchref[2]{\ref{#1}\subref{#2}}
\newcommand{\dbgacronym}{{\em \underline{D}egree-\underline{B}ased \underline{G}rouping}}
\newcommand{\dbgfullname}{Degree-Based Grouping\xspace}
\newcommand{\dbg}{DBG\xspace}
\newcommand{\gorder}{Gorder\xspace}
\newcommand{\samacronym}{{\em GRASP -- \underline{GRA}ph-\underline{SP}ecialized }}
\newcommand\sam{GRASP\xspace}
\newcommand\samnospace{GRASP}

\newcommand{\floor}[1]{\left\lfloor #1 \right\rfloor}

\newcommand\maxdegree{{$\mathbb{M}$}\xspace}
\newcommand\avgdegree{{$\mathbb{A}$}\xspace} %
\newcommand\constant{{$\mathbb{C}$}\xspace} %
\newcommand\infinity{{$\infty$}\xspace}
\usepackage{pifont}%
\newcommand{\cmark}{\ding{51}}%
\newcommand{\xmark}{\ding{55}}%

\newcommand\silentdrop[1]{}

\newcommand\slightlylarge[0]{\fontsize{12}{0}\selectfont}
\newcommand\vspacexs[0]{\vspace{0.1\baselineskip}}
\newcommand\vspaces[0]{\vspace{0.25\baselineskip}}

\newcommand\vspacel[0]{\vspace{1\baselineskip}}

\definecolor{amber}{rgb}{1.0, 0.49, 0.0}
\definecolor{amethyst}{rgb}{0.6, 0.4, 0.8}
\ifdefined\RELEASE
\else
\fi

\newcommand\noindenttitle[1]{\vspaces \noindent {\bf #1}}

\newcommand\noindentsubsectiontitle[1]{\vspacexs \noindent {\bf #1}}
\newcommand\noindentnotitle[0]{\vspaces \noindent {}}
\newcommand\indentnotitle[0]{\vspaces {}}

\newcommand\visiblespace{\vspace{0.2\baselineskip}}

\newcommand\num[1]{\footnotesize{\numprint{#1}}}
\newcommand\tw{\emph{tw}\xspace}
\newcommand\fr{\emph{fr}\xspace}
\newcommand\wl{\emph{wl}\xspace}
\newcommand\kr{\emph{kr}\xspace}
\newcommand\lj{\emph{lj}\xspace}
\newcommand\pl{\emph{pl}\xspace}
\newcommand\pld{\emph{pl}\xspace}
\newcommand\sd{\emph{sd}\xspace}
\newcommand\mpi{\emph{mp}\xspace}
\newcommand\uni{\emph{uni}\xspace}
\newcommand\pr{PR\xspace}
\newcommand\prd{PRD\xspace}
\newcommand\sssp{SSSP\xspace}
\newcommand\bc{BC\xspace}
\newcommand\radii{Radii\xspace}
\newcommand\ie[0]{i.e., }
\newcommand\eg[0]{e.g., }
\newcommand\etal[0]{et al. }
\usepackage{tikz}
\newcommand*\smalltextcircled[1]{\tikz[baseline=(char.base)]{
    \node[shape=circle,draw,inner sep=1pt] (char) {#1};}}

\newcommand\citeall[0]{\cite{PseudoLRU,dip,rrip,gippr,tadip,segmentedLRU,timekeeping,timekeeping2,counter,sampler,ship,mdpp,perceptron,multi,hawkeye,harmony,pd,EAF,shepherd,pseudo,mlp,pipp,lrfu,doubledip,adaptivecaches}}
\newcommand\citeallsoft[0]{\cite{PseudoLRU,dip,rrip,gippr,tadip,segmentedLRU,timekeeping,timekeeping2,counter,sampler,ship,mdpp,perceptron,multi,hawkeye,harmony,pd,EAF,shepherd,pseudo,mlp,pipp,lrfu,doubledip,adaptivecaches,cache-hints1,cache-hints1-soft-assisted,cache-hints2,cache-hints3, cache-hints5-pacman,xmem}}

\newcommand\citeprob[0]{\cite{dip,rrip,tadip,gippr,segmentedLRU,doubledip,adaptivecaches}}
\newcommand\citehistory[0]{\cite{timekeeping,timekeeping2,counter,sampler,ship,mdpp,perceptron,multi,hawkeye,harmony}}

\newcommand\citepriordbp[0]{\cite{counter,sampler,mdpp,ship,perceptron,hawkeye}}
\newcommand\citepcandcrc[0]{\cite{counter,sampler,mdpp,ship,perceptron,multi,hawkeye,harmony,crc2-1shippp,crc2-2lime,crc2-3multi,crc2-4expected,c-crc2-5reuse,crc2-6hawkeye,crc2-7red}}

\newcommand\citesoftware[0]{\cite{cache-hints1, cache-hints1-soft-assisted, cache-hints2, cache-hints3, cache-hints5-pacman, xmem}}
\newcommand\citepc[0]{\cite{counter,sampler,mdpp,ship,perceptron,multi,hawkeye,harmony}} %
\newcommand\citecrconly[0]{\cite{crc2-1shippp,crc2-2lime,crc2-3multi,crc2-4expected,c-crc2-5reuse,crc2-6hawkeye,crc2-7red}}

\title{Addressing Variability in Reuse Prediction for Last-Level Caches}

\author{Priyank Faldu}

\submityear{2019}

\abstract {

Last-Level Cache (LLC) represents the bulk of a modern CPU processor's transistor budget and is essential for application performance as LLC enables fast access to data in contrast to much slower main memory. Problematically, technology constraints make it infeasible to scale LLC capacity to meet the ever-increasing working set size of the applications. Thus, future processors will rely on effective cache management mechanisms and policies to get more performance out of the scarce LLC capacity.

Applications with large working set size often exhibit streaming and/or thrashing access patterns at LLC. As a result, a large fraction of the LLC capacity is occupied by \emph{dead blocks} that will not be referenced again, leading to inefficient utilization of the LLC capacity. To improve cache efficiency, the state-of-the-art cache management techniques employ prediction mechanisms that learn from the past access patterns with an aim to accurately identify as many dead blocks as possible. Once identified, dead blocks are evicted from LLC to make space for potentially high reuse cache blocks.

In this thesis, we identify {\em variability} in the reuse behavior of cache blocks as the key limiting factor in maximizing cache efficiency for state-of-the-art predictive techniques. Variability in reuse prediction is inevitable due to numerous factors that are outside the control of LLC. The sources of variability include control-flow variation, speculative execution and contention from cores sharing the cache, among others. Variability in reuse prediction challenges existing techniques in reliably identifying the end of a block's useful lifetime, thus causing lower prediction accuracy, coverage, or both. To address this challenge, this thesis aims to design robust cache management mechanisms and policies for LLC in the face of variability in reuse prediction to minimize cache misses, while keeping the cost and complexity of the hardware implementation low. To that end, we propose two cache management techniques, one domain-agnostic and one domain-specialized, to improve cache efficiency by addressing variability in reuse prediction.

In the first part of the thesis, we consider domain-agnostic cache management, a conventional approach to cache management, in which the LLC is managed fully in hardware, and thus the cache management is transparent to the software. In this context, we propose {\em Leeway}, a novel domain-agnostic cache management technique. Leeway introduces a new metric, {\em Live Distance}, that captures the largest interval of temporal reuse for a cache block, providing a conservative estimate of a cache block's useful lifetime. Leeway implements a robust prediction mechanism that identifies dead blocks based on their past Live Distance values. Leeway monitors the change in Live Distance values at runtime and dynamically adapts its reuse-aware policies to maximize cache efficiency in the face of variability.

In the second part of the thesis, we identify applications, for which existing domain-agnostic cache management techniques struggle in exploiting the high reuse due to variability arising from certain fundamental application characteristics. Specifically, applications from the domain of graph analytics inherently exhibit high reuse when processing natural graphs. However, the reuse pattern is highly irregular and dependent on graph topology; a small fraction of vertices, {\em hot vertices}, exhibit high reuse whereas a large fraction of vertices exhibit low- or no-reuse. Moreover, the hot vertices are sparsely distributed in the memory space. Data-dependent irregular access patterns, combined with the sparse distribution of hot vertices, make it difficult for existing domain-agnostic predictive techniques in reliably identifying, and, in turn, retaining hot vertices in cache, causing severe underutilization of the LLC capacity.

In this thesis, we observe that the software is aware of the application reuse characteristics, which, if passed on to the hardware efficiently, can help hardware in reliably identifying the most useful working set even amidst irregular access patterns. To that end, we propose a holistic approach of software-hardware co-design to effectively manage LLC for the domain of graph analytics. Our software component implements a novel lightweight software technique, called {\em Degree-Based Grouping (DBG)}, that applies a coarse-grain graph reordering to segregate hot vertices in a contiguous memory region to improve spatial locality. Meanwhile, our hardware component implements a novel domain-specialized cache management technique, called {\em Graph Specialized Cache Management (GRASP)}. GRASP augments existing cache policies to maximize reuse of hot vertices by protecting them against cache thrashing, while maintaining sufficient flexibility to capture the reuse of other vertices as needed. To reliably identify hot vertices amidst irregular access patterns, GRASP leverages the DBG-enabled contiguity of hot vertices. Our domain-specialized cache management not only outperforms the state-of-the-art domain-agnostic predictive techniques, but also eliminates the need for any storage-intensive prediction mechanisms. 

}

\begin{document}

\begin{preliminary}

\maketitle

\begin{laysummary}
Over the past few decades, technological advancements in the semiconductor industry have made the processors and the main memory significantly faster. However, the main memory has been getting faster at a much slower rate than the processors, widening the gap between the speed of the processor and the main memory. 
Consequently, slow access time of the main memory is one of the major performance bottlenecks in modern computer systems as the processor needs to access data items (\ie program instructions and data) from the main memory to perform computations.

To avoid accessing the main memory for every data item the processor needs, the computer systems employ multiple caches between the processor and the main memory. A cache is a form of memory, which is significantly faster (and closer to the processor) than the main memory and thus retrieving a data item from the cache is much faster than retrieving it from the main memory. However, a cache is significantly more expensive (in dollars per byte) than the main memory. Consequently, caches tend to have considerably smaller capacity in comparison to the main memory, warranting judicious use of the precious cache capacity. To that end, the goal of a cache management technique is to decide which data items to store in the cache to minimize the number of accesses to the main memory.

For cache management, Last-Level Cache (LLC) is of particular interest as it offers the largest capacity among all caches. 
Cache management for LLC controls which data items are stored in the LLC. As application executes and accesses more data items, cache management predicts which data items are more likely to be reused in the near future, and thus should be stored in the LLC. Meanwhile, when the cache is full, cache management also predicts which data items are unlikely to be reused in the near future, and thus can be removed.
Naturally, the more accurate the reuse predictions, the better the cache efficiency.

State-of-the-art cache management techniques for LLC observe cache access patterns of the data items over time and utilize this information to predict the future reuse of data items. 
In this thesis, we show that the LLC observes inconsistent access patterns for many data items due to numerous factors that are outside the control of the LLC. Thus, data items inevitably exhibit {\em variability} in the reuse behavior at LLC, limiting existing techniques in making accurate predictions. In response, this thesis aims to design robust cache management mechanisms and policies for LLC to minimize cache misses in the face of variability in reuse prediction, while keeping the cost and complexity of the hardware implementation low. To that end, we propose two new cache management techniques incorporating various variability-tolerant features.

\end{laysummary}

\begin{acknowledgements}

It is impossible to get admitted to the PhD program of a world class university, let alone graduate from it, without the constant help, support and guidance from family, friends and teachers. Naming all of them is not possible, but I sincerely thank each and every one of them from the bottom of my heart. Below, I specifically acknowledge a selected group of people without whom this thesis wouldn't have been possible.

First and foremost, my sincerest gratitude to Prof. Boris Grot, who has been truly a remarkable advisor throughout my PhD program. Boris has always been open for discussions and brainstorming, and has also given me the freedom to explore new problems on my own. Boris not only helped me improve my research skills, but also ensured my all round development; whether it was encouraging me to mentor students in their projects, trusting me with the teaching and tutoring duties, nominating me for the organizing committee of ISCA, enabling me to network with the wider research community or even patiently helping me with my writing skills, his contributions have been enormous. His critical thinking, great attention to details, and above all, his compassionate attitude towards his students make him the perfect advisor one could ask for. I am very grateful to Boris for all the guidance and support throughout my PhD, and also privileged to be his first PhD student. 

The other most important person to whom I am indebted is my wife, Kruti, for her unconditional love and unwavering support. She is the one who encouraged me to pursue PhD, even if that meant getting off of the driver's seat of her career. Words are not enough to describe her contributions as she took all the responsibilities upon herself to ensure I can focus on my research. Kruti has made several sacrifices for me to be able to complete my PhD, and for that she deserves an equal credit, if not more, for this thesis. She has been the source of encouragement during the tough times of paper rejections, and the perfect companion to celebrate every milestone on the way, little or big. I can safely say that Kruti has made me a better researcher, and more importantly a better person.

I thank Oracle for the internship opportunity, and my mentors over there, Dr. Jeff Diamond and Dr. Avadh Patel, for making my internship an enriching experience. The work that I started during the internship, and expanded in the subsequent years, turned out to be a stepping stone for my thesis, spanning three out of four technical chapters.

I am fortunate to have had the opportunity to interact with and learn from Prof. Vijay Nagarajan, Prof. Björn Franke, Prof. Daniel Sorin, Prof. Babak Falsafi, Prof. Timothy Pinkston, Prof. Murali Annavaram, Prof. Daniel Jiménez, Prof. Rajeev Balasubramonian, my thesis examiners Prof. Michael O’Boyle and Dr. Gabriel Loh, and the anonymous reviewers from the Computer Architecture community. Learning from the very best of the field has been a privilege, and has made a far reaching impact on me.

Special thanks to my friends in the School of Informatics and my academic siblings, Artemiy Margaritov and Amna Shahab, without whom the days would have passed far more slowly. They provided valuable feedback and suggestions to improve my ideas. Endless discussions, sometimes technical but more often not, provided much needed break during those intense days before the submissions. We have been through each others ups and downs together.

I thank the faculty members, support staff and students of the School of Informatics for their help and support. I would like to specially thank Antonios Katsarakis, Arpit Joshi, Cheng-Chieh Huang, Dmitrii Ustiugov, Rakesh Kumar, Saumay Dublish, Siavash Katebzadeh and Vasilis Gavrielatos, for their valuable help and support, both technical and otherwise.

My time in Edinburgh has been made special with the friendship of Supriya and Sidharth Kashyap. I thank them for their company and providing such a rich source of conversation, education and entertainment. They have been the family away from home.

Finally, last but not the least, I would like to thank my parents, Popatlal and Jyotsana, and sisters, Urvi and Ronak, for their endless love. I would not be who I am today without their enormous support and sacrifices throughout my life.

As the submission of this thesis turns a new chapter in my life, I thank God for the perfectly timed wonderful gift in the form of my little son, Mivaan. With him on my side, I look forward to embark upon a new journey \ldots

\end{acknowledgements}

\dedication{\em \large{Dedicated to my wife, Kruti.}}

\tableofcontents

\end{preliminary}

\ifdefined\RELEASE
\chapter{Introduction\label{ch:intro}}
The microprocessor industry has enjoyed four decades of exponentially growing transistor budgets, enabling complex core microarchitectures, multi-core processors, and cache capacities reaching into tens of megabytes (MB) for commodity processors. 
The looming reality, however, is that Moore's law is nearing its limits both in terms of physics and economics. 
Combined with the end of voltage scaling, the semiconductor industry is entering a new phase where transistors become a limited resource and a new technology generation cannot be counted on to double them. 
This calls for a new regime in computer systems, one in which every transistor counts.

Last-Level Cache (LLC) represents the bulk of a modern processor's transistor budget and is
an essential feature of modern processors. 
Fig.~\ref{fig:back:die} shows die photos of two modern processors showing LLC (labeled L3) occupying nearly the same area as the processor cores.
LLC has been instrumental in bridging the gap in the speed of processor and memory via ever-larger capacities, providing performance gains across processor generations.
In the future, however, further increases in cache capacity may become a difficult proposition due to technology constraints. 
Thus, future processors will rely on effective cache management mechanisms and policies to get more performance out of the scarce LLC capacity and minimize long latency memory accesses.

\begin{figure}[!t]
    \subfloat[Intel Broadwell E Core i7-6950X featuring 10 cores and 25MB shared L3 (2.5MB L3 slice per core)~\cite{inteldie}.]{\label{fig:back:inteldie}{\includegraphics[width=0.45\linewidth,valign=b]{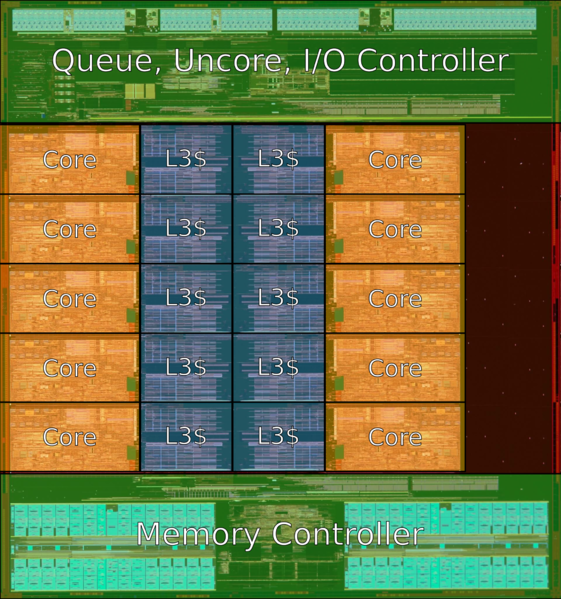}}}
    \hfill
    \subfloat[AMD Zen microarchitecture Core Complex (CCX) featuring 4 cores and 8MB shared L3 (2 MB L3 slice per core)~\cite{amddie}.]{\label{fig:back:amddie}{\includegraphics[width=0.5\linewidth,valign=b]{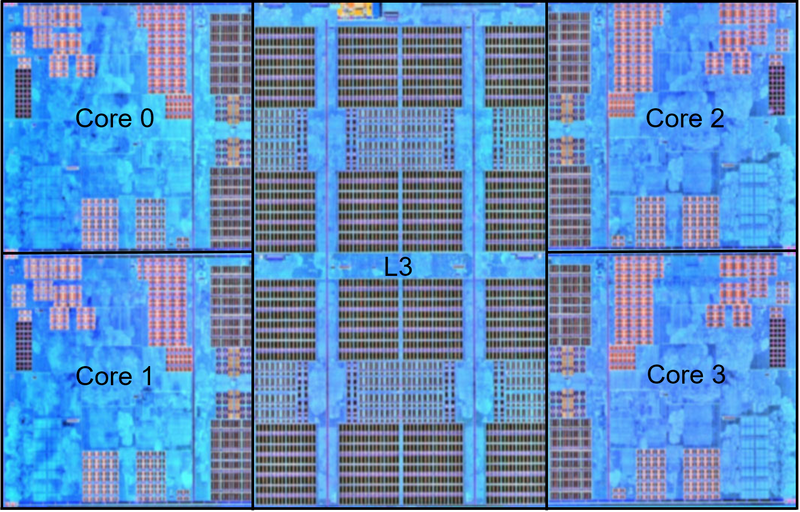}}}
    \caption{\label{fig:back:die}Die photos of modern processors highlighting floor area devoted to different components.}
\end{figure}

\section{The Problem}
Applications with large working set size often exhibit thrashing and/or streaming access patterns at LLC, leading to premature evictions of useful cache blocks that are likely to be referenced in the near future. Meanwhile, a large fraction of LLC capacity is occupied by {\em dead blocks} that will eventually be evicted without incurring further hits, leading to inefficient utilization of the LLC capacity.
Cache efficiency can be improved significantly by identifying dead blocks and discarding them immediately after their last use, thereby providing an opportunity for cache blocks with long temporal reuse distances to persist in the cache longer and accumulate more hits.

The state-of-the-art cache management techniques employ prediction mechanisms that learn from the past access patterns with an aim to correctly identify as many dead blocks as possible. Effectiveness of these predictors hinges on the stability of application behavior with respect to the metric used for determining whether the block is dead. Naturally, the more consistent the reuse behavior across the block's lifetimes (also called {\em generations}) in the cache, the more accurate the predictions.

In practice, applications exhibit {\em variability} in the reuse behavior of cache blocks. The sources of variability are numerous such as microarchitectural noise (e.g., speculation), control-flow variation, cache pressure from other threads and inherent application characteristics.
These sources of variability are outside the control of LLC, making variability in the reuse behavior an inevitable challenge for a cache management technique. 
Variability in reuse prediction challenges existing techniques in reliably identifying the end of a block's useful lifetime, thus causing lower prediction accuracy, coverage or both. A wrong prediction may either cause premature eviction of a useful cache block, leading to an additional cache miss or cause delay in eviction of a dead block, leading to wastage of cache capacity.
This calls for cache management mechanisms and policies that can tolerate variability in the reuse behavior of cache blocks to maximize cache efficiency.

\section{Our Proposals}
Aim of this thesis is:
\begin{center}
{\em \slightlylarge To design robust cache management mechanisms and policies for LLC that~minimize cache misses in the face of variability in the reuse behavior of cache~blocks, while keeping the cost and complexity of the hardware implementation~low}.
\end{center}

To that end, we propose two cache management techniques, one domain-agnostic and one domain-specialized, that introduce robust mechanisms and policies to address variability in reuse prediction. The rest of the chapter provides a brief overview of both proposals.

\subsection{Domain-Agnostic Cache Management}
In this part of the thesis, we consider a conventional approach to cache management, namely domain-agnostic cache management, in which the LLC is managed completely in hardware.
Such approach is quite attractive in practice as the cache management remains fully transparent to the application software. 
There has been a rich history of works that proposed various domain-agnostic techniques to improve cache efficiency~\citeall{}.

The state-of-the-art techniques employ prediction mechanisms that seek to correctly identify as many dead blocks as possible and evict them immediately after their last use to reduce cache thrashing.
These predictors all rely on some metric of temporal reuse to make their decisions regarding the end of a given block's useful life. 
Previous works have suggested hit count~\cite{counter}, last-touch PC~\cite{sampler}, and number of references to the block's set since the last reference~\cite{pd}, among others, as metrics for determining whether the block is dead at a given point in time. 
However, we observe that existing metrics limit the accurate identification of dead blocks in the face of variability. For example, when the number of hits to a cache block is inconsistent across generations, a technique relying on this metric (\ie hit count) would either prematurely classify the cache block dead or may not classify the cache block dead altogether until its eviction, both of which lead to cache inefficiency.
This calls for robust metrics and policies that can tolerate inconsistencies.

To that end, we propose {\em Live Distance}, a new metric of temporal reuse based on stack distance; stack distance of a given access to a cache block is defined as the number of unique cache blocks accessed since the previous access to the cache block~\cite{stack-distance}. For a given generation of a cache block (from allocation to eviction), live distance is defined as the largest observed stack distance in the generation.
Live distance provides a conservative estimate of a cache block's useful lifetime.

We introduce Leeway, a new domain-agnostic cache management technique that uses live distance as a metric for dead block predictions. 
Leeway uses code-data correlation to associate live distance for a group of blocks with a PC that brings the block into the cache. 
While live distance as a metric provides a high degree of resilience to variability, the per-PC live distance values themselves may fluctuate across generations.  
To correctly train live distance values in the face of fluctuation, we observe that an individual application's cache behavior tends to fall in one of two categories: streaming (most allocated blocks see no hits) and reuse (most allocated blocks see one or more hits). Based on this simple insight, we design a pair of corresponding policies that steer updates in live distance values either toward zero (for bypassing) or toward the maximum recently-observed value (to maximize reuse). For each application, Leeway picks the best policy dynamically based on the observed cache reuse behavior.

To avoid the need to access specialized external structures (e.g, prediction tables) upon each LLC access, Leeway embeds its prediction metadata (i.e., Live Distance) directly with cache blocks. 
This is in contrast with prior predictors~\cite{sampler, mdpp, hawkeye, perceptron}, which need to access a dedicated predictor table upon every single LLC access. Because modern multi-core processors feature distributed LLC, accesses to dedicated prediction tables introduce detrimental latency and energy overheads in traversing the on-chip interconnect to query such structures.

\subsection{Domain-Specialized Cache Management}

In this part of the thesis, we identify applications for which existing domain-agnostic cache management techniques struggle in exploiting the high reuse due to variability arising from certain fundamental application characteristics. Specifically, we explore applications from the domain of graph analytics processing natural graphs.
For natural graphs, the vertex degrees follow a skewed power-law distribution, in which a small fraction of vertices have many connections while the majority of vertices have relatively few connections~\cite{power-law,power-law-internet,powergraph,fc,hubcluster}.
Such graphs are prevalent in a variety of domains, including social networks, computer networks, financial networks, semantic networks, and airline networks.

The power-law skew in the degree distribution means that a small set of vertices with the largest number of connections is responsible for a major share of off-chip memory accesses. The fact that these richly-connected vertices, {\em hot vertices}, comprise a small fraction of the overall footprint while exhibiting high reuse makes them prime candidates for caching. Meanwhile, the rest of the vertices, {\em cold vertices}, comprise a large fraction of the overall footprint while exhibiting low or no reuse.

Despite the high reuse inherent in accesses to the hot vertices, graph applications exhibit poor cache efficiency due to the following two reasons:

\noindenttitle{\smalltextcircled{1} Lack of spatial locality: } hot vertices are sparsely distributed throughout the memory space, exhibiting a lack of spatial locality.  When hot vertices share the cache block with cold vertices, valuable cache space is underutilized. 

\noindenttitle{\smalltextcircled{2}  Difficult to exploit temporal locality: } hot vertices inherently exhibit high temporal reuse. However, the reuse patterns of graph-analytic applications is highly irregular and is dependent on graph topology, which cause severe cache thrashing when processing large graphs.
Accesses to a large number of cold vertices are responsible for thrashing, often forcing hot vertices out of the cache.

\noindentnotitle

Both problems are orthogonal in nature as solving one problem does not solve the other.
Overcoming the former problem requires improving cache block utilization by focusing on intra-block reuse, whereas the latter problem requires retaining high reuse cache blocks in the LLC by focusing on inter-block reuse. 

The former problem is outside the scope of any cache management technique as it stems from the fact that vertex properties usually require just 4 to 16 bytes in comparison to 64 or 128 bytes of a cache block size in modern processors. Thus, the effective spatial locality is completely dictated by the vertex layout in memory for a given graph dataset, which is in complete control of the software.

The latter problem is what a cache management technique targets. However, long reuse distances along with irregular access patterns impede learning mechanisms of the state-of-the-art domain-agnostic cache management techniques, rendering them deficient for the entire application domain.

We observe that the software not only has the knowledge of crucial application semantics such as vertex degrees, but also controls the placement of vertices in memory. Thus, cache management for graph analytics can be significantly improved by leveraging software support.

To that end, we propose a holistic approach of software-hardware co-design to improve cache efficiency for the domain of graph analytics processing natural graphs. 
Our software component implements a novel lightweight software technique, called Degree-Based Grouping (DBG), that applies a coarse-grain graph reordering to segregate hot vertices in a contiguous memory region to improve spatial locality. %

Our hardware component implements Graph Specialized Cache Management (GRASP).  
GRASP augments existing cache insertion and hit-promotion policies to provide preferential treatment to cache blocks containing hot vertices to shield them from thrashing.
To cater to the variability in the reuse behavior, GRASP policies are designed to be flexible to cache other blocks exhibiting reuse, if needed. %

GRASP relies on lightweight software support to accurately pinpoint hot vertices amidst irregular access patterns, in contrast to the state-of-the-art domain-agnostic techniques that rely on storage-intensive prediction mechanisms. By leveraging contiguity among hot vertices (enabled by DBG), GRASP employs a lightweight software-hardware interface comprising of only a few configurable registers, which are programmed by software using its knowledge of the graph data structure.

The strength and novelty of our co-design lies in the interplay between software (DBG) and hardware (GRASP). Software aids hardware in pinpointing hot vertices via a lightweight interface, thus eliminating the need for storage-intensive cache metadata required by the state-of-the-art domain-agnostic techniques.
Meanwhile, hardware is responsible for exploiting temporal locality in presence of cache thrashing, allowing software to focus only on inducing spatial locality, enabling low-overhead software reordering compared to high-overhead complex software-only vertex reordering techniques that target both spatial and temporal locality.
A holistic software-hardware co-design enables high cache efficiency for graph analytics while keeping both software and hardware components simple.

\section{Published Work}
\label{sec:intro:publications}

Some of the contents of this thesis have appeared in the following publications:

\vspacel
\noindent
The publications appearing in \emph{Chapter~\ref{ch:leeway}} :
\begin{itemize}
\item P. Faldu and B. Grot. {``LLC Dead Block Prediction Considered Not Useful''}. {\em In International Workshop on Duplicating, Deconstructing and Debunking (WDDD), co-located with ISCA}. 2016.~\cite{c-dead}

\item P. Faldu and B. Grot. {``Reuse-Aware Management for Last-Level Caches''}. {\em In International Workshop on Cache Replacement Championship (CRC), co-located with ISCA}. 2017.~\cite{c-crc2-5reuse}

\item P. Faldu and B. Grot. {``Leeway: Addressing Variability in Dead-Block \mbox{Prediction} for Last-Level Caches''}. {\em In Proceedings of the International Conference on Parallel Architectures and Compilation Techniques (PACT)}. 2017.~\cite{c-leeway}
\end{itemize}

\vspacel
\noindent
The publication appearing in \emph{Chapter~\ref{ch:dbg}} :
\begin{itemize}
\item P. Faldu, J. Diamond and B. Grot. {``A Closer Look at Lightweight Graph Reordering''}. {\em In Proceedings of the International Symposium on Workload Characterization (IISWC)}. 2019.~\cite{c-dbg}
\end{itemize}

\vspacel
\noindent
The publications appearing in \emph{Chapter~\ref{ch:grasp}} :
\begin{itemize}
\item P. Faldu, J. Diamond and A. Patel. {``Cache Memory Architecture and Policies for Accelerating Graph Algorithms''}. U.S. Patent 10417134. Oracle International Corporation. 2019.~\cite{c-grasp-patent}

\item P. Faldu, J. Diamond and B. Grot. {``POSTER: Domain-Specialized Cache Management for Graph Analytics''}. {\em In Proceedings of the International Conference on Parallel Architectures and Compilation Techniques (PACT)}. 2019.~\cite{c-grasp-poster}

\item P. Faldu, J. Diamond and B. Grot. {``Domain-Specialized Cache Management for Graph Analytics''}. {\em In Proceedings of the International \mbox{Symposium} on High-Performance Computer Architecture (HPCA)}. 2020.~\cite{c-grasp} .
\end{itemize}

\section{Thesis Organization}
Rest of the thesis is organized as follows: Chapter~\ref{ch:background} presents the necessary background on cache management techniques to understand the limitations of the state-of-the-art techniques. Chapter~\ref{ch:leeway} presents the design and evaluation of Leeway, our domain-agnostic cache management technique.

Chapter~\ref{ch:motivation} highlights the limitations of domain-agnostic cache management techniques for the domain of graph analytics and motivates the need for a software-hardware co-design to manage LLC for graph analytics. The next two chapters present software and hardware components of the proposed co-design: Chapter~\ref{ch:dbg} presents DBG, a new software vertex reordering technique to improve spatial locality and Chapter~\ref{ch:grasp} presents GRASP, a domain-specialized cache management that leverages DBG to further improve cache efficiency for graph analytics. Finally, we conclude our proposals in Chapter~\ref{ch:conclusion} and provide potential future directions of research for cache management.

\chapter{Background\label{ch:background}}

In typical desktop and server computers, the memory hierarchy is organized as several levels of memories of different speeds and sizes. Each level of memory is bigger and cheaper per byte, but slower than the previous higher-level that is closer to the processor.
Fig.~\ref{fig:cache_hierarchy} shows a three-level cache hierarchy along with the adjacent levels, including their typical access times and sizes. Fig.~\ref{fig:back:nuca} shows a typical layout of a cache hierarchy in a modern multi-core processor. L1 and L2 caches are private per core whereas L3, also called {\em Last-Level Cache (LLC)}, is shared across processors.
While for the purpose of caching, L3 can be logically seen as a single structure, physically, L3 is organized as multiple {\em Non-Uniform Cache Accesses (NUCA)} slices~\cite{nuca} as shown in the figure. 

\begin{figure}[!b]
    \centering
    \includegraphics[width=0.75\linewidth]{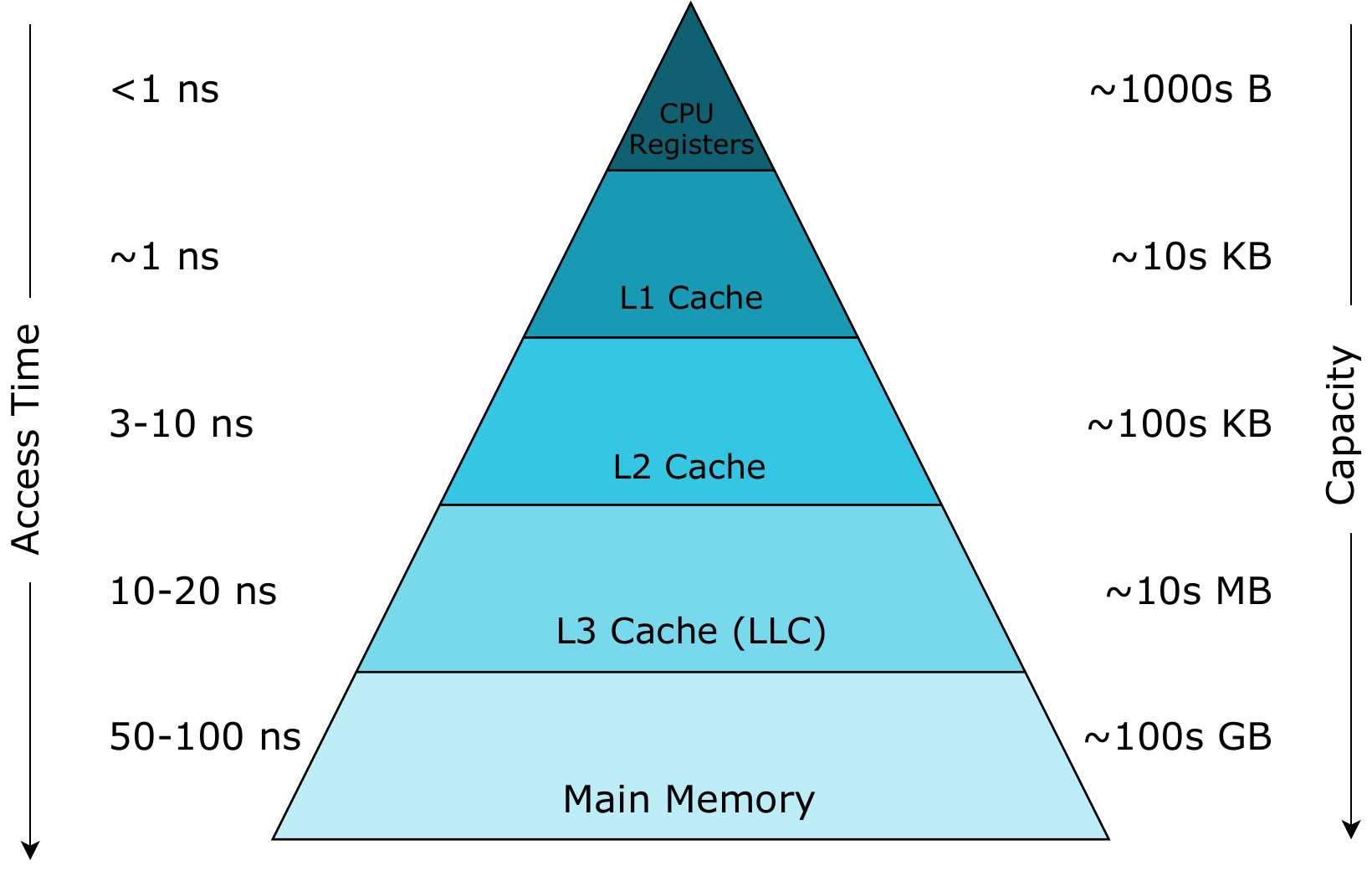}
    \caption{A typical memory hierarchy containing three levels of caches, including typically access times (on left) and typical sizes (on right)~\cite{hp}.}
    \label{fig:cache_hierarchy}
\end{figure}

A cache hierarchy can be maintained as {\em fully-inclusive}, {\em fully-exclusive} or {\em non-inclusive non-exclusive}. A fully-inclusive level of cache must contain all the cache blocks that are present in the previous higher-level cache. Conversely, a fully-exclusive level of cache must {\em not} contain any cache block that is present in the previous higher-level cache. Finally, a non-inclusive non-exclusive level of cache does not observe any such constrains, and it may or may not contain the cache blocks that
are present in the previous higher-level cache. Meanwhile, the main memory is inclusive of all the cache levels, meaning memory stores all addresses regardless of whether they are present in any of the cache levels.

During execution, a CPU core first queries the L1 cache for the data item (\ie a program instruction or an application data) needed to perform computations. 
If the data item is found (\ie a cache hit), L1 responds to the request with the necessary data. Meanwhile, if the data item is not found (\ie a cache miss), next lower-level cache is queried. The process is repeated until the data item is found in one of the caches. If the data item is not found in any caches, it will be retrieved from the main memory.

 \begin{figure}[!t]
    \centering
    \includegraphics[width=0.6\linewidth]{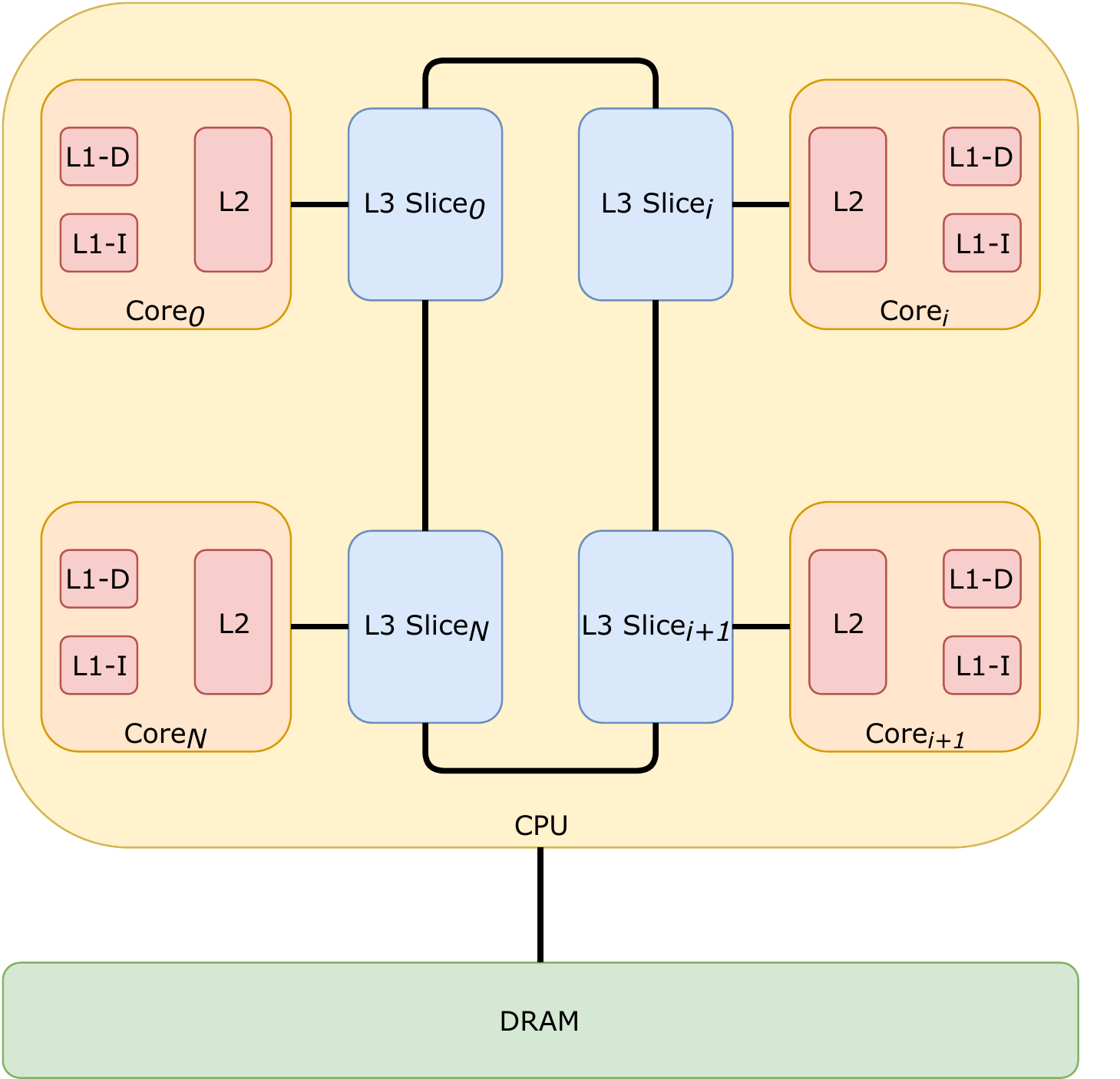}
    \caption{\label{fig:back:nuca}A typical layout of a modern multi-core processor with three levels of the cache hierarchy.}
\end{figure}

Last-Level Cache (LLC) (\ie L3 for a three-level cache hierarchy or L2 for a two-level cache hierarchy) is of particular interest as it acts as an on-chip frontier, miss to which requires a long latency memory access. LLC offers the largest capacity among all the on-chip caches, and thus can store the largest fraction of the working set of an application. However, the LLC capacity offered by modern processors is significantly smaller than the working set size of emerging applications. Fortunately, most applications do not access all data items uniformly, meaning some data items are likely to be reused more frequently than others, due to an application property known as locality as discussed below.

\section{Principle of Locality\label{sec:locality}}
Caches are designed to exploit the {\em principle of locality} observed in most applications. Two different types of locality have been observed:

\noindenttitle{\smalltextcircled{1} Spatial locality} refers to locality in space, which states that the data items whose addresses are near one another tend to be referenced close together in time.

\noindenttitle{\smalltextcircled{2} Temporal locality} refers to locality in time, which states that recently accessed data items are likely to be accessed in near future.

\indentnotitle{}
To exploit spatial locality, caches operate at a granularity of a unit called {\em cache block} (or {\em cache line}), which consists of several bytes (typically, 64 or 128 bytes). When moving data between caches, an entire cache block containing the data item is transferred, in anticipation that the other nearby data items will be accessed soon due to spatial~locality.

Temporal locality is exploited by caching the most recently accessed cache blocks.
A widely popular cache management technique that achieves this is called {\em Least Recently Used (LRU)}. LRU maintains the cache blocks in a cache set as a {\em recency stack}. 
Cache blocks are ordered in the stack based on how recently they were accessed with the {\em Most Recently Used (MRU)} cache block at the top of the stack and the {\em Least Recently Used (LRU)} cache block at the bottom of the stack. When a cache set is full and a new block must be inserted into this set, a cache block at the LRU position is evicted, in anticipation that other, more recently accessed, cache blocks will be accessed soon due to temporal locality. Fig.~\ref{fig:lru} depicts the functionality of LRU cache management technique for three events: insertion, eviction and hit.

\begin{figure}[!t]
    \centering
    \includegraphics[width=0.6\linewidth]{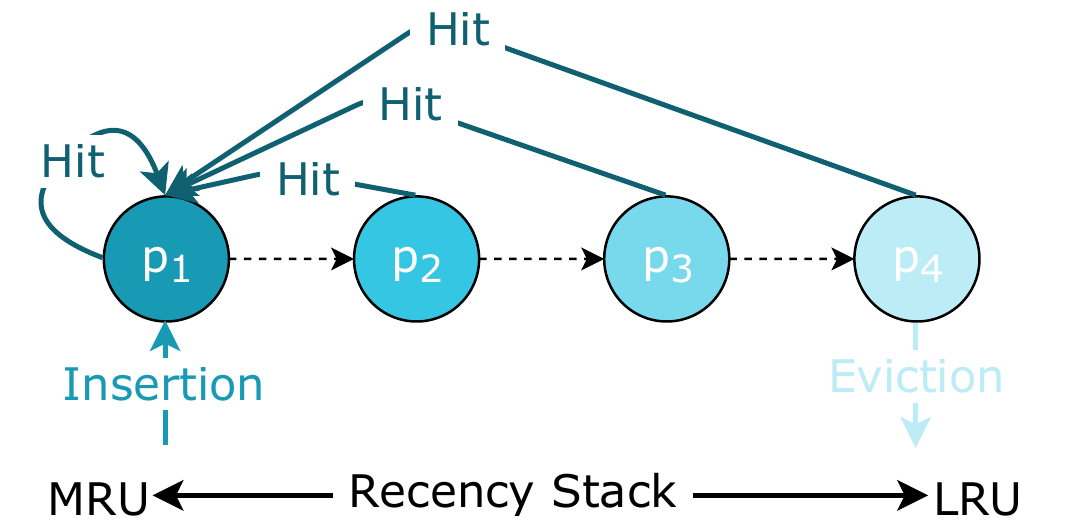}
    \caption{LRU cache management for a 4-way associative cache. Circles labeled $p_i$ show positions of the cache blocks in the recency stack with position $p_1$ for the Most Recently Used (MRU) cache block and $p_4$ for the Least Recently Used (LRU) cache block. The solid arrows point to the new positions for cache blocks for a given cache event, while the dotted arrows point to the new positions of cache blocks when other blocks are placed into their positions.}
    \label{fig:lru}
\end{figure}

\section{Cache Access Patterns\label{sec:cache-access-patterns}}
LRU can effectively exploit temporal locality. However, LRU is not an efficient cache management technique for LLC as temporal locality of an application is often filtered by the higher-level caches. As such, not all access patterns observed at LLC confirm to the principle of locality. Prior works listed three most common access patterns observed at LLC, which are summarized in Table~\ref{tab:access_patterns}~\cite{dip,rrip}.

\begin{table}[!b]
    \small
    \begin{tabularx}{0.9\linewidth}
        {|>{\raggedright\arraybackslash\hsize=0.25\hsize}X| 
          >{\raggedright\arraybackslash\hsize=0.75\hsize}X|
        }
        \hline
        \makecell{Access Pattern} & \makecell{Stream of cache accesses $a_i$ to a given cache set} \\
        \hline
        \hline
        Recency-friendly & ($a_1$, $a_2$, ..., $a_{k-1}$, $a_k$, $a_k$, $a_{k-1}$, ..., $a_2$, $a_1$)$^N,$ for $k$ > 0 and N > 0 \\ \hline
        Streaming & ($a_1$, $a_2$, ..., $a_k$), for $k$ > 0 \\ \hline
        Thrashing &  ($a_1$, $a_2$, ..., $a_k$)$^N,$ for $k$ > associativity and N > 1 \\ \hline
    \end{tabularx}
    \caption{Common cache access patterns at LLC.}
    \label{tab:access_patterns}
\end{table}

\noindenttitle{\smalltextcircled{1} Recency-friendly access pattern} exhibits good temporal locality as the recently accessed cache blocks are more likely to be accessed soon, making LRU perfectly suitable for such patterns.

\noindenttitle{\smalltextcircled{2} Streaming access pattern} has no temporal locality in its references. For strictly streaming access patterns, LRU is no worse than any other cache management technique as replacement decisions are irrelevant. However, LRU is inefficient when LLC observes a mix access pattern that is a combination of streaming and some other access patterns. Amidst the mix patterns, LRU inserts all cache blocks at the MRU position. The cache blocks exhibiting streaming accesses are
gradually propagated to the LRU position, all the while occupying cache space, until eventually evicted from the LLC without incurring any cache hit, wasting valuable cache capacity. In contrast, the optimal cache management technique may insert all these cache blocks in the LRU position or may bypass their cache insertions altogether and directly forward them to the higher-level caches.

\begin{figure}[!t]
    \centering
    \includegraphics[width=1\linewidth]{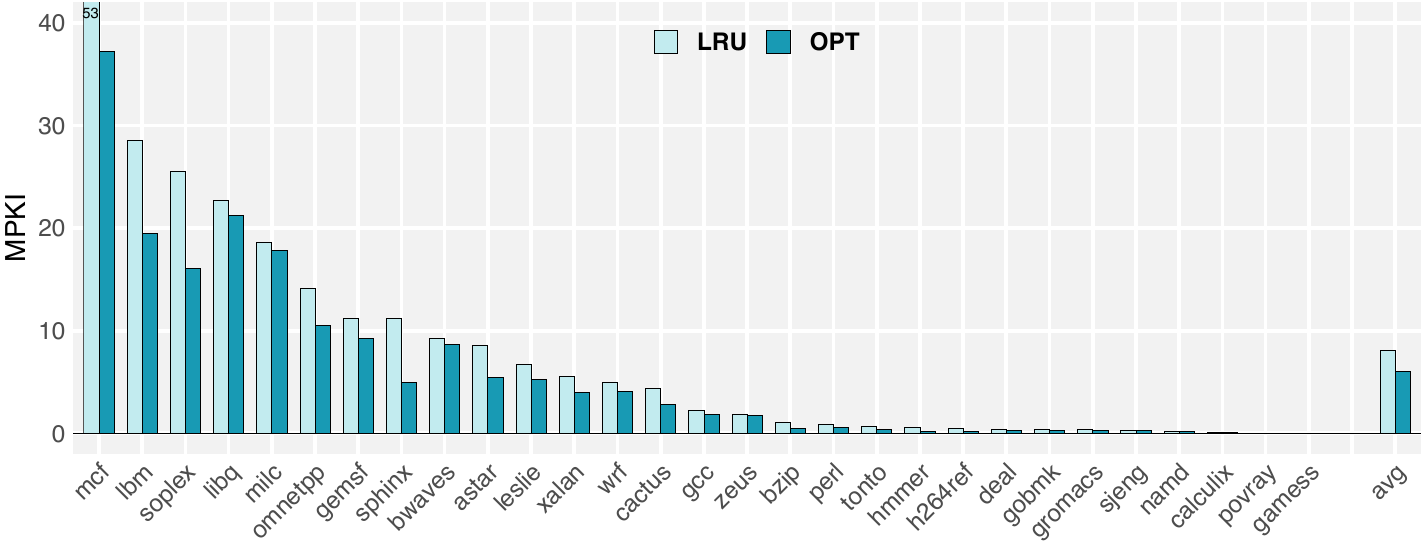}
    \caption{Misses Per Kilo Instructions (MPKI) for SPEC CPU 2006 applications under LRU and OPT cache management techniques for 16-way associative 2MB LLC. 
    Applications on x-axis are sorted by the MPKI under LRU.}
    \label{fig:opt-lru}
\end{figure}

\noindenttitle{\smalltextcircled{3} Thrashing access pattern} is a cyclic access pattern of length $k$, when $k$ is greater than the associativity of a cache. LRU is inadequate for such an access pattern as LRU receives zero cache hit for such patterns. These access patterns present a pathological case for LRU as LRU tries to retain the entire working set in the cache, and ends with zero hit. In contrast, the optimal cache management technique may retain a partial working set in the cache and may observe cache hits for a fraction of cache accesses.

\noindentnotitle
In practice, applications exhibit access patterns that are some combination of the above access patterns, thus offering significant room for improving cache efficiency over a traditional cache management technique like LRU.

To quantify the maximum opportunity in eliminating misses over LRU, we simulate LLC under Belady's {\em OPT}~\cite{opt}, an offline optimal replacement technique that has the perfect knowledge of the future. OPT replaces a cache block whose next reference is farthest in the future among the cache blocks in a given set.
While OPT is impractical to implement, it provides a theoretical upper bound on the number of misses a cache management technique can eliminate. %
Fig.~\ref{fig:opt-lru} plots the {\em Misses Per Kilo Instructions (MPKI)} for OPT as well as the baseline LRU for all 29 SPEC CPU 2006 applications.
OPT is able to eliminate 26\% (max 67\%) of misses on average over LRU, highlighting a significant opportunity in improving the cache efficiency over LRU. In the following sections, we explain the basics of cache management techniques 
followed by a discussion on the most relevant prior cache management techniques.

\section{Basics of Cache Management}
The goal of a cache management technique is to decide which cache blocks to retain in the cache in order to minimize cache misses (or equivalently, maximize cache hits).
Therefore, the efficiency of a cache management technique depends on how effectively it answers the following question:
{\em Which cache block in a given cache set is the least likely to be accessed soon, and thus should be replaced when a new cache block is inserted in the set?} An offline technique like OPT can provide the optimal answer by looking into the future accesses. However, a practical cache management technique does not know the future LLC accesses, and thus relies on a heuristic that predicts reuse of cache blocks by analyzing the past LLC accesses.

A typical cache management technique maintains relative priorities of the cache blocks in a given cache set. Priority of a cache block reflects how likely it is going to be reused in the near future under a given heuristic.
Priorities may be adjusted on certain cache events such as cache hits or cache misses.
Overall, every cache management technique implements three policies, each defining how to adjust the priorities of cache blocks for a corresponding cache event.

\noindenttitle{\smalltextcircled{1} Insertion policy} is responsible for assigning priority of a new cache block, when inserted in the cache due to a miss. Meanwhile, insertion policy may also adjust the priorities of other cache blocks already present in the cache set. In some cases, the insertion policy may choose to bypass the insertion altogether by forwarding data directly to the higher-level caches, if the existing cache blocks in the set are more likely to be reused in comparison to the new cache block. %

For example, the insertion policy of LRU assumes that the application exhibits a recency-friendly access pattern and thus, a newly inserted cache block is likely to be accessed soon. 
Based on this assumption, LRU never bypasses the insertion and always assigns the highest priority to a new cache block by inserting it at the MRU position.
Before inserting a new cache block, insertion policy shifts every cache block by one position towards the LRU position in the recency stack as shown using the dotted arrows in Fig.~\ref{fig:lru}.

\noindenttitle{\smalltextcircled{2} Eviction policy} is responsible for choosing which cache block to replace for a case when the insertion policy decides to insert a new cache block in the cache set and the set is full.
If a technique supports multiple cache blocks to have the same priority, the eviction policy also defines a tie-breaker logic.

For example, as LRU maintains cache blocks in the recency stack using a total order, no two cache blocks can have the same priority. Thus, the eviction policy of LRU simply chooses a cache block at the LRU position as a replacement candidate.

\noindenttitle{\smalltextcircled{3} Hit-promotion policy} is responsible for adjusting the priority of a cache block upon hit. Meanwhile, hit-promotion policy may also adjust the priorities of other cache blocks already present in the cache set.

For example, the hit-promotion policy of LRU assumes that the application exhibits a recency-friendly access pattern and thus, a recently accessed cache block is likely to be accessed soon.
Based on this assumption, LRU promotes the cache block to the MRU position, regardless of its current position in the recency stack. Meanwhile, the cache blocks between the MRU position and the position of the cache block before the promotion are shifted one position towards LRU.

\section{Prior Cache Management Techniques\label{sec:background:prior}}

There has been a rich history of cache management techniques to improve cache efficiency~\citeallsoft{}. %
Based on the amount of state maintained by the heuristic employed by a cache management technique and how the state is updated, existing cache management techniques can be broadly classified into the following four categories.

\noindenttitle{\smalltextcircled{1} Static techniques} apply static policies for insertion, eviction and hit-promotion. Such techniques maintain a local state per cache block by augmenting each cache block with a few bits.
Local state (\eg recency state under LRU) is used to maintain relative priorities of cache blocks within a cache set under some heuristic.
The local state of a cache block is only relevant during its current generation, which is defined as the time between insertion and eviction of the cache block; the local state is reset when the cache block is replaced with a new cache block. 
The static techniques provide fundamental building blocks for more advanced cache management techniques as discussed next.

\noindenttitle{\smalltextcircled{2} Lightweight dynamic techniques} apply dynamic policy for at least one of three cache events -- insertion, eviction or hit-promotion. Such techniques are built on top of static cache management techniques, and thus, like static techniques, maintain a local cache state in the LLC for each cache block. Additionally, these techniques also maintain some state outside the cache, which is referred to as the external state. The external state is usually minimal, and hence the name lightweight.

\noindenttitle{\smalltextcircled{3} History-based predictive techniques} apply dynamic policies for cache management based on historical access patterns. In addition to the local state for each cache block, these techniques also record information pertaining to the reuse of the cache blocks beyond their current generations in some external structure(s). As a result, these techniques require significantly more storage than the lightweight dynamic techniques.

\noindenttitle{\smalltextcircled{4} Software-aided techniques} apply dynamic policies for cache management, which rely on software to identify high-reuse cache blocks. For each cache access, software provides some sort of a reuse hint for hardware to make policy decisions. 

\noindenttitle{}
In the rest of the chapter, we discuss each of these classes in detail.

\subsection{Static Techniques}
Static cache management techniques employ static policies for insertion, eviction and hit-promotion, which disregard the reuse of cache blocks in their previous generations. 
A static technique may maintain a local state for each cache block during its current generation, which is reset when the cache block is evicted and replaced by another cache block. 

LRU is a classic example of static techniques, which maintains how recently a cache block is accessed relative to the other cache blocks in a given cache set and makes policy decisions exclusively based on that information. For example, the insertion policy of LRU always assigns a new cache block the highest priority by inserting it at the MRU position, regardless of its reuse in the previous generations. Similarly, the hit-promotion policy promotes a cache block to the MRU position on a hit regardless of the number of hits the cache block may have observed in the current or previous generations. Finally, the eviction policy always evicts a cache block at the LRU position, regardless of the number of hits incurred by the cache block in the current generation or previous generations. Other example of static techniques include PseudoLRU~\cite{PseudoLRU}, LIP~\cite{dip}, SRRIP~\cite{rrip}, Static GIPPR~\cite{gippr} and Static MDPP~\cite{mdpp}, among others.

\subsubsection{Storage}
Static techniques typically maintain between $1$ and $log_{2}n$ bits of a local state (usually a recency state) per cache block, where $n$ is the associativity of the cache.
Fig.~\ref{fig:back:static_state} shows a logical organization of LLC along with the storage devoted to a recency state, tag and data for each cache block.
Static techniques typically require the least amount of state per cache block, as other techniques are built on top of a static technique(s).%
\begin{figure}[!t]
    \centering
    \includegraphics[width=0.8\linewidth]{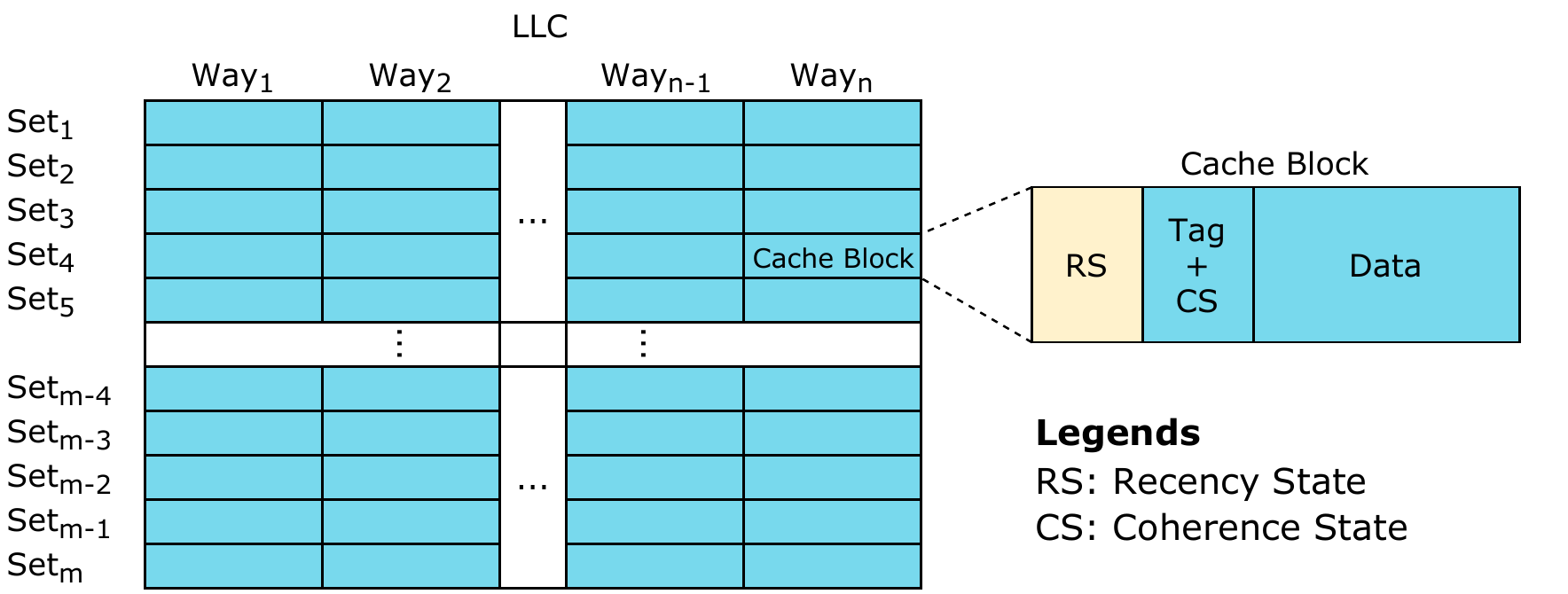}
    \caption{A set-associative cache with $n$ ways and $m$ sets. A static technique requires $k$-bits per cache block to maintain recency state, where $k$ is typically between $1$ and $log_{2}n$. In comparison, D bytes (typically 64 or 128 bytes) are allotted for data whereas Tag requires $A-log_{2}m-log_{2}D$ bits, where $A$ is the number of bits needed to represent an address. }
    \label{fig:back:static_state}
\end{figure}

\subsubsection{Limitations}
Static cache management techniques target specific access patterns, and cannot adapt to application behavior due to the static nature of their policies. For example, LRU targets recency-friendly access patterns. However, LRU is not suitable to address thrashing or streaming access patterns as explained in Sec.~\ref{sec:cache-access-patterns}.

Prior work proposed {\em LRU Insertion Policy (LIP)}~\cite{dip}, that makes a simple modification to the insertion policy of LRU to target streaming access patterns. LIP is identical to LRU except for its insertion policy, which assigns a new cache block the lowest priority by inserting it at the LRU position, in anticipation of streaming access patterns. Under LIP management, the cache blocks that do not exhibit any reuse are evicted from the cache soon after their insertion, thus minimizing cache pollution for applications dominated by streaming access patterns. However, due to the static nature of the policies, LIP, as a standalone technique, is not suitable for recency-friendly access patterns.

\subsection{Lightweight Dynamic Techniques}
Lightweight dynamic cache management techniques employ dynamic policy for at least one of the three cache events of insertion, hit-promotion and eviction~\citeprob{}. A lightweight dynamic technique may maintain some external state, in addition to maintaining a local state per cache block.
Policy decisions are influenced by a combination of the local state for the cache blocks in a given set and the external state. Therefore, two cache blocks with an identical recency state may be treated differently based on an external state.

A lightweight dynamic technique is typically constructed by composing a few techniques, each of which is either a static technique or another lightweight dynamic technique. Thus, unlike static techniques, lightweight dynamic techniques can adapt to application behavior.

For example, {\em Bimodal Insertion Policy (BIP)}~\cite{dip} is composed of two static techniques, LRU and LIP. 
BIP dynamically selects between LRU and LIP probabilistically, wherein LRU is chosen with a low probability. Thus, BIP inserts a new cache block at the MRU position with a low probability and at the LRU position with a high probability. Thus, a new cache block's insertion priority is dynamically decided based on the external state (e.g., a pseudo random number generator or a saturating counter) at the time of insertion.

BIP is able to target certain thrashing access patterns for which neither of its constituent techniques (\ie LRU and LIP) alone is suitable. Consider an access pattern to a particular set of the form ($a_1$, $a_2$, ..., $a_{k-1}$, $a_k$)$^N$ followed by ($b_1$, $b_2$, ..., $b_{k-1}$, $b_k$)$^N$, where k is greater than the cache associativity and N is greater than 1.
For such access patterns, LRU is not suitable for either of the streams and would incur zero hit for both the streams. LIP also struggles as it won't be able to adapt to the change in the working set from the stream $a_i$ to the stream $b_i$ and would incur zero hit for the second stream of accesses as all the new cache blocks from the stream $b_i$ will be inserted at the LRU position and evicted immediately after their insertion without incurring any hits. In contrast, BIP can adapt to the change in the working set by dynamically switching between LRU and LIP. For BIP, some cache blocks of the stream $b_i$ are inserted at the MRU position, thus allowing them to persist in the cache longer to incur further hits. Meanwhile, the rest of the cache blocks are inserted at the LRU position, thus reducing cache thrashing.

Another example of a lightweight dynamic technique is {\em Dynamic Insertion Policy (DIP)}~\cite{dip}, which is composed of LRU, a static technique, and BIP, a lightweight dynamic technique. 
DIP chooses between LRU and BIP based on the observed access pattern, and thus DIP is suitable for application exhibiting any of the three -- recency-friendly, streaming and thrashing -- access patterns.

\begin{figure}[!t]
    \centering
    \includegraphics[width=0.8\linewidth]{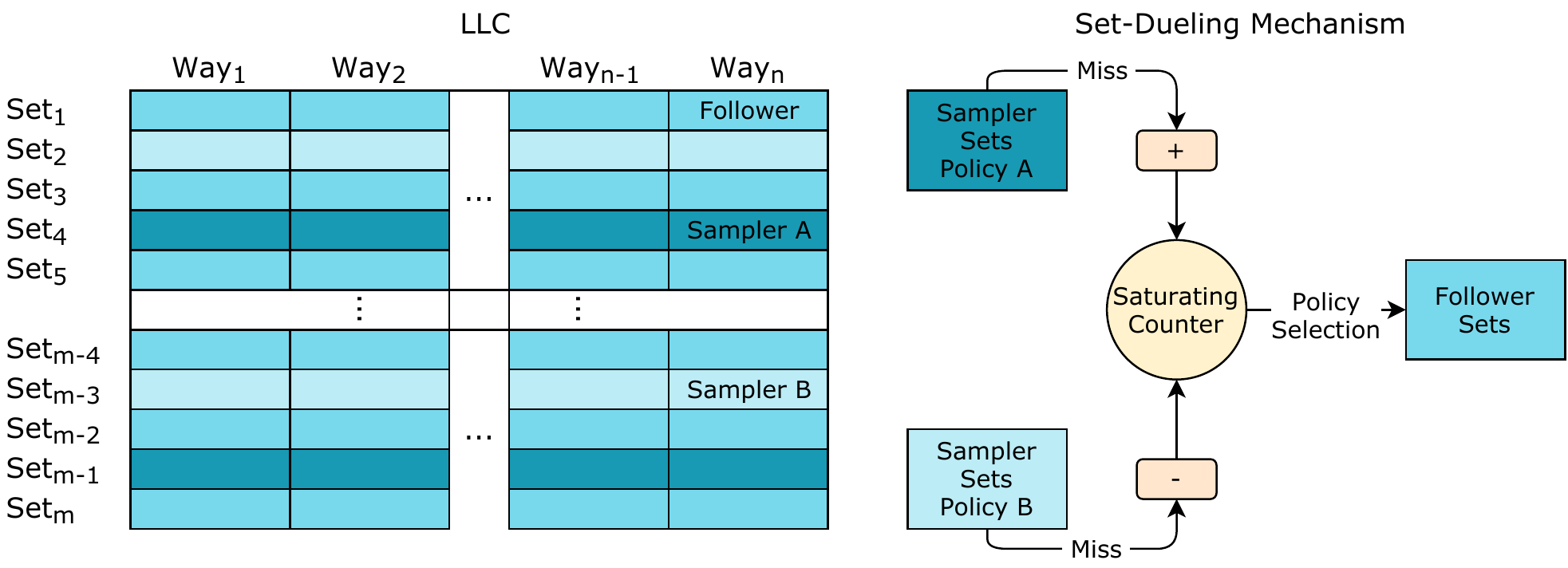}
    \caption{%
    The figure shows a dynamic technique composed of two techniques, A and B. A small number of sampled sets, Sampler Sets, implement technique A and an equal number of some other sets implement technique B. Remaining sets, Follower Sets, implement the winning technique based on the value of the saturating counter.
    }
    \label{fig:back:dynamic_state}
\end{figure}

DIP introduced a set-dueling mechanism for the policy selection. DIP allocates a small number of cache sets (called {\em sampler sets}) which are exclusively managed under the LRU technique. An equal number of other sampled sets are exclusively managed under the BIP technique. DIP maintains a saturating counter (\ie an external state) outside the cache to track the difference in misses due to each technique.
DIP dynamically selects a technique that causes fewer misses, and manages the rest of the sets (called {\em follower sets}) using the most effective technique for a given access pattern.

RRIP is the state-of-the-art lightweight dynamic technique~\cite{rrip}. RRIP is fundamentally very similar to DIP. However, RRIP is practically more attractive than DIP as RRIP does not rely on LRU for a base static technique. RRIP maintains the cache blocks in a given set in only $k$ unique recency classes ($k$ is typically smaller than the associativity $n$), thus requiring $log_{2}k$ bits per cache block. In comparison, LRU maintains the cache blocks in a set in total order ($n$ unique recency classes), which requires $log_{2}n$ bits per cache block.

\subsubsection{Storage}
As a lightweight dynamic technique is composed of static techniques, it also maintains a local state per cache block as required by the base static techniques. 
Finally, a technique also maintains an external state that guides the dynamic policy selection. 
For example, the set-dueling mechanism of DIP requires a saturating counter to keep track of the winning policy between LRU and BIP as shown in Fig.~\ref{fig:back:dynamic_state}.

\subsubsection{Limitations}
A lightweight technique can adapt to application behavior and dynamically select the policy best suited for the application at a given time. However, due to minimal external state, a lightweight technique cannot provide fine-grain cache management for different streams, when each of the stream exhibits diverging access patterns.
Consider an example of two streams $a_i$ and $b_i$, wherein $a_i$ exhibits a streaming access pattern and $b_i$ exhibits a recency-friendly access pattern. Also assume that the accesses from both streams are interleaved. 
A lightweight technique may apply a policy that is suitable for the access pattern that dominates the cache misses (e.g, apply LIP for both streams if $a_i$ dominates or apply LRU for both streams if $b_i$ dominates). Consequently, a lightweight technique is unable to manage individual streams. In contrast, the optimal technique may apply policy individually for each stream (e.g., by managing $a_i$ under LIP and $b_i$ under LRU), showing significant opportunity in improving cache efficiency by applying fine-grain cache management for individual streams.

\subsection{History-Based Predictive Techniques\label{sec:back:history}}
History-based predictive techniques implement dynamic policies that identify dead blocks (or conversely, useful blocks) based on historical access patterns~\citehistory{}.
These techniques encode reuse information of cache blocks beyond their current generations in some external structure, for subsequent recall when the cache blocks are accessed again. External state maintained by these techniques is often non-trivial, unlike that of lightweight dynamic techniques.

Majority of history-based techniques encode reuse information in an external structure called history table(s). To avoid the prohibitive storage costs of tracking individual cache blocks, these techniques use a single entry in the history table to encode reuse information for a set of cache blocks that are likely to exhibit homogeneous reuse. For example, prior works have used different correlating features such as the sequence of memory
access instruction addresses (PCs) leading to a block's access, the single PC accessing a cache block and starting address of a fixed size memory region containing a cache block~\cite{lasttouch,reftrace,sampler,ship}.

History-based techniques can provide fine-grain cache management 
by adapting their policies for individual access-streams as we explain below using the example of three state-of-the-art history-based predictive techniques.

\noindenttitle{SHiP~\cite{ship}} leverages PC-correlating reuse behavior by adapting its policies at per-PC granularity. Each PC is classified as Streaming-PC or Reuse-PC. If cache blocks inserted by a particular PC are evicted without incurring any reuse, the PC is classified as Streaming-PC. Any other PC is classified as Reuse-PC. For Streaming-PCs, SHiP applies policy suitable for streaming access patterns. For Reuse-PCs, SHiP applies policy suitable for recency-friendly access patterns.

\noindenttitle{Sampling Dead Block Predictor (SDBP)~\cite{sampler}} leverages PC-correlating reuse behavior by aiming to detect the last access to a cache block, \ie the instance at which a cache block becomes {\em dead}. Each PC is classified as Last-PC or Not-a-Last-PC. If cache blocks accessed by a particular PC are evicted without incurring a further reuse, the PC is classified as Last-PC. All other PCs are classified as Not-a-Last-PC. A cache block accessed by any Last-PC is predicted dead
and its priority is set to the lowest to make it the immediate candidate for eviction; if a cache access by a Last-PC leads to a cache miss, the corresponding cache insertion may be bypassed by forwarding data directly to the higher-level caches. Meanwhile, cache blocks accessed by Not-a-Last-PC are managed under a simple static cache management technique.

\noindenttitle{Hawkeye~\cite{hawkeye}} is the state-of-the-art technique that relies on PC-correlating reuse behavior. Hawkeye simulates Belady's OPT~\cite{opt} on past cache accesses and based on the policy decisions taken by OPT, it classifies each PC as cache-averse or cache-friendly. Cache blocks accessed by PCs tagged as cache-averse are made the immediate candidates for eviction. Meanwhile, other cache blocks are managed under a simple static cache management technique.

All three history-based techniques discussed above exploit some form of PC-correlating reuse, which is one of the most commonly used correlating features among prior history-based techniques. We also note that SHiP also proposed leveraging memory region as another correlating feature, in which SHiP adapts its policies at per region granularity and all cache blocks belonging to the same memory region are managed under the same policy.

\begin{figure}[!t]
    \centering
    \includegraphics[width=0.8\linewidth]{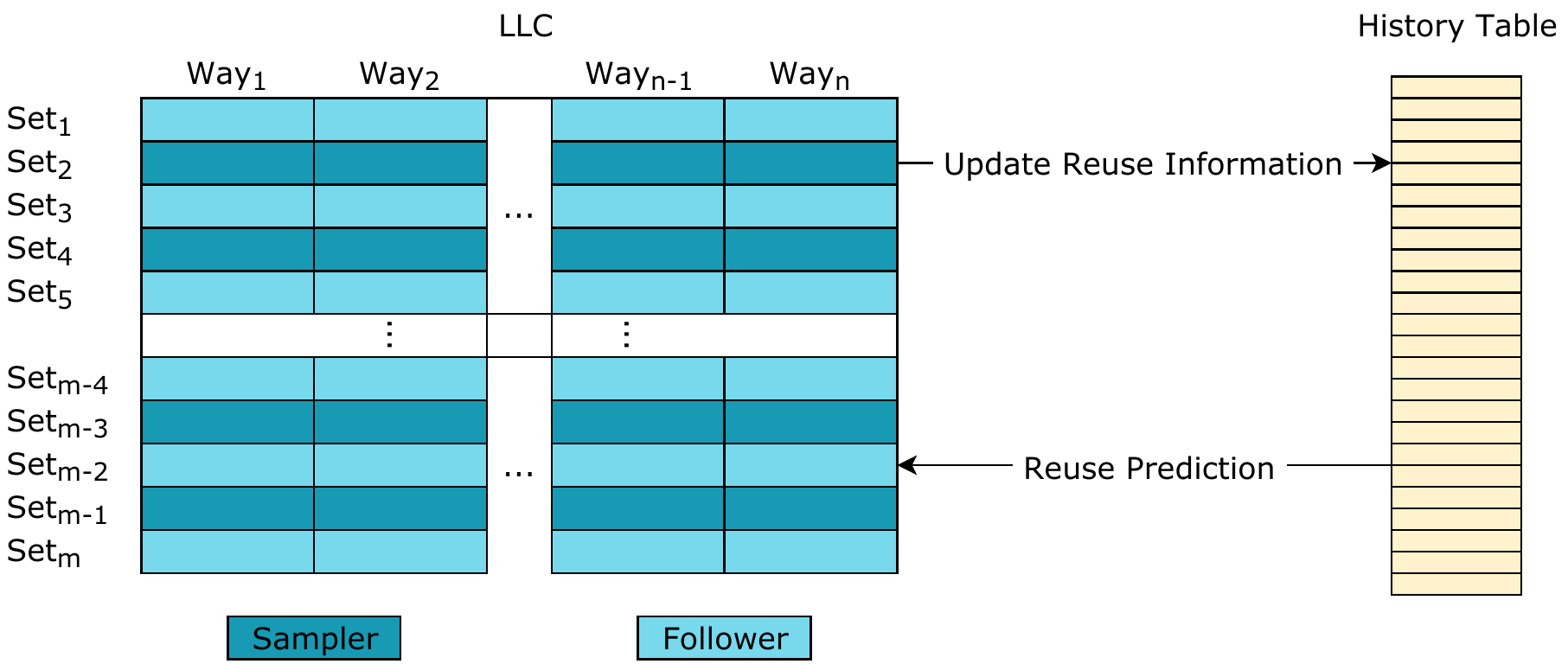}
    \caption{%
    A history-based predictive technique employs history table that encodes reuse information of cache blocks. History table is updated with the reuse information observed for cache blocks (potentially, for only cache blocks from the sampler sets). History table is queried to make reuse prediction for a cache block from any set.
    }
    \label{fig:back:history_state}
\end{figure}

\subsubsection{Storage}
\noindenttitle{Local state:}
As history-based predictive techniques are typically built on top of simple static or lightweight dynamic techniques, these also maintain a local state per cache block as required by the base technique.

\noindenttitle{Reuse information:}
The cache blocks may be augmented with additional state needed to encode reuse information, using which the history table is trained. To reduce the need to update history table frequently, only cache blocks belonging to a small number of preconfigured cache sets (called {\em sampler sets}) may be used to train the history table. Thus, the only cache blocks that belong to the sampler sets require additional storage.

\noindenttitle{Embedded prediction metadata:}
Some history-based techniques may also embed prediction metadata in every cache block. Prediction metadata is updated for a cache block on insertion and, potentially, on every subsequent hits. 
For example, SDBP uses 1-bit per cache block to indicate if a cache block is predicted dead, which is updated on every access to the cache block.

\noindenttitle{External prediction metadata:}
History-based techniques encode prediction metadata in some external structure, usually, history table, as shown in Fig.~\ref{fig:back:history_state}.  
For example, prior techniques employ history tables with 10s of KB of storage per core for 1MB LLC~\cite{ship, sampler, hawkeye}.

\subsubsection{Limitations}
\noindenttitle{PC-based reuse correlation:} A large fraction of history-based techniques rely on PC-based correlation to make reuse predictions as code-data correlation generally enables higher accuracy predictions than other features~\citepc{}. Indeed, all seven techniques~\citecrconly{} presented at the Cache Replacement Championship'17~\cite{crc2} rely on some form of a PC-based reuse correlation. Therefore, these techniques need to pass PCs through the load-store queue and all the levels of a cache hierarchy, requiring extra logic, wiring and energy consumption. This is partially mitigated by storing only a hash of a PC, which requires only a fraction of bits compared to the whole PC (e.g., 14-bits for a PC hash vs 48-bits for a full PC address). Nevertheless, it still poses a significant challenge for commercial processors to implement PC-based techniques~\cite{kpc}.

\noindenttitle{Reuse correlating features:} History-based techniques use correlating features (e.g., PC-based reuse correlation) to reduce the storage cost of history tables. Use of a correlating feature also helps train history table faster, when the reuse behavior for all cache blocks mapped to the same history table entry is similar. 
However, when the reuse behavior diverges for the cache blocks mapped to the same entry, it leads to a pathological case for prediction. Consider an example when a non-trivial fraction of cache blocks accessed by a PC exhibit high reuse, but the rest of the cache blocks accessed by the same PC exhibit no reuse. In such a case, history-based technique that relies on PC-correlating reuse may struggle in reliably identifying high-reuse cache blocks from the no-reuse cache blocks.

\noindenttitle{History table look-ups:}
SHiP relies on history-based predictions for only its insertion policy. For SHiP, every new cache block is inserted in the cache after querying the history table. In contrast, SDBP and Hawkeye rely on history-based predictions for insertion as well as the hit-promotion policy. Thus, the history table is queried on all cache accesses (including cache hits), which puts history table look-ups on the critical path as a look-up may increase the latency of a cache hit. Such critical path look-ups are even more undesirable in a modern multi-core processor with a NUCA LLC (as shown in Fig.~\ref{fig:back:nuca}), as each LLC hit under these techniques require accessing the PC-indexed history table that might be located elsewhere on a chip, incurring latency, energy, and traffic overheads due to the need to traverse the on-chip network.

\subsection{Software-Aided Techniques}

Software-aided cache management techniques rely on software hints to identify which cache blocks are likely to exhibit high reuse~\citesoftware{}. 
For cache management, these techniques typically rely on a lightweight dynamic technique. However, the policy selection is guided by the software, unlike lightweight dynamic techniques that rely on some hardware mechanisms such as set-dueling. 

For example, Pacman~\cite{cache-hints5-pacman} communicates a 1-bit hint with every memory access to guide whether a cache block should be inserted at the MRU position or the LRU position.
Pacman optimizes loop code using runtime profiling over multiple training runs as follows. 
During training, Pacman analyzes access patterns of memory addresses within a loop that are dependent on the loop index variable and attempts to find a correlation between the loop index and the reuse distance of the memory accesses. If it finds a linear correlation, the loop is split into two with all memory accesses in one loop are tagged with a non-temporal hint (e.g., LRU hint) and all memory accesses from the other loop are tagged with a temporal hint (e.g., MRU hint). Overall, the cache management under Pacman is very similar to that of DIP, except that Pacman relies on software to select between LRU or MRU insertion position whereas DIP relies on a lightweight hardware mechanism.

XMem~\cite{xmem}, a recently proposed software-aided technique, relies on pinning-based cache management for applications that benefit from cache tiling. Pinned cache blocks are protected from eviction until explicitly unpinned by the software, usually done when the tile is fully processed. XMem dedicates 75\% of LLC capacity for pinning cache blocks that belong to the tile whereas the remaining capacity is managed by some other hardware-only cache management technique.%

\subsubsection{Storage}
As software-aided techniques are typically built on top of lightweight dynamic techniques, these also maintain a local state per cache block.
Additionally, these techniques require nominal additional state, if any. For example, Pacman does not require any additional state whereas XMem requires 1-bit per cache block to identify whether a cache block is pinned.

\subsubsection{Limitations}
\noindenttitle{Custom interface:}
Software-aided techniques, unlike other techniques discussed so far, are not completely transparent to the software, and thus require additional hardware support for software to communicate hints. 
For example, Pacman proposed changes in the {\em Instruction Set Architecture (ISA)} by embedding load/store instructions with 1-bit reuse hint to guide cache management policies. Meanwhile, XMem proposed region-based interface as follows: XMem supports custom cache management for $n$ different memory regions. For each memory region, XMem hardware exposes a pair of registers, with each pair is required to be populated by software with the bounds of the region of interest. Software also sets the reuse hint for each memory region it populates to indicate whether the cache blocks from a given region should be pinned.

\noindenttitle{Limited scope:}
The majority of prior software-aided techniques rely on compiler analysis and/or runtime profiling to provide software hints. For example, Pacman only optimizes loops with regular access patterns, and thus may not be effective for applications dominated by irregular access patterns (e.g., indirect memory accesses of graph analytics), making such techniques difficult to apply for a broad spectrum of applications.

\section{Summary}
In this chapter, we provided background on cache management techniques necessary to understand our contributions in the following chapters. We also provided a broad classification of existing cache management techniques depending on the state needed by their heuristics, which is summarized in  Table~\ref{tab:back:state_summary}.%

Static techniques require the least amount of state, a few bits per cache block, among all classes of techniques. While standalone static techniques are the least effective in addressing complex access patterns at LLC, these techniques serve as building blocks for more advanced dynamic techniques. 

Lightweight dynamic techniques build on top of static techniques and require nominal additional state. These techniques provide significant value-addition over static techniques by dynamically adapting to the the observed access patterns.
However, due to limited state, lightweight techniques are unable to provide fine-grain cache management for individual access-streams.

History-based predictive techniques are the state-of-the-art in cache management that provide fine-grain cache management by adapting their policies according to the access patterns of individual access-streams. However, these techniques require non-trivial storage to maintain state in external structure(s), whose accesses may fall on the critical path of cache accesses.

Software-aided techniques can provide more accurate identification of high-reuse cache blocks as opposed to the hardware-only techniques for some applications. However, these techniques may require changes in the existing ISA. Finally, existing proposals target a set of applications with specific properties (e.g., tile-based algorithms or loops with regular access patterns).

Overall, history-based techniques and software-aided techniques generally manage LLC more efficiently than static or lightweight dynamic techniques. Unsurprisingly, to provide higher efficiency, these techniques also require more hardware (e.g., history table or new ISA extensions). However, the cost of additional hardware is usually insignificant in comparison to the LLC. For example, the storage requirement of a history table is less than 2\% of the LLC for the state-of-the-art history-based techniques~\cite{sampler,ship,hawkeye}.

\begin{table}[!t]
    \centering
    \small
    \begin{tabularx}{1\linewidth}
    {|>{\centering\arraybackslash\hsize=0.2\hsize}X| 
      >{\centering\arraybackslash\hsize=0.3\hsize}X|
      >{\centering\arraybackslash\hsize=0.25\hsize}X|
      >{\centering\arraybackslash\hsize=0.25\hsize}X|
      }
        \hline
        \makecell{Technique} & \makecell{State Within Cache} & External State & Software Support? \tn
        \hhline{|=|=|=|=|}
        \makecell{Static} & Recency State  & - & -   \tn
        \hline
        \makecell{Lightweight \\ Dynamic} & Recency State  & Nominal & -   \tn
        \hline
        \makecell{History-based} & \makecell{Recency State +  \\ Reuse Information + \\ Embedded Prediction \\ Metadata}  & History Table(s) & -   \tn
        \hline
        \makecell{Software-aided} & Recency State & - & ISA Extension \tn
        \hline
    \end{tabularx}
    \caption{Overview of state required for various classes of cache management techniques.}
    \label{tab:back:state_summary}
\end{table}

\chapter{Leeway -- Domain-Agnostic Cache Management\label{ch:leeway}}
\section{Introduction\label{sec:leeway:intro}}

History-based predictive techniques (also known as {\em Dead Block Predictors or DBP}) have been shown to be effective in improving LLC efficiency through better utilization of existing capacity~\citepriordbp{}. These schemes all rely on some metric of temporal reuse to make their decisions regarding the end of a given block's useful life. Previous works have suggested hit count~\cite{counter}, last-touch PC~\cite{sampler}, and number of references to the block's set since the last reference~\cite{pd}, among others, as metrics for determining whether the block is dead at a given point in time. By identifying and evicting dead blocks in a timely and accurate manner, these schemes allow other blocks (that have not exhausted their useful life) to persist in the cache and see further~hits. 

The task of a DBP is complicated by the fact that applications often exhibit {\em variability} in the reuse behavior of cache blocks. The sources of variability are numerous, stemming from microarchitectural noise (e.g., speculation), control-flow variation, cache pressure from other co-running applications, etc. The variability manifests itself as an inconsistent behavior of the individual cache blocks from one cache generation (from allocation to eviction) to the next. This inconsistency challenges DBPs in reliably identifying the end of a block's useful lifetime, thus resulting in lower prediction accuracy, coverage, or both.

A DBP requires metrics and policies that can tolerate inconsistencies. To that end, we propose {\em Live Distance}, a new metric of temporal reuse based on {\em Stack Distance}. 
Stack distance for a cache reference to a given cache block is defined as the number of unique cache blocks accessed since the previous reference to the cache block~\cite{stack-distance}.
For a given generation of a cache block, live distance is then defined as the largest observed stack distance in the generation.
Live distance is an efficient way to represent a block's range of temporal use and, as we argue in Sec.~\ref{sec:leeway:live-dist}, has a number of useful properties that make it attractive for dead block prediction in the face of variability.

We introduce Leeway, a new DBP that uses live distance as a metric for prediction. Leeway uses code-data correlation to associate live distance for a group of blocks with a PC that brings the block into the cache. While live distance as a metric provides a high degree of resilience to variability by conservatively capturing a block's temporal reuse, the per-PC live distance values themselves may fluctuate across generations.  To correctly train live distance values in the face of fluctuation, we observe that individual applications' cache behavior tends to fall in one of two categories: streaming (most allocated blocks see no hits) and reuse (most allocated blocks see one or more hits). Based on this simple insight, we design a pair of corresponding policies that steer updates in live distance values either toward zero (for bypassing) or toward the maximum recently-observed value (to maximize reuse). For each application, Leeway dynamically picks the best policy based on the observed reuse behavior at LLC.

To avoid the need to access specialized external structures (e.g, predictor or history table) upon each LLC access, Leeway embeds its prediction metadata (i.e., live distance) directly with cache blocks. 
This is in contrast with prior predictors~\cite{sampler, mdpp, hawkeye, perceptron}, which need to access a dedicated history table upon every single LLC access. Because modern multi-core processors feature distributed NUCA LLC, accesses to dedicated history tables introduce detrimental latency and energy overheads in traversing the on-chip interconnect to query such structures. 

\noindentnotitle
We study cache management techniques on various deployment configurations, and make the following contributions:
\begin{itemize}
    \item We propose Leeway, a dead block predictor for LLC that introduces a new metric, Live Distance, to track a block's useful lifetime in the cache.
    To provide high performance in the face of variability, 
    Leeway deploys novel reuse-aware update policies that steer live distance values to maximize either bypass or reuse opportunities based on the application preference.
    \item Leeway embeds prediction metadata in the cache, and thus accesses history table only on misses, keeping the table look-ups off the critical path. This is in contrast to prior DBPs that access history tables on all cache accesses (including cache hits).
    \item We compare Leeway to prior cache management techniques for LLC, demonstrating that Leeway consistently provides good performance that generally matches or exceeds that of state-of-the-art approaches. 
\end{itemize}

\section{Motivation}
\label{sec:leeway:motivation}

\subsection{Variability in the Reuse Behavior of Cache Blocks}
\label{sec:leeway:variation}

DBPs aim to improve cache behavior by identifying dead blocks and discarding them shortly after their last use, thereby providing an opportunity for blocks with long temporal reuse distances to persist. Effectiveness of a dead block prediction hinges on the stability of application behavior with respect to the metric used for determining whether the block is dead. Naturally, the more consistent the reuse behavior across the block's generations in the cache, the more accurate the predictions. %

In practice, there are many reasons for why a block's live time may vary across generations, including:

\noindenttitle{Control flow variation:} When the memory reference instruction is predicated on a condition whose behavior varies at runtime, the corresponding cache block might be referenced a different number of times across generations based on the predicate.

\noindenttitle{Microarchitectural noise:} This includes references on a mispredicted control flow path and hits in lower-level caches due to conflicts in higher-level caches.

\noindenttitle{Shared data:} When a block is shared by multiple threads, it might see different reference patterns due to runtime dynamics and scheduler decisions. 

\noindenttitle{Cache pressure:} An application behavior may be consistent but due to cache pressure in the presence of co-running applications, a block may be prematurely evicted. As a result, the block would observe fewer references in a prematurely terminated generation than it would~otherwise. 

\noindenttitle{Application characteristics:} An application may inherently exhibit irregular behavior, leading to inconsistent access patterns for cache blocks. For example, for graph processing applications, reuse patterns of accesses to vertices are dependent on the graph topology. Specifically, the number of times a vertex is accessed depends on the number of edges connected to the vertex and the reuse distance of an access depends on the number of other vertices and edges accessed since the previous access to the same vertex.

Our insight is that the ability of a DBP to tolerate {\em inconsistency across generations} hinges on the choice of the metric used for making the predictions. Spurred by the observation, we next use a simple taxonomy to understand the space of metrics.

\subsection{Metrics for Dead Block Prediction}
\label{sec:leeway:classification}

Fundamentally, all DBPs require a metric for determining when a block has reached the end of its useful life. Existing metrics can be classified broadly into two categories: {\em direct} and {\em indirect}.

\noindenttitle{\smalltextcircled{1} Direct metrics:} Also known as {\em event-based metrics}, these rely on monitoring accesses to the block in order to detect the final access based on previously observed behavior. Reference count~\cite{counter}, trace signature of instructions referencing a block~\cite{reftrace,lasttouch}, and last-touch PC~\cite{sampler} are all examples of direct metrics used by previously proposed DBPs. An advantage of direct metrics is that a block's fate is determined exclusively by accesses to itself, thereby shielding the decision-making mechanism from noise due to accesses to other blocks. 

{
\centering
\begin{lstlisting}[float=t,mathescape,frame=single,caption=A code snippet showing potential  variability in the reuse behavior of reference X due to a data-dependent branch.,label=lst:leeway:snippet,basicstyle=\small,captionpos=b]
  PC$_i$: Ld X            
  . . .                                 
  PC$_v$: Beq cond, SKIP  
  PC$_w$: Ld X            
  SKIP:                                
\end{lstlisting}
}

The downside of direct metrics is their inflexibility in the face of {\em inconsistent} behavior, which we define as any variation from one generation of a block to the next. Consider a simple code snippet shown in Listing~\ref{lst:leeway:snippet}, which shows a reference to a cache block holding the variable X, followed by a predicated second reference to X.

Assuming that the second reference occurs only a fraction of the time due to the data-dependent nature of the predicate, predictors that rely on direct metrics are faced with two choices: (1) predict the block dead after the first reference, incurring a miss if the predicate resolves to False; or (2) predict the block dead after the second reference, which may never occur if the predicate resolves to True, and thus the prediction is never made.
Alas, none of the options are satisfying, as they reduce either accuracy or coverage of the predictions.

\begin{figure}[!t]
    \centering
    \includegraphics[width=0.8\linewidth]{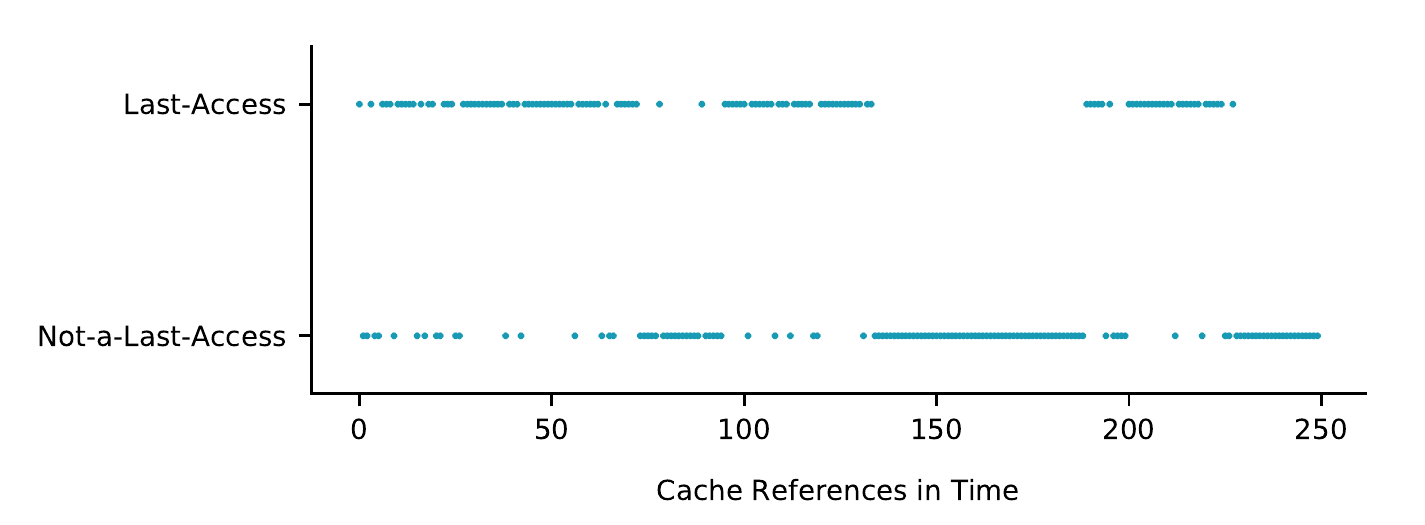}
    \caption{Variability for a PC being the last touch or not in {\em h264ref}} 
    \label{fig:leeway:last-pc}
\end{figure}

Fig.~\ref{fig:leeway:last-pc} demonstrates such behavior for the last-PC metric used by SDBP~\cite{sampler} in {\em h264ref}, one of the SPEC CPU 2006 applications, for a PC responsible for 37\% of the misses. The behavior captured in the figure is representative of the entire execution; for clarity, however, the figure shows only a sample of 250 consecutive cache references by that PC (X axis). For each reference, the Y axis shows whether the reference is, indeed, the last access to the block or not under the LRU cache management technique. For the last-PC metric to be useful in identifying dead blocks upon a last access to them, this behavior should be consistent, with all points falling on either the Last-Access (indicating dead blocks) or Not-a-Last-Access (indicating live blocks) line. Meanwhile, the fluctuation shown in the figure indicates that the predictor using last-PC metric may struggle in accurately determining the end of a useful lifetime for blocks touched by this PC.

\noindenttitle{\smalltextcircled{2} Indirect metrics:} Also known as {\em age-based metrics}, these rely on an external reference signal to inform the prediction mechanism of the block's age. 
A block's age increases with some notion of time, which is reset upon a hit.
The age can be computed in 
number of cycles~\cite{timekeeping}, number of accesses to the cache~\cite{timekeeping2}, or number of accesses to the set~\cite{pd, counter}. 
When a block's age crosses a set threshold (e.g., the maximum observed age from the previous generations), the block may be predicted~dead.

A major advantage of indirect metrics is their inherent ability to tolerate uncertainty in a block's behavior. Coming back to the code snippet in Listing~\ref{lst:leeway:snippet}, a carefully chosen age threshold may allow the block to stay in the cache long enough to see the second hit, if any, while ensuring that the block won't greatly overstay its likely useful lifetime. 

The drawback of existing indirect metrics is their imprecision and susceptibility to noise.
Because the prediction is made based on events unrelated to the block itself (e.g., the count of all cache accesses), the age used for deciding whether the block is dead must have some tolerance to fluctuation built into it. This tolerance inevitably increases the block's dead time, even for highly predictable blocks, potentially causing the block to stay in the cache long after its last access while waiting for the age to reach the conservatively set threshold. 

\subsection{Toward a Better Metric}
\label{sec:leeway:live-dist}

{\em Stack distance} for a reference to a given cache block is defined as the number of unique cache blocks accessed since the previous reference to the cache block~\cite{stack-distance}. Stack distance provides a useful way to reason about a block's reuse behavior: blocks that have short reuse intervals will have short stack distances, while blocks with long reuse intervals will see larger stack distances over their lifetime in the cache. In practice, a short stack distance means that a block is likely to experience a hit when it is near the top of the LRU stack (i.e., close to the MRU position). Conversely, a long stack distance means that a hit may come near the LRU position, or, if the stack distance exceeds the associativity of the cache, will result in a miss to the block. By predicting dead blocks early, DBPs aim to keep blocks with long stack distances in the cache long enough for them to see a hit. 

\begin{figure}[!t]
    \centering
    \includegraphics[width=0.8\linewidth]{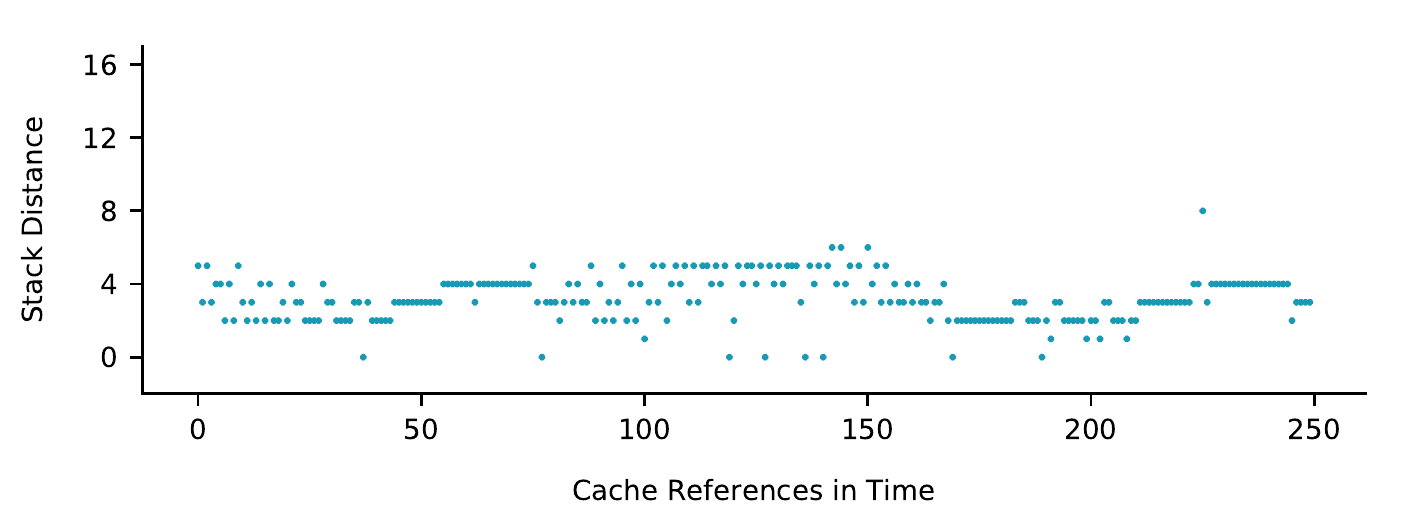}
    \caption{Stack Distances for one PC in {\em GemsFDTD} for 16-way set-associative cache. For a cache hit, a stack distance ranges from 1--16. A cache block that is evicted with zero hits is shown to have a stack distance of 0.}
    \label{fig:leeway:example}
\end{figure}

We make the observation that stack distance can be turned into a powerful metric for dead block prediction. 
Fig.~\ref{fig:leeway:example} provides the intuition. The figure shows the observed stack distances for a sample of 250 cache references for all blocks allocated by a single PC which is responsible for the highest number of LLC misses in {\em GemsFDTD}. 
The key take-away is that despite significant variability across references, the stack distance is largely confined to 5.

Based on this insight, we define {\em Live Distance} as the maximum observed stack distance during a block's generation (from insertion to eviction). Live distance is a good indicator of the block's temporal reuse limit, so when the block's position within the LRU stack exceeds its known live distance, the block is unlikely to be referenced again and can be predicted dead.
To obtain stack distance values, we exploit the fact that LRU-based policies implicitly track stack distances of cache-resident blocks. In true LRU, when a block hits, its current LRU stack position corresponds to its stack distance. 
For policies that deviate from true LRU, such as multi-bit NRU (see Sec.~\ref{sec:leeway:nru} for details), a block's stack position upon a hit only approximates the true stack distance. Nevertheless, it provides an efficient heuristic to approximate stack distance and, correspondingly, live distance.

\begin{table}[!t]
\centering
\small
\begin{tabular}{|c||l|c|c|c|}
\hline
Ref \#  & Reference Pattern        & Stack Distance   & Live Distance  & Cache Event      \\
\hline \hline
1    & \texttt{X A X}           &   2   & 2   &   Hit   \\
\hline
2    & \texttt{X A B X}         &   3    & 3  &   Hit   \\
\hline
3    & \texttt{X A A A B B B A X}   &   3    & 3  &   Hit   \\
\hline
4    & \texttt{X F X}           &   2    & 3  &   Hit   \\
\hline
5    & \texttt{X A B C P Q R S T X}    &   $\infty$ ($>$8)  & 3 &   Miss   \\
\hline
\end{tabular}
\caption{Stack Distance \& Live Distance for block X in 8-way set for a reference pattern \texttt{X A X A B X A A A B B B A X F X A B C P Q R S T X}. Assuming LRU policy, X incurs 4 cache hits in a generation that starts with a cache fill of the first instance of X in Ref \#1 and ends when X is evicted in the Ref \#5 upon an access to T. Last instance of X in Ref \# 5 misses in the cache, which starts another generation with a cache fill of X.}
\label{fig:leeway:stack-distance}
\end{table}

Table~\ref{fig:leeway:stack-distance} demonstrates how stack and live distance is determined for a block {\texttt X} for various reference patterns in a 8-way set. In this example, the largest observed stack distance is 3, yielding a live distance of 3 and indicating that 
{\texttt X} can be predicted dead after the reference to {\texttt C} in reference pattern \#5. 

Live distance combines the best properties of both direct and indirect metrics, making it more effective than ``pure'' approaches. Specifically, to determine if a block is dead, live distance uses an indirect signal, which is the block's place within the LRU stack. This signal is indirect, since the block ages as a result of hits to other blocks within the set. Crucially, however, live distance for a block X is trained only upon hits to X (same as direct metrics), which demarcate the range of the block's temporal reuse within the LRU stack. Because of this combination, live distance can naturally tolerate variability across generations as long as the reuse interval for the block falls within the previously observed range. At the same time, live distance provides an efficient mechanism for rapidly identifying blocks that have exceeded their typical reuse window and can therefore be predicted dead.

Compared to other indirect metrics, live distance has an additional attractive property. By relying on stack distance, which only grows as a result of hits to {\em unique} blocks, live distance provides a degree of dampening to noise resulting from variability in access patterns to recently-accessed blocks. Because most recently accessed blocks are the ones likely to receive future hits, suppressing variability in these hit counts is beneficial ~\cite{burst}. For instance, consider reference patterns \#2 and \#3 in Table ~\ref{fig:leeway:stack-distance}. 
When trying to learn the reuse distance for X, counting the number of all accesses, unique or not, to the set between references to X as proposed in prior work~\cite{pd} produces an inconsistent distance. In contrast, the stack distance for X in both reference patterns is unaffected by variability in the number of accesses to blocks A and B, resulting in a consistent live distance value.

\section{Leeway Design}
\label{sec:leeway:design-llc}

We introduce Leeway, a history-based predictive cache management technique that uses live distance as its underlying metric. 
We first explain the Leeway basics and features that make it robust against variability in the context of LLC. We then show how Leeway works with a low-cost 2-bit NRU cache management technique. We then discuss microarchitectural details and compare its cost and complexity with prior techniques. Later we extend Leeway to a multi-core setup.

\subsection{Overview}
\label{sec:leeway:overview}
LRU-based Leeway uses a full LRU stack and records the maximum observed hit position (i.e., live distance) during a block's residency in the cache. At eviction time, the live distance is recorded in a separate structure, {\em Live Distance Predictor Table (LDPT)}, for subsequent recall when the block is allocated again. Leeway uses the live distance learned in the block's previous generations to infer when the block may have exceeded its useful lifetime and predicts it dead.  To avoid the prohibitive storage costs of tracking individual cache blocks in the LDPT, Leeway exploits code-data correlation and associates all cache blocks allocated by the same PC with one PC-indexed LDPT~entry.

The functionality of Leeway can be divided into three categories -- {\em Learning, Prediction} and {\em Update}. Learning is a continuous process for cache-resident blocks that involves checking a block's position in the LRU stack upon each hit and, if the current position exceeds the past maximum, updating the live distance. Prediction is triggered during victim selection on a miss to a set. Any block that has moved past its predicted live distance in the LRU stack is predicted dead. Update occurs upon a block's eviction from the cache, propagating the latest live distance to the LDPT. To effectively handle variability in live distance across generations of a given block and across blocks tracked by a single PC-indexed LDPT entry, the update process is conditional as explained in the next section.

Leeway implements set-sampling, similar to~\cite{ship}, to learn the blocks' live distances by observing their behavior in a small number of sampler sets. 
Sampling significantly reduces Leeway's storage requirement as the only blocks belonging to the sampler sets need to be augmented with storage needed for learning.

\subsection{Adapting to Variability}
\label{sec:leeway:adapt}

As explained in Sec.~\ref{sec:leeway:variation}, a block's observed reuse behavior may fluctuate in time even if its fundamental reuse characteristics are not changing. While the live distance metric provides a degree of protection from intra-generation noise, Leeway must contend with inevitable fluctuation in live distance across generations and across different blocks allocated by the same PC. In particular, it must separate unrepresentative live distance values from actual shifts in the reuse behavior. This observation points to the need for an intelligent update policy for Leeway's  live distance values.

To design a variability-tolerant update policy, we study SPEC CPU 2006 applications to understand their reuse behavior. Our analysis reveals that applications tend to fall in one of two categories in terms of their reuse behavior affecting LLC management. 

The first category is dominated by streaming cache blocks that do not observe any LLC hits and should be bypassed. For example, in {\em mcf}, over 90\% of cache blocks are not reused after allocation in LLC under LRU. In many cases, however, we find that blocks allocated by certain streaming PCs will occasionally observe one or more hits. Fig.~\ref{fig:leeway:bypass} shows one such PC responsible for 21\% of the misses in {\em mcf}. Moreover, such behavior sometimes occurs in clusters, forcing a shift in cache management policy from bypassing to keeping blocks on chip. Such a shift is generally undesirable, as the behavior tends to quickly revert back to streaming. 
A multi-bit hysteresis threshold may be effective in delaying a shift in policy; however, the high threshold is counter-productive when the behavior reverts back to streaming as it will lead to blocks being allocated in LLC rather than be bypassed.

\begin{figure}[t!]
    \centering
    \includegraphics[width=0.8\linewidth]{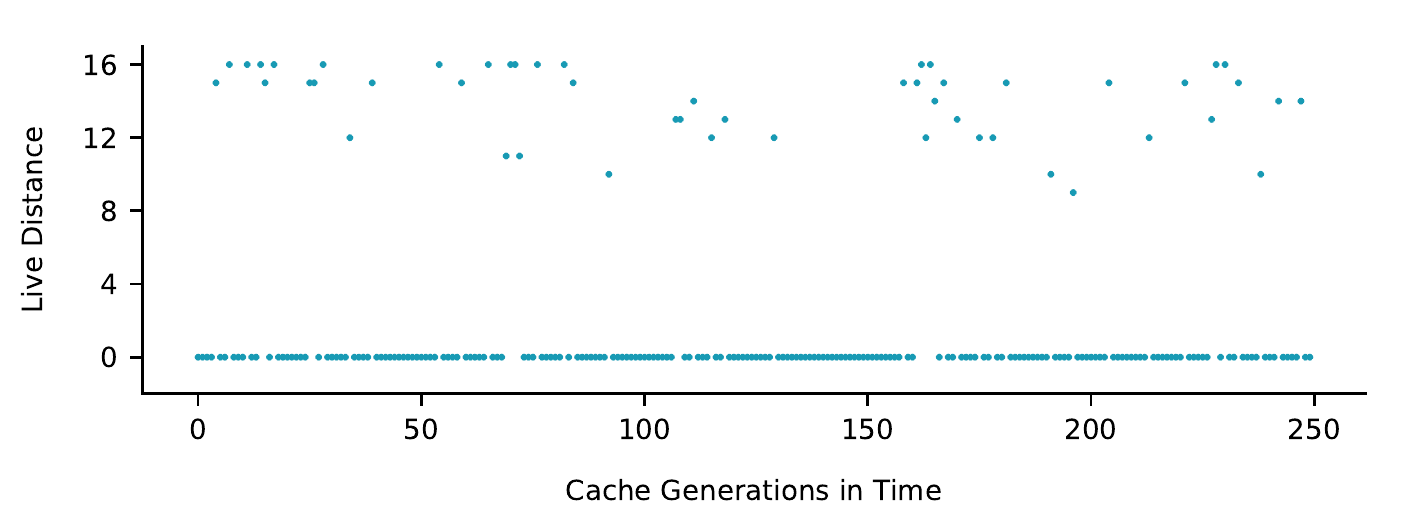}
    \caption{Variability in live distance with a bias of streaming for a PC in {\em mcf}. A Live Distance of 0 indicates a bypass opportunity.\label{fig:leeway:bypass}}
    \end{figure}
    \begin{figure}
    \includegraphics[width=0.8\linewidth]{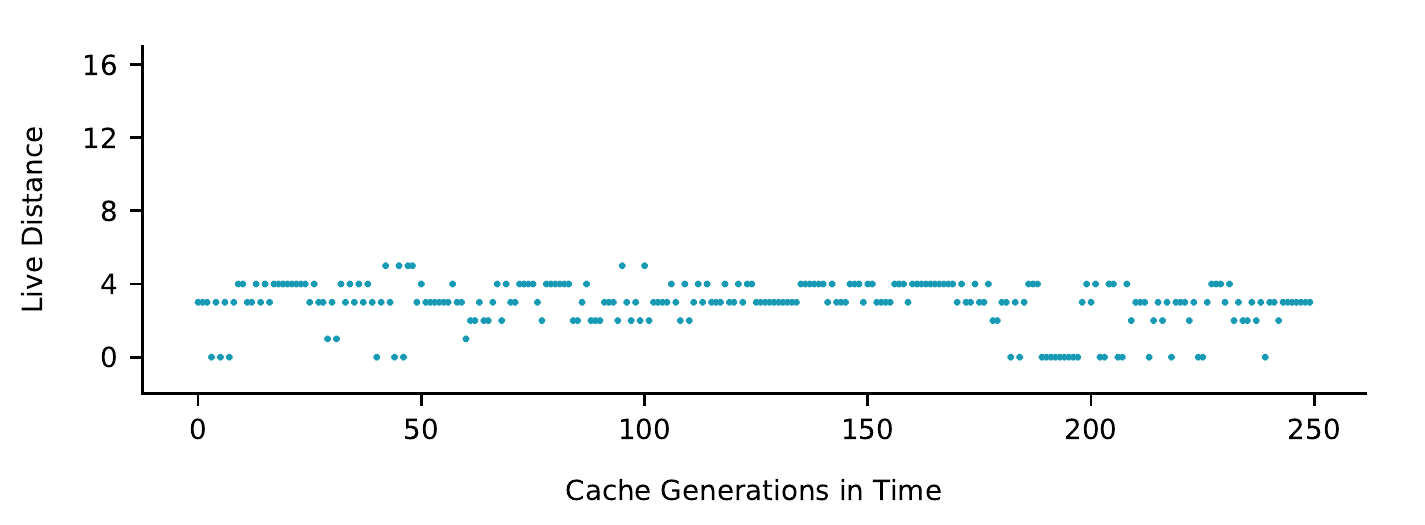}
    \caption{Variability in live distance with a bias of reuse for a PC in {\em calculix}.\label{fig:leeway:reuse}}
\end{figure}

The second category of applications is dominated by blocks that do see reuse prior to being evicted from the LLC. For example, in {\em calculix}, more than 60\% blocks are reused at least once after their allocation in LLC under LRU. We observe considerable variability in live distance for many PCs that allocate blocks exhibiting reuse. 
Fig.~\ref{fig:leeway:reuse} shows one such PC responsible for 29\% of the misses in {\em calculix}. This observation is consistent with our work that observed that the blocks exhibiting reuse are more prone to variability in inter-generational behavior than the streaming blocks, thus posing a challenge for DBPs~\cite{c-dead}. Given the uncertainty in the reuse behavior, such blocks should be kept longer to maximize opportunity for reuse.

The two types of behavior naturally lead to a pair of policies designed to maximize bypass opportunities for streaming applications and reuse opportunities for others.%

\noindenttitle{\smalltextcircled{1} Bypass-Oriented Policy (BOP):} This policy seeks to maximize opportunities for bypass by being slow to increase the live distance and fast in dropping it back towards 0, in the face of variability in live distance values. An incoming block with a predicted live distance of 0 is bypassed, unless it maps to a sampler set (see Sec.~\ref{sec:leeway:leeway-actions} for details).

\noindenttitle{\smalltextcircled{2} Reuse-Oriented Policy (ROP):} To maximize reuse opportunities for allocated blocks when there is a fluctuation in live distance values, this policy is quick to increase the live distance and slow to decrease it. Since Leeway does not evict blocks that have not reached their live distance value in the LRU or multi-bit NRU stack, a larger live distance enables a longer temporal window for a block to uncover reuse.

\noindenttitle{Enabling the policies:}
The two policies call for diametrically opposite behavior: whereas the Bypass-Oriented policy is slow to increase the live distance values in LDPT but fast to decrease them, the Reuse-Oriented policy is fast to increase live distance values but slow to decrease them. To satisfy the demand for separate policies in increasing and decreasing live distance in the LDPT, Leeway deploys two {\em Variability Tolerance Thresholds (VTTs)} that control the rate at which live distance values are adjusted based on workload behavior and the direction of change in live distance.

In order to choose the preferred policy for a running application, Leeway leverages Set-Dueling~\cite{dip} and implements both policies (Bypass- and Reuse-Oriented) simultaneously on separate sampler sets. The rest of the cache follows the policy that minimizes the misses.

\subsection{Leeway with Cost-Efficient NRU}
\label{sec:leeway:nru}

So far, we have considered Leeway on top of true LRU, which may be unattractive for highly-associative caches.
In this section, we explain the minimal modifications required to make Leeway work with a low-cost multi-bit {\em Not Recently Used (NRU)} family of techniques. 

NRU uses 1-bit per cache block to keep track of blocks that have not been used recently with respect to some time frame in the past. 
Multi-bit NRU is an extension of NRU that uses two or more bits per cache block to indicate a partial relative order of LRU stack positions. For instance, a 2-bit NRU policy keeps blocks in a set in one of four equivalence classes as a function of their relative stack positions, with class 1 for MRU blocks and class 4 for LRU ones. During victim selection, a block in class 4 is evicted (ties are broken through random selection). If no block is found in class 4, every block is moved to the next class and the process is repeated. Both RRIP~\cite{rrip} and SHiP~\cite{ship} use 2-bit NRU.

Leeway implementation over (1-bit or multi-bit) NRU, {\em Leeway-NRU}, relies on the partial relative order maintained by NRU to make dead block predictions. It uses a block's NRU value to approximate its stack distance, and in turn, live distance. It cannot differentiate between the relative order of blocks in the same recency class.

In general, Leeway can be implemented with any base technique which maintains (1) a partial relative order of blocks based on their relative reference time and (2) a monotonically non-decreasing order for a given block's position between re-references or until eviction.

\subsection{Microarchitecture}
\label{sec:leeway:uarch}

\begin{figure}[!t]
    \centering
    \includegraphics[width=1\linewidth]{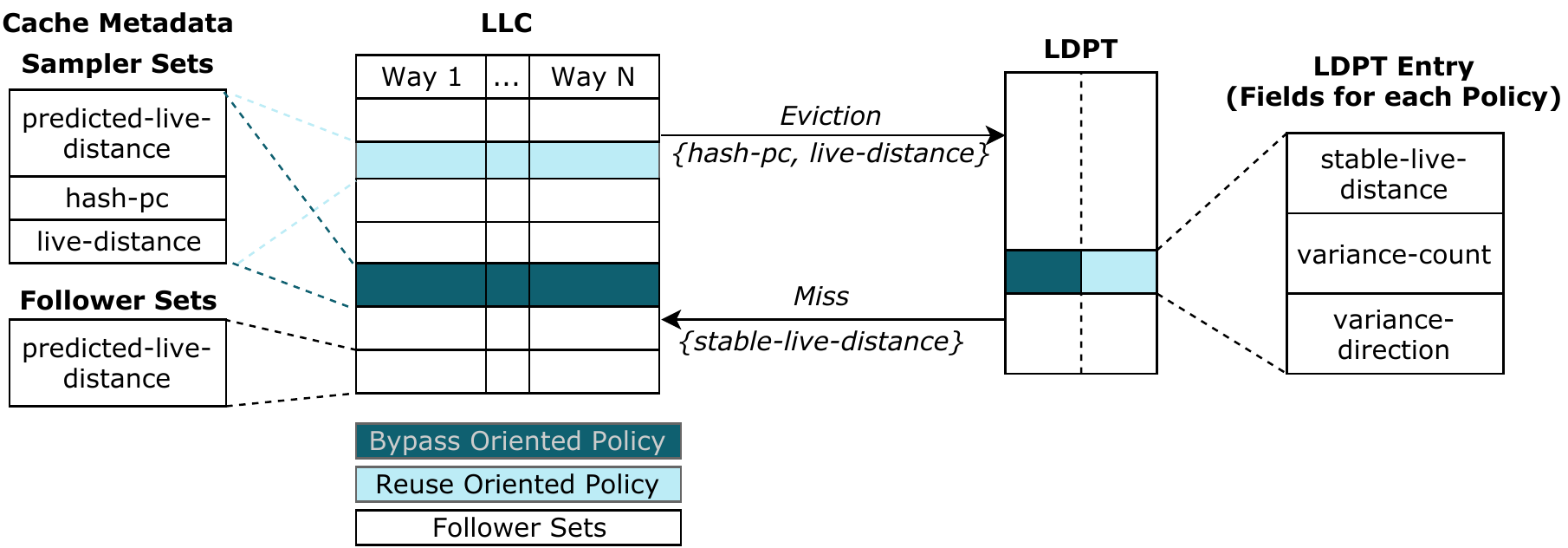}
    \caption{Schematic of Leeway for LLC}
    \label{fig:leeway:structures}
\end{figure}

\subsubsection{Physical Fields and Structures}
\label{sec:leeway:fields}
Fig.~\ref{fig:leeway:structures} summarizes key elements of the design.

\noindenttitle{LDPT:} Each PC-indexed LDPT entry contains a {\em stable-live-distance} field that indicates the current live distance based on most recent history. Updates to stable-live-distance are controlled by VTTs and two additional LDPT fields: 
(1) {\em variance-count} is a counter for tracking the number of consecutively evicted cache lines whose live distance differs from  the stored value, and (2) {\em variance-direction} is a bit indicating the direction of the change. Once the count matches the value of a VTT for a given direction, the value of {\em stable-live-distance} is updated. To avoid additional storage for transient live distance values, the new {\em stable-live-distance} value is taken from the evicted block that triggers the update.

\noindenttitle{VTTs:} To enable Bypass- and Reuse-Oriented policies, Leeway uses a pair of Variability Tolerance Thresholds that control the rate at which {\em stable-live-distance} values are updated (Sec.~\ref{sec:leeway:adapt}). 
Empirically, we find that a 3-bit VTT is sufficient, and use the maximum value for the slow update (i.e., requiring 7 consecutive evictions with a live distance different, and in the same direction, from the {\em stable-live-distance}) and a value of 1 for the aggressive threshold.
Thus, the two valid VTT configurations are either \{7,1\} (for the Bypass-Oriented policy, with a slow increase and fast decrease) and \{1,7\} (for the Reuse-Oriented policy with a fast increase and slow decrease).

\noindenttitle{LLC:} Leeway requires all LLC blocks to carry a field, {\em predicted-live-distance}, which is read from the LDPT at block allocation time and is subsequently used for dead block prediction. As this field is embedded in the cache, dead block prediction can be done locally in cache just by comparing a block's LRU stack position with the value of its predicted-live-distance field.
Meanwhile, the cache blocks from the sampler sets carry two additional fields: {\em live-distance} \& {\em hash-pc}. These are used for learning, allowing evicted blocks to index the LDPT and, if necessary, update its fields as explained above.

\subsubsection{Leeway in Action}
\label{sec:leeway:leeway-actions}

\noindenttitle{\smalltextcircled{1} Cache miss:} On an LLC miss, the LDPT is indexed using a hash of the miss PC to recall the {\em stable-live-distance}, which is then transferred to the incoming block's {\em predicted-live-distance} field. If {\em stable-live-distance} is 0, the block is expected to have no reuse and is bypassed to the higher-level caches. Since bypassed blocks have no opportunity to retrain, Leeway inserts them into the sampler sets with a small probability (1\% for Bypass-Oriented Policy and 3\% for Reuse-Oriented Policy) to enhance learning.

\noindenttitle{\smalltextcircled{2} Cache hit (Learning):} On a hit to a sampler set, the block's {\em live-distance} field is updated if 
its current stack position is greater than the value of the {\em live-distance} field. 
Meanwhile, for all sets (sampler as well as the follower sets), the block's {\em predicted-live-distance} is also updated if its current stack position is greater than the value of the {\em predicted-live-distance} field. Note that the {\em predicted-live-distance} field is never used to update LDPT, and thus the change remains local and protects the only block for which the {\em predicted-live-distance} is increased.

\noindenttitle{\smalltextcircled{3} Eviction (Prediction and Update):} To find victim, Leeway searches for a dead block by comparing each block's LRU or NRU position to its {\em predicted-live-distance} field. 
If more than one blocks are found dead, a block with the minimum {\em predicted-live-distance} value is picked for replacement. If no block is found dead, the LRU block is evicted. If the evicted block resides in the sampler set (dead or not), 
its {\em live-distance} and {\em hash-pc} fields are forwarded to the LDPT for a potential update.

\subsubsection{Mechanism for Policy Selection}
\label{sec:leeway:mechanism:setdueling}
To dynamically choose between Bypass- and Reuse-Oriented policies, Leeway relies on a set-dueling mechanism~\cite{dip}. Thus, two separate groups of sampler sets are used, with each group implementing one of the two policies. To support simultaneous implementation of policies, the LDPT must be extended to support two sets of \{{\em stable-live-distance, variance-count, variance-direction}\} fields per entry. While the sampler sets always access their dedicated fields based on a static mapping, the rest of the sets read the {\em stable-live-distance} from the winning policy.

To determine the winning policy, Leeway maintains two saturating miss counters, one for each policy. The counters are incremented on a miss to a sampler set of a respective policy. Periodically, the miss counters are sampled and the winning policy is selected based on the counter with the lowest value.

Often, the winning policy remains the same throughout the application's execution. In some cases, however, the winning policy may change due to changes in the application's phase or its co-runner(s). 
In theory, a policy change requires reloading {\em predicted-live-distance} for all cache blocks using the {\em stable-live-distance} of the new winning policy in LDPT.
In practice, we find that policy change is infrequent, indicating that the simplest way to deal with it is to leave existing blocks untouched, potentially incurring a handful of poor decisions but minimizing microarchitectural complexity.

\subsection{Cost and Complexity Analysis}
\label{sec:leeway:cost-complexity}

\noindenttitle{Storage cost:}
We analyze storage requirements for a 16-way 2MB LLC with 64B blocks. We find that a 16K-entry LDPT per core is sufficient and is not affected by destructive aliasing, thus affording a tagless design. For LRU-based Leeway, each LDPT entry of each of two Leeway policies has 8 bits: 4 for {\em stable-live-distance}, 3 for {\em variance-count} and 1 for {\em variance-direction}. The resulting cost of LDPT is thus 32KB.

We use a 64-set sampler per policy. Each block in the sampler carries a 4-bit {\em live-distance} and 14-bit {\em hash-pc} fields, requiring 4.5KB of storage in total. All cache blocks, including the sampler, include a 4-bit {\em predicted-live-distance}, totaling 16KB storage. The total storage storage of Leeway is thus 68.5KB (52.5KB overhead + 16KB of LRU state), or 2.3\% of the LLC storage.
Using 2-bit NRU instead of LRU further reduces the storage by 36\% to 44KB, or 1.4\% of the LLC storage, by lowering live distance storage costs from 4 to 2 bits. 

\setlength\tabcolsep{5pt}
\begin{table}[!t]
        \centering
        \small
        \begin{tabular}{|l|N{4}{2}|N{4}{2}|N{4}{2}|N{4}{2}|c|}
        \hline
        \multirow{2}{*}{Technique}  &   \multicolumn{1}{c|}{Recency}    &   \multicolumn{2}{c|}{Predictor State (KB)}    &   \multicolumn{1}{c|}{Total}   & \multicolumn{1}{c|}{When is History} \\ \cline{3-4}
                                    &   \multicolumn{1}{c|}{State (KB)}               &   \multicolumn{1}{c|}{Within LLC} & \multicolumn{1}{c|}{External to LLC} & \multicolumn{1}{c|}{(KB)} & \multicolumn{1}{c|}{Table accessed?} \\ 
        \hline \hline
        SDBP~\cite{sampler}        &   16     &     4  &   18.75       &   38.75   &   Hits + Misses       \\ \hline            
        SHiP~\cite{ship}        &   8       &   3.75 &   6        &   17.75    &   Misses*      \\ \hline
        Hawkeye~\cite{hawkeye}     &   12  & \multicolumn{1}{c|}{-}       &   19          &   31      &   Hits + Misses       \\ \hline
        Leeway-LRU  &   16  & 20.5  &   32        &   68.5      & Misses        \\ \hline
        \rowcolor{lightgray} Leeway-NRU  &   8   &  12  &  24           &   44    &   Misses      \\ \hline
        \end{tabular}
        \caption{Storage cost (excluding tag and data) for 16-way 2MB LLC, 128 sampler sets, and 16K-entry Predictor Table for history-based predictive techniques. (*For SHiP, cache hits to the follower sets do not access the history table. Meanwhile, cache hits to the sampler sets do update the history table; however, the updates to the table can be pipelined and taken off the critical path.)}
        \label{tab:leeway:llc-storage}
\end{table}
\setlength\tabcolsep{6pt}

Table~\ref{tab:leeway:llc-storage} compares the storage requirements of Leeway to those of prior techniques. SHiP~\cite{ship}, an insertion technique, has the lowest storage cost at the expense of not predicting blocks that are reused. Among dead block predictors that also predict reused blocks, 
the preferred Leeway-NRU configuration requires 44KB of storage in total (including NRU bits), compared to 38.75KB for SDBP~\cite{sampler} and 31KB for Hawkeye~\cite{hawkeye}, considering the same number of sampler sets and predictor table entries for all techniques.
While Leeway is slightly more expensive, we observe that the storage requirements for all techniques are in a similar range of several tens of KBs. Such modest storage requirements are dwarfed by the size of the LLC.

\noindenttitle{Complexity:}
Operations performed by Leeway at various stages are limited to simple additions and comparisons, which are quite hardware friendly. Additionally, Leeway embeds the metadata necessary for the prediction (i.e., {\em live distance}) with the cache blocks. As a result, LLC hits and replacement decisions never access remote metadata. The only time Leeway accesses its prediction table (LDPT) is upon cache misses, when {\em stable-live-distance} is read and possibly updated. These accesses are entirely off the critical path, since they do not involve state updates to a live cache block.

In contrast, state-of-the-art predictive techniques, such as SDBP~\cite{sampler} and Hawkeye~\cite{hawkeye}, use a PC-indexed prediction table that is probed on every LLC access (including a cache hit) to inform the block's eviction priority. For example, Hawkeye incurs 2.3$x$ accesses to its prediction table when compared to Leeway (SPEC average). Such frequent accesses to the prediction table are particularly undesirable in a modern multi-core processor
with a NUCA LLC, 
as each LLC hit requires state-of-the-art predictive techniques to access the PC-indexed prediction table located elsewhere on a chip, incurring latency, energy, and traffic overheads due to the need to traverse the on-chip network.

\subsection{Leeway for Multi-Core}
\label{sec:leeway:multi-core-strategies}

Leeway can naturally be extended to multi-core deployments. The only notable difference is in determining the winning policy for each individual core. When extended to multi-core, the sampler sets for a given core, referred to as the {\em owner core}, are shared with other {\em follower cores} that will use them as followers of their respective
(and potentially different) policies.
Thus, the cache policy for each core seeks to minimize the {\em total misses across all applications.}
Note that a core may select a policy which may not work best for its own application but reduces overall misses.

\noindenttitle{Microarchitectural extensions:} 
For a multi-core setup, LDPT is implemented as a per-core private structure. Thus, when a core initiates a memory instruction, LDPT that is private to the core is accessed using the PC of the memory instruction.
As with single-core implementation, Leeway requires two saturating counters per core (one each for Bypass- and Reuse-Oriented policies) for tracking aggregate misses in a sampling interval.

\section{Methodology\label{sec:leeway:methodology}}

\subsection{Workloads and Simulation Infrastructure}

We evaluate the performance of 
SPEC CPU 2006 applications using a modified version of CMP\$im~\cite{cmpsim} provided with the JILP Cache Replacement Championship \cite{championship} %
Table~\ref{tab:leeway:spec-method} summarizes the features of the simulated processor.

For each SPEC application, we use {\em SimPoint} \cite{simpoints} to identify up to six simpoints of one billion instructions each representing a different phase of an application. 
We use SimPoint tool to generate the weights for each simpoint that are then used to calculate the overall performance. Each program is run with the first {\em ref} input provided by {\em runspec} command. For each run, the simpoint is used to warm microarchitectural structures for 200M instructions, then it measures and reports the result for the subsequent one billion instructions. The result reported for each benchmark is the weighted average of the results for the individual simpoints. 

\begin{table}[!t]
\centering
\small
\begin{tabular}{|l||c|}
    \hline
    \multirow{1}{*}{Core Model}             &   OoO: 4-wide pipeline, 128-entry ROB\\
    \hline
    \multirow{1}{*}{L1 Caches}              &   Private, Split, 8-ways 32KB\\
    \hline
    \multirow{1}{*}{L2 Cache}               &   Private, Unified, 8-ways 256KB\\
    \hline
    \multirow{2}{*}{L3 Cache}               &   Shared, Unified, 16-ways 2MB per core\\
                                            &   Non-Inclusive Non-Exclusive \\
    \hline
    \multirow{1}{*}{Memory}             &   200-cycle access latency\\
    \hline
\end{tabular}
\caption{System parameters for simulations.}
\label{tab:leeway:spec-method}
\end{table}

For multi-core applications, we use 100 multi-programmed mixes, with each individual application for a mix is randomly selected from 23 (of 29) SPEC applications whose performance is sensitive to cache replacement decisions. 
For each application in the mix, we use the highest weighted simpoint. Each mix is run on a quad-core system for 1 billion instructions following a warmup of 200 million instructions. Applications which finish before others are restarted to maintain the cache pressure until the slowest one has finished. We report the weighted speed-up over LRU. To compute it, we run every application in isolation with 8MB LLC under LRU to calculate $SingleIPC_i$. We then calculate Weighted IPC as $\sum_{i=1}^{N} (IPC_i$ / $SingleIPC_i$), where $IPC_i$ is the application's IPC in the presence of~co-runners.

\subsection{Evaluated Cache Management Techniques}

\noindenttitle{RRIP}~\cite{rrip} is the state-of-the-art lightweight dynamic technique that does not depend on history-based learning. We implement RRIP based on the source code from the cache replacement championship~\cite{championship} for RRIP.

\noindenttitle{Sampling Dead Block Predictor} (SDBP)~\cite{sampler} is a dead block predictor that correlates ``last touch'' to the block with the PC of the memory instruction making the touch. We use source code from the cache replacement championship~\cite{championship} for SDBP. We use default
settings provided for SPEC workloads except for increasing the number of sampler sets from 32 to 128.

\noindenttitle{Signature-based Hit Predictor} (SHiP)~\cite{ship} is an insertion policy which builds on RRIP~\cite{rrip}. It learns and records whether a block is re-referenced after insertion and uses this information to guide insertion placement. We implement SHiP with 2-bit RRIP as a baseline technique and 14-bit PC signature. Each predictor table entry contains a 3-bit saturating counter which is updated by the 128 sampled sets.

\noindenttitle{Hawkeye}~\cite{hawkeye} learns a block's behavior by simulating Belady's optimal algorithm~\cite{opt} and trains the predictor that, on each cache access, updates the block's eviction priority. The authors kindly provided the source code of their technique, which we use for the evaluation.

\noindenttitle{Leeway:} For learning, Leeway uses 64 sets per core for each policy. Leeway uses set-dueling to find the preferred policy (Sec.~\ref{sec:leeway:mechanism:setdueling}).
Miss counters are sampled every 200M instructions or 100K cache accesses in the sampler sets, whichever occurs first. The LDPT has 16K entries per core. Finally, for the configurations that enable data prefetchers in the higher-level caches, Leeway always uses Bypass-Oriented Policy for the cache blocks inserted by prefetch requests. Leeway implementations are referred to as Leeway-LRU or Dynamic Leeway-LRU for LRU-based implementations and Leeway-NRU or Dynamic Leeway-LRU for NRU-based implementations. Leeway-NRU uses 2-bit NRU as the base technique, unless specified otherwise.

\section{Evaluation\label{sec:leeway:eval-spec}}
In this section, we evaluate Leeway and state-of-the-art cache management techniques on four different machine configurations -- single-core with data prefetchers off, single-core with data prefetchers on, quad-core with data prefetchers off and quad-core with data prefetchers on. We first provide average speed-ups for all techniques for each configuration. Next, we analyze performance for both quad-core configurations in Sec.~\ref{sec:leeway:eval-quad-core}, followed by a detailed analysis for a single-core configuration in Sec.~\ref{sec:leeway:eval-single-core}.

Fig.~\ref{fig:leeway:perf-summary} shows average speed-up for SPEC applications on all four deployment configurations. 
For each configuration, the speed-up is reported over the baseline implementing LRU-managed cache on the same configuration. While we below discuss the speed-up for different techniques on each configuration, it is worth noting that the baseline configurations with data prefetchers by themselves outperform the respective configuration without the data prefetchers for LRU, 39.1\% for single-core and 33.0\% for multi-core, which is not shown in this figure.

When data perfetchers are off, both Leeway implementations achieve good performance for both single-core and quad-core configurations. On a single-core configuration, Leeway-LRU and Leeway-NRU both yield an average speed-up of 6.5\% over LRU vs 3.9\% for RRIP, 4.3\% for SDBP, 4.5\% for SHiP and 6.4\% for Hawkeye. On a quad-core configuration, Leeway-LRU and Leeway-NRU yield an average speed-up of 7.5\% and 8.0\%, respectively, vs 4.0\% for RRIP, 6.9\% for SDBP, 8.0\% for SHiP and 9.7\% for Hawkeye.

When the data perfetchers in the higher-level caches are on, average speed-ups for prior techniques significantly drops whereas both Leeway implementations continue to achieve good performance. On a single-core configuration, Leeway-LRU and Leeway-NRU yield an average speed-up of 4.5\% and 4.8\%, respectively, vs 1.9\% for RRIP, 1.0\% for SDBP, 2.1\% for SHiP and 1.7\% for Hawkeye. Similarly, on a quad-core configuration, Leeway-LRU and Leeway-NRU outperform prior techniques with an average speed-up of 7.7\% and 7.8\% over LRU, respectively, vs 2.7\% for RRIP, 4.1\% for SDBP, 4.8\% for SHiP and 0.8\% for Hawkeye. Note that Hawkeye, which provides the highest average performance among prior techniques in the absence of data prefetchers, is among the least effective techniques in the presence of data prefetchers.

\begin{figure}[!t]
    \centering
    \includegraphics{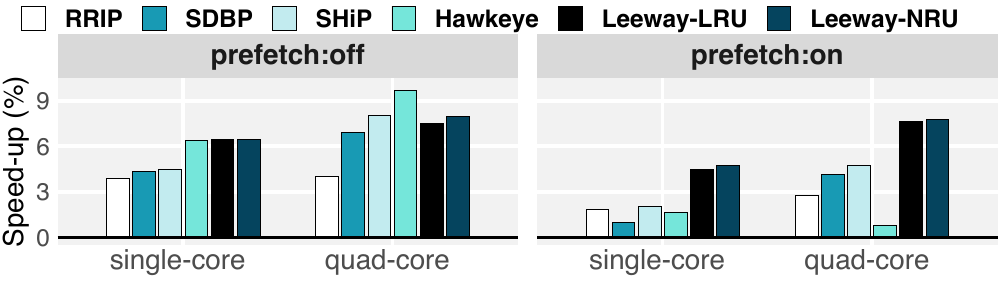}
    \caption{Average speed-up for SPEC applications on four machine configurations.}
    \label{fig:leeway:perf-summary}
\end{figure}

A quad-core configuration with data prefetchers is the most representative of a real-world deployment scenario. The performance trend on this configuration shows that history-based predictive techniques (except for Hawkeye) outperform RRIP (state-of-the-art lightweight dynamic technique) and LRU (a recency-friendly static technique), corroborating prior works~\cite{sampler, ship}. Surprisingly, Hawkeye provides the least performance improvements, which is a new result as the prior work evaluated Hawkeye in the absence of data prefetchers~\cite{hawkeye}. 

\subsection{Performance on Quad-Core Configurations\label{sec:leeway:eval-quad-core}}
In this section, we evaluate the effectiveness of Leeway-NRU and three state-of-the-art history-based predictive techniques (SDBP, SHiP and Hawkeye) for both quad-core configurations. 
We omit the results for RRIP and Leeway-LRU from the subsequent studies for brevity.

\noindenttitle{In the absence of prefetchers}, all techniques provide similar average speed-up, with SDBP providing the lowest (6.9\%) and Hawkeye providing the highest (9.7\%) average speed-up as shown in Fig.~\ref{fig:leeway:quad-core-prefetch-off}.
Hawkeye's effectiveness can be attributed to its learning mechanism. Like other techniques, Hawkeye also relies on a PC-based reuse correlation. However, unlike other techniques, Hawkeye's learning mechanism simulates optimal replacement on past LLC accesses, and thus provides more accurate reuse predictions.

\noindenttitle{In the presence of prefetchers}, variability in the reuse behavior of cache blocks increases as prefetchers speculatively load cache blocks in the higher-level caches, some of which are bound to be inaccurate, leading to extra LLC accesses that would not have occurred in the absence of prefetchers. As shown in Fig.~\ref{fig:leeway:quad-core-prefetch-on}, Leeway-NRU is the most effective in tolerating prefetcher-induced variability by yielding an average speed-up of 7.8\% over LRU. In comparison, SDBP and SHiP yield an average speed-up of 4.1\% and 4.8\% respectively. Hawkeye provides the least performance with an average speed-up of 0.8\%, in stark contrast to its performance without the prefetchers.

\begin{figure}[!t]
    \centering
    \includegraphics[width=1\linewidth]{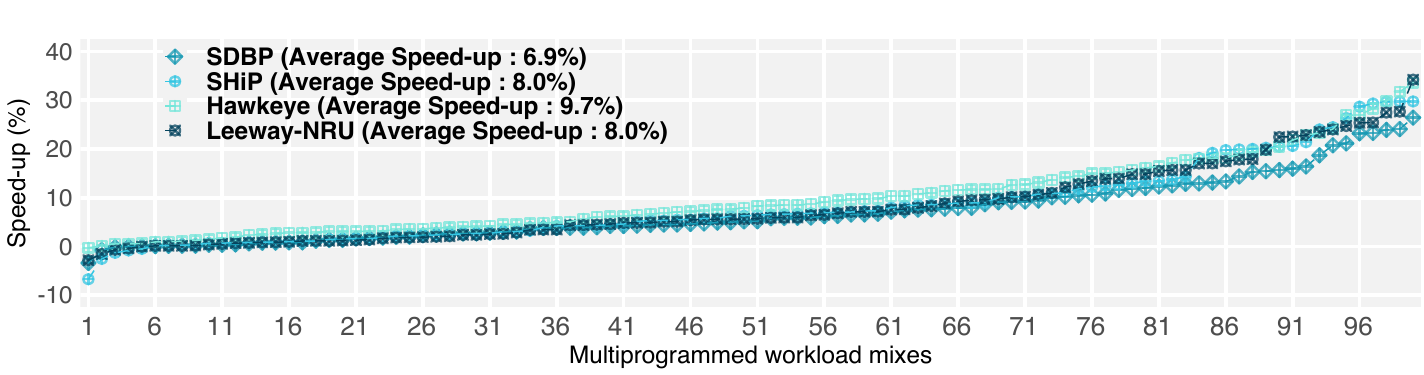}
    \caption{Weighted speed-up for multi-programmed SPEC mixes when prefetchers are off. The speed-ups for mixes are sorted for each technique~individually.}
    \label{fig:leeway:quad-core-prefetch-off}
\end{figure}
\begin{figure}[!t]
    \centering
    \includegraphics[width=1\linewidth]{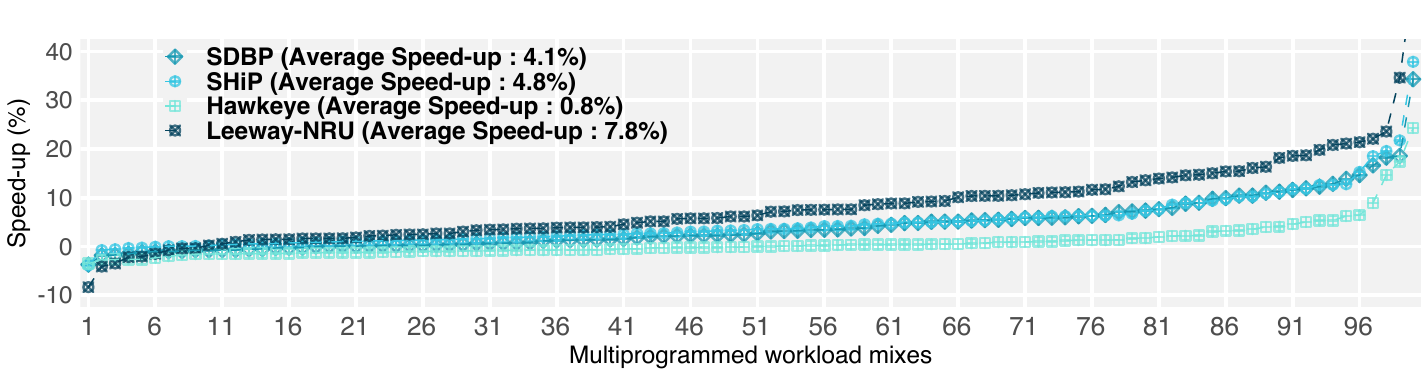}
    \caption{Weighted speed-up on multi-programmed SPEC mixes when prefetchers are on. The speed-ups for mixes are sorted for each technique~individually.}
    \label{fig:leeway:quad-core-prefetch-on}
\end{figure}

When compared to the prior techniques, Leeway-NRU achieves an average speed-up of 3.5\% over SDBP, 2.9\% over SHiP, 6.9\% over Hawkeye and 7.8\% over LRU. Of the 100 evaluated mixes, on 78 mixes Leeway-NRU provides higher performance than any of the prior techniques, while outperforming SDBP on 85 mixes, SHiP on 79 mixes and Hawkeye on 93 mixes.

\subsection{Performance Analysis on a Single-Core Configuration\label{sec:leeway:eval-single-core}}
In this section, We provide a detailed performance analysis of various techniques for a single-core configuration with data prefetchers off as this configuration has the minimum noise in access patterns. In other configurations, the reuse behavior of cache blocks is significantly affected by prefetchers or cache pressure from the co-located workloads sharing LLC.

To better understand the effects of all cache management techniques, we classify SPEC applications into three categories:
(1) {\em High Opportunity}, if performance improves by at least 10\% over LRU with any one technique; (2) {\em No Opportunity} if performance doesn't vary by more than 0.5\% for all techniques; (3) {\em Mix Opportunity} for the rest.

\noindenttitle{High opportunity applications: }
\begin{figure}[!t]
{
    \centering
    \subfloat[Miss Reduction over LRU]{\label{fig:leeway:high-miss}{\includegraphics{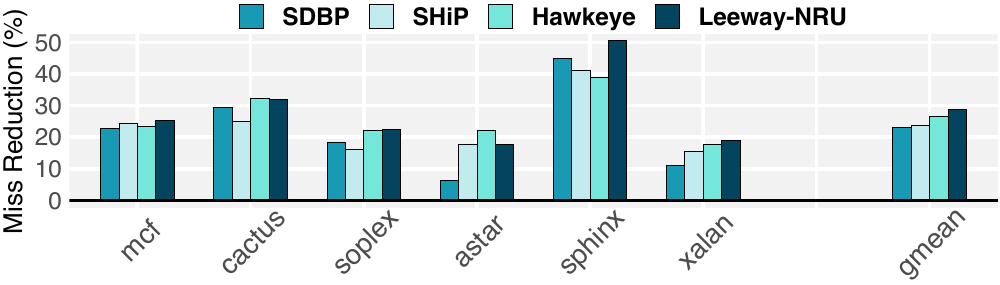}}} \vfill
    \subfloat[Speed-up over LRU]{\label{fig:leeway:high-perf}{\includegraphics{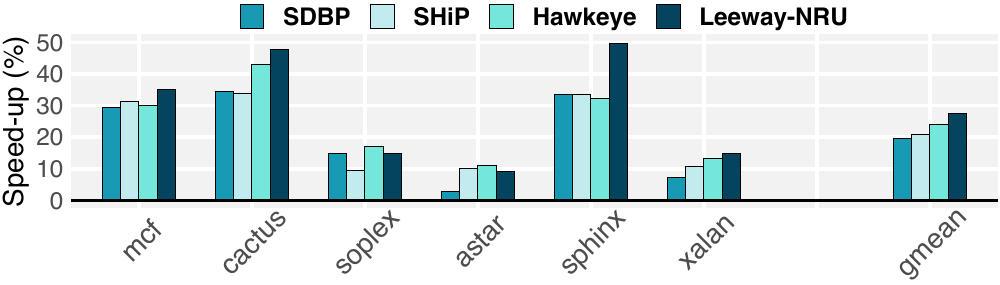}}}
    \caption{Evaluation of various cache management techniques for the {\em High Opportunity} SPEC CPU 2006 applications. Name of some applications are shortened as follows: cactus for cactusADM, sphinx for sphinx3 and xalan for xalancbmk.\label{fig:leeway:high}
    }
}
\end{figure}
Fig.~\stitchref{fig:leeway:high}{fig:leeway:high-miss} shows the reduction in LLC misses and Fig.~\stitchref{fig:leeway:high}{fig:leeway:high-perf} shows the improvement in performance compared to the baseline LRU for the high opportunity applications.
Overall all techniques are highly effective on these applications with Leeway-NRU reducing the most misses on average (28.9\% over LRU), vs 23.2\% for SDBP, 23.9\% for SHiP and 26.5\% for Hawkeye.
The performance of all techniques generally correlate well with the miss reduction, with Leeway-NRU achieving the highest average speed-up (27.6\% over LRU), vs 19.7\% for SDBP, 21.0\% for SHiP and 24.0\% for Hawkeye.

\noindenttitle{Mix opportunity applications: }
\begin{figure}[!t]
{
    \centering
    \subfloat[Miss Reduction over LRU]{\label{fig:leeway:mix-miss}{\includegraphics[width=1\linewidth]{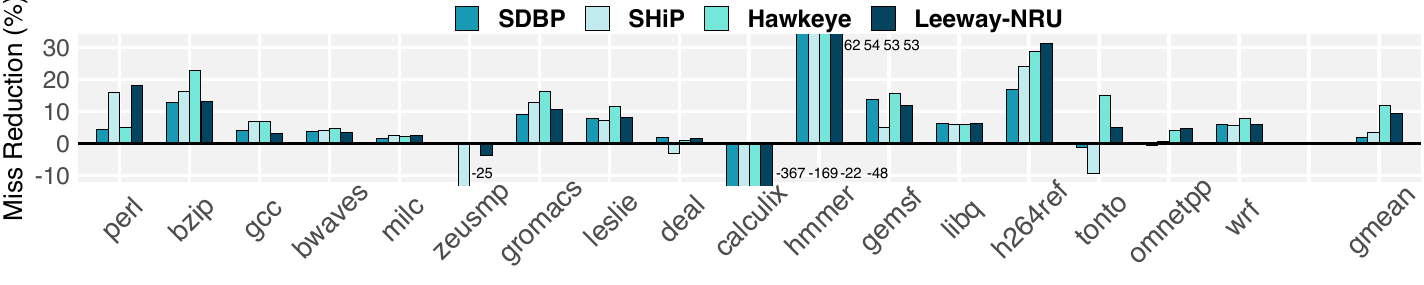}}} \vfill
    \subfloat[Speed-up over LRU]{\label{fig:leeway:mix-perf}{\includegraphics[width=1\linewidth]{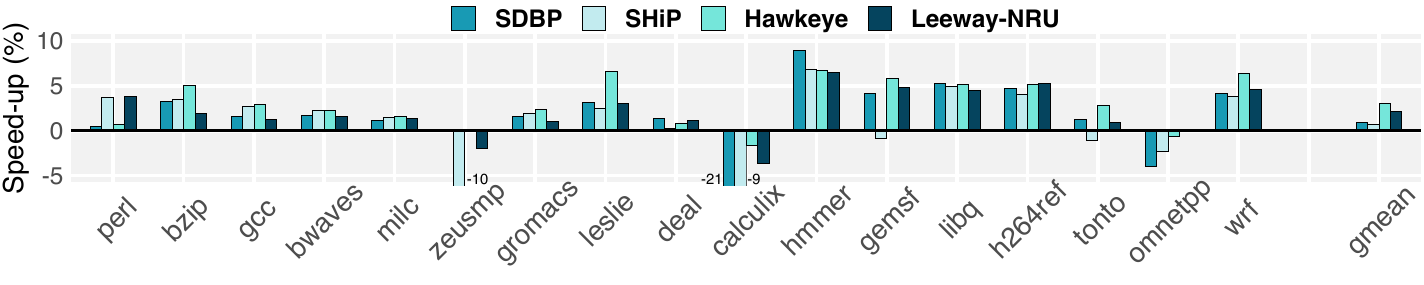}}}
    \caption{Evaluation of various cache management techniques for the {\em Mix Opportunity} SPEC CPU 2006 applications. Name of some applications are shortened as follows: perl for perlbench, bzip for bzip2, leslie for leslie3d, deal for dealII, gemsf for GemsFDTD and libq for libquantum.\label{fig:leeway:mix}
    }
}
\end{figure}
Fig.~\stitchref{fig:leeway:mix}{fig:leeway:mix-miss} shows the reduction in LLC misses and Fig.~\stitchref{fig:leeway:mix}{fig:leeway:mix-perf} shows the improvement in performance compared to the baseline LRU for the mix opportunity applications.
Overall, Hawkeye and Leeway-NRU are far more effective than SDBP and SHiP on the mix opportunity applications with 12.0\% average miss reduction for Hawkeye and 9.5\% for Leeway-NRU vs only 2.0\% for SDBP and 3.4\% for SHiP. 

For four applications ({\em zeusmp}, {\em calculix}, {\em tonto} and {\em omnetpp}), at least one of the techniques incurs more misses than the baseline LRU. For two of these applications, Leeway-NRU also increases misses, but the miss reduction is relatively small. For example, on {\em zeusmp}, Leeway-NRU increases misses by 3.7\% vs 25.5\% for SHiP. Similarly, on {\em calculix}, Leeway-NRU increases misses by 47.7\% vs 366.6\% for SDBP and	168.7\% for SHiP. On {\em tonto} and {\em omnetpp}, SDBP and SHiP increase misses (1\%-9\%) whereas Leeway-NRU manages to {\em reduce} misses (4\%-5\%) over LRU.

The performance of all techniques generally correlate well with the miss reduction with Hawkeye and Leeway-NRU achieving an average speed-up of 3.0\% and 2.1\%, respectively vs 0.9\% for SDBP and 0.7\% for SHiP. Leeway-NRU slows down the fewest applications, {\em zeusmp} and {\em calculix},  with the maximum slowdown of 3.6\%. In comparison, SDBP slows down 3 applications (max slowdown of 20.6\%), SHiP slows down 5 applications (max slowdown of 10.3\%) and Hawkeye slows down 3 applications (max slowdown of 1.7\%).

\noindenttitle{No opportunity applications: }
\begin{figure}[!t]
{
    \centering
    \subfloat[Miss Reduction over LRU]{\label{fig:leeway:no-miss}{\includegraphics{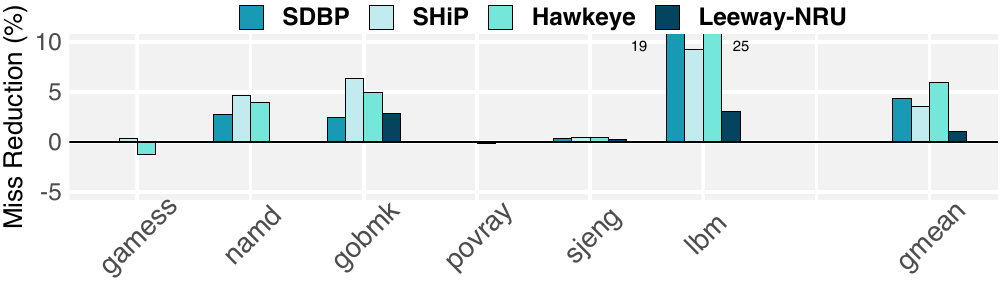}}} \vfill
    \subfloat[Speed-up over LRU]{\label{fig:leeway:no-perf}{\includegraphics{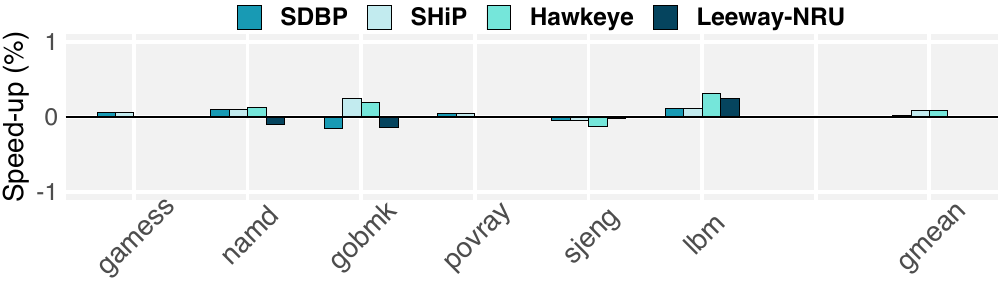}}}
    \caption{Evaluation of various cache management techniques for the {\em No Opportunity} SPEC CPU 2006 applications.\label{fig:leeway:no}
    }
}
\end{figure}
Fig.~\stitchref{fig:leeway:no}{fig:leeway:no-miss} shows the reduction in LLC misses and Fig.~\stitchref{fig:leeway:no}{fig:leeway:no-perf} shows the improvement in performance compared to the baseline LRU for the no opportunity applications. An average miss reduction for all techniques range between 1\%-6\%. However, the performance for these applications is not sensitive to replacement decisions and the change in performance due to any technique is at most 0.5\% over LRU.

\subsection{Dissecting Performance of Hawkeye}
Hawkeye's learning mechanism simulates optimal replacement (OPT) on past LLC accesses, unlike Leeway (as well as SDBP and SHiP) that relies on baseline LRU or NRU for learning. Thus, Hawkeye, in theory, can provide more accurate reuse predictions.
For example, between two cache blocks, each having a reuse distance greater than the associativity, OPT can identify a cache block having a smaller reuse distance accurately, in contrast to LRU-like techniques. Thus, Hawkeye is more likely to retain a cache block with a smaller reuse distance in the presence of thrashing than Leeway.

To quantitatively support this hypothesis, we study prediction coverage and accuracy for Hawkeye and Leeway-NRU. Coverage is measured as a percentage of total evictions that are predicted dead by a cache management technique. Accuracy is measured as a percentage of predicted evictions that are correct.~\footnote{While comparing coverage and accuracy of different techniques, it should be noted that both are self normalized metrics; if the total evictions under two techniques are significantly different for the same application, analyzing coverage and accuracy metrics in isolation may lead to wrong conclusions.}

Table~\ref{tab:leeway:coverage} shows prediction coverage and accuracy for Hawkeye and Leeway-NRU, averaged across SPEC applications (excluding the no opportunity applications). 
Hawkeye' prediction coverage is nearly the same as Leeway-NRU. However, Hawkeye has a higher prediction accuracy (78.4\% vs 72.3\% for Leeway-NRU), thanks to the OPT-based learning.

In the presence of data prefetch, however, effectiveness of Hawkeye reduces significantly. Amidst the prefetcher-induced variability, Hawkeye takes a conservative approach and makes far less predictions, reducing the opportunity to evict dead blocks. Prediction coverage for Hawkeye averages 71.2\% (vs 80.3\% without prefetch) and accuracy also drops to 74.3\% (vs 78.4\% without prefetch), explaining Hawkeye's poor performance in the presence of data prefetchers.

\begin{table}[!t]
    \centering
    \small
    \begin{tabular}{|c||c|c|}
    \hline   
    Single-core Configuration  & Coverage  & Accuracy \\ \hline   
    \hline   
    Hawkeye &   80.3\%  &   78.4\%  \\ \hline   
    Leeway-NRU & 82.8\% & 	72.3\% \\ \hline
    \end{tabular}
    \caption{Prediction coverage and accuracy, averaged across SPEC applications (excluding the no opportunity applications) on a single-core configuration in the absence of data prefetchers.}
    \label{tab:leeway:coverage}
\end{table}

\subsection{Adaptivity of Leeway\label{sec:leeway:eval-adapt}}

\noindenttitle{Reuse-aware update policies:} To understand the effect of Leeway's policy choice, we compare the performance of individual static policies (Bypass- and Reuse-Oriented) with an adaptive scheme (Dynamic Leeway or simply Leeway) that dynamically chooses one of the static policies at runtime (Sec.~\ref{sec:leeway:adapt}). Dynamic Leeway was used throughout the evaluation. Fig.~\ref{fig:leeway:adapt} presents the results for SPEC applications on a single-core configuration without data prefetcher. 
No opportunity applications are not shown for clarity.

Applications benefiting more from the Bypass-Oriented Policy (BOP) are shown in the Fig.~\ref{fig:leeway:adapt}. Such applications include four of the six high opportunity applications (left group of Fig.~\stitchref{fig:leeway:adapt}{fig:leeway:adapt:bop}) and several mixed opportunity ones (left group of Fig.~\stitchref{fig:leeway:adapt}{fig:leeway:adapt:rop}).
For these applications, the access pattern is dominated by bypassable blocks. For example, for these applications, on average, only 7.7\% (max 26.3\% for {\em deal}) of blocks inserted in the cache incur at least one hit under the OPT replacement policy. The Reuse-Oriented Policy conservatively increases the live distance in the face of variability. Predicting high live distance for such blocks only contributes in increasing the dead time, which, in turn, lowers the cache efficiency. 

\begin{figure}[!t]
{
    \centering
    \subfloat[Speed-up over LRU for High Opportunity Applications]{\label{fig:leeway:adapt:bop}{\includegraphics[width=1\linewidth]{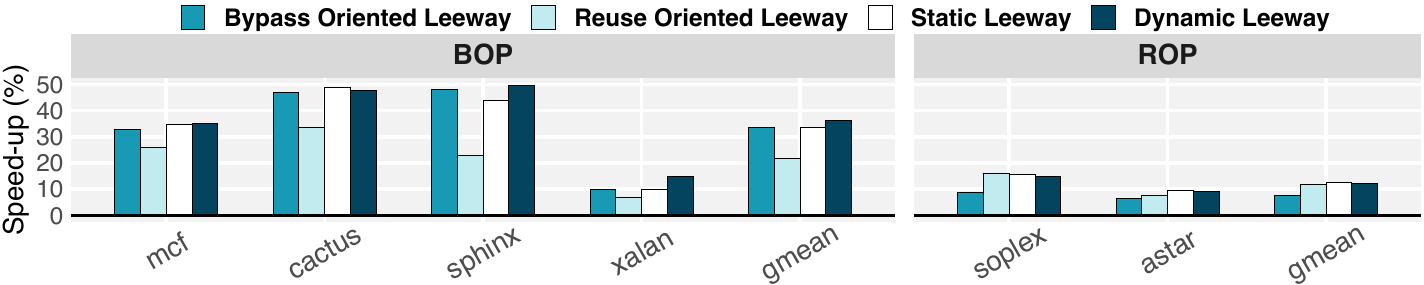}}} \vfill
    \subfloat[Speed-up over LRU for Mix Opportunity Applications]{\label{fig:leeway:adapt:rop}{\includegraphics[width=1\linewidth]{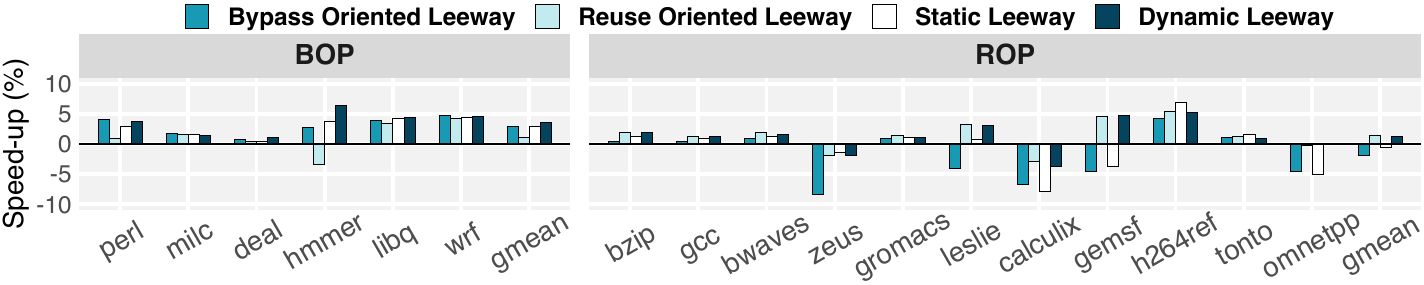}}}
    \caption{Evaluation of various Leeway-NRU configurations (all using 2-bit NRU as the base policy).\label{fig:leeway:adapt}
    }
}
\end{figure}

Right side of Fig.~\stitchref{fig:leeway:adapt}{fig:leeway:adapt:bop} and Fig.~\stitchref{fig:leeway:adapt}{fig:leeway:adapt:rop} respectively show two high opportunity applications and several mixed opportunity applications that benefit more from the Reuse-Oriented Policy (ROP).
For most of these applications, none of the techniques are very effective. The culprit is high incidence of blocks with reuse and inter-generational variability.
For example, for these applications, on average, 33.9\% (max 74.7\% for {\em tonto}) of blocks inserted in the cache incur at least one hit under the OPT replacement policy.
In the case of Leeway, the Reuse-Oriented policy generally proves beneficial by steering the live distance toward the recently-observed maximum in order to boost the opportunity for reuse. For instance, this proves particularly beneficial on {\em omnetpp}, on which Leeway-NRU is the only technique to avoid a slowdown (see Fig.~\stitchref{fig:leeway:mix}{fig:leeway:mix-perf}).

To understand how BOP and ROP makes predictions, we compare the coverage and accuracy for both Static BOP and Static ROP policies. Fig.~\ref{fig:leeway:coverage} shows prediction coverage and accuracy, averaged across all SPEC applications (excluding the no opportunity ones). The figure also shows data for {\em mcf}, which prefers BOP and {\em calculix}, which prefers ROP as representative examples. On {\em mcf}, ROP reduces coverage to 86.5\% from 99.5\% for BOP. However, that only marginally increases accuracy to 96.1\% from 95.5\% for BOP. The end result is the loss of opportunity for ROP in making predictions (indicated by low coverage), which hurts performance. On {\em calculix}, ROP reduces coverage to 92.4\% from 97.9\% for BOP. However, that significantly increases accuracy to 64.5\% from 46.7\% for BOP, providing higher performance for ROP. The results show that BOP, in general, trades coverage for accuracy, which is beneficial for applications that are dominated by bypassable blocks as likelihood of making wrong prediction is already low to begin with. In contrast, ROP trades accuracy for coverage, which is beneficial for applications that exhibit significant amount of inter-generational~variability.

\begin{figure}[!t]
    \centering
    \includegraphics{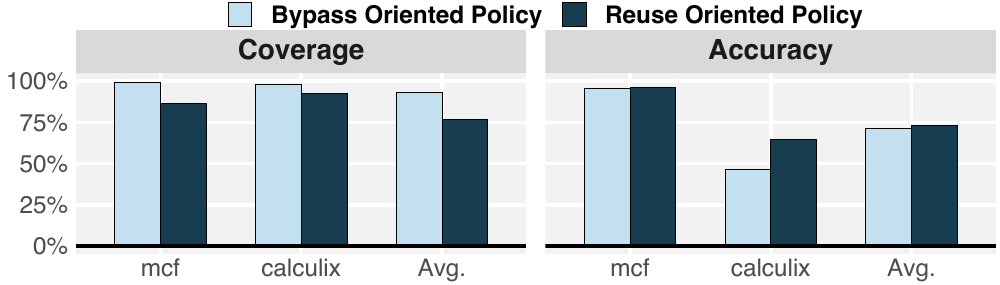}
    \caption{Prediction Coverage and Accuracy for Leeway-NRU static policies.}
    \label{fig:leeway:coverage}
\end{figure}

Finally, we show that at runtime, Dynamic Leeway generally selects the policy that is most suited for a given application. Recall Fig.~\ref{fig:leeway:adapt}, which shows that Dynamic Leeway effectively selects between two static policies, ROP and BOP, for all applications with Dynamic Leeway matching the performance of the best performing static policy.
Moreover, Leeway can adapt to phase behavior within a single application, as demonstrated on three applications ({\em mcf}, {\em hmmer} and {\em xalan}) that have distinct cache behavior across phases. On these applications, dynamic Leeway outperforms the best static policy by over 2\%.

\noindenttitle{Reuse-unaware static Leeway:} To isolate the performance due to dead block predictions using live distance as a metric from the reuse-aware dynamic update policies, we evaluate {\em Static Leeway-NRU}. Static Leeway-NRU employs a static VTT value of 7 in both directions, and thus does not require set-dueling for policy selection, requiring only 32K of total storage (vs 44KB for Dynamic Leeway-NRU).

Fig.~\ref{fig:leeway:adapt} shows the performance for Static Leeway on SPEC applications. Overall, Static Leeway provides an average speed-up of 5.3\%; however, due to its reuse-unaware design, it underperforms the dynamic Leeway-NRU (6.5\%) for almost all applications, thus justifying the additional storage cost in LDPT for the Dynamic Leeway design.

\subsection{Sensitivity of Leeway-NRU on Number of NRU Bits}
In this section, we evaluate sensitivity of performance for Leeway-NRU on the number of bits used by the baseline NRU technique. So far, we have used Leeway-NRU with 2-bits per cache block. Fig.~\ref{fig:leeway:nru-bits} shows average speed-up for Leeway-NRU (1-4 bits per cache block) for all four different configurations. The figure also shows performance for Leeway-LRU as reference. Overall, Leeway-NRU (2b), which was used throughout the evaluation, consistently provides good performance across the configurations.

Leeway-LRU uses LRU as the baseline technique, which maintains precise recency state for the cache blocks in a set. However, this is largely beneficial only for applications that benefit more from Reuse-Oriented Policy. For example, on a single-core configuration in the absence of data perfetchers, across applications that benefit more from ROP, Leeway-LRU provides 3.2\% average speed-up vs 2.8\% for the best performing Leeway-NRU. Meanwhile, for applications that benefit more from BOP, Leeway-LRU achieves an average speed-up of 10.1\% vs 10.9\% for the best performing Leeway-NRU. As explained in Sec.~\ref{sec:leeway:eval-adapt}, for applications that benefit more from BOP, are dominated by bypassable blocks. For these applications, maintaining precise recency state is not required (and sometimes counterproductive) as live distance for most of the blocks is zero.

\begin{figure}[!t]
    \centering
    \includegraphics{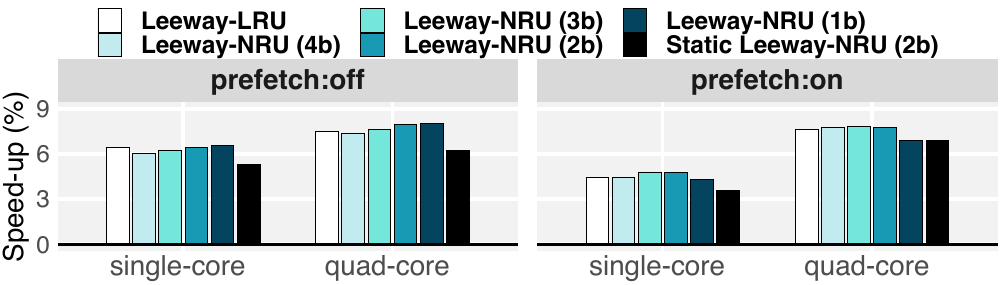}
    \caption{speed-up for various Leeway configurations.}
    \label{fig:leeway:nru-bits}
\end{figure}

\subsection{Measuring the Number of History Table Look-Ups\label{sec:leeway:cost-complexity-eval}}
\begin{table}[!t]
    \centering
    \small
    \begin{tabular}{|c||c|c|c|c|}
    \hline   
    Technique  & SDBP  & SHiP & Hawkeye & Leeway-NRU \\ \hline   
    Table Lookups    & 2.5$x$  &  1.1$x$  &   2.3$x$  & 1.0$x$\\ \hline   
    \end{tabular}
    \caption{History table look-ups, normalized to Leeway-NRU, averaged over SPEC applications (excluding no opportunity ones).}
    \label{tab:leeway:lookups}
\end{table}
As explained in Sec.~\ref{sec:leeway:cost-complexity}, prior techniques such as Hawkeye access their history tables on every cache access, increasing on-chip traffic. Table~\ref{tab:leeway:lookups} compares the number of table look-ups across techniques. Overall, SDBP and Hawkeye require 2.3 and 2.5 times the table look-ups when compared to that of Leeway-NRU. Also note that almost half the number of look-ups for SDBP and Hawkeye are during cache hits, and thus are on the critical path. In contrast, Leeway not just requires significantly fewer table look-ups, but also performs all these look-ups only during cache misses, which are off the critical path.

\subsection{Reducing Storage Cost for Leeway}

\begin{table}[!b]
    \centering
    \small
    \begin{tabular}{|c|c|c|}
        \hline
        Leeway-NRU  & Metadata  & Avg. Speed-up \\ 
        Implementation    &  Storage & (Core:Quad-Prefetch:On)\\ \hline
         \hline
        \rowcolor{lightgray} Dynamic Leeway-NRU (2-bits) &  44KB & 7.8\% \\ \hline
        Dynamic Leeway-NRU (1-bit) & 32KB & 6.9\% \\ \hline
        Static Leeway-NRU (2-bits) & 32KB & 7.0\% \\ \hline
    \end{tabular}
    \caption{Per-core storage cost (assuming 2MB of LLC) for different Leeway-NRU implementations. Fig.~\ref{fig:leeway:nru-bits} shows the performance for the other configurations.}
    \label{tab:leeway:reduce-storage}
\end{table}

Leeway trades more storage for fewer table look-ups by embedding prediction metadata in the cache, because of which, Leeway-NRU requires slightly more storage than the prior techniques as shown in Sec.~\ref{sec:leeway:cost-complexity}. While the storage requirement is relatively small in comparison to the LLC, there is some room for reducing storage for Leeway by changing the Leeway-NRU configuration as follows: (1) By reducing NRU bits from 2 to 1, storage requirement for Leeway-NRU drops from 44KB to 32KB. (2) Other option is to use Static (\ie Reuse-unaware) version of Leeway-NRU, which reduces the storage requirement of LDPT by half, reducing the total storage from 44KB to 32KB. However, those configurations also lead to lower performance as shown in Table~\ref{tab:leeway:reduce-storage}.

\section{Evaluation of Concurrent Techniques\label{sec:leeway:crc2-evel}}

In this section, we provide an evaluation summary of concurrent techniques submitted to the Cache Replacement Championship (CRC2)~\cite{crc2}. Each competing technique was allowed to utilize the maximum storage of 32KB.
We evaluate top five ranked techniques -- LIME~\cite{crc2-2lime}, MPP~\cite{crc2-3multi}, RED~\cite{crc2-7red}, SHiP++~\cite{crc2-1shippp} and Hawkeye++~\cite{crc2-6hawkeye} -- from the fifteen techniques competed in CRC2. 
SHiP++ and Hawkeye++ are improved, prefetch-aware, implementations of SHiP~\cite{ship} and Hawkeye~\cite{hawkeye}, respectively.
We evaluate techniques using the methodology used in CRC2, which is very similar to the methodology used in the evaluation so far (Sec.~\ref{sec:leeway:methodology}) except for two major differences as follows: (1) CRC2 uses the ChampSim~\cite{champsim} cycle-accurate simulator instead of CMP\$im~\cite{cmpsim}. (2) CRC2 evaluates five Cloudsuite~\cite{cloudsuite} applications -- {\em media streaming, web search, software testing, data serving, map reduce} -- as a representative benchmark suite for the server applications executing in the data-centers, in addition to single-core SPEC CPU 2006 applications and quad-core multi-programmed SPEC applications.

We note that the implementation of Leeway-NRU submitted in CRC2 had a bug, because of which, live distance values were not read correctly from the LDPT. We use the updated version from \url{https://github.com/faldupriyank.com/leeway} in this evaluation. This implementation is identical to the Dynamic Leeway-NRU implementation evaluated so far, except that we reduced the number of LDPT entries to bring the total storage under 32KB.

\begin{figure}[!t]
    \centering
    \includegraphics{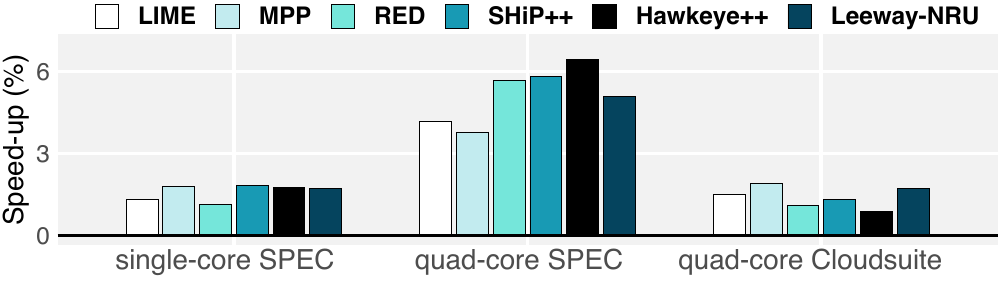}
    \caption{Average speed-up for three benchmark suites -- single-core and quad-core multi-programmed SPEC applications, and quad-core Cloudsuite applications.}
    \label{fig:leeway:crc2-perf-summary}
\end{figure}

Fig.~\ref{fig:leeway:crc2-perf-summary} shows an average speed-up over LRU for all three application benchmark suites with data prefetchers kept on for all simulations. On single-core SPEC application benchmarks, Leeway provides an average speed-up of 1.7\% vs 1.8\% for the best performing techniques (MPP, SHiP++ and Hawkeye++). Leeway achieves a higher average speed-up than LIME (1.3\%) and RED (1.2\%).

On quad-core SPEC benchmarks, Leeway yields an average speed-up of 5.1\% over LRU vs 6.4\% for Hawkeye++, the best performing technique. Leeway achieves a higher average speed-up than LIME (4.2\%) and MPP (3.8\%).

Finally, on Cloudsuite applications, Leeway achieves an average speed-up of 1.7\% vs 1.9\% for MPP, the best performing technique. Leeway achieves a higher average speed-up than all but MPP while Hawkeye++ achieves the least average speed-up~(0.9\%).

The results show that Leeway consistently provides good performance across the benchmark suites, which is on par with the best performing concurrent techniques. While all these techniques utilize the same storage of 32KB for predictions, note that the results do not factor in the hardware complexities. For example, Hawkeye++, the winner of the CRC2, has fundamentally similar design as Hawkeye, and thus requires history table look-ups on every cache access. Recall from Table~\ref{tab:leeway:lookups} that Hawkeye requires significantly higher number of history table look-ups than Leeway-NRU. Moreover, about half of the history table look-ups for Hawkeye are on cache hits and thus on the critical path. In contrast, Leeway look-ups are exclusively on cache misses, thus are off the critical path, making Leeway more attractive from the implementation point of view.

\section{Related Work}
\label{sec:related}

Duong \etal introduced a DBP based on the notion of Protected Distance (PD)~\cite{pd}. PD leverages reuse distance, an indirect metric that counts non-unique references to a set. A single PD is used for an entire application. If a block is not referenced beyond the application's PD, it is predicted dead. 
While conceptually PD sounds similar to Leeway, 
Leeway has two key advantages over PD. First, PD maintains a single Protected Distance for an entire application, whereas Leeway maintains a Live Distance per PC that is continuously trained throughout the application's execution. This maximizes Leeway's adaptivity while minimizing dead time of blocks prior to prediction. Secondly, Live Distance relies on stack distance, and thus naturally ``filters'' non-unique references to the set. In contrast, PD counts all references to the set, which can inflate PD values and lead to increased dead time for cache blocks. 
Indeed, our evaluation of PD shows that it is generally inferior to both Leeway and other recent cache management techniques. On SPEC, average performance improvement for Leeway-NRU is 6.5\% versus 4.4\% for PD for a single-core configuration without data prefetchers, and 4.8\% versus 1.1\% (in favor of Leeway-NRU) with the prefetchers.%

Others have also suggested using stack distance or reuse distance for cache replacement or modeling~\cite{rd1,rd2,rd-daniel,timekeeping2,pd, reusedist}. Doing so requires maintaining a Reuse Distance Distribution (RDD) for an application, which itself can be storage intensive as it involves keeping separate counter for different reuse distances maintained. Further, turning this RDD into a useful metric is challenging and computationally intensive. For example,~\cite{pd} proposes dedicated compute logic while ~\cite{rd-daniel} relies on a software framework that runs on a core. In contrast, Leeway monitors the readily-available stack position within a set, which is already maintained by the base replacement policy. Deriving a block’s live distance is then as simple as taking the max of observed stack positions upon hits in its lifetime. Thus, live distance fundamentally enables a very efficient hardware implementation within this general class of metrics.

Teran \etal~\cite{perceptron} proposed perceptron learning based predictor for LLC.
Instead of correlating cache block behavior with just a single feature like load-PC, the predictor combines multiple features for predicting block's reuse behavior.
To do so, the predictor maintains a separate predictor table for each feature, for a total of six tables. Each of these predictor tables need to be accessed on every cache access (including hits) which makes this design difficult to scale for multi-core processors as explained in Sec.~\ref{sec:leeway:cost-complexity}. 

Contrary to the traditional recency stack, {\em Pseudo-LIFO}~\cite{pseudo} manages the LLC as a fill stack.
The approach dynamically learns the preferred eviction positions within the fill stack, and prioritizes the blocks close to the top of the stack for eviction. It learns the preferred positions for an application based on the combined behavior of all the cache blocks, lacking fine-granularity adaptation that state-of-the-art approaches, including Leeway, use.

We primarily used DBPs for efficient cache management of LLC. Prior works have proposed DBPs for other use cases.
Lai \etal used dead block prediction to optimize coherence protocol~\cite{lasttouch}. They proposed predicting the last access to a cache block on a core and self-invalidating the block after its last access; consequently, a subsequent access to the same cache block in other core do not incur the invalidation latency, improving the performance for applications dominated by coherence communications.
Lai \etal used DBPs at L1D and used dead blocks as prefetch targets, obviating the need for auxiliary prefetch buffers~\cite{reftrace}. 
Prior works have explored using dead block prediction to dynamically turn-off cache blocks at LLC that are predicted dead to reduce leakage power~\cite{turnoffdead, cachedecay, letcachesdecay}.
Khan \etal used dead block prediction to implement virtual victim cache~\cite{dbpvictimcache}. They used dead blocks to hold blocks evicted from other sets, thus forming a pool of dead blocks as a virtual victim cache.

\section{Conclusion}
\label{sec:conclusion}

In this chapter, we showed that variability in the reuse behavior of cache blocks limits state-of-the-art history-based predictive techniques in achieving high performance.
In response, we argued for variability-tolerant mechanisms and policies for cache management.
As a step in that direction, we proposed Leeway, a history-based predictive technique employing two variability-tolerant features.
First, Leeway introduces a new metric, Live Distance, that captures the largest interval of temporal reuse for a cache block, providing a conservative estimate of a cache block's useful lifetime.
Second, Leeway implements a robust prediction mechanism that identifies dead blocks based on their past Live Distance values. To maximize cache efficiency in the face of variability, Leeway monitors the change in Live Distance values at runtime using its reuse-aware policies to adapt to the observed access patterns.
Meanwhile, Leeway embeds prediction metadata with cache blocks in order to avoid critical path history table look-ups on cache hits and reduce the on-chip network traffic, in contrast to the state-of-the-art techniques that access history table on every cache access (including cache hits).
On a variety of applications and deployment scenarios, Leeway consistently provides good performance that generally matches or exceeds that of state-of-the-art techniques.

\chapter{A Case for Domain-Specialized Cache Management\label{ch:motivation}}

In the previous chapter, we showed that history-based predictive techniques provide significant performance improvement over simple static and lightweight dynamic techniques for a broad range of applications. 
However, these history-based predictive techniques struggle in exploiting the high reuse for certain applications for which variability arises due to fundamental application characteristics. 
In this chapter, we specifically analyze the suitability of domain-agnostic predictive techniques  for the applications from the domain of graph analytics.
We qualitatively and quantitatively explain why these domain-agnostic techniques are fundamentally deficient for an important domain of graph analytics and motivate the need for software-hardware co-design in managing LLC for graph analytics.

The chapter is organized as follows. First, in Sec.~\ref{sec:motivation:skew}, we discuss two important properties of graph datasets that influence cache efficiency. Next, we explain the basics of data-structures used in graph processing, followed by cache access patterns of individual data-structures (Secs.~\ref{sec:motivation:graph-processing-basics} \& \ref{sec:motivation:general-caching}). Finally, we highlight the challenges in improving cache efficiency for graph analytics and discuss the limitations of prior software and hardware techniques in addressing those challenges (Secs.~\ref{sec:motivation:compute-array}, \ref{sec:motivation:prior:sw} \& \ref{sec:motivation:prior:hw}).

\section{Properties of Real-World Graphs\label{sec:motivation:graph-properties}}
Graph analytics is an exciting and rapidly growing field with applications spanning diverse areas such as uncovering latent relationships (e.g., for recommendation systems), pinpointing influencers in social graphs (e.g., for marketing purposes), among others. Real-world graphs from these areas often
have two distinguishing properties, skew in their degree distribution and community structure, that influence cache efficiency while processing graphs.

\subsection{Skew in Degree Distribution\label{sec:motivation:skew}}
A distinguishing property of graph datasets common in many graph-analytic applications is that the vertex degrees follow a skewed {\em power-law} distribution, in which a small fraction of vertices, {\em hot vertices}, have many connections while the majority of vertices, {\em cold vertices}, have relatively few connections~\cite{power-law,power-law-internet,powergraph,fc,hubcluster}.
Graphs characterized by such a distribution are known as {\em natural} or {\em scale-free} graphs and are prevalent in a variety of domains, including social networks, computer networks, financial networks, semantic networks, and airline networks.

Table~\ref{tab:motivation:skew-hot-vertices} quantifies the skew for the datasets evaluated in this thesis (Sec.~\ref{sec:dbg:method} of Chapter~\ref{ch:dbg} contains more details of the datasets). %
For example, in the \sd dataset, 11\% of total vertices are classified as hot vertices in terms of their in-degree (13\% for out-degree) distribution. These Hot vertices are connected to 88\% of all in-edges (88\% of all out-edges) in the graph. Similarly, in other datasets, 9\%-26\% of vertices are classified as hot vertices, which are connected to 80\%-94\% of all edges.

\begin{table}[!t]
    \centering
    \small
    \begin{tabularx}{1\linewidth}
        {|>{\centering\arraybackslash\hsize=0.25\hsize}X|
        >{\raggedleft\arraybackslash\hsize=0.35\hsize}X|
        >{\raggedleft\arraybackslash\hsize=0.05\hsize}X|
        >{\raggedleft\arraybackslash\hsize=0.05\hsize}X|
        >{\raggedleft\arraybackslash\hsize=0.05\hsize}X|
        >{\raggedleft\arraybackslash\hsize=0.05\hsize}X|
        >{\raggedleft\arraybackslash\hsize=0.05\hsize}X|
        >{\raggedleft\arraybackslash\hsize=0.05\hsize}X|
        >{\raggedleft\arraybackslash\hsize=0.05\hsize}X|
        >{\raggedleft\arraybackslash\hsize=0.05\hsize}X|
        }
    \hline
        & & \kr & \pl & \tw & \sd & \lj & \wl & \fr & \mpi \\ \hline \hline
        In & Hot Vertices (\%) & 9 &  16 &  12 &  11 &  25 &  12 &  24 &  10 \\
        Edges & Edge Coverage (\%) & 93 & 83 &  84 &  88 &  81 &  88 &  86 &  80 \\ \hline
        Out & Hot Vertices (\%) & 9 & 13 &  10 &  13 &  26 &  20 &  18 &  12 \\
        Edges & Edge Coverage (\%) & 93 & 88 &  83 &  88 & 82 & 94 & 92 & 81 \\ \hline
    \end{tabularx}
    \caption{\label{tab:motivation:skew-hot-vertices} {Rows \#2 and \#4 show the percentage of vertices having degree equal or greater than the average (\ie hot vertices), with respect to in-edges and out-edges, respectively; the higher the skew, the lower the percentage.
    Rows \#3 and \#5 show the percentage of in-edges and out-edges connected to the hot vertices, respectively; the higher the skew, the higher the percentage.}}
\end{table}

\subsection{Community Structure\label{sec:motivation:community}} 
Real-world graphs often feature clusters of highly interconnected vertices such as communities of common friends in a social graph~\cite{community1,community2}. 
Such community structure is often captured by vertex ordering within a graph dataset by placing vertices from the same community nearby in the memory space.
At runtime, vertices that are placed nearby in memory are typically processed within a short time window of each other. Thus, by placing vertices from the same community nearby in memory, both temporal and spatial locality is improved at the cache block level for such datasets.%

\section{Graph Processing Basics}
\label{sec:motivation:graph-processing-basics}

\begin{figure}[!t]
    \centering
    \subfloat{\label{fig:motivation:basic-graph-a}{\transparent{1}\includegraphics[width=1px,height=1px]{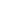}}}
    \subfloat{\label{fig:motivation:basic-graph-b}{\includegraphics[width=0.9\linewidth]{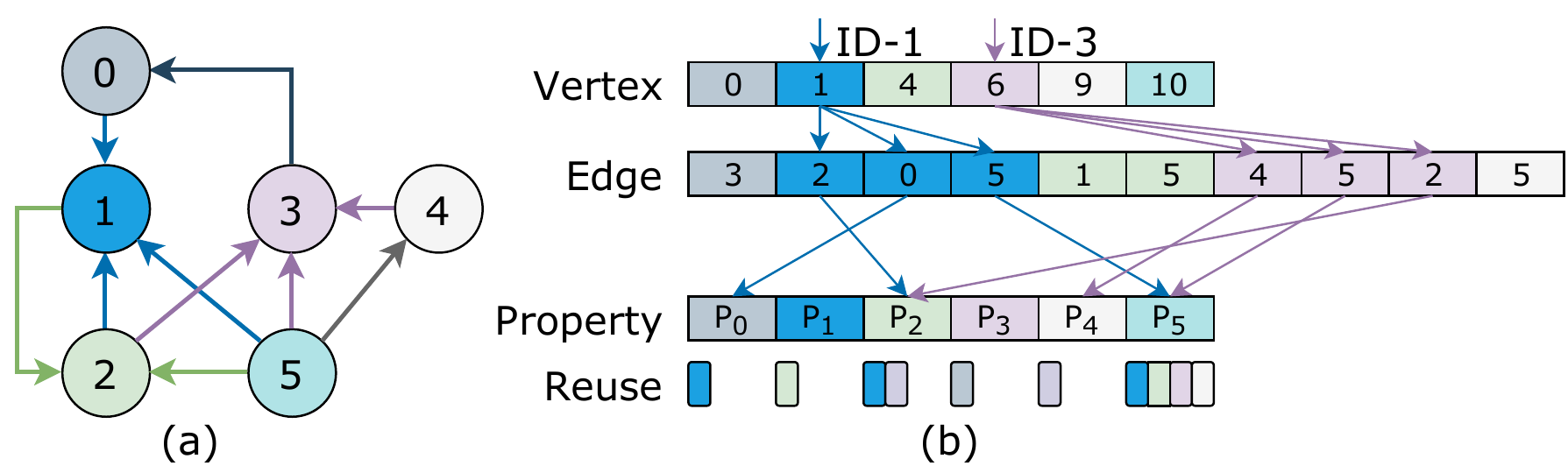}}}
    \subfloat{\label{fig:motivation:basic-graph-c}{\transparent{1}\includegraphics[width=1px,height=1px]{chapter_motivation/pdfs/transparent.png}}}  
    \caption{(a) An example graph. (b) CSR format encoding in-edges. Elements of the same colors across the arrays, correspond to the same destination vertex. Number of bars (labeled \emph{Reuse}) below each element of the Property Array shows the number of times an element is accessed in one full iteration, where the color of a bar indicates the vertex making an access.
    \label{fig:motivation:basic-graph}}
 \end{figure}

The majority of shared-memory graph frameworks are based on a vertex-centric model, in which an application computes some information for each vertex based on the properties of its neighboring vertices~\cite{ligra, galois, gap, graphmat, graphlab, graphchi}. Applications may perform pull- or push-based computations. In pull-based computations, a vertex pulls updates from its in-neighbors. In push-based computations, a vertex pushes updates to its out-neighbors. This process may be iterative, and all or only a subset of vertices may participate in a given~iteration.

The \emph{Compressed Sparse Row (CSR)} format is commonly used to represent graphs in a storage-efficient manner. CSR uses a pair of arrays, \emph{Vertex} and \emph{Edge}, to encode the graph. 
CSR encodes in-edges for pull-based computations and out-edges for push-based computations. 
In this discussion, we focus on pull-based computations and note that the observations hold for push-based computation.
For every vertex, the Vertex Array  maintains an index that points to its first in-edge in the Edge Array. 
The Edge Array stores all in-edges, grouped by destination vertex ID. For each in-edge, the Edge Array entry stores the associated source vertex ID. %

The graph applications use an additional \emph{Property} Array(s) to hold partial or final results for every vertex. For example, the \emph{PageRank} application maintains two ranks for every vertex; one computed from the previous iteration and one being computed in the current iteration. Implementation may use either two separate arrays (each storing one rank per vertex) or may use one array (storing two ranks per vertex).
Fig.~\stitchref{fig:motivation:basic-graph}{fig:motivation:basic-graph-a}~and~\stitchref{fig:motivation:basic-graph}{fig:motivation:basic-graph-b} respectively show a simple graph and its CSR representation for pull-based computations, along with one Property Array.

\section{Cache Behavior in Graph Analytics\label{sec:motivation:general-caching}}

At the most fundamental level, a graph application computes a property for a vertex based on the properties of its neighbors.
To find the neighboring vertices, an application traverses the portion of the Edge Array corresponding to a given vertex, and then accesses elements of the Property Array corresponding to these neighboring vertices.
Fig.~\stitchref{fig:motivation:basic-graph}{fig:motivation:basic-graph-b} highlights the elements accessed during computations for vertex ID-1 and ID-3.

As the figure shows, each element in the Vertex and Edge Arrays is accessed exactly once during an iteration, exhibiting no temporal locality at LLC. These arrays may exhibit high spatial locality, which is filtered by the L1-D cache, leading to a streaming access pattern in the~LLC.

In contrast, the Property Array does exhibit temporal reuse. However, reuse is not consistent for all elements. %
Specifically, reuse is proportional to the number of out-edges for pull-based algorithms. Thus, the elements corresponding to high out-degree vertices exhibit high reuse. Fig.~\stitchref{fig:motivation:basic-graph}{fig:motivation:basic-graph-b} shows the reuse for high out-degree (\ie hot) vertices P$_2$ and P$_5$ of the Property Array assuming pull-based computations; other elements do not exhibit reuse. The same observation applies to high in-degree vertices in push-based algorithms.

\begin{figure}[!t]  
    \includegraphics{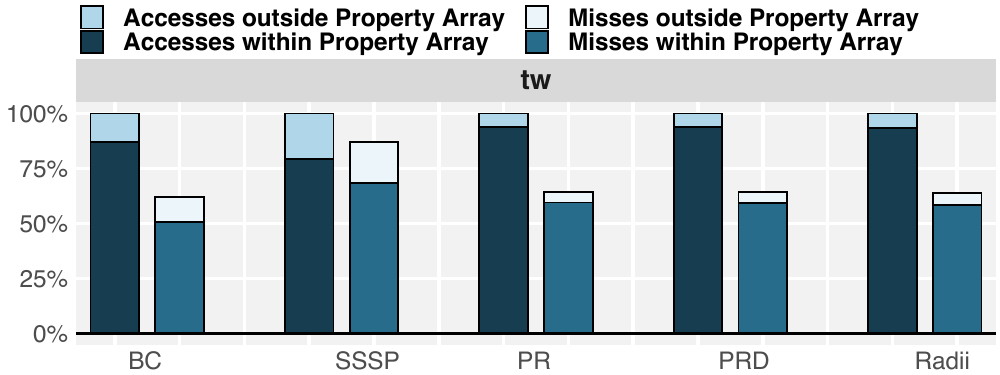}
    \caption{Classification of LLC accesses and misses (normalized to total accesses) for five graph applications when processing the \tw{} dataset.\label{fig:motivation:breakup}
    }
\end{figure}

Finally, Fig.~\ref{fig:motivation:breakup} quantifies the LLC behavior of various graph applications (Sec.~\ref{sec:dbg:method} of Chapter~\ref{ch:dbg} contains more details of the applications) on the \tw{} dataset as a \mbox{representative} example of real-world graph datasets. The figure differentiates all LLC accesses and misses as falling either within or outside the Property Array. 
Unsurprisingly, the Property Array accounts for 78-93\% of all LLC accesses. However, despite the high reuse, the Property Array is also responsible for large fraction of LLC misses, the reasons for which are explained next.

\section{Challenges in Caching the Property Array\label{sec:motivation:compute-array}}

As discussed in the previous section, elements in the Property Array corresponding to the hot vertices exhibit high reuse. Unfortunately, on-chip caches struggle in capitalizing on the high reuse for the two reasons: lack of spatial locality and difficult to exploit temporal locality.

\subsection{Lack of Spatial Locality\label{sec:motivation:low-spatial-locality}} 
A cache block is typically comprised of multiple vertices as the properties associated with a vertex are much smaller than the size of a cache block. Moreover, hot vertices constitute a relatively smaller fraction of all vertices and are sparsely distributed throughout the memory space of the Property Array. 
Thus, inevitably, hot vertices share space in a cache block with cold vertices, leading to low spatial locality for hot vertices.
Even when a cache block holding a hot vertex is retained in the cache, it leads to underutilization of cache capacity as a considerable fraction of the cache block is occupied by cold vertices that exhibit low or no reuse.  

\begin{table}[!b]
    \centering
    \small
    \begin{tabularx}{0.8\linewidth}
        {|>{\centering\arraybackslash\hsize=0.2\hsize}X|
        >{\centering\arraybackslash\hsize=0.1\hsize}X|
        >{\centering\arraybackslash\hsize=0.1\hsize}X|
        >{\centering\arraybackslash\hsize=0.1\hsize}X|
        >{\centering\arraybackslash\hsize=0.1\hsize}X|
        >{\centering\arraybackslash\hsize=0.1\hsize}X|
        >{\centering\arraybackslash\hsize=0.1\hsize}X|
        >{\centering\arraybackslash\hsize=0.1\hsize}X|
        >{\centering\arraybackslash\hsize=0.1\hsize}X|
        }
    \hline
        Dataset & \kr & \pl & \tw & \sd & \lj & \wl & \fr & \mpi \\ \hline
        Avg.    & 1.3 & 1.6 & 1.5 & 1.8 & 3.5 & 3.1 & 2.7 & 2.6 \\
     \hline
    \end{tabularx}
    \caption{\label{tab:motivation:avg-hot-vertex}Average number of hot vertices per cache block. 
    \mbox{Calculation} assumes 8 bytes per vertex and 64 bytes per cache block, and counts only cache blocks that contain at least one hot vertex. As a result, any cache block can contain between 1--8 hot vertices. %
    }
\end{table}

Table~\ref{tab:motivation:avg-hot-vertex} shows the average number of hot vertices per cache block, assuming typical values of 8 bytes per vertex and 64 bytes per cache block. While, at best, 8 hot vertices can be packed together in a cache block, in practice, only 1.3 to 3.5 hot vertices are found per cache block across the datasets. As the footprint (\ie number of cache blocks) to store hot vertices is inversely proportional to the average number of hot vertices per cache block, the data shows significant opportunity in reducing the cache footprint of hot vertices, and in turn, improving cache efficiency.

\subsection{Difficult to Exploit Temporal Locality} 
\begin{figure}[!t]
    \centering
    \includegraphics{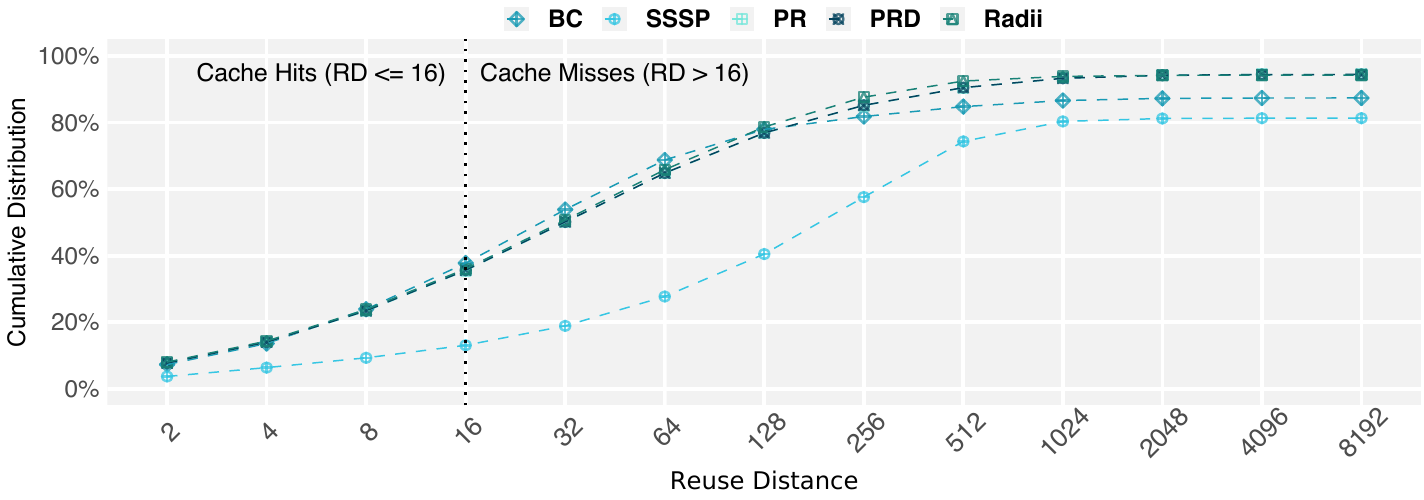}
    \caption{Reuse Distance Distribution on 16-way set-associative 16MB LLC for five graph applications, each processing the \tw dataset. Vertical dotted line at reuse distance of 16 shows hit-rate under LRU management. Remaining percentage of LLC accesses after reuse distance of 8192 are corresponding to cold misses and thus, have infinite reuse distances.}
    \label{fig:motivation:rdd}
\end{figure}
The access pattern to the Property Array is highly irregular, being heavily dependent on both graph structure and application. Between a pair of accesses to a given hot vertex in the Property Array, a number of other, low-/no-reuse, data elements (e.g, cold vertices or elements of the Vertex and Edge Arrays) may be accessed, increasing reuse distance of the accesses to the hot vertices.
Any block allocated by these low-/no-reuse data elements will trigger evictions at the LLC, potentially displacing cache blocks holding hot vertices.

Fig.~\ref{fig:motivation:rdd} shows the cumulative reuse distance distribution of LLC accesses for five graph applications processing the \tw dataset. For a stream of accesses to a given cache set, the reuse distance of a cache block access is calculated as the number of unique LLC accesses in the set since the previous LLC access to the same cache block.
Thus, any LLC access with reuse distance less than or equal to the associativity of the cache (16 in the study) would result in a cache hit under LRU. 
As the figure shows, at most 38\% of LLC accesses have reuse distance less than or equal to 16 (shown using a vertical dotted line). Meanwhile, 19\%-54\% of all LLC accesses have reuse distance greater than 64 (\ie 4$x$ the associativity). Long reuse distances, along with irregular access patterns, lead to severe cache thrashing at LLC, making it difficult for domain-agnostic techniques to capitalize on the high reuse inherent in accesses to hot vertices.

We next discuss the most relevant state-of-the-art techniques in both software and hardware that attempt to address the above mentioned challenges for graph analytics.

\section{Prior Software Techniques\label{sec:motivation:prior:sw}}

The order of vertices in memory is under the control of a graph application. Thus, the application can reorder vertices in memory before processing a graph to improve cache locality. To accomplish this, researchers have proposed various reordering techniques~\cite{gorder, rabbit, recall, LDG, METIS, SlashBurn, rcm, CHDFS, fc, hubcluster}. Reordering techniques only relabel vertices (and edges), which does not alter the graph itself and does not require any changes to the graph algorithms.
Following the relabeling, vertices (and edges) are reordered in memory based on the new vertex IDs.

The most powerful reordering techniques like Gorder~\cite{gorder} leverage community structure, typically found in real-world graphs, to improve spatio-temporal locality. Gorder{} \mbox{comprehensively} analyzes the vertex connectivity and reorders vertices such that vertices that share common neighbors, and thus are likely to belong to the same community, are placed nearby in memory. While Gorder is effective at reducing cache misses, it requires a staggering reordering time that is often multiple orders of magnitude higher than the total application runtime, rendering Gorder impractical~\cite{hubcluster}.

To keep the reordering cost affordable, we argue for limiting the scope of vertex reordering to induce spatial locality only while leaving the task of exploiting temporal locality to a hardware cache management technique. 
We collectively refer to such techniques as {\em skew-aware} reordering techniques.
Unlike Gorder, skew-aware reordering techniques require lightweight analysis as these reorder vertices solely based on vertex degrees, and thus can speed-up applications even after accounting for the reordering time~\cite{fc, hubcluster}.

Existing skew-aware reordering techniques seek to induce spatial locality among hot vertices by segregating them into a contiguous memory region. As a result, the cache footprint of hot vertices is reduced, which, in turn, improves cache efficiency.
However, as a side-effect of reordering, these may destroy a graph's community structure, which could negate the performance gains achieved from the reduced footprint of hot vertices. Thus, there exists a tension between reducing the cache footprint of hot vertices and preserving graph structure when reordering vertices, which must be addressed by a skew-aware technique in order to maximize cache efficiency.

\section{Prior Hardware Techniques\label{sec:motivation:prior:hw}}

In the previous section, we argued for exploiting temporal locality of hot vertices through a hardware cache management technique to keep software reordering lightweight. 
In this section, we discuss how effective existing hardware cache management techniques are in exploiting temporal locality, specifically in context of graph analytics.

\noindenttitle{\smalltextcircled{1} Lightweight techniques} (\ie static and lightweight dynamic techniques from Sec.~\ref{sec:background:prior}) use simple heuristics to manage LLC. RRIP~\cite{rrip} is the state-of-the-art technique in this category that relies on a probabilistic approach to classify a cache block as low or high reuse at the time of inserting a new block in the cache. As these techniques do not exploit the reuse behavior of cache blocks from their past generations, these are limited in accurately identifying high-reuse blocks. 

\begin{figure}[!t]
    \centering
    \includegraphics{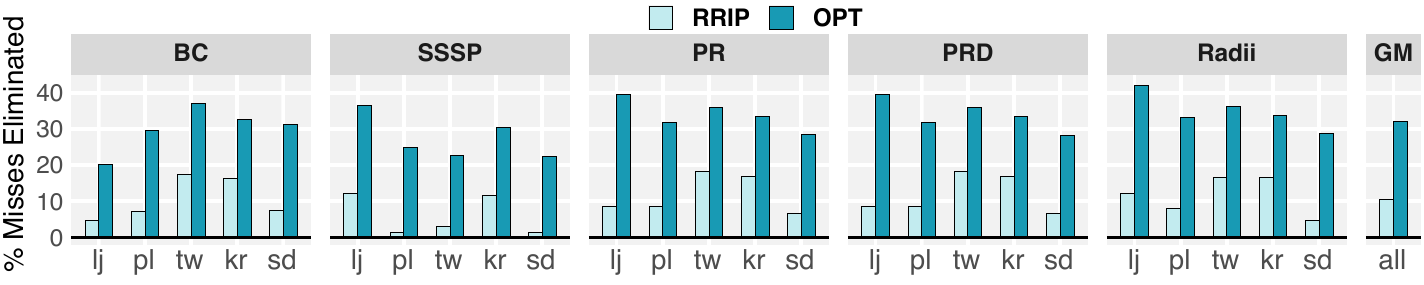}
    \caption{Percentage of misses eliminated by RRIP and OPT over LRU on 16MB LLC. Trace for each application-dataset pair consists of up to 2 billion LLC accesses.}
    \label{fig:motivation:belady_orig}
\end{figure}

We quantify the effectiveness of RRIP over LRU using a trace-based study on a set of graph applications processing various high-skew datasets as shown in Fig.~\ref{fig:motivation:belady_orig}.
The figure plots the percentage of misses eliminated by RRIP over LRU, along with misses eliminated by OPT~\cite{opt} to show the maximum opportunity for any cache management technique.
RRIP consistently reduces misses over LRU across datapoints with an average miss reduction of 10.5\%. Meanwhile, OPT shows that on an average, 32.3\% of misses can be eliminated over LRU, showing a significant opportunity in improving cache efficiency over RRIP.

\noindenttitle{\smalltextcircled{2} History-based predictive techniques} such as the state-of-the-art Hawkeye~\cite{hawkeye} and many others ~\cite{counter,sampler,ship,mdpp,harmony,perceptron,multi} learn past reuse behavior of cache blocks by employing sophisticated storage-intensive prediction mechanisms. A large body of recent works focus on history-based predictive techniques as these generally provide higher performance than the lightweight techniques for a wide range of applications as shown in Sec.~\ref{sec:leeway:eval-spec} of Chapter~\ref{ch:leeway}. Meanwhile, for graph analytics, we find that graph-dependent irregular access patterns, combined with long reuse distances, prevent these predictive techniques from correctly learning which cache blocks to preserve. 
For example, as explained in Sec.~\ref{sec:back:history} of Chapter~\ref{ch:background}, most history-based predictive techniques rely on a PC-based correlation to learn which set of PC addresses access high-reuse cache blocks to prioritize these blocks for caching over others.
However, we observe that the reuse for elements of the Property Array, which are the prime target for LLC caching in graph analytics (Sec~\ref{sec:motivation:general-caching}), does not correlate with the PC because the same PC accesses hot and cold vertices alike.

We quantify the performance of three state-of-the-art history-based predictive techniques -- SHiP-MEM, Hawkeye\footnote{We use an improved, prefetch-aware, version of Hawkeye from CRC2 (\ie Hawkeye++ from Sec.~\ref{sec:leeway:crc2-evel} of Chapter~\ref{ch:leeway}).} and Leeway. Hawkeye and Leeway rely on a PC-based reuse correlation whereas SHiP-MEM, a variant of SHiP, exploits a region-based correlation.
Fig.~\ref{fig:motivation:orig-perf} plots application speed-up for these techniques over RRIP for five graph applications, each processing five graph datasets. We use RRIP as a new, stronger, baseline as RRIP consistently reduces more misses than LRU as shown in Fig.~\ref{fig:motivation:belady_orig}.

The results show that all predictive techniques on average cause slowdown over the RRIP baseline. 
Irregular access patterns, combined with long reuse distance accesses, impede learning of these predictive techniques, rendering them deficient for the whole domain of graph analytics. 
As expected, Leeway tolerates variability in the reuse behavior the most by causing an average slowdown of 0.8\% only vs 5.7\% for SHiP-MEM and 14.8\% for Hawkeye. Alas, Leeway causes a slowdown nonetheless.
The results highlight that existing domain-agnostic cache management techniques are unable to exploit temporal locality despite a significant opportunity. 

\begin{figure}
    \centering
    \includegraphics{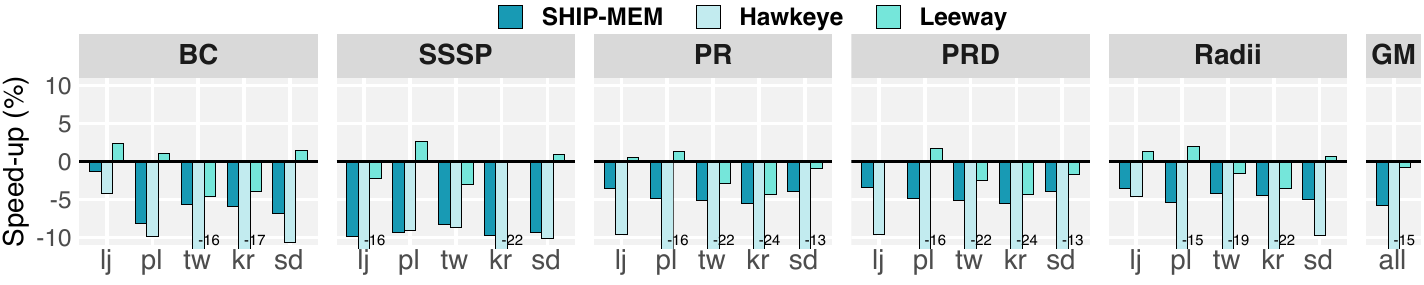}
    \caption{Performance evaluation for state-of-the-art domain-agnostic cache management techniques over RRIP.}
    \label{fig:motivation:orig-perf}
\end{figure}

\noindenttitle{\smalltextcircled{3} Software-aided techniques} use compiler analysis, runtime profiling or domain-knowledge of the programmers to identify high-reuse cache blocks. 
The majority of these techniques target regular access patterns, making them infeasible for graph applications that are dominated by irregular access patterns.

Techniques such as XMem~\cite{xmem} dedicate partial or full cache capacity by \emph{pinning} high-reuse blocks to cache.
Hardware ensures that the pinned blocks cannot be evicted by other cache blocks and thus are protected from cache thrashing. Such an approach is only feasible when the high-reuse working set fits in the available cache capacity. Unfortunately, for large graph datasets, even with high skew, it is unlikely that all hot vertices will fit in the LLC; recall from Table~\ref{tab:motivation:skew-hot-vertices} that hot vertices account for up to 26\% of the total vertices. 
Moreover, some of the colder vertices might also exhibit short-term temporal reuse, particularly in graphs with community structure.

\noindentnotitle
These observations call for a new LLC management technique that employs (1) a reliable mechanism to identify hot vertices amidst irregular access patterns and (2) flexible cache policies that maximize reuse among hot vertices by protecting them in the cache without denying colder vertices the ability to be cache resident if they exhibit reuse.

\section{Solution: Software-Hardware Co-Design}

Graph analytics on natural graphs exhibit poor cache efficiency due to low spatial locality and difficult to exploit temporal locality. Existing domain-agnostic hardware cache management techniques are limited in addressing both these challenges. First, hardware alone cannot enforce spatial locality, which is dictated by vertex placement in the memory space and is under software control. Second, domain-agnostic hardware cache management techniques struggle in pinpointing hot vertices under cache thrashing due to long reuse distance accesses and irregular access patterns endemic of graph analytics. 

Both of these challenges can be addressed by leveraging a lightweight software support. 
First, a skew-aware lightweight software technique can induce spatial locality by segregating hot vertices in a contiguous memory region.
Second, software has the knowledge of the memory locations of hot vertices. Utilizing software knowledge can enable a reliable mechanism for hardware to identify hot vertices amidst irregular access patterns.

Based on these observations, we propose a holistic software-hardware co-design to improve cache efficiency for graph analytics. Our software component is responsible for inducing spatial locality of hot vertices. The software component also facilitates our hardware's task of pinpointing the cache blocks containing hot vertices. While the software informs hardware, the hardware is ultimately in control of deciding which vertices to evict and which to preserve based on available cache capacity and temporal access patterns, thus relinquishing software from any additional runtime overhead. The end result is software that incurs minimal runtime overhead, and simple hardware that
reliably identifies cache blocks that are likely to exhibit high reuse.

In the following chapters, we discuss each of these components in detail. In Chapter~\ref{ch:dbg}, we present DBG, a new skew-aware vertex reordering technique. In Chapter~\ref{ch:grasp}, we introduce GRASP, domain-specialized cache management for graph analytics.

\chapter{DBG -- Lightweight Vertex Reordering\label{ch:dbg}}

\section{Introduction}
\label{sec:dbg:intro}

For a typical graph application, a cache block
contains multiple vertices, as vertex properties usually require just 4 to 16 bytes whereas a cache block size in modern processors is typically 64 or 128 bytes.
Since hot vertices are sparsely distributed in memory, and are smaller in number, 
they inevitably share 
cache blocks with cold vertices, leading to underutilization of a considerable fraction of useful cache capacity.

Skew-aware techniques reorder vertices in memory such that hot vertices are adjacent to each other in a contiguous memory region.
As a result, each cache block is comprised of exclusively hot or cold vertices, reducing the total footprint (\ie number of cache blocks) required to store hot vertices.
Blocks that are exclusively comprised of hot vertices are far more likely to be retained in the cache due to higher aggregate hit rates, leading to
higher utilization of existing cache capacity.

A straightforward way to pack vertices with similar degree into each cache block
is to %
apply {\em Sort Reordering}, which sorts vertices based on their degree.
However, Sort is not always beneficial, %
because many real-world graph datasets exhibit a strong structure, \eg clusters of webpages within the same domain in a web graph, or communities of common friends in a social graph~\cite{community1,community2}.
In such datasets, vertices within the same community are accessed together, and often reside nearby in memory, exhibiting spatio-temporal locality that should be preserved. Fine-grain vertex reordering, such as Sort and Hub Sorting~\cite{fc}, destroys the spatio-temporal locality, which limits the effectiveness of such reordering on datasets that exhibit~structure. 

In this chapter, we quantify potential performance loss due to disruption of graph structure on various datasets.
We further characterize locality at all three levels of the cache hierarchy, and show that all skew-aware techniques are generally effective at reducing LLC misses. However, techniques employing fine-grain reordering significantly disrupt graph structure, increasing misses in higher-level caches. %
Our results highlight a tension between reducing the cache footprint of hot vertices and preserving graph structure, limiting the effectiveness of prior skew-aware~techniques.

To overcome the limitations of prior techniques, we propose \dbgacronym, a novel reordering technique that largely preserves graph structure while 
reducing the cache footprint of hot vertices.
Like prior skew-aware techniques, \dbg segregates hot vertices from the cold ones. However, to preserve existing graph structure, \dbg employs {\em coarse-grain} reordering.
\dbg partitions vertices into a small number of groups based on their degree but maintains the original relative order of vertices within each group.
As \dbg \emph{does not sort} vertices within any group to minimize structure disruption, \dbg also incurs a very low reordering overhead.%

\noindentnotitle
To summarize, we make the following contributions:
\begin{itemize}
    \item We study existing skew-aware reordering techniques on a variety of multi-threaded graph applications processing varied datasets.
       Our characterization reveals the inherent tension between reducing the cache footprint of hot vertices and preserving graph structure.
    \item We propose \dbg, a new skew-aware reordering technique that employs lightweight coarse-grain reordering to largely preserve existing graph structure while reducing the cache footprint of hot vertices. %

    \item Our evaluation on a real machine shows that \dbg \mbox{outperforms} existing skew-aware techniques.
    Averaging across 40 datapoints, \dbg yields a speed-up of 16.8\%, vs 11.6\% for the best-performing existing skew-aware \mbox{technique} over the baseline with no reordering.%
\end{itemize}

\section{Skew-Aware Reordering Techniques\label{sec:dbg:lwr}}

\subsection{Objectives for High Performance Reordering\label{sec:dbg:objectives}}

In order to provide high performance for graph applications, skew-aware reordering techniques should achieve all of the following three objectives:

\visiblespace

\noindenttitle{O1. Low Reordering Time:} Reordering time plays a crucial role in deciding whether a technique is viable in providing end-to-end application performance after accounting for the reordering time. Lower reordering time facilitates amortizing the reordering overhead in a fewer graph traversals.

\noindenttitle{O2. High Cache Efficiency:} As explained in Sec.~\ref{sec:motivation:low-spatial-locality} of Chapter~\ref{ch:motivation}, a cache block is comprised of multiple vertices. Problematically, hot vertices are sparsely distributed throughout the memory space, which leads to cache blocks containing vertices with vastly different degrees. %
To address this, vertex reordering should ensure that hot vertices are placed adjacent to each other in the memory space, thus reducing the cache footprint of hot vertices, and in turn, improving cache efficiency.

\noindenttitle{O3. Structure Preservation:} As explained in Sec.~\ref{sec:motivation:community} of Chapter~\ref{ch:motivation}, many real-world graph datasets have vertex ordering that results in high spatio-temporal cache locality. For such datasets, vertex reordering should ensure that the original structure is preserved as much as possible. If structure is not preserved, reordering may adversely affect the locality, negating performance gains achieved from the reduced footprint of hot vertices.

\subsection{Implications of Not Preserving Graph Structure\label{sec:dbg:random}}

In this section, we characterize how important it is to preserve graph structure for different datasets.
To \mbox{quantify} the potential performance loss due to reduction in spatio-temporal locality arising from reordering, we randomly reorder vertices, which decimates any existing structure.
Randomly \mbox{reordering} all vertices would cause a slowdown for two potential \mbox{reasons}: 
(1) By destroying graph structure, thus reducing spatio-temporal locality.
(2) By further scattering hot vertices in memory, thus \mbox{increasing} the cache footprint of hot vertices.
To isolate performance loss due to the former, we also evaluate random reordering at a cache block granularity. In such a reordering, cache blocks (not individual vertices) are randomly reordered in memory, which means that the vertices within a cache block are moved as a group. As a result, the cache footprint of hot vertices is unaffected, and any change in performance can be directly attributed to a change in graph structure.
Fig.~\stitchref{fig:dbg:reordering-existing-example}{fig:dbg:reordering-existing-example-a} illustrates vertex placement in memory after Random Reordering at a vertex and at a cache block~granularity.

\begin{figure}[!t]
    \centering
    \includegraphics{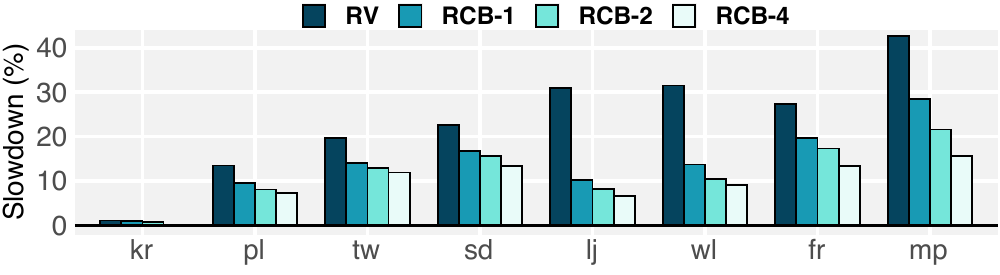}
    \caption{Application slowdown after random reordering at different granularity for the Radii application. The lower the bar, the better the application performance.\label{fig:dbg:random}}
\end{figure}

Fig.~\ref{fig:dbg:random} shows performance slowdown for Random Reordering for the Radii application on all datasets listed in Table~\ref{tab:dbg:datasets}.
The figure shows four configurations, Random Vertex (RV) that reorders at a granularity of one vertex and Random Cache Block-$n$ (RCB-$n$) that reorders at a granularity of $n$ cache blocks, where $n$ is 1, 2 or 4.

Performance difference between RV and RCB-1 is very large for the four right-most datasets. Recall from Table~\ref{tab:motivation:avg-hot-vertex} of Chapter~\ref{ch:motivation} that these datasets have relatively high number of hot vertices per cache block. RV scatters the hot vertices in memory, incurring large slowdowns for these datasets.

Performance slowdown for RCB-1 is significant on all real-world datasets (\ie all but \kr), and ranges from 9.6\% to 28.5\%.
This slowdown can be attributed to disruption in spatio-temporal locality for the real-world datasets, confirming existence of community structure in the original ordering of the datasets. 
As reordering granularity increases, disruption in graph structure reduces, which also reduces the slowdown.
For example, on the \mpi dataset, the most affected dataset by the Random Reordering among all, performance slowdown is 28.5\% for RCB-1, which 
reduces to 21.6\% for RCB-2 and 15.6\% for RCB-4.

Results for \kr{}, the only synthetic dataset in the mix, are in stark contrast with that of the real-world datasets.
As \kr{} is generated synthetically, \kr does not have any structure in the original ordering.
Thus, the performance on the \kr{} dataset is largely oblivious to random reordering at any granularity. 

\begin{figure}[!t]
    \centering
    \subfloat{\label{fig:dbg:reordering-existing-example-a}{\transparent{1}\includegraphics[width=1px,height=1px]{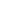}}}
    \subfloat{\label{fig:dbg:reordering-existing-example-b}\includegraphics[width=0.995\linewidth]{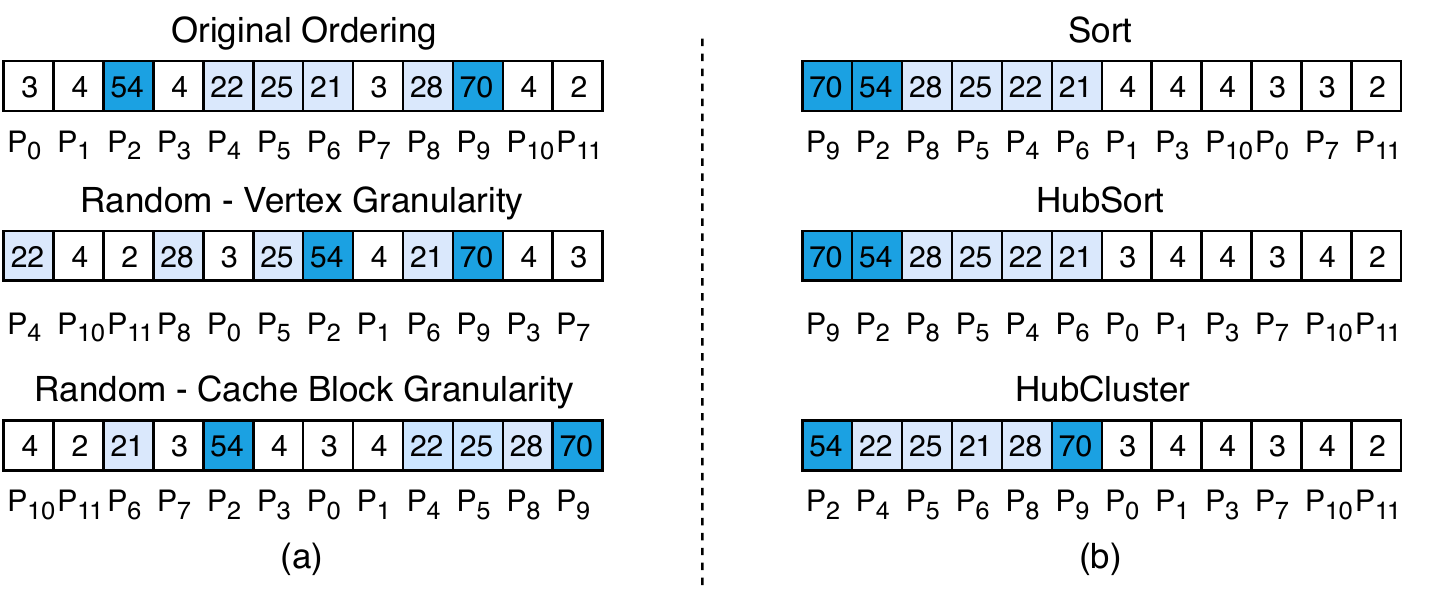}}
    \caption{Vertex ordering in memory for different techniques.
    Vertex degree is shown inside the box while original vertex ID is shown below the box.
    Hot vertices (degree $\ge$ 20) are shown in color. Hottest among the hot vertices (degree $\ge$ 40) are shown in a darker shade. Finally, Random (Cache Block Granularity) assumes two vertices per cache block.}
    \label{fig:dbg:reordering-existing-example}
\end{figure}

The results show that the real-world graph datasets exhibit some structure in their original ordering, which, if not preserved, is likely to adversely affect the performance.
The results also indicate that structure can be largely preserved by applying reordering at a coarse granularity.

\subsection{Limitations of Prior Skew-Aware Reordering Techniques\label{sec:dbg:existing-skew-aware}}

This section describes the existing skew-aware techniques and how they fare in achieving the three objectives listed in Sec.~\ref{sec:dbg:objectives}. As skew-aware techniques solely rely on vertex degrees for reordering, they all incur relatively low reordering time, achieving objective O1. However, for the two remaining objectives, reducing the cache footprint of hot vertices and preserving existing graph structure, existing techniques trade one for the other, hence failing to achieve at
least one of the two objectives.

\noindenttitle{Sort} reorders vertices based on the descending order of their degree.
Sort requires the least possible number of cache blocks to store hot vertices without explicitly classifying individual vertices as hot or cold. 
However, sort reorders \emph{all} vertices, which completely destroys the original graph structure. 
Fig.~\stitchref{fig:dbg:reordering-existing-example}{fig:dbg:reordering-existing-example-b} shows vertex placement in memory after the Sort Reordering.

\noindenttitle{Hub Sorting}~\cite{fc} (also known as Frequency-based Clustering) was proposed as a variant of Sort that aims to preserve some structure while reducing the cache footprint of hot vertices.
Hub Sorting uses an average degree of the dataset as a threshold to classify vertices as hot or cold, and only sorts the hot vertices.%

Hub Sorting does preserve partial structure by not sorting the cold vertices, but problematically, the hot vertices are fully sorted. While hot vertices constitute a smaller fraction compared to the cold ones, recall from Table~\ref{tab:motivation:skew-hot-vertices} of Chapter~\ref{ch:motivation} that 
hot vertices account for up to 26\% of the total vertices. Moreover, hot vertices are connected to the vast majority of edges (80\%-94\%), and thus are responsible for the majority of reuse. Consequently, preserving structure for hot vertices is also important, at which Hub Sorting fails.

\noindenttitle{Hub Clustering}~\cite{hubcluster} is a variant of Hub Sorting that only segregates hot vertices from the cold ones but does not sort them. While Hub Clustering was proposed as an alternative to Hub Sorting that has lower reordering time, we note that Hub Clustering is also better than Hub Sorting at preserving graph structure as Hub Clustering does not sort any vertices. However, by not sorting hot vertices, Hub Clustering sacrifices significant opportunity in
improving cache efficiency as discussed~next.

\begin{table}[t]
    \centering
    \small
    \begin{tabularx}{1\linewidth}
        {|>{\centering\arraybackslash\hsize=0.4\hsize}X|
          >{\raggedleft\arraybackslash\hsize=0.075\hsize}X|
          >{\raggedleft\arraybackslash\hsize=0.075\hsize}X|
          >{\raggedleft\arraybackslash\hsize=0.075\hsize}X|
          >{\raggedleft\arraybackslash\hsize=0.075\hsize}X|
          >{\raggedleft\arraybackslash\hsize=0.075\hsize}X|
          >{\raggedleft\arraybackslash\hsize=0.075\hsize}X|
          >{\raggedleft\arraybackslash\hsize=0.075\hsize}X|
          >{\raggedleft\arraybackslash\hsize=0.075\hsize}X|
          }
    \hline
        Per-Vertex Property & \kr & \pl & \tw & \sd & \lj & \wl & \fr & \mpi \\ \hline \hline
        8 Bytes & 44 & 51 & 56 & 80 & 9 &  16 & 115 & 39 \\ \hline
        16 Bytes & 88 & 102 & 112 & 160 & 18 & 32 & 230 & 78 \\ \hline
    \end{tabularx}
    \caption{\label{tab:dbg:fp-hot-vertices} Cache size (MB) needed to store {\em all hot vertices}, assuming 8 and 16 bytes per property, respectively. Vertex is classified hot if its degree is equal or greater than the average degree of the dataset.}  
\end{table}

For large graph datasets, it is unlikely that all hot vertices fit in the LLC. For example, the \sd dataset requires at least 80MB to store all hot vertices assuming only 8 bytes per vertex (refer to Table~\ref{tab:dbg:fp-hot-vertices} for requirements of the remaining datasets). The required capacity significantly exceeds a typical LLC size of commodity server processors. As a result, all hot vertices compete for the
limited LLC capacity, causing cache thrashing.

Fortunately, not all hot vertices have similar reuse, as vertex degree varies vastly among hot vertices. 
Table~\ref{tab:dbg:degree-range} shows the degree distribution for just the hot vertices of the \sd dataset. 
Each column in the table represents a degree range as a function of \avgdegree, the average degree of the dataset. For instance, the first column covers vertices whose degree ranges from \avgdegree to 2\avgdegree; these are the lowest-degree vertices among the hot ones (recall that a vertex is classified as hot if its degree is equal or greater than \avgdegree). 
For a given range, the table shows number of vertices (as a percentage of total hot vertices) whose degree is within that range. The table also shows cache capacity needed for those many vertices assuming 8 bytes per vertex property.
Unsurprisingly, given the power-law degree distribution, the table shows that the least-hot vertices are the most numerous, representing 45\% of all hot vertices and requiring 35.8MB capacity, yet likely exhibiting the least reuse among hot vertices. 
In contrast, vertices with degree above 8\avgdegree (three right-most columns) are the hottest of all, constituting just 12\% of total hot vertices ($<10$MB footprint).
Naturally, these hottest vertices are the ones that should be retained in the cache.
However, by not sorting hot vertices, Hub Cluster fails to differentiate between the most- and the least-hot vertices, hence denying the hottest vertices an opportunity to stay in the cache in the presence of cache thrashing.

\begin{table}[!t]
    \centering
    \small
    \begin{tabularx}{1\linewidth}
        {|>{\centering\arraybackslash\hsize=0.22\hsize}X|
          >{\raggedleft\arraybackslash\hsize=0.12\hsize}X|
          >{\raggedleft\arraybackslash\hsize=0.12\hsize}X|
          >{\raggedleft\arraybackslash\hsize=0.125\hsize}X|
          >{\raggedleft\arraybackslash\hsize=0.135\hsize}X|
          >{\raggedleft\arraybackslash\hsize=0.145\hsize}X|
          >{\raggedleft\arraybackslash\hsize=0.135\hsize}X|
          }
        \hline 
        Degree Range & {\footnotesize[1\avgdegree,2\avgdegree)} & {\footnotesize[2\avgdegree,4\avgdegree)} & {\footnotesize[4\avgdegree,8\avgdegree)} & {\footnotesize[8\avgdegree,16\avgdegree)} & {\footnotesize[16\avgdegree,32\avgdegree)} & {\footnotesize[32\avgdegree,\infinity)} \\ \hline
        Vertices (\%) & 45\% & 28\% & 15\% & 7\% & 3\% & 2\%  \\ \hline
        Footprint & 35.8 & 22.3 & 12.0 & 5.7 & 2.2 & 1.8 \\ \hline  
    \end{tabularx}
    \caption{\label{tab:dbg:degree-range}
    Degree distribution of hot vertices for the \sd dataset, whose Average Degree (\avgdegree) is 20. Row \#2 shows percentage of total hot vertices while row \#3 shows the footprint requirement in MB, assuming 8 bytes per property.
    }
\end{table}

\noindentnotitle
To summarize, Sort achieves the maximum reduction in the cache footprint of hot vertices. However, in doing so, Sort completely decimates existing graph structure. Hub Sorting and Hub Clustering both classify vertices as hot or cold based on their degree and preserve the structure for cold vertices. However, in dealing with hot vertices, they resort to inefficient extremes. 
At one extreme, Hub Sorting employs fine-grain reordering that sorts all hot vertices, destroying existing graph structure.
At the other extreme, Hub Clustering does not apply any kind of reordering among hot vertices, sacrificing significant opportunity in improving cache efficiency. %

\section{Degree-Based Grouping (DBG)\label{sec:db:spot}}

\begin{lstlisting}[caption=\dbg algorithm. Degree can be in-degree or out-degree or sum of both., label=listing:dbg:spot-listing, float, floatplacement=t, aboveskip=0.65em, 
morekeywords={D,M,K,P,Q,G,V,E,Group},basicstyle=\small,captionpos=b]
G(V, E) where Graph G has V vertices and E edges.
%*\emph{Input}*): Degree Distribution D[], where D[v] is degree of vertex v.
%*\emph{Output}*): Mapping M[], where M[v] is the new ID of vertex v.
%*\emph{\dbg}*): Binning algorithm to reorder vertices into K groups (K $>$ 0). 

1: Assign contiguous range [P$_k$, Q$_k$) to every Group$_k$ such that,
       Q$_1$ > max(D[])    &    
       P$_K$ $\leq$ min(D[])    &
       Q$_{k+1}$ = P$_k$  < Q$_k$,    for every k < K
2: For every vertex v from 1 to V
       Append v to the Group$_k$ for which D[v] $\in$ [P$_k$, Q$_k$).
3: Assign new IDs to all vertices as follows:
       id := 1
       For every Group$_k$ from 1 to K
            For every vertex v in Group$_k$
                M[v] := id++, where v is the original ID
\end{lstlisting}

To address the limitations of prior skew-aware reordering techniques, we propose \emph{\dbgfullname (\dbg)}, a novel skew-aware technique that applies coarse-grain reordering such that each cache block is comprised of vertices with similar degree, and in turn, similar hotness, while also preserving graph structure at~large.

Unlike Hub Sorting and Hub Clustering, which rely on a single threshold to classify vertices as hot or cold, \dbg employs a simple binning algorithm to \emph{coarsely} partition vertices into different groups (or bins) based on their hotness level.
Groups are assigned exclusive degree ranges such that the degree of any vertex falls within a degree range of exactly one group.
Within each group, \dbg maintains the original relative order of vertices to preserve graph structure at large.
To keep the reordering time low, \dbg maintains only a small number of groups and does not sort vertices within any group. 
Listing~\ref{listing:dbg:spot-listing} presents the formal \dbg algorithm.

\begin{figure}[!t]
    \centering
    \includegraphics[width=1\linewidth]{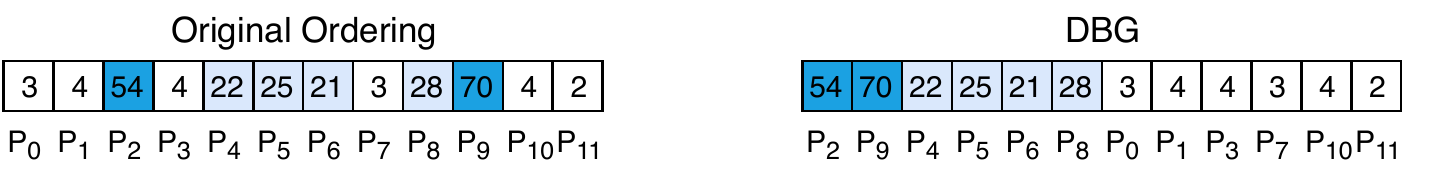}
    \caption{Vertex ordering in memory after DBG. In this example, \dbg partitions vertices into three groups with degree ranges [0, 20), [20, 40) and [40, 80). \dbg maintains a relative order of vertices within a group. As a result, many vertices are placed nearby the same vertices as before the reordering such as vertex sets (P$_4$, P$_5$, P$_6$), (P$_0$, P$_1$) and (P$_{10}$, P$_{11}$).}
    \label{fig:dbg:reordering-example}
\end{figure}

To assign degree ranges to different groups, \dbg leverages the power-law distribution of vertex connectivity in natural graphs. %
For example, recall Table~\ref{tab:dbg:degree-range}, which shows distribution of hot vertices across different degree ranges. 
Vertices with the smallest degree range constitute the largest fraction of hot vertices. As degree range doubles, the number of vertices are roughly halved, exhibiting the power-law distribution.
Thus, geometrically-spaced degree ranges provide a natural way to segregate vertices with different levels of hotness. 
At the same time, using such wide ranges to partition vertices facilitates reordering at a very coarse granularity, preserving structure at large. 
Meanwhile, by not sorting vertices within any group, \dbg incurs a very low reordering time.  
Thus, \dbg successfully achieves all three objectives listed in Sec.~\ref{sec:dbg:objectives}.
Fig.~\ref{fig:dbg:reordering-example} shows vertex placement in memory after the \dbg Reordering, for a synthetic example.

\begin{table}[!t]
    \small
    \centering
    \begin{tabularx}{1\linewidth}
        {|>{\raggedright\arraybackslash\hsize=0.2\hsize}X| %
          >{\centering\arraybackslash\hsize=0.2\hsize}X|
          >{\raggedright\arraybackslash\hsize=0.6\hsize}X|
          }
        \hline
        Reordering & \#Groups & \centering Degree Range \tabularnewline \hline
        \hline
        Sort & \maxdegree{}$+1$ & [$n$, $n+1$) where n $\in$ [0, \maxdegree{}] \\ \hline
        Hub  Sorting   & \multirow{1}{*}{\maxdegree{}-\avgdegree{}$+2$} & [0, \avgdegree), [$n$, $n+1$) where n $\in$ [\avgdegree{}, \maxdegree{}]\\ \hline
        Hub Clustering & \multirow{1}{*}{2} & [0, \avgdegree{}), [\avgdegree{}, \maxdegree{}] \\ \hline
        \multirow{1}{*}{\dbg}    &  \multirow{1}{*}{$\floor{log_{2}{\frac{\mathbb{M}}{\mathbb{C}}}} + 2$}  & [0, \constant), [2$^n$\constant, 2$^{n+1}$\constant) where n $\in$ \bigg[0, $\floor{log_{2}{\frac{\mathbb{M}}{\mathbb{C}}}}\bigg]$ \\ \hline
    \end{tabularx}
    \caption{Implementation of various skew-aware techniques using \dbg algorithm. 
    \avgdegree is the average and \maxdegree is the maximum degree of the dataset.
    For \dbg, \constant is some threshold such that 0 $<$ \constant $<$ \maxdegree. 
    \label{tab:dbg:spot-bin}}
\end{table}

Finally, we note that the \dbg algorithm (Listing~\ref{listing:dbg:spot-listing}) provides a general framework to understand trade-offs between reducing the cache footprint of hot vertices and preserving graph structure just by varying a number of groups and their degree ranges. Indeed, Table~\ref{tab:dbg:spot-bin} shows how different skew-aware techniques can be implemented using the \dbg algorithm. 
For example,
Hub Clustering can be viewed as an implementation of \dbg algorithm with two groups, one containing hot vertices and another one containing cold vertices.
Similarly,
Sort can be seen as an implementation of \dbg algorithm with as many number of groups as many unique
degrees a given dataset has. 
Consequently, for a given unique degree, the associated group contains all vertices having the same degree, effectively sorting vertices by their degree.
In general, as the number of groups is increased, the degree range gets narrower and vertex reordering gets finer, causing more disruption to existing structure.
Table~\ref{tabl:dbg:spot-cost} qualitatively compares \dbg to prior techniques.

\begin{table}[!b]
    \centering
    \small
    \begin{tabularx}{1\linewidth}
        {|>{\raggedright\arraybackslash\hsize=0.4\hsize}X| %
          >{\centering\arraybackslash\hsize=0.215\hsize}X|
          >{\centering\arraybackslash\hsize=0.2\hsize}X|
          >{\centering\arraybackslash\hsize=0.22\hsize}X|
        }
    \hline
    \multirow{2}{*}{Technique} &
    \multicolumn{1}{c|}{Structure} &
    \multicolumn{1}{c|}{Reordering} &
    \multicolumn{1}{c|}{Net} \\
    
    &
    \multicolumn{1}{c|}{Preservation} &
    \multicolumn{1}{c|}{Time} &
    \multicolumn{1}{c|}{Performance} \\ \hline
    \hline
       
    Sort &
    \xmark &
    \cmark &
    \cmark \\ \hline
    
    Hub Sorting~\cite{fc} &
    \cmark &
    \cmark &
    \cmark \\ \hline
    
    Hub Clustering~\cite{hubcluster} &
    \cmark\cmark &
    \cmark\cmark &
    \cmark  \\ \hline
    
    \rowcolor{lightgray}
    \dbg (proposed) &
    \cmark\cmark &
    \cmark\cmark &
    \cmark\cmark \\ \hline

    \gorder~\cite{gorder} &
    \cmark\cmark &
    \xmark &
    \xmark \\ \hline
 
    \end{tabularx}
    \caption{\label{tabl:dbg:spot-cost}Qualitative performance of different reordering techniques for graph analytics on natural graphs.}
\end{table}

\section{Methodology}
\label{sec:dbg:method}
\begin{table}[!t]
    \small
    \centering
    \begin{tabularx}{1\linewidth}
        {|>{\centering\arraybackslash\hsize=0.24\hsize}X| 
          >{\raggedright\arraybackslash\hsize=0.76\hsize}X|
        }
        \hline
        Application    & \multicolumn{1}{c|}{Brief Description} \\ \hline
        \hline
        {Betweenness Centrality ~~~~~~~~~~~~(BC)} & finds the most central vertices in a graph by using a BFS kernel to count the number of shortest paths passing through each vertex from a given root vertex. \\ \hline
        {Single Source Shortest Path (SSSP)} & computes shortest distance for vertices in a weighted graph from a given root vertex using the Bellman Ford algorithm. \\ \hline
        {PageRank ~~~~~~~~~~~~~~~~~~(PR)}  & is an iterative algorithm that calculates ranks of vertices based on the number and quality of incoming edges to them~\cite{pagerank}. \\ \hline
        {PageRank-Delta (PRD)} &is a faster variant of PageRank in which vertices are active in an iteration only if they have accumulated enough change in their PageRank score. \\ \hline %
        {Radii Estimation (Radii)} & estimates the radius of each vertex by performing multiple parallel BFS's from a small sample of vertices~\cite{radii}. \\ \hline
    \end{tabularx}
    \caption{\label{tab:dbg:graph-worklads}A list of evaluated graph applications.}
\end{table}
\begin{table}[!t]
    \centering
    \small
    \begin{tabularx}{1\linewidth}
        {|>{\raggedright\arraybackslash\hsize=0.15\hsize}X|
          >{\raggedright\arraybackslash\hsize=0.18\hsize}X| 
          >{\raggedleft\arraybackslash\hsize=0.17\hsize}X| 
          >{\raggedleft\arraybackslash\hsize=0.3\hsize}X| 
          >{\centering\arraybackslash\hsize=0.2\hsize}X|
        }
        \hline
         &  & \multicolumn{2}{c|}{Per-Vertex Property Size (Bytes)} & \centering{Degree}    \tabularnewline \cline{3-4}
        \centering{Graph Application} &\centering{Computation Type} & \centering{All Properties} & \centering{Only Properties with Irregular Accesses} & \centering{Type used for Reordering} \tabularnewline \hline \hline
        BC & pull-push & 17 & 8 & out \tabularnewline \hline 
        SSSP & push-only & 8 & 8 & in \tabularnewline \hline 
        PR & pull-only & 20 & 12 & out \tabularnewline \hline 
        PRD & push-only & 20 & 8 & in \tabularnewline \hline 
        Radii & pull-push & 20 & 8 & out \tabularnewline \hline 
    \end{tabularx}
    \caption{Properties of graph applications. 
    In addition to the vertex properties, all graph applications require 4 bytes to encode a vertex and 8 bytes to encode an edge. 
    }
    \label{tab:dbg:graph-workloads-properties}
\end{table}

\subsection{Graph Processing Framework, Applications and Datasets}
\label{sec:dbg:framework}

For the evaluation, we use \emph{Ligra}~\cite{ligra}, a widely used shared-memory graph processing framework that supports both pull- and push-based computations, including switching from pull to push (and vice versa) at the start of a new iteration. We evaluate various reordering techniques using five iterative graph applications listed in Table~\ref{tab:dbg:graph-worklads}, on eight graph datasets listed in Table~\ref{tab:dbg:datasets}, resulting in 40 datapoints for each technique.
Table~\ref{tab:dbg:graph-workloads-properties} lists various properties for the Ligra implementation of the evaluated graph applications.

\begin{table}[!t]
    \centering
    \small
    \begin{tabularx}{1\linewidth}
    {|>{\raggedright\arraybackslash\hsize=0.33\hsize}X| 
      >{\raggedleft\arraybackslash\hsize=0.1\hsize}X|
      >{\raggedleft\arraybackslash\hsize=0.11\hsize}X| 
      >{\raggedleft\arraybackslash\hsize=0.11\hsize}X| 
      >{\centering\arraybackslash\hsize=0.14\hsize}X| 
      >{\centering\arraybackslash\hsize=0.21\hsize}X| 
    }
    \hline
        { \centering \multirow{2}{*}{Dataset} }                              &  \centering Vertex    & \centering Edge   & \centering Avg.& \centering \multirow{2}{*}{Type}  & \centering Original \tabularnewline
        {  }                                                &  \centering Count     & \centering Count  & \centering Degree &                   & \centering Ordering  \tabularnewline
    \hline
    \hline 
        {Kron ({\em kr})~\cite{gap}}                        & \num{67}{M}           & \num{1323}{M}     & 20 & Synthetic            & Unstructured                   \tabularnewline \hline
        {PLD ({\em pl})~\cite{pld}}                         & \num{43}{M}           & \num{623}{M}      & 15 & Real              & Unstructured          \tabularnewline \hline
        {Twitter ({\em tw})~\cite{twitter}}                 & \num{62}{M}           & \num{1468}{M}     & 24 & Real              & Unstructured          \tabularnewline \hline
        {SD ({\em sd})~\cite{pld}}                          & \num{95}{M}           & \num{1937}{M}     & 20 & Real              & Unstructured          \tabularnewline \hline
        {LiveJournal ({\em lj})~\cite{snapnets}}            & \num{5}{M}            & \num{68}{M}       & 14 & Real              & Structured            \tabularnewline \hline
        {WikiLinks ({\em wl})~\cite{konect-wl}}       & \num{18}{M}           & \num{172}{M}      & 9 & Real              & Structured            \tabularnewline \hline
        {Friendster ({\em fr}) \cite{konect-friendster}}    & \num{64}{M}           & \num{2147}{M}     & 33 & Real              & Structured            \tabularnewline \hline
        {MPI ({\em mp})~\cite{twitter_mpi}}                 & \num{53}{M}           & \num{1963}{M}     & 37 & Real              & Structured            \tabularnewline \hline
    \end{tabularx}
    \caption{\label{tab:dbg:datasets}Properties of the evaluated graph datasets. 
    We empirically label those datasets as {\em sturctured} on which Random Reordering (RV) causes more than 25\% slowdown (Fig.~\ref{fig:dbg:random}).
    }
\end{table}
\begin{table}[!t]
    \centering
    \small
    \begin{tabularx}{1\linewidth}
    {|>{\raggedright\arraybackslash\hsize=0.52\hsize}X| 
      >{\raggedleft\arraybackslash\hsize=0.11\hsize}X|
      >{\raggedleft\arraybackslash\hsize=0.11\hsize}X| 
      >{\raggedleft\arraybackslash\hsize=0.11\hsize}X| 
      >{\centering\arraybackslash\hsize=0.15\hsize}X| 
    }
    \hline
        { \centering \multirow{2}{*}{Dataset} } &  \centering Vertex    & \centering Edge   & \centering Avg.& \centering \multirow{2}{*}{Type}  \tabularnewline
        &  \centering Count     & \centering Count  & \centering Degree & \tabularnewline
    \hline
    \hline 
        {Uniform ({\em uni})~\cite{PaRMAT}} & \num{50}{M} & \num{1000}{M} & 20.0 & Synthetic \tabularnewline \hline
        {USA Road Network ({\em road})~\cite{road}} & \num{24}{M} & \num{29}{M} & 1.2 & Real \tabularnewline \hline
    \end{tabularx}
    \caption{\label{tab:dbg:no-skew-datasets}Properties of the no-skew graph datasets. The {\em uni} dataset is generated using R-MAT~\cite{rmat} methodology with parameter values of A=B=C=25.
    }
\end{table}

We obtained the source code for the graph applications from Ligra~\cite{ligra}. Implementation of the graph applications is unchanged except for an addition of an array to keep a mapping between the vertex ID assignments before and after the reordering. The mapping is needed to ensure that root-dependent traversal applications running on the reordered graph datasets use the same root as the baseline execution running on the original graph dataset. We compile the applications using g++-6.4 with O3
optimization level on Ubuntu 14.04.1 booted with Linux kernel 4.4.0-96-lowlatency and use OpenMP for parallelization. To utilize memory bandwidth from both sockets, we run every application under NUMA interleave memory allocation policy.

\subsection{Evaluation Platform and Methodology\label{sec:dbg:soft-eval}}

Evaluation is done on a dual-socket server with two Broadwell based {\em Intel Xeon CPU E5-2630}~\cite{xeon}, each with 10 cores clocked at 2.2GHz and a 25MB shared LLC. Hyper-threading is kept on, exposing 40 hardware execution contexts across both CPUs. Server has 128GB of DRAM provided by eight DIMMs clocked at 2133MHz. Applications use 40 threads, and the threads are pinned to avoid performance variations due to OS scheduling. To further reduce
sources of performance variation, DVFS features are disabled. Finally, \emph{Transparent Huge Pages} is kept on to reduce TLB misses.

We evaluate each reordering technique on every combination of graph applications and graph datasets 11 times, and record the average runtime of 10 executions, excluding the timing of the first execution to allow the caches to warm up. We report the speed-up over the entire application runtime (with and without reordering cost) but exclude the graph loading time from the disk. For iterative applications, \pr{} and \prd{}, we run them until convergence and consider the aggregate runtime over all
iterations. For root-dependent traversal applications, \sssp{} and \bc{}, we run them from eight different root vertices for each input dataset and consider the aggregate runtime over all eight traversals. Finally, we note that the application runtime is relatively stable across executions. For each reported datapoint, coefficient of variation is at most 2.3\% for PRD and at most 1.6\% for other applications.

\subsection{Evaluated Reordering Techniques}

We evaluate \dbg and compare it with all three existing skew-aware techniques described in Sec.~\ref{sec:dbg:existing-skew-aware} (Sort, HubSort~\cite{fc} and HubCluster~\cite{hubcluster}) along with \gorder{}~\cite{gorder}, the state-of-the-art structure-aware reordering technique.

We use the source code available from \url{https://github.com/datourat/Gorder} for \gorder{}. As \gorder is only available in a single-thread implementation, while reporting the reordering time of \gorder for a given dataset, we optimistically divide the reordering time by 40 (maximum number of threads supported on the server) to provide a fair comparison with skew-aware techniques whose reordering implementation is fully parallelized.

For \dbg{}, we use 8 groups with the ranges [32\avgdegree, \infinity), [16\avgdegree, 32\avgdegree), [8\avgdegree, 16\avgdegree), [4\avgdegree, 8\avgdegree), [2\avgdegree, 4\avgdegree), [1\avgdegree, 2\avgdegree), [\avgdegree/2, \avgdegree) and [0, \avgdegree/2), where \avgdegree{} is the average degree of the graph dataset. Note that we also partition cold vertices into two groups. We developed a multi-threaded implementation of \dbg, which is available at \url{https://github.com/faldupriyank/dbg}.

\begin{figure}[!t]
    \centering
    \includegraphics{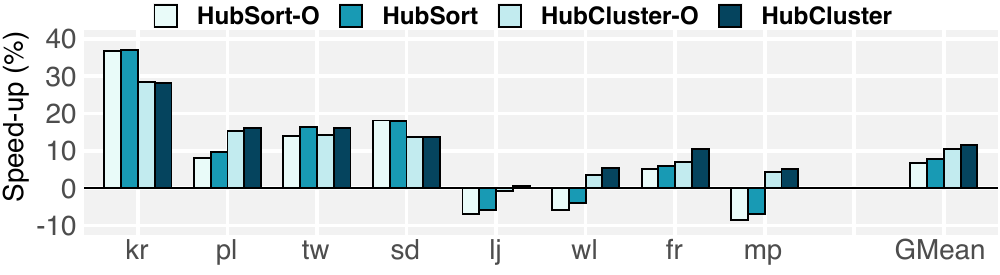}
    \caption{Application speed-up over the baseline with no reordering. 
    Techniques with suffix O use their original implementations whereas techniques without any suffix are implemented using \dbg algorithm as per Table~\ref{tab:dbg:spot-bin}.
    The bars for the datasets show geometric mean of speed-ups across five applications for a dataset.
    }
    \label{fig:dbg:hub-s-perf}
\end{figure}
\begin{table}[!t]
    \centering
    \small
    \begin{tabularx}{1\linewidth}
        {|>{\raggedright\arraybackslash\hsize=0.28\hsize}X| 
          >{\centering\arraybackslash\hsize=0.09\hsize}X|
          >{\centering\arraybackslash\hsize=0.09\hsize}X|
          >{\centering\arraybackslash\hsize=0.09\hsize}X|
          >{\centering\arraybackslash\hsize=0.09\hsize}X|
          >{\centering\arraybackslash\hsize=0.09\hsize}X|
          >{\centering\arraybackslash\hsize=0.09\hsize}X|
          >{\centering\arraybackslash\hsize=0.09\hsize}X|
          >{\centering\arraybackslash\hsize=0.09\hsize}X|
        }
        \hline
        Technique	    & \kr	& \pld	& \tw	& \sd	& \lj	& \wl	& \fr	& \mpi \tabularnewline \hline \hline
        HubSort-O	        &1.02	&1.04	&1.01	&1.02	&1.09	&0.79	&1.04	&1.01 \tabularnewline 
        \hline \rowcolor{lightgray}
        HubSort	    &0.80	&0.82	&0.84	&0.84	&0.87	&0.91	&0.90	&0.89 \tabularnewline \hline \hline
        HubCluster-O	    &0.78	&0.79	&0.81	&0.81	&0.78	&0.56	&0.88	&0.87 \tabularnewline 
        \hline \rowcolor{lightgray}
        HubCluster	&0.77	&0.74	&0.81	&0.78	&0.76	&0.81	&0.84	&0.82 \tabularnewline \hline
    \end{tabularx}
    \caption{Reordering time for existing skew-aware techniques, normalized to that of Sort. Lower is better.}
    \label{tab:dbg:reordering-time-norm}
\end{table}

Finally, we implement HubSort and HubCluster using the \dbg algorithm as shown in Table~\ref{tab:dbg:spot-bin}. We found our implementations to be more effective than the original implementations (referred to as HubSort-O and \mbox{HubCluster-O}) provided by the authors of HubCluster.
Fig.~\ref{fig:dbg:hub-s-perf} shows application speed-up over the baseline with no reordering. Table~\ref{tab:dbg:reordering-time-norm} shows reordering time normalized to that of Sort.  As our implementation of both techniques provides better speed-ups and lower reordering time, we use our implementations in the main evaluation.

\section{Evaluation\label{sec:dbg:eval}}

In this section, we evaluate the effectiveness of \dbg against the state-of-the-art reordering techniques. In Sec.~\ref{sec:dbg:sw-excluding-cost}, we compare the application speed-up for these techniques {\em without considering the reordering time}. In Sec.~\ref{sec:dbg:pull-dominated} and Sec.~\ref{sec:dbg:push-dominated}, we  analyze different levels of cache hierarchy to understand the sources of performance variation. Subsequently, to understand  the effect of the reordering time on end-to-end performance, we
compare the application speed-up {\em after accounting for the reordering time} in Sec.~\ref{sec:dbg:sw-including-cost}.

\subsection{Performance Excluding Reordering Time\label{sec:dbg:sw-excluding-cost}}

Fig.~\ref{fig:dbg:all-app-perf} shows application speed-up excluding reordering time for various datasets. Averaging across all 40 datapoints (combining all structured and unstructured), \dbg provides 16.8\% speed-up over the baseline with no reordering, outperforming all existing skew-aware techniques: Sort (8.4\%), HubSort (7.9\%) and HubCluster (11.6\%). \gorder, which comprehensively analyzes graph structure, yields 18.6\% average speed-up, marginally higher than that of \dbg. We
next analyze performance variations across datasets and applications.

\subsubsection{\textbf{\textit{Unstructured vs Structured}}}
As shown in Fig.~\stitchref{fig:dbg:all-app-perf}{fig:dbg:all-app-perf-unstructured}, on unstructured datasets, all reordering techniques provide positive speed-ups for all applications except for PRD.
Sec.~\ref{sec:dbg:push-dominated} explains the reasons for slowdowns for the PRD application.
Among skew-aware techniques, \dbg provides the highest average speed-up of 28.1\% in comparison to 22.1\% for Sort, 19.8\% for HubSort and 18.3\% for HubCluster.

\begin{figure}[!t]
    \centering
    \subfloat[\em Unstructured datasets.]{\label{fig:dbg:all-app-perf-unstructured}\includegraphics[width=1\linewidth]{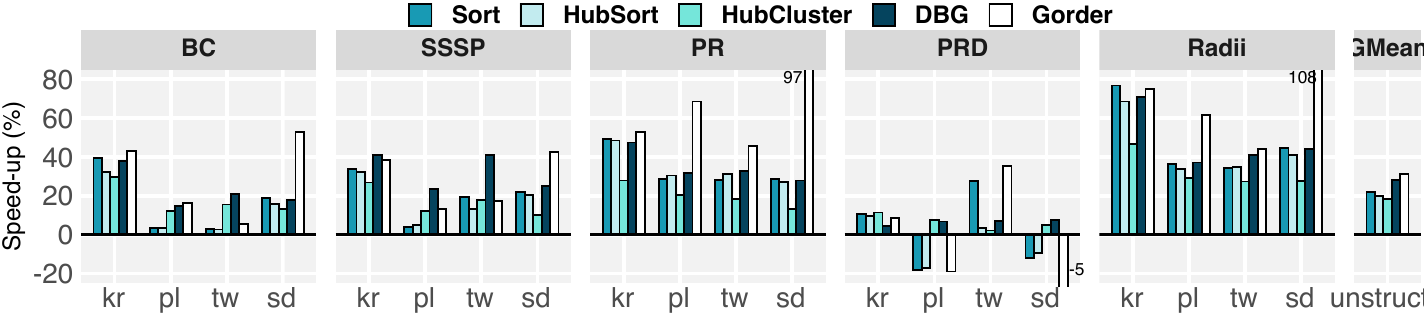}}\\
    \subfloat[\em Structured datasets.]{\label{fig:dbg:all-app-perf-structured}\includegraphics[width=1\linewidth]{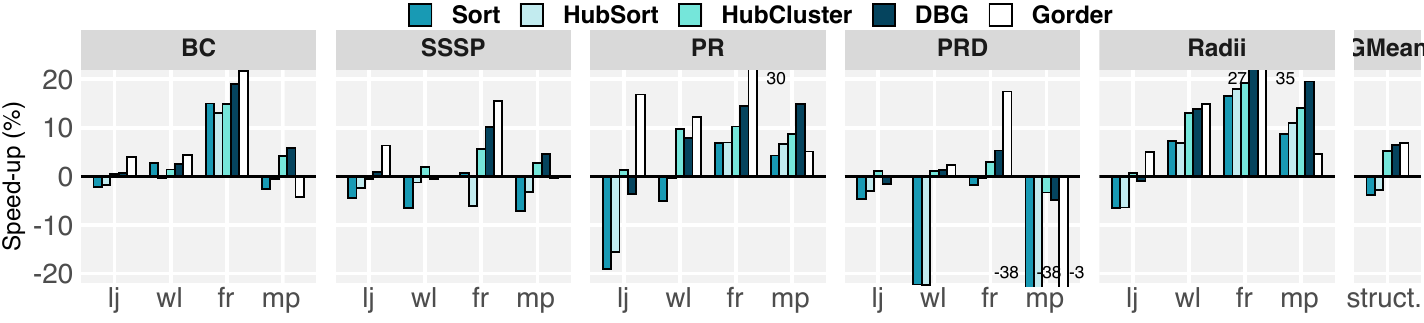}}
    \caption{\label{fig:dbg:all-app-perf}Application speed-up (excluding reordering time) for reordering techniques over the baseline with no reordering.}
\end{figure}

On synthetic dataset \kr, all techniques except HubCluster provide similar speed-ups as \kr is largely insensitive to structure preservation.
Similarly, on other unstructured datasets, as hot vertices are relatively more scattered in memory (see Table~\ref{tab:motivation:avg-hot-vertex} of Chapter~\ref{ch:motivation}), the benefit of vertex packing outweighs potential slowdown due to structure disruption. Thus, Sort, despite completely decimating the original graph structure, outperforms HubSort and HubCluster on more than half datapoints.
Meanwhile, \dbg, which also preserves graph structure while reducing the cache footprint of hot vertices, provides higher performance than Sort on more than half~datapoints.%

Overall, \dbg provides more than 30\% speed-up over the baseline on half datapoints.
\dbg outperforms or matches existing skew-aware techniques on nearly all datapoints.
Over the best performing prior skew-aware technique, \dbg provides the highest performance improvements on the SSSP application, with maximum speed-up of 18.0\% on the \tw dataset.

Structured datasets exhibit high spatio-temporal locality in their original ordering. Thus, any technique that does not preserve the graph structure is likely to yield only a marginal speed-up, if any. 
Among skew-aware techniques, \dbg provides the highest average speed-up of 6.5\% in comparison to -3.7\% for Sort, -2.8\% for HubSort and 5.3\% for HubCluster.

On structured datasets, performance gains from the reduction in footprint of hot vertices are negated by the disruption in graph structure.
Thus, Sort and HubSort, which preserve graph structure the least, cause slowdown (up to 38.4\%) on more than half datapoints. \dbg, in contrast, successfully avoids slowdown on almost all datapoints and causes a marginal slowdown (up to 4.9\%) only on 4 datapoints.

\subsubsection{\textbf{\textit{\dbg vs \gorder}}}
\gorder comprehensively analyzes vertex connectivity to improve cache locality whereas \dbg reorders vertices solely based on their degrees. Thus, it is expected for \gorder to outperform \dbg (and other skew-aware techniques). 
On average, \gorder yields a speed-up of 31.5\% (vs 28.1\% for \dbg) for unstructured datasets and 6.9\% (vs 6.5\% for \dbg) for structured datasets. 

Specifically, difference in speed-ups for \dbg and \gorder is very small for datasets \kr, \tw, \wl and \mpi. These datasets have relatively small clustering coefficient compared to other datasets~\cite{hats}, which makes it difficult for \gorder to approximate suitable vertex ordering. On other datasets, \gorder provides significantly higher speed-ups than any skew-aware technique. Problematically, \gorder incurs staggering reordering overhead, and thus causes severe slowdowns when accounted for its reordering time (see Sec.~\ref{sec:dbg:sw-including-cost}), making it impractical.

\subsubsection{\textbf{\textit{Reordering on No-Skew Graphs}}}
In this section, we evaluate the effect of reordering techniques on graph datasets that have no skew. Skew-aware techniques are not expected to provide significant speed-up for these datasets due to lack of skew in their degree distribution. More importantly, these techniques are also not expected to cause any significant slowdown due to a nearly complete lack of locality in the baseline ordering to begin with.

Fig.~\ref{fig:dbg:uni} shows speed-ups for reordering techniques on two datasets, {\em uni} and {\em road}, listed in Table~\ref{tab:dbg:no-skew-datasets}. As expected, all skew-aware techniques have a relatively neutral effect, with an average change in execution time within 1.2\% on the {\em uni} dataset and within {0.4\%} on the {\em road} dataset. Meanwhile, Gorder yields slightly more speed-up (3.5\% on both {\em uni} and {\em road} datasets), as it can exploit fine-grain spatio-temporal locality, which is not entirely skew dependent.

\begin{figure}
    \centering
    \includegraphics{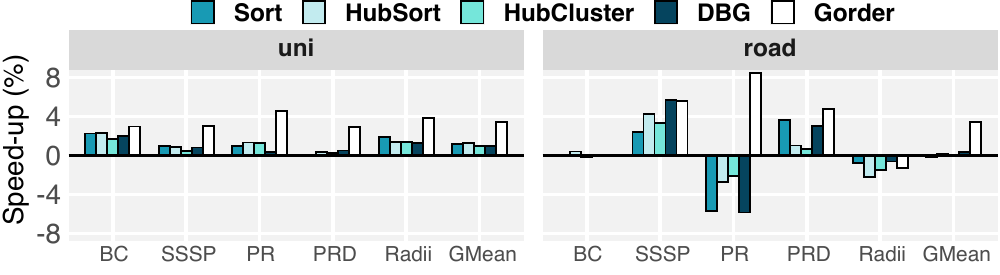}
    \caption{Effect of reordering techniques on graph datasets having no skew.}
    \label{fig:dbg:uni}
\end{figure}

\subsection{MPKI Across Cache Levels\label{sec:dbg:pull-dominated}}
In this section, we explain the sources of performance variations for different reordering techniques by analyzing their effects on all three levels of the cache hierarchy.
Fig.~\ref{fig:dbg:mpki-pr} plots {\em Misses Per Kilo Instructions (MPKI)} for L1, L2 and L3 cache, measured using hardware performance counters, for the PR application as a representative example.

In the baseline with the original ordering, on all datasets except \lj and \wl, L1 MPKI is more than 100 (\ie at least 1 L1 miss for every 10 instructions on average), which confirms the memory intensive nature of graph applications.
For the original ordering, L2 MPKI is only marginally lower than L1 MPKI across datasets, which shows that almost all memory accesses that miss in the L1 cache also miss in the L2 cache. As L3 cache is significantly larger than L2 cache, L3 MPKI is much lower than L2 MPKI;
nonetheless, L3 MPKI is very high for the original ordering, ranging from 56.2 to 82.9 across large datasets (excluding \lj and \wl).

While all skew-aware techniques target L3 cache, we observe that analyzing the effect of reordering on all three cache levels is necessary to understand application performance.
For example, for \wl{} dataset, Sort yields 5.5\% reduction in L3 MPKI over the baseline and yet causes a slowdown of 5.1\%. 
In fact, the slowdown is caused by 15.3\% and 19.6\% increase in L1 and L2 MPKI, respectively, over the baseline.

\begin{figure}[!t]
    \centering
    \subfloat{\label{fig:dbg:mpki-pr-l1}{\transparent{1}\includegraphics[width=1px,height=1px]{chapter_dbg/pdfs/transparent.png}}}
    \subfloat{\label{fig:dbg:mpki-pr-l2}\includegraphics[width=0.99\linewidth]{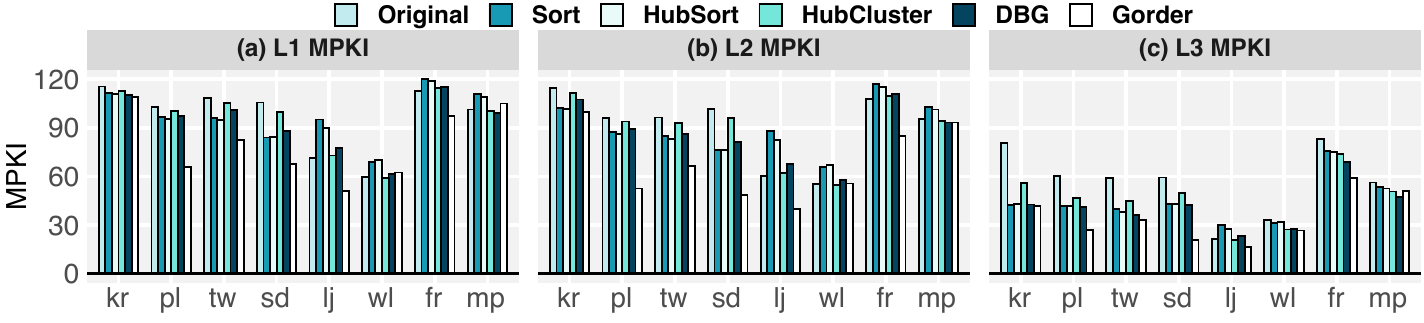}}
    \subfloat{\label{fig:dbg:mpki-pr-l3}{\transparent{1}\includegraphics[width=1px,height=1px]{chapter_dbg/pdfs/transparent.png}}}
    \caption{\label{fig:dbg:mpki-pr} Misses Per Kilo Instructions (MPKI) for the PR application across datasets. Lower is better.}
\end{figure}

All skew-aware techniques are generally effective in reducing L3 MPKI on all datasets but \lj. On unstructured datasets (the left-most four datasets), all skew-aware techniques reduce L1 and L2 MPKI, with the highest reduction on the \sd dataset. Meanwhile, on structured datasets (the right-most four datasets), Sort and HubSort, which do not preserve graph structure, significantly {\em increase} L1 and L2 MPKI (increase of 5.7 to 27.6 over original ordering). In contrast, HubCluster and
\dbg, which largely preserve existing structure, only marginally increase L1 and L2 MPKI (difference of -2.0 to 7.5) on structured datasets.

\subsection{Performance Analysis of Push-Dominated Applications\label{sec:dbg:push-dominated}}
\begin{figure}[!t]
    \centering
    \subfloat{\label{fig:dbg:push-false-sharing-orig}{\transparent{1}\includegraphics[width=1px,height=1px]{chapter_dbg/pdfs/transparent.png}}}
    \subfloat{\label{fig:dbg:push-false-sharing-spot}\includegraphics[width=0.995\linewidth]{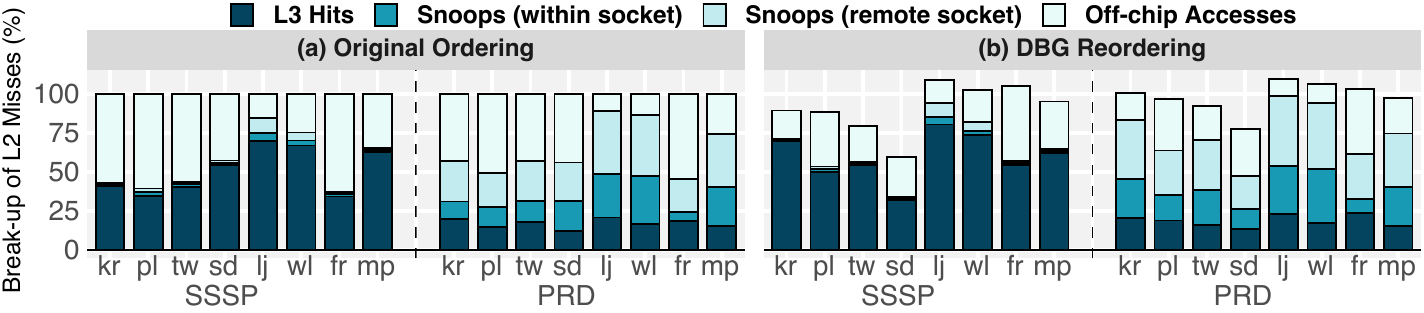}}
    \caption{\label{fig:dbg:push-false-sharing} Break-up of L2 misses for the push-dominated applications (SSSP and PRD) for datasets with original and \dbg ordering, normalized to the L2 misses of the original ordering.}
\end{figure}

As seen in Fig.~\ref{fig:dbg:all-app-perf}, all reordering techniques slowdown the PRD application on many datasets, the cause of which can be attributed to the push-based computation model employed by PRD. 
In push-based computations, when a vertex pushes an update through the out-edges, it generates scattered or irregular write accesses (as opposed to irregular read accesses in pull-based computations). As different threads may concurrently update the same vertex (true sharing) or update different vertices in the same cache block (false sharing), the push-based model leads to read-write or write-write sharing, hence generating on-chip coherence traffic.

Fig.~\ref{fig:dbg:push-false-sharing} quantifies coherence traffic on both push-dominated applications, SSSP and PRD.
The figure shows the break-up of L2 misses into four categories -- L3 Hits (served by L3 without requiring any snoops to other cores), Snoops to other cores within the same socket, Snoops to another socket and off-chip accesses. 
For the first three categories, data is served by an on-chip cache whereas for the last category, data is served from the main memory.

The two push-dominated applications have strikingly different fraction of coherence traffic while processing the datasets with the original ordering (middle two stacked bars in Fig.~\stitchref{fig:dbg:push-false-sharing}{fig:dbg:push-false-sharing-orig}). For SSSP, a relatively small fraction of L2 misses (14.5\% for \lj{} and below 9\% for other datasets) required snoops whereas for PRD, a considerable fraction of L2 misses (from 26.9\% for \fr{} to 69.4\% for \wl{}) required snoops.
 
While processing a vertex using push-based computations, an application pushes updates (writes) to some or all destination vertices of the out-edges. In the case of PRD, it unconditionally pushes an update (\ie a PageRank score) to all destination vertices while processing a vertex. In contrast, SSSP pushes an update to an out-edge only if it finds a shorter path through that edge. Thus, SSSP has much fewer number of irregular writes, and in turn, less coherence traffic, in comparison to PRD.

Fig.~\stitchref{fig:dbg:push-false-sharing}{fig:dbg:push-false-sharing-spot} shows a similar break-up for SSSP and PRD on the datasets after \dbg reordering.
For PRD, \dbg consistently reduces off-chip accesses (top stacked bar) across datasets, thus, a significantly higher fraction of requests are served by on-chip caches. However, most of these requests (37.8\% to 77.0\% of L2 misses) incur a snoop latency. For example, for \dbg, while processing the \pl{} dataset, 65.4\% (vs 49.2\% for the original ordering) of L2 misses are served by on-chip caches (bottom three stacked bars combined). However, most of these on-chip hits required snooping to other cores, incurring high access
latency. Specifically, only 18.9\% (vs 14.8\% for the original ordering) of total L2 misses are served without requiring snooping. For most of the datasets, increase in L3 hits (\ie no snooping) due to \dbg is relatively small despite a significant reduction in off-chip accesses, which explains the marginal speed-up for \dbg for the PRD application (Fig.~\ref{fig:dbg:all-app-perf}).

For SSSP, most of the savings in off-chip accesses directly translate to L3 hits (\ie no snooping) as the application does not exhibit high amount of coherence traffic even in the baseline. Thus, \dbg is highly effective on SSSP, despite being dominated by push-based computations.

\begin{figure}[!t]
    \centering
    \includegraphics[width=1\linewidth]{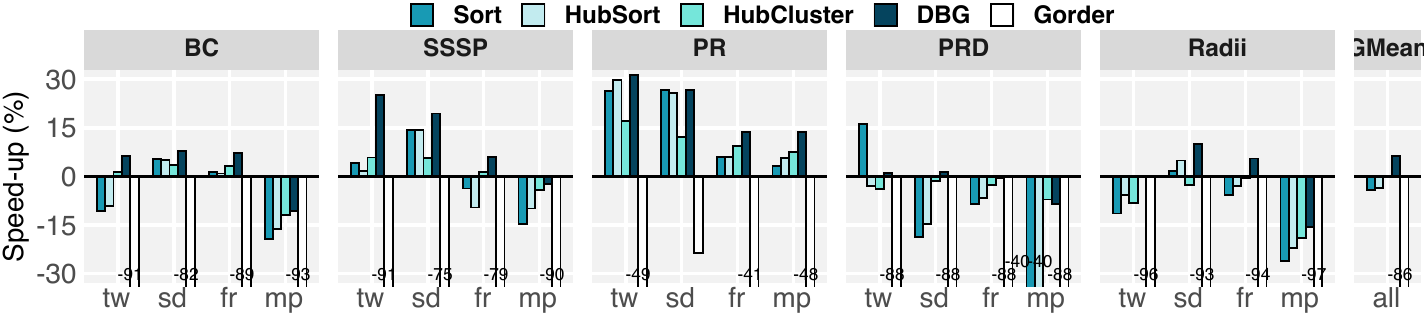}
    \caption{\label{fig:dbg:all-cost-perf}Net speed-up for software reordering techniques over the baseline with original ordering of datasets. GMean shows geometric mean across speed-ups for all five applications on four datasets.}
\end{figure}

\subsection{Performance Including Reordering Time\label{sec:dbg:sw-including-cost}}

Fig.~\ref{fig:dbg:all-cost-perf} shows end-to-end application speed-up for different reordering techniques after accounting for the reordering time. Without loss of generality, we show four datasets (two largest unstructured and two largest structured datasets). 

\gorder{}, while more effective at improving application speed-up (Fig.~\ref{fig:dbg:all-app-perf}), when accounted for its reordering time, causes severe slowdowns (up to 96.5\%) across datasets, corroborating prior work~\cite{hubcluster}. In contrast, all skew-aware techniques provide a net speed-up on at least some datapoints.

\dbg outperforms all prior techniques on 17 out of 20 datapoints. \dbg provides a net speed-up (up to 31.4\%) on 14 out of 20 datapoints, even after accounting for its reordering time. On the remaining 6 datapoints, \dbg reduces slowdown when compared to prior techniques, with maximum slowdown of 15.6\% for the Radii application on the \mpi dataset and below 10\% for others.
In contrast, existing skew-aware techniques cause slowdown of up to 40.2\% on half datapoints. Overall, DBG is the only technique that yields an average net speed-up (6.2\%) by providing high performance while incurring low reordering overhead.

We next study how 
long it takes to amortize the reordering cost for an iterative application (PR) and a root-dependent traversal application (SSSP).

\subsubsection{Amortization Point for PR}
\begin{table}[!t]
    \centering
    \small
    \begin{tabularx}{0.7\linewidth}
        {|>{\centering\arraybackslash\hsize=0.15\hsize}X| 
          >{\centering\arraybackslash\hsize=0.16\hsize}X|
          >{\centering\arraybackslash\hsize=0.16\hsize}X|
          >{\centering\arraybackslash\hsize=0.21\hsize}X|
          >{\centering\arraybackslash\hsize=0.16\hsize\columncolor{lightgray}}X|
          >{\centering\arraybackslash\hsize=0.16\hsize}X|
        }
       
        \hline
            Dataset & \centering Sort    & \centering HubSort   & \centering HubCluster    & \centering \dbg    & \centering \gorder \tabularnewline \hline \hline
            \tw  &  3.3  & 2.4  & 3.5  & 1.9  & 258.6 \tabularnewline \hline
            \sd  &  3.7  & 3.0  & 5.0  & 2.4  & 112.2 \tabularnewline \hline
            \fr  &  8.6  & 7.4  & 4.7  & 3.2  & 254.9 \tabularnewline \hline
            \mpi  &  18.2  &    10.3  &    7.5  & 4.4  & 1359.4 \tabularnewline \hline

    \end{tabularx}
    \caption{Minimum number of iterations needed for the PR application to amortize the reordering time of different reordering techniques.}
    \label{tab:dbg:min-iter-amortize}
\end{table}
The PR application has the largest runtime among all five applications for any given dataset, thus all skew-aware techniques are highly effective for the PR application and yield a net speed-up on all four datasets. 
Averaging across four datasets for the PR application, \dbg outperforms all reordering techniques with 21.2\% speed-up vs 15.1\% for Sort, 16.3\% for HubSort, 11.6\% for HubCluster and -41.3\% for~\gorder. %

Table~\ref{tab:dbg:min-iter-amortize} lists the minimum number of iterations needed for the PR application to amortize the cost of different reordering techniques. For all four datasets, \dbg is quickest in amortizing its reordering time, providing a net speed-up for all four datasets after just 2-5 iterations. 

\subsubsection{Amortization Point for SSSP}

We now evaluate net performance sensitivity to the number of successive graph traversals for different techniques for the SSSP application.
The runtime for root-dependent applications depends on the number of traversals (or queries) performed from different roots. The exact number of traversals required depends on the specific use case. Thus, we perform a sensitivity analysis by varying the number of traversals from 1 to 32 in multiples of 8. 

As shown in Fig.~\ref{fig:dbg:net-cost-sssp}, with the increase in the number of traversals, performance for each technique also increases, as the reordering needs to be applied only once and its cost is amortized over multiple graph traversals. Thus, a single traversal is the worst-case scenario, with all techniques causing slowdown due to their inability to amortize the reordering cost. Of all the techniques, DBG causes the minimum slowdown (20.6\% on average vs 27.7\% for the next best) and is the quickest in amortizing the reordering cost, providing an average speed-up of 11.5\% (vs 2.1\% for the next best) with as few as 8 graph traversals.

\begin{figure}[!t]
    \centering
    \includegraphics[width=1\linewidth]{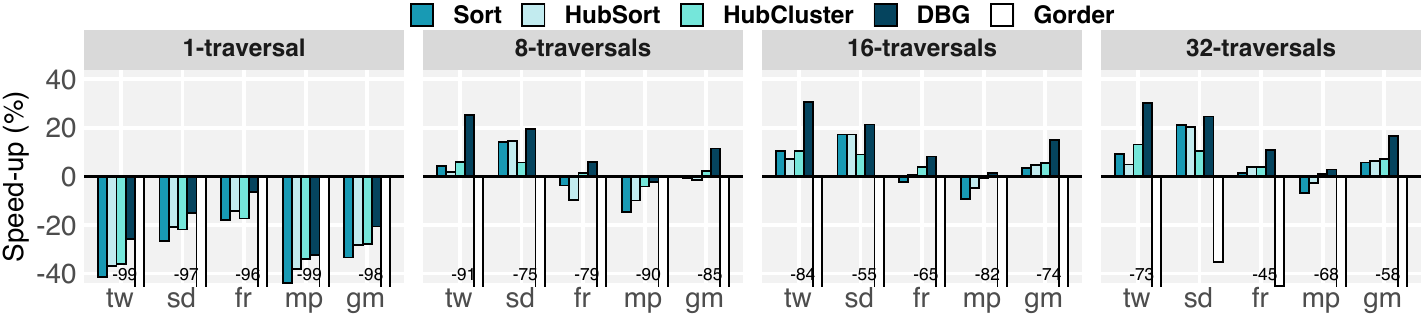}
    \caption{\label{fig:dbg:net-cost-sssp}Net speed-up for reordering techniques over the baseline with no reordering for SSSP with different number of traversals.}
\end{figure}

\section{Related Work}
\label{sec:dbg:related}

A significant amount of research has focused on designing high performance software frameworks for graph applications (e.g., ~\cite{ligra, galois,gap, graphmat, graphlab, graphchi}). In this section, we highlight the most relevant works that focus on improving cache efficiency for graph applications. %

\noindenttitle{Graph slicing:}
Researchers have proposed graph slicing that slices the graph in LLC-size partitions and processes one partition at a time to nullify the effect of irregular memory accesses~\cite{fc,graphicionado,graphmat}.
While generally effective, slicing has two important limitations. First, it requires invasive framework changes to form the slices (which may include replicating vertices to avoid out-of-slice accesses) and manage them at runtime. Secondly, for a given cache size, the number of slices increases with the size of the graph, resulting in greater processing overheads in creating and maintaining partitions for larger graphs.
In comparison, \dbg only requires a preprocessing pass over the graph dataset to relabel vertex IDs and does not require any change in the graph algorithms.

\noindenttitle{Traversal scheduling:}
Mukkara \etal proposed Bounded Depth-First Scheduling (BDFS) to exploit cache locality for graphs exhibiting community structure~\cite{hats}.
Problematically, the software implementation of BDFS introduces significant book-keeping overheads, causing slowdowns despite improving cache efficiency. To avoid software overheads, the authors propose an accelerator that implements BDFS scheduling in hardware. In comparison, \dbg is a software technique that can improve application performance without any additional hardware~support.

\section{Conclusion}
\label{sec:dbg:conclusion}

In this chapter, we studied existing skew-aware reordering techniques that seek to improve cache efficiency for graph analytics by reducing the cache footprint of hot vertices. 
We demonstrated the inherent tension between reducing the cache footprint of hot vertices and preserving original graph structure, which limits the effectiveness of existing skew-aware reordering techniques. In response, we proposed Degree-Based Grouping (DBG), a lightweight vertex reordering software technique that employs
coarse-grain reordering to preserve graph structure while reducing the cache footprint of hot vertices. 
On a variety of graph applications and datasets, DBG achieves higher average performance than all existing skew-aware techniques and nearly matches the average performance of the state-of-the-art complex reordering technique.

\chapter{GRASP -- Domain-Specialized Cache Management\label{ch:grasp}}

\section{Introduction}
\label{sec:grasp:intro}

Almost all prior works on hardware cache management targeting cache thrashing are {\em domain-agnostic}~\citeall{}. 
These hardware techniques aim to perform two tasks: (1) identify cache blocks that are likely to exhibit high reuse, and (2) protect high reuse cache blocks from cache thrashing. To accomplish the first task, these techniques deploy either probabilistic or prediction-based hardware mechanisms~\cite{dip,rrip,counter,sampler,ship,mdpp,hawkeye,perceptron,harmony}. However, as we showed in Chapter~\ref{ch:motivation}, graph-dependent irregular access patterns, combined with long reuse distance of accesses, prevent these techniques from correctly learning which cache blocks to preserve, rendering them deficient for the broad domain of graph analytics. Meanwhile, to accomplish the second task, recent work proposed \emph{pinning} of high-reuse cache blocks in LLC to ensure that these blocks are not evicted~\cite{xmem}. However, we find that pinning-based techniques are overly rigid and result in sub-optimal utilization of cache capacity.

To overcome the limitations of existing hardware cache \mbox{management} techniques, we propose \samacronym cache management at the LLC. 
To the best of our knowledge, this is the first work to introduce domain-specialized cache management for the domain of graph analytics. 
\sam augments existing cache insertion and hit-promotion policies to provide preferential treatment to the cache blocks containing hot vertices to shield them from thrashing.
To cater to the irregular access patterns, \sam policies are designed to be flexible to cache other blocks exhibiting reuse.
By not relying on pinning, \sam maximizes cache efficiency based on observed access patterns.

\sam relies on lightweight software support to \mbox{accurately} pinpoint hot vertices amidst irregular access patterns, in contrast to history-based predictive techniques that rely on storage-intensive hardware mechanisms. By leveraging vertex reordering techniques such as DBG, \sam enables a lightweight software-hardware interface comprising of only a few configurable registers, which are programmed by software using its knowledge of the graph data~structures.

\sam requires minimal changes to the existing microarchitecture as \sam only augments existing cache policies and its interface is lightweight.
\sam does not require additional metadata in the LLC or storage-intensive prediction tables. 
Thus, \sam can easily be integrated into commodity server processors, enabling domain-specific acceleration for graph analytics at minimal hardware cost.

\noindentnotitle
To summarize, our contributions are as follows:

\begin{itemize}
    \item We qualitatively and quantitatively show that a wide range of prior domain-agnostic hardware cache management techniques, despite their sophisticated prediction mechanisms, are inefficient for the domain of graph analytics.

    \item We propose {\em \samnospace}, graph-specialized LLC management for graph analytics on natural graphs. \sam augments existing cache policies to protect hot vertices against thrashing while also maintaining flexibility to capture reuse in other cache blocks. \sam employs a lightweight software interface to pinpoint hot vertices amidst irregular accesses, which eliminates the need for metedata storage at the LLC, keeping the existing cache structure largely~unchanged.

    \item Our evaluation on several multi-threaded graph applications operating on large, high-skew datasets shows that \sam outperforms state-of-the-art domain-agnostic techniques on all datapoints, yielding an average speed-up of 4.2\% (max 9.4\%) over the best-performing prior technique.
    \sam is also robust on low-/no-skew datasets whereas prior techniques consistently cause a slowdown.

\end{itemize}

\section{\sam: Caching In on the Skew}
\label{sec:grasp:graphinitydesign}

This chapter introduces \sam, graph-specialized cache management at LLC for graph analytics processing natural graphs. 
\sam augments existing cache management techniques with simple modifications to their insertion and hit-promotion policies that provide preferential treatment to the cache blocks containing hot vertices to protect them from thrashing. 
\sam policies are sufficiently flexible to capture reuse of other blocks as needed.

\samnospace's domain-specialized design is influenced by the following two challenges faced by existing hardware cache management techniques. First, hardware alone cannot enforce spatial locality, which is dictated by vertex placement in the memory space and is under software control. Second, domain-agnostic hardware cache management techniques struggle in pinpointing hot vertices under cache thrashing due to irregular access patterns endemic of graph analytics. 

\begin{figure}[!t]
{
    \centering
    \subfloat{\label{fig:grasp:graphinity-a}{\transparent{1}\includegraphics[width=1px,height=1px]{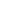}}}
    \subfloat{\label{fig:grasp:graphinity-b}{\transparent{1}\includegraphics[width=1px,height=1px]{chapter_grasp/pdfs/transparent.png}}}
    \subfloat{\label{fig:grasp:graphinity-c}{\includegraphics[width=0.8\linewidth]{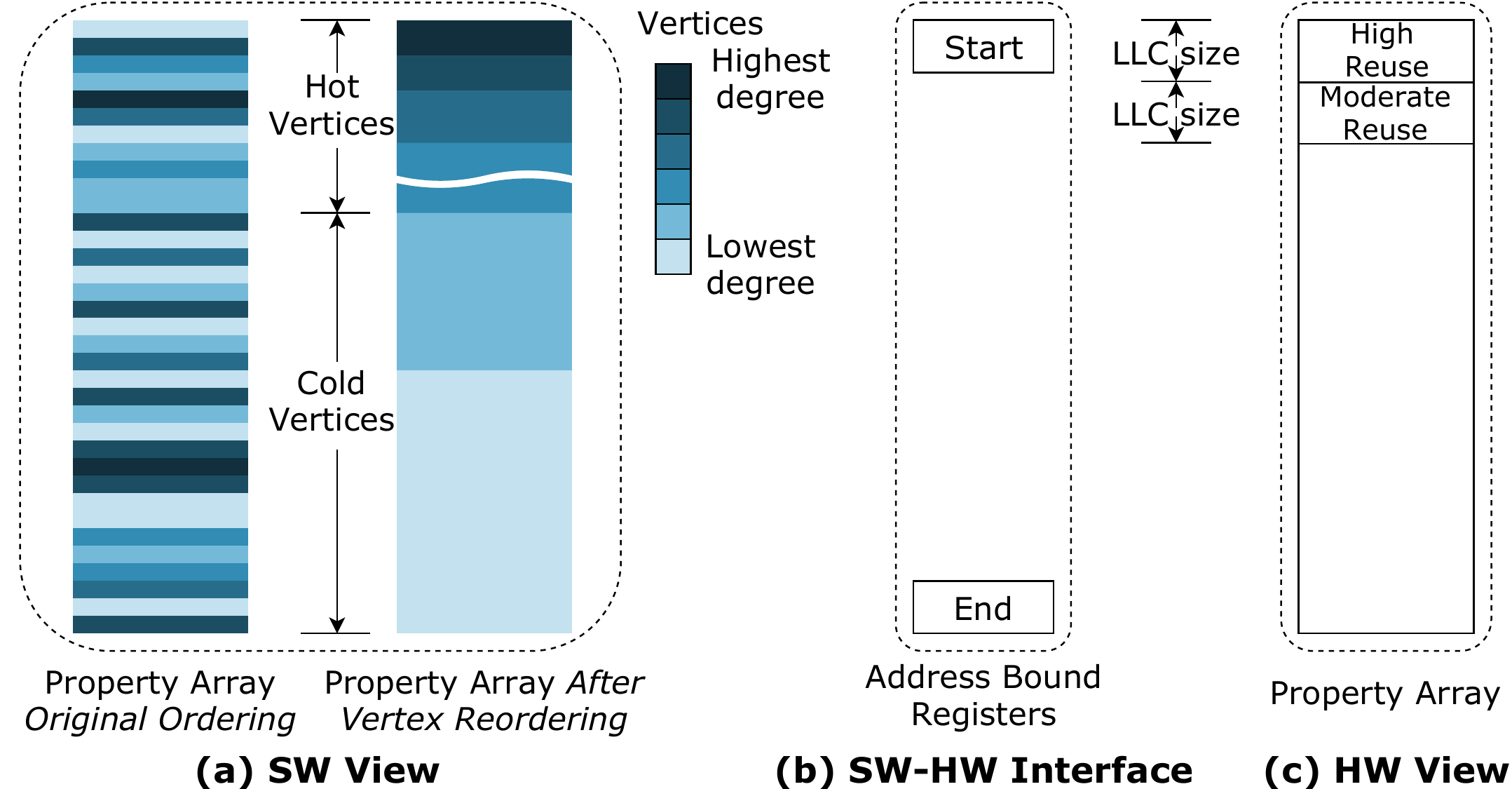}}}
    \caption{\label{fig:grasp:graphinity}\sam overview. (a) Software applies vertex reordering, which segregates hot vertices at the beginning of the array. (b) \sam interface exposes an ABR pair per Property Array to be configured with the bounds of the array. (c) \sam identifies regions exhibiting different reuse based on an LLC size.}
}
\end{figure}

To overcome both challenges, \sam relies on skew-aware reordering techniques to induce spatial locality by segregating hot vertices in a contiguous memory region. While these techniques offer different trade-offs in terms of reordering cost and their ability to preserve graph structure, they all work by isolating hot vertices from the cold ones. Fig.~\stitchref{fig:grasp:graphinity}{fig:grasp:graphinity-a} shows a logical view of the placement of hot vertices in the Property Array after reordering by such a technique. 
\sam subsequently leverages the contiguity among hot vertices in the memory space to (1) pinpoint them via a lightweight interface and (2) protect them from thrashing. \sam design consists of three hardware components as follows.

\noindenttitle{\smalltextcircled{A} Software-hardware interface:} \sam interface is minimal, consisting of a few configurable registers that software populates with the bounds of the Property Array during the initialization of an application (see Fig.~\stitchref{fig:grasp:graphinity}{fig:grasp:graphinity-b}). Once populated, \sam does not rely on any further intervention from software. 

\noindenttitle{\smalltextcircled{B} Classification logic:} \sam logically partitions the Property Array into different regions  based on expected reuse.
(See Fig.~\stitchref{fig:grasp:graphinity}{fig:grasp:graphinity-c}). \sam implements simple comparison-based logic, which, at runtime, checks whether a cache request belongs to any one of these regions.

\noindenttitle{\smalltextcircled{C} Specialized cache policies:} \sam specializes cache policies for each region to ensure hot vertices are protected from thrashing while retaining flexibility in caching other blocks. The classification logic guides which policy to apply to a given cache block.

\noindentnotitle
Fig.~\ref{fig:grasp:sam-block-diagram} shows how \sam interacts with other hardware components in the system. In the following sections, we describe each of \samnospace's components in detail.

\begin{figure}[!t]
{
    \centering
    \includegraphics[width=0.8\linewidth]{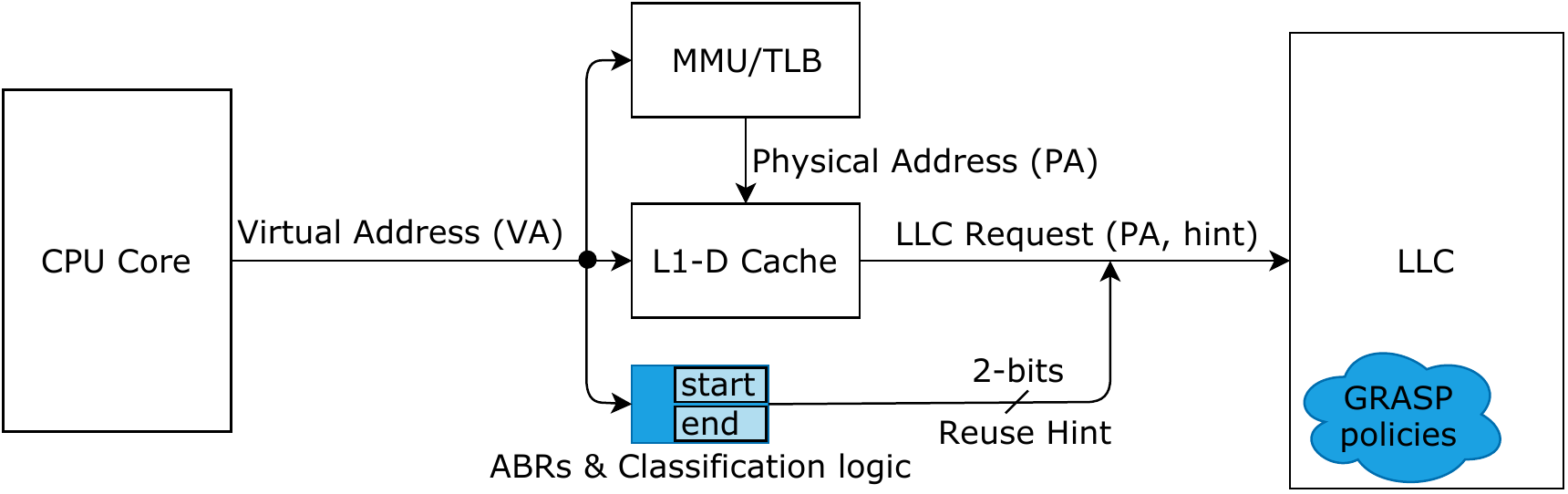}%
	\caption{\label{fig:grasp:sam-block-diagram}Block diagram of \sam and other hardware components with which it interacts. \sam components are shown in color. For brevity, the figure shows only one CPU core.}
}
\end{figure}

\subsection{Software-Hardware Interface\label{sec:grasp:interface}}

\samnospace's interface consists of one pair of \emph{Address Bound Registers (ABR)} per Property Array; recall from Sec.~\ref{sec:motivation:graph-processing-basics} of Chapter~\ref{ch:motivation} that an application may maintain more than one Property Array, each of which requires a dedicated ABR pair. ABRs are part of an application context and are exposed to the software.
At application start-up, the graph framework populates each ABR pair with the start and end virtual address of the entire Property Array (Fig.~\stitchref{fig:grasp:graphinity}{fig:grasp:graphinity-b}). 
Setting these registers activates the custom cache management for graph analytics. 
When the ABRs are not set by the software (\ie the default case for other applications), specialized cache management is essentially disabled.

The use of virtual addresses keeps the GRASP interface independent of the existing TLB design, allowing GRASP to perform address classification (described next) in parallel with the usual virtual-to-physical address translation carried out by TLB (see Fig.~\ref{fig:grasp:sam-block-diagram}).
Prior works have used similar approaches to pass data-structure bounds to aid microarchitecture mechanisms~\cite{prefetch-data-structure,barren,whirlpool,xmem}.

\subsection{Classification Logic}

This component of \sam is responsible for reliably identifying cache blocks containing hot vertices in hardware by leveraging the bounds of the Property Array(s) available in the ABRs as explained in the following sections:

\noindenttitle{Identifying hot vertices: } In theory, all hot vertices should be cached. In practice, it is unlikely that all hot vertices will fit in the LLC for large datasets as shown in Table~\ref{tab:dbg:fp-hot-vertices} of Chapter~\ref{ch:dbg}.
In such a case, providing preferential treatment to all hot vertices is {\em not} beneficial as they can thrash each other in the LLC.
To avoid this problem, \sam prioritizes cache blocks containing only a subset of hot vertices, comprised of only the hottest vertices based on the available LLC capacity. Conveniently, the hottest vertices are located at the beginning of the  Property Array in a contiguous region thanks to the application of skew-aware reordering as shown in Fig.~\stitchref{fig:grasp:graphinity}{fig:grasp:graphinity-a}.

\noindenttitle{Pinpointing the High Reuse Region: }
\sam labels two LLC-sized sub-regions within the Property Array: 
The LLC-sized memory region at the start of the Property Array is labeled as \emph{High Reuse Region}; another LLC-sized memory region starting immediately after the High Reuse Region is labeled as the \emph{Moderate Reuse Region} (Fig.~\stitchref{fig:grasp:graphinity}{fig:grasp:graphinity-c}). Finally, if an application specifies more than one Property Array, \sam divides LLC-size by the number of Property Arrays before labeling the~regions.

\noindenttitle{Classifying LLC accesses: }
At runtime, \sam classifies a memory address making an LLC access as \emph{High-Reuse} if the address belongs to the {High Reuse Region} of any Property Array; \sam determines this by comparing the address with the bounds of the High Reuse Region of each Property Array. Similarly, an address is classified as \emph{Moderate-Reuse} if the address belongs to the {Moderate Reuse Region}.
All other LLC accesses are classified as \emph{Low-Reuse}. 
For non-graph applications, the ABRs are not initialized and all accesses are classified as \emph{Default}, effectively disabling domain-specialized cache management. \sam encodes the classification result (High-Reuse, Moderate-Reuse, Low-Reuse or Default) as a 2-bit \emph{Reuse Hint}, and forwards it to the LLC along with each cache request, as shown in Fig.~\ref{fig:grasp:sam-block-diagram}, to guide specialized insertion and hit-promotion policies as described next.

\subsection{Specialized Cache Policies\label{sec:grasp:sam}}%

This component of \sam implements specialized cache policies that protect the cache blocks associated with High-Reuse LLC accesses against thrashing. One naive way of doing so is to pin the High-Reuse cache blocks in the LLC. However, pinning would sacrifice any opportunity in exploiting temporal reuse that may be exposed by other cache blocks (\eg Moderate-Reuse cache blocks).

To overcome this challenge, \sam adopts a flexible approach by augmenting an existing cache replacement policy with a specialized insertion policy for LLC misses and a hit-promotion policy for LLC hits. \samnospace's specialized policies provide preferential treatment to High-Reuse blocks while maintaining flexibility in exploiting temporal reuse in other cache blocks, as discussed next.

\noindenttitle{Insertion policy:} Accesses tagged as {High-Reuse}, comprising the set of the hottest vertices belonging to the High Reuse Region, are inserted in the cache at the MRU position to protect them from thrashing. 
Accesses tagged as {Moderate-Reuse}, likely exhibiting lower reuse when compared to the High-Reuse region, are inserted near the LRU position. Such insertion policy allows Moderate-Reuse cache blocks an opportunity to experience a hit without causing thrashing. Finally, accesses tagged as {Low-Reuse}, comprising the rest of the graph dataset, including the long tail of the Property Array containing cold vertices, are inserted at the LRU position, thus making them immediate candidates for replacement while still providing them with an opportunity to experience a hit and be promoted using the specialized policy described~next. 

\noindenttitle{Hit-promotion policy:} Cache blocks associated with {High-Reuse} LLC accesses are immediately promoted to the MRU position on a hit to protect them from thrashing. LLC hits to blocks classified as {Moderate-Reuse} or {Low-Reuse} make for an interesting case. 
On the one hand, the likelihood of these blocks having further reuse is quite limited, which means they should not be promoted directly to the MRU position. On the other hand, by experiencing at least one hit, these blocks have demonstrated temporal locality, which cannot be completely ignored. \sam takes a middle ground for such blocks by gradually promoting them towards MRU position on every hit.

\noindenttitle{Eviction policy:} \samnospace's eviction policy does not differentiate among blocks at replacement time; hence, it is unmodified from the baseline technique. This is a key factor that keeps the cache management flexible for \samnospace. By not prioritizing candidates for eviction, \sam ensures that blocks classified as {High-Reuse} but not referenced for a long time can yield cache space to other blocks that do exhibit reuse. Because the unchanged eviction policy does not need to differentiate between blocks with High-Reuse and other hints, cache blocks do \emph{not} need to explicitly store the Reuse Hint as additional LLC metadata.

\visiblespace

\begin{table}[!t]
    \small
    \centering
    \begin{tabularx}{0.8\linewidth}
    {|>{\raggedright\arraybackslash\hsize=0.30\hsize}X| 
      >{\centering\arraybackslash\hsize=0.35\hsize}X|
      >{\centering\arraybackslash\hsize=0.35\hsize}X|
    }
        \hline
        Reuse Hint & \multicolumn{1}{c|}{Insertion Policy} & \multicolumn{1}{c|}{Hit Policy} \\ \hline
        \hline
        High-Reuse  & RRPV = 0  & RRPV = 0    \\ \hline
        Moderate-Reuse & RRPV = 6 & if RRPV $>$ 0:    \\ \cline{1-2}
        Low-Reuse   & RRPV = 7 & \multicolumn{1}{c|}{RRPV - -}    \\ \hline
        Default   & RRPV = 6 or 7 & RRPV = 0    \\ \hline
    \end{tabularx}
    \caption{\label{tab:grasp:sam-rrip}Policy columns show how \sam updates per-block 3-bit RRPV counter of RRIP (base technique) for a given Reuse Hint. Higher RRPV value indicates higher eviction priority.}
\end{table}

Table~\ref{tab:grasp:sam-rrip} shows the specialized cache policies for all Reuse Hints under \samnospace. While the table, and our evaluation, assumes RRIP~\cite{rrip} as the base replacement technique, we note that \sam is not fundamentally dependent on RRIP and can be implemented over many other techniques including, but not limited to, LRU, Pseudo-LRU and DIP~\cite{dip}. %

\subsection{Benefits of \sam over Prior Techniques}
The state-of-the-art history-based predictive techniques \cite{sampler,mdpp,ship,perceptron,hawkeye,harmony} require intrusive modifications to the cache structure in form of embedded metadata in cache blocks and/or dedicated predictor tables. These techniques also require propagating a PC signature through the core pipeline all the way to the LLC, which so far has hindered their commercial adoption. 

In comparison, \sam is implemented within the same hardware structure required by the base technique (e.g., RRIP). 
\sam propagates only a \mbox{2-bit} {Reuse Hint} to the LLC on each cache access to guide cache policy decisions. %
By relying on lightweight software support, \sam reliably pinpoints hot vertices in hardware without requiring costly prediction tables and/or additional per-cache-block metadata. 

When compared to pinning-based techniques, \sam policies protect hot vertices from thrashing while remaining flexible to capture reuse of other blocks as needed.
Combining robust cache policies with minimal hardware modifications makes \sam feasible for commercial adoption while also providing higher LLC~efficiency.

\section{Methodology}
\label{sec:grasp:method}

\begin{table}[!t]
    \centering
    \small
    \begin{tabularx}{0.8\linewidth}
    {|>{\raggedright\arraybackslash\hsize=0.4\hsize}X| 
      >{\raggedleft\arraybackslash\hsize=0.2\hsize}X|
      >{\raggedleft\arraybackslash\hsize=0.2\hsize}X| 
      >{\raggedleft\arraybackslash\hsize=0.2\hsize}X| 
    }
    \hline
    { Dataset } & 
    \multicolumn{1}{c|}{Vertex Count} & 
    \multicolumn{1}{c|}{Edge Count}  & 
    \multicolumn{1}{c|}{Avg. Degree} \\

    \hline
    \hline
    { LiveJournal ({\lj})~\cite{snapnets}}
    & \num{5}{M}
    & \num{68}{M}
    & 14 \\ \hline 
    
    { PLD ({\pld})~\cite{pld}}            
    & \num{43}{M}             
    & \num{623}{M}       
    & 15 \\ \hline
    
    {Twitter ({\tw})~\cite{twitter}}      
    & \num{62}{M}             
    & \num{1468}{M}       
    & {24} \\ \hline

    { Kron ({\kr})~\cite{gap}}            
    & \num{67}{M}             
    & \num{1323}{M}       
    & {20} \\ \hline
    
    { Sd1-arc ({\sd})~\cite{pld}}            
    & \num{95}{M}             
    & \num{1937}{M}       
    & {20} \\ \hline \hline

    {Friendster ({\fr}) \cite{konect-friendster}}   & 
    \num{64}{M} & 
    \num{2147}{M} & 
    {33} \\ \hline
    
    {Uniform ({\uni}) \cite{rmat}}   & 
    \num{50}{M} & 
    \num{1000}{M} & 
    {20} \\
    
    \hline
    \end{tabularx}
    \caption{\label{tab:grasp:datasets}Properties of the graph datasets. Top five datasets are used in the main evaluation whereas the bottom two datasets are used as adversarial datasets.}%
\end{table}

\subsection{Graph Processing Framework}
\label{sec:grasp:Applications}

For the evaluation, we use the same set of applications as we did in the Chapter~\ref{ch:dbg} (see Table~\ref{tab:dbg:graph-worklads}).
We combine these five applications -- BC, SSSP, PR, PRD and Radii -- with the five high-skew graph datasets listed in Table~\ref{tab:grasp:datasets}, resulting in 25 benchmarks. To test the robustness of \sam to adversarial workloads, we use two additional datasets with low-/no-skew.

We obtained the source code for the graph applications from Ligra~\cite{ligra} 
and applied a simple data-structure optimization to improve locality in the baseline implementation as follows.
As explained in Sec.~\ref{sec:motivation:general-caching} of Chapter~\ref{ch:motivation}, graph applications exhibit irregular accesses for the Property Array, with applications potentially maintaining more than one such array. When multiple Property arrays are used, elements corresponding to a given vertex may need to be sourced from all of the arrays. We
merge these arrays (\ie Structure of Arrays to Array of Structure transformation) to induce spatial locality, which reduces number of misses, and in turn, improves performance on all datasets for \pr{}, \prd{} and \sssp{} (see Table~\ref{tab:grasp:opt}). 
We use the optimized implementation of these three applications as a stronger baseline for our evaluation. The optimized applications are available at \url{https://github.com/faldupriyank/grasp}. We do note that GRASP does not mandate merging arrays as GRASP design can accommodate multiple arrays. Nevertheless, merging does reduce the number of arrays needed to be tracked.

For \prd, two versions of the algorithm are provided with Ligra: push-based and pull-push. In the baseline \mbox{implementation}, the push-based version is faster. However, after merging the Property Arrays, the pull-push variant performs better, and is what we use for the evaluation. 

\begin{table}[!t]
    \small
    \centering
    \begin{tabularx}{0.8\linewidth}
        {|>{\centering\arraybackslash\hsize=0.30\hsize}X|
          >{\centering\arraybackslash\hsize=0.35\hsize}X|
          >{\centering\arraybackslash\hsize=0.35\hsize}X|
        }
        \hline
        \multicolumn{1}{|c|}{Application} &  \multicolumn{1}{c|}{Merging Opportunity?} & \multicolumn{1}{c|}{Speed-up} \\
        \hline \hline
        \bc{} & No & - \\ \hline
        \sssp{} & Yes & 3-8\%\\ \hline
        \pr{} & Yes & 40-52\%\\ \hline
        \prd{} & Yes & 14-49\%\\ \hline
        \radii{} & No & - \\ \hline
        
    \end{tabularx}
    \caption{Effect of our optimization on the original Ligra implementation for different applications. PR applies pull-based computations whereas SSSP applies push-based computations throughout the execution; the rest of the applications switch between pull or push based on a number of active vertices in a given iteration.}
    \label{tab:grasp:opt}
\end{table}

\subsection{Methodology for Software Evaluation \label{sec:grasp:sw-eval-method}}

Methodology for the evaluation of software reordering techniques -- Sort, Hub Sorting, DBG and Gorder -- is identical to the methodology used in the previous Chapter (see \ref{sec:dbg:method} of Chapter~\ref{ch:dbg}).

\subsection{Methodology for Hardware Evaluation\label{sec:grasp:hw-eval-method}}

\noindenttitle{Simulation infrastructure:} We use the {\em Sniper}~\cite{sniper} simulator modeling 8 OoO cores. Table~\ref{tab:grasp:sim-params} lists the parameters of the simulated system. The applications are evaluated in a multi-threaded mode with 8-threads.

We find that the graph applications spend significant fraction (86\% on average in our evaluations) of time in push-based iterations for \sssp{} or pull-based iterations for all other evaluated applications. Thus, we simulate the \emph{Region of Interest (ROI)} covering only push- or pull-based iterations (whichever one dominates) for the respective applications.
Because simulating all iterations of a graph-analytic application in a detailed microarchitectural simulator is prohibitive, time-wise, we instead simulate one iteration that has the highest number of active vertices. To validate the soundness of our methodology, we also simulated one more randomly chosen iteration for each application-dataset pair with at least 20\% of vertices active and observed trends similar to the ones reported in the paper.

\begin{table}[!t]
    \centering
    \small
    \begin{tabularx}{0.9\linewidth} %
        {|>{\centering\arraybackslash\hsize=0.2\hsize}X| 
          >{\raggedright\arraybackslash\hsize=0.8\hsize}X|
        }
        \hline
        Core        &   OoO @ 2.66GHz, 4-wide front-end\\ \hline%
        \multirow{2}{*}{L1-I/D Cache} &   4/8-ways 32KB, 4 cycles access latency   \\
                    & stride-based prefetchers with 16 streams \\ \hline
        L2 Cache    &   Unified, 8-ways 256KB, 6 cycles access latency  \\ \hline            
        \multirow{2}{*}{L3 Cache}    &   16-ways 16MB NUCA (2MB slice per core), Non-Inclusive \\ & Non-Exclusive, 10 cycles bank access latency\\ \hline
        NOC         &   Ring network with 2 cycles per hop \\ \hline            
        Memory      &   50ns latency, 2 on-chip memory controllers \\ \hline
    \end{tabularx}
    \caption{\label{tab:grasp:sim-params}Parameters of the simulated system for evaluation of the hardware techniques.}
\end{table}

\noindenttitle{Evaluated cache management techniques:} 
We evaluate \sam and compare it with the state-of-the-art thrash-resistant cache management techniques described below.

\noindentsubsectiontitle{\bf RRIP}~\cite{rrip} is the state-of-the-art technique among static and lightweight dynamic techniques that do not depend on history-based learning. RRIP is the most appropriate comparison point given that \sam builds upon RRIP as the base technique (Sec.~\ref{sec:grasp:sam}). We implement RRIP (specifically, {\em DRRIP}) based on the source code from the cache replacement championship~\cite{championship} for RRIP, and use a 3-bit counter per cache block. We use RRIP as high performance baseline and report speed-up for all hardware techniques over the RRIP baseline (except for the studies in Sec~\ref{sec:grasp:oracle} that use LRU baseline).

\noindentsubsectiontitle{\bf Signature-based Hit Predictor (SHiP)}~\cite{ship} is the state-of-the-art insertion policy which builds on RRIP~\cite{rrip}. Due to the shortcomings of PC-based correlation for graph applications as explained in Sec.~\ref{sec:motivation:prior:hw} of Chapter~\ref{ch:motivation}, we evaluate a SHiP-MEM variant that correlates a block's reuse with the block's memory region. We evaluate 16KB memory regions as in the original proposal. The predictor table is provisioned with an {\em unlimited number of entries} to assess the maximum potential of this technique.
Every entry in the predictor table contains a 3-bit saturating counter that tracks the re-reference behavior of cache blocks of the memory region associated with that entry.

\noindentsubsectiontitle{\bf Hawkeye}~\cite{hawkeye} is the state-of-the-art cache management technique and winner of the recent cache replacement championship (CRC2)~\cite{crc2}. Hawkeye trains its predictor table by simulating OPT~\cite{opt} on past LLC accesses to infer block's cache friendliness. 
We use an improved, prefetch-aware, version of Hawkeye from CRC2 (\ie Hawkeye++ from Sec.~\ref{sec:leeway:crc2-evel} of Chapter~\ref{ch:leeway})
We appropriately scale the number of sampling sets and predictor table entries for a 16MB cache.

\noindentsubsectiontitle{\bf Leeway}
(specifically, Leeway-NRU from Chapter~\ref{ch:leeway}) is a history-based predictive cache management technique that applies dead block predictions based on a metric called Live Distance, which conservatively captures the reuse interval of a cache block. 
We appropriately scale the number of sampling sets and predictor table entries for a 16MB cache.

\noindentsubsectiontitle{\bf XMem}~\cite{xmem} is a pinning-based technique proposed for algorithms that benefit from {\em cache tiling}. Once pinned, a cache block cannot be evicted until explicitly unpinned by the software, usually done when the processing of a tile is complete. 
In the original proposal, XMem reserves 75\% of LLC capacity to pin tile data whereas the remaining capacity is managed by the base replacement technique for the rest of the data. In this work, we explore four configurations of XMem, labeled PIN-X, where X refers to the percentage (25\%, 50\%, 75\% or 100\%) of LLC capacity reserved for pinning.
We adopt XMem design for graph analytics and identify the cache blocks from the high reuse region that benefit from pinning using the \sam interface.
Finally, XMem requires an additional 1-bit for every cache block to identify whether a cache block is pinned, along with an additional mechanism to track how much of the capacity is used by the pinned cache blocks at any given time.

\noindentsubsectiontitle{\bf \sam} is the proposed domain-specialized cache \mbox{management} technique for graph analytics. We instrument the applications to communicate the address bounds of the Property Arrays to the simulated \sam hardware. For the evaluated applications, we needed to instrument at most two arrays.
Finally, \sam uses RRIP as the base cache policy with a 3-bit saturating counter and does {\em not} add any further storage to per-block metadata.

\section{Evaluation\label{sec:grasp:eval}}

We first evaluate hardware cache management techniques on top of a software skew-aware reordering technique (Sec.~\ref{sec:grasp:eval-predictive} \&~\ref{sec:grasp:eval-pin}). Due to long simulation time, evaluating all hardware techniques on top of all reordering techniques would be prohibitive. Thus, without loss of generality, we evaluate hardware techniques on top of DBG, which consistently outperforms other reordering techniques (Sec.~\ref{sec:grasp:eval-reordering-techniques}). In Sec.~\ref{sec:grasp:sam-compatibility}, we evaluate \sam with other reordering techniques to show GRASP's generality.

\begin{figure}[!t]
    \centering
    \includegraphics[width=1\linewidth]{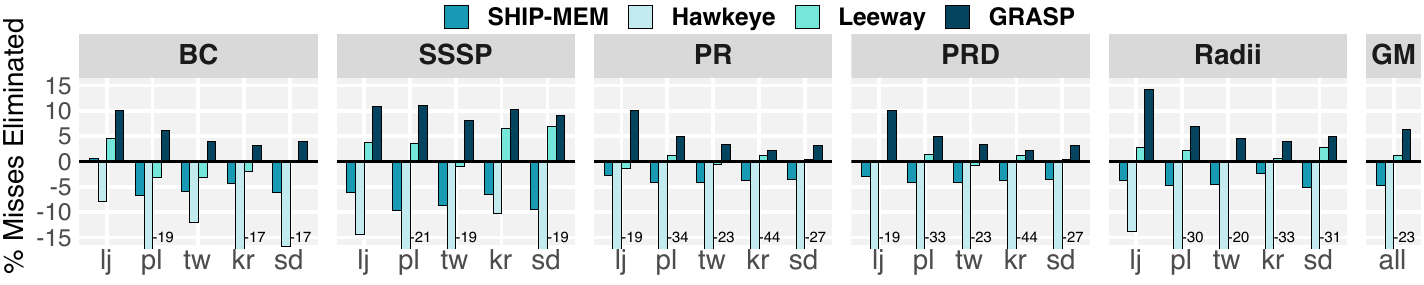}
    \caption{\label{fig:grasp:sam-main1-miss} LLC miss reduction for \sam and state-of-the-art history-based predictive techniques over the RRIP baseline.}
\end{figure}
\begin{figure}[!t]
    \centering
    \includegraphics[width=1\linewidth]{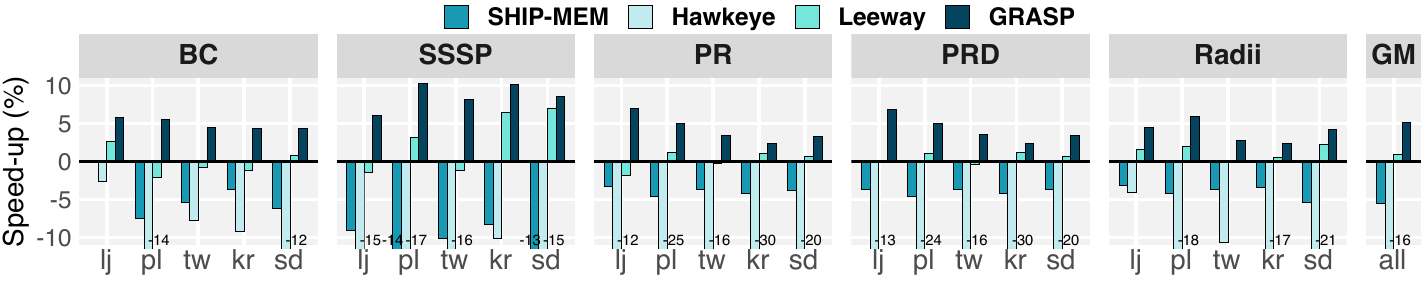}
    \caption{\label{fig:grasp:sam-main1-perf} Speed-up for \sam and state-of-the-art history-based  predictive cache management techniques over the RRIP baseline.}
\end{figure}

\subsection{History-Based Predictive Techniques\label{sec:grasp:eval-predictive}}

In this section, we compare GRASP with the state-of-the-art hardware techniques, SHiP-MEM~\cite{ship}, Hawkeye~\cite{hawkeye} and Leeway.
As we showed in Chapter~\ref{ch:motivation}, RRIP consistently outperforms LRU across the datapoints, we use RRIP as a stronger baseline.
Finally, we use DBG as the software baseline; thus, all speed-ups reported in this section are \emph{over and above DBG}.

\noindenttitle{Miss reduction:} Fig.~\ref{fig:grasp:sam-main1-miss} shows the miss reduction over the RRIP baseline. \sam consistently reduces misses on all datapoints, eliminating 6.4\% of LLC misses on average and up to 14.2\% in the best case (on \lj{} dataset for the Radii application). The domain-specialized design allows \sam to accurately identify the high-reuse working set (\ie hot vertices), which \sam is able to retain in the cache through its specialized policies, effectively exploiting the temporal reuse.%

Among prior techniques, Leeway is the only technique that reduces misses, albeit marginal, with an average miss reduction of 1.1\% over the RRIP baseline. The other two techniques are not effective for graph applications, with SHiP-MEM and Hawkeye \emph{increasing} misses across the datapoints, with an average miss reduction of -4.8\% and -22.7\%, respectively, over the baseline. This is a new result as prior works show that Hawkeye and SHiP-MEM outperform RRIP on a wide range of applications~\cite{ship,hawkeye}.%
The result indicates that the learning mechanisms of the state-of-the-art domain-agnostic techniques are deficient in retaining the high reuse working set (\ie hot vertices) for graph applications, which ends up hurting application performance as discussed next.

\noindenttitle{Application speed-up:}
Fig.~\ref{fig:grasp:sam-main1-perf} shows the speed-up for hardware techniques over the RRIP baseline. Overall, performance correlates well with the change in LLC misses; 
\sam consistently provides a speed-up across datapoints with an average speed-up of 5.2\% and up to 10.2\% in the best case (on \pld{} dataset for SSSP application) over the baseline. When compared to the same baseline, SHiP-MEM and Hawkeye consistently cause slowdown with an average speed-up of -5.5\% and -16.2\%, respectively whereas Leeway yields a marginal speed-up of 0.9\%.
Finally, when compared to prior works directly, \sam yields 4.2\%, 5.2\%, 11.2\% and 25.5\% average speed-up over Leeway, RRIP, SHiP-MEM and Hawkeye, respectively, while not causing slowdown on any datapoints. %

Recall from Chapter~\ref{ch:motivation}, in which we also evaluated prior techniques without applying any vertex reordering. As shown in  Fig.~\ref{fig:motivation:orig-perf}, Leeway, SHiP-MEM and Hawkeye yield an average speed-up of -0.8\%, -5.7\% and \mbox{-14.8\%}, respectively, over RRIP on the datasets with no reordering. 

\noindenttitle{Dissecting performance of SHiP-MEM:} SHiP-MEM is a predictive technique that predicts reuse of a cache block based on the fine-grained memory region it belongs to. Thus, SHiP-MEM relies on a homogeneous cache behavior for all blocks belonging to the same memory region. In theory, DBG should allow SHiP-MEM to identify memory regions containing hottest of vertices (corresponding to High Reuse Region from Fig.~\stitchref{fig:grasp:graphinity}{fig:grasp:graphinity-c}). In practice, however, irregular access patterns to these regions and thrashing by cache blocks from other regions impede learning.
Thus, despite leveraging software and utilizing a sophisticated storage-intensive prediction mechanism in hardware, SHiP-MEM underperforms domain-specialized \samnospace.

\noindenttitle{Dissecting performance of Hawkeye:} Hawkeye is the state-of-the-art predictive technique that uses PC-based correlation to predict whether a cache block has a cache-friendly or cache-averse behavior based on past LLC accesses.
Thus, Hawkeye fundamentally relies on homogeneous cache behavior for all blocks accessed by the same PC address.
When Hawkeye is employed for graph analytics, Hawkeye struggles to learn the behavior of cache blocks in the Property Array as hot vertices exhibit cache-friendly behavior while cold vertices exhibit cache-averse behavior, yet all vertices are accessed by the same PC address.
To make matters worse, if a block incurs a hit and Hawkeye predicts the PC making the access as cache-averse, the cache block is prioritized for eviction instead of promoting the block to MRU as is done in the baseline. 
Thus, Hawkeye performs even worse than the baseline for all combinations of graph applications and datasets.
While not evaluated, other prior PC-based techniques (e.g.,~\cite{sampler,ship}) that rely on a PC-based correlation would also struggle on graph applications for the same reason.

\noindenttitle{Dissecting performance of Leeway:}
Leeway, like Hawkeye, also relies on a PC-based reuse correlation, and thus is not expected to provide significant speed-ups for graph-analytics. However, Leeway successfully avoids the slowdown on 10 of the 25 datapoints and significantly limits the slowdown on the rest of the datapoints (max slowdown of 2.1\% vs 13.6\% for SHiP-MEM and 30.2\% for Hawkeye). The reasons why Leeway perfroms better than prior PC-based techniques can be attributed to (1) the conservative nature of the Live Distance
metric, which Leeway uses to determine if a cache block is dead, and (2) adaptive reuse-aware policies that control the rate of predictions based on the observed access patterns. Because of these two factors, performance of Leeway remains close the the base replacement technique in the presence of variability in the reuse behavior of cache blocks.

\noindenttitle{Dissecting performance of GRASP:}
Performance of GRASP over its base technique, RRIP, can be attributed to three features: software hints, insertion policy and hit-promotion policy. 
Fig.~\ref{fig:grasp:grasp-policies} shows the performance impact due to each of these features. 
RRIP inserts every new cache block at one of the two positions (as specified in the Default Reuse Hint of Table~\ref{tab:grasp:sam-rrip}); a cache block is inserted at the LRU position with a high probability or near the LRU position with a low probability.
RRIP+Hints is identical to RRIP except for how a new cache block is assigned these positions. RRIP+Hints uses software hints (similar to GRASP) to guide the insertion. A cache block with High-Reuse hint is inserted near the LRU position and all other blocks are inserted at the LRU position.
GRASP (Insertion-Only) refers to the technique that applies insertion policy of GRASP as specified in Table~\ref{tab:grasp:sam-rrip} but the hit-promotion policy is unchanged from RRIP. 
Finally, GRASP (Hit-Promotion) refers to the technique that applies hit-promotion policy of GRASP along with its insertion policy, which is essentially the full GRASP design. Note that each successive technique adds a new feature on top of the features incorporated by the previous ones. For example, GRASP (Insertion-Only) features a new insertion policy in addition to the software hints.

\begin{figure}[!t]
    \centering
    \includegraphics[width=1\linewidth]{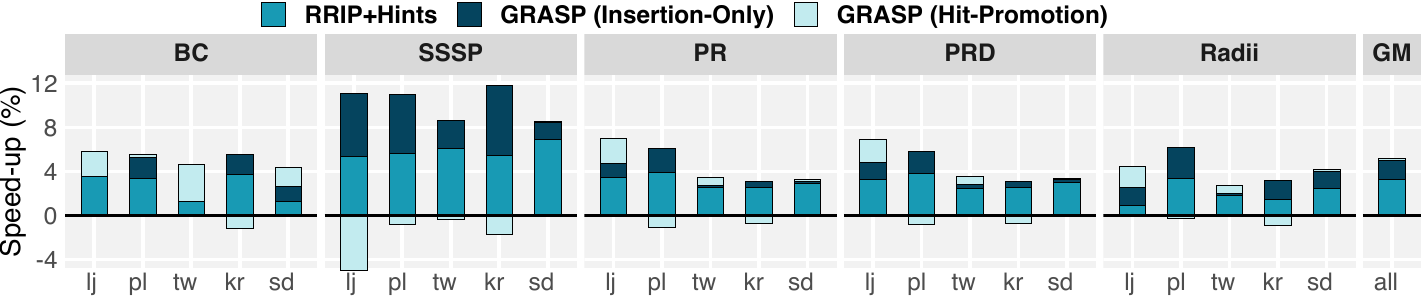}
    \caption{Impact of GRASP features on performance.}
    \label{fig:grasp:grasp-policies}
\end{figure}

As the figure shows, RRIP+Hints yields an average speed-up of 3.3\% over probabilistic RRIP, confirming the utility of software hints. 
GRASP (Insertion-Only) further increases performance by yielding an average speed-up of 5.0\%. 
GRASP (Insertion-Only) provides additional protection to the High-Reuse cache blocks in comparison to RRIP+Hints by inserting High-Reuse cache blocks directly at the MRU position.
Finally, GRASP (Hit-Promotion) yields an average speed-up of 5.2\%. Difference between GRASP (Hit-Promotion) and GRASP (Insertion-Only) is marginal as the hit-promotion policy of GRASP has negative effect on slightly less than half datapoints. 
The results are inline with the observations from our work that showed that the value-addition of hit-promotion policies over insertion policies is low in presence of cache thrashing~\cite{c-dead}.

\noindenttitle{Summary:} Hardware cache management is an established difficult problem, which is reflected in the small average speed-ups (usually 1\%-5\%) achieved by state-of-the-art techniques over the prior best techniques~\cite{rrip,sampler,ship,mdpp,hawkeye,perceptron,harmony}. Our work shows that graph applications present a particularly challenging workload for these techniques, in many cases leading to significant performance slowdowns. 
In this light, \sam is quite successful in improving performance of graph applications by yielding an average speed-up of 5.2\% (max 10.2\%) over a high performing software and hardware baseline, while not causing slowdown on any datapoint. Moreover, unlike state-of-the-art techniques, \sam achieves this without requiring storage-intensive~metadata.

\subsection{Pinning-Based Techniques\label{sec:grasp:eval-pin}}

\begin{figure}[!t]
    \centering
    \includegraphics[width=1\linewidth]{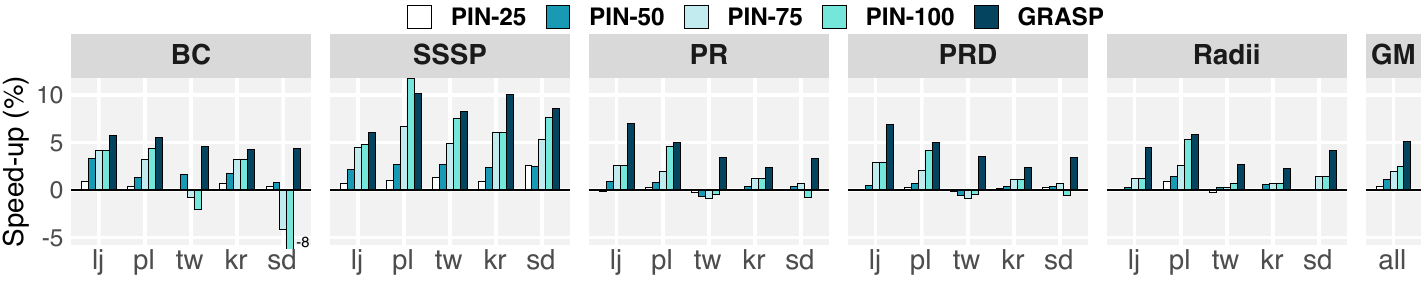}
    \caption{\label{fig:grasp:sam-main2-perf} Speed-up for \sam and pinning-based techniques over the RRIP baseline on high-skew datasets.}
\end{figure}

In this section, we show the benefit of flexible \sam policies over pinning-based rigid approaches. 
We first present the results on the high-skew datasets and then on the low-/no-skew datasets to test their resilience in adversarial scenarios.

\noindenttitle{High-skew datasets:} Fig.~\ref{fig:grasp:sam-main2-perf} shows speed-ups for four XMem configuration (PIN-25, PIN-50, PIN-75 and PIN-100) and GRASP over the RRIP baseline on high-skew datasets. 
\sam outperforms all XMem configurations on 24 of 25 datapoints with an average speed-up of 5.2\%. 
In comparison, PIN-25, PIN-50, PIN-75 and PIN-100 yield 0.4\%, 1.1\%, 2.0\% and 2.5\%, respectively.

PIN-100 outperforms the other three XMem configurations as for those configurations, significant fraction of the capacity can still be occupied by cold vertices, which causes thrashing in the unreserved capacity.
Nevertheless, PIN-100 causes slowdown on many datapoints (\eg for BC, PR and PRD applications on \tw{} and \sd{} datasets). Moreover, PIN-100 cannot capitalize on reuse from Moderate Reuse Region as pinned vertices cannot be evicted even when they stopped exhibiting reuse.
Thus, PIN-100 provides only a marginal speed-up on many datapoints (\eg Radii application on \lj{}, \tw{} and \kr{} datasets).

PIN-75 and PIN-100, (two of the high performing XMem configurations), while yield only marginal speed-ups, still outperform the state-of-the-art domain-agnostic techniques -- SHiP-MEM, Leeway and Hawkeye -- (Figs.~\ref{fig:grasp:sam-main1-perf} \& \ref{fig:grasp:sam-main2-perf}) which confirms that utilizing software knowledge for cache management is a promising direction over a storage-intensive domain-agnostic design for the challenging access patterns of graph~analytics.

\noindenttitle{Low-/No-skew datasets:} Next, we evaluate the robustness of \sam and pinning-based techniques (PIN-75 and PIN-100) for adversarial datasets with low-/no-skew. Naturally, these techniques are not expected to provide a significant speed-up in the absence of high skew; however, a robust technique would reduce/avoid the slowdown. Fig.~\ref{fig:grasp:sam-no-skew} shows the speed-up for a low-skew dataset \fr{} and a no-skew dataset \uni{} for these techniques over the RRIP baseline.

\begin{figure}[!t]
    \centering
    \includegraphics{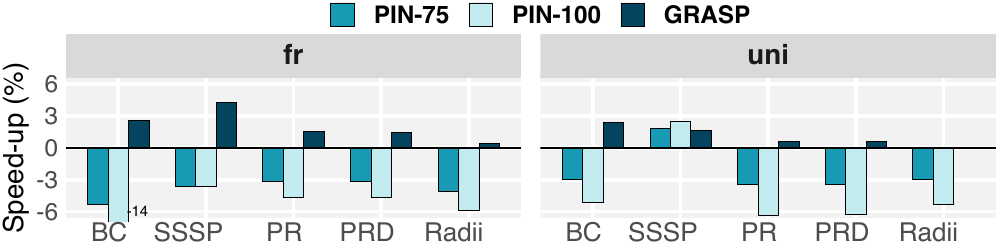}
    \caption{\label{fig:grasp:sam-no-skew} Speed-up over the RRIP baseline on \fr{}, a low-skew dataset and \uni, a no-skew dataset.}
\end{figure}

\sam provides a positive speed-up on 9 out of 10 datapoints even for low-/no-skew datasets. On the low-skew dataset \fr, \sam yields a speed-up between 0.4\% and 4.3\% whereas on the no-skew dataset \uni, \sam yields a speed-up between -0.1\% and 2.4\%. In contrast, PIN-75 and PIN-100 cause slowdown on almost all datapoints. 

In the absence of high-skew, cache blocks belonging to the High Reuse Region do not dominate the overall LLC accesses. Thus, pinning these blocks throughout the execution is counter-productive for PIN-75 and PIN-100. In contrast, \sam adopts a flexible approach, wherein the high priority cache blocks from High Reuse Region can make way for other blocks that observe some reuse, as needed. Thus, \sam successfully limits slowdown, and even provides reasonable speed-up on some datapoints, for such highly adversarial datasets.

Finally, combining results on all 7 datasets (5 datasets from Fig.~\ref{fig:grasp:sam-main2-perf} and 2 from Fig.~\ref{fig:grasp:sam-no-skew}), GRASP yields an average speed-up of 4.1\%. In comparison, PIN-75 and PIN-100 provide a marginal speed-up of only 0.5\% and 0.1\%, respectively.
PIN-75 and PIN-100 cause slowdown of up to 5.3\% and 14.2\% whereas max slowdown for GRASP is only 0.1\%.

\subsection{Reordering Techniques and GRASP}
Thus far, we evaluated GRASP on graph applications \mbox{processing} datasets that are reordered using DBG. In this section, we compare performance of vertex reordering techniques, followed by an evaluation of GRASP on top of these techniques, demonstrating GRASP's generality.

\subsubsection{Effectiveness of Reordering Techniques\label{sec:grasp:eval-reordering-techniques}}
\begin{figure}[!t]
    \centering
    \subfloat[\small Net speed-up for existing software reordering techniques after accounting for their reordering cost on a real machine.]{\label{fig:grasp:sw-perf}{\includegraphics{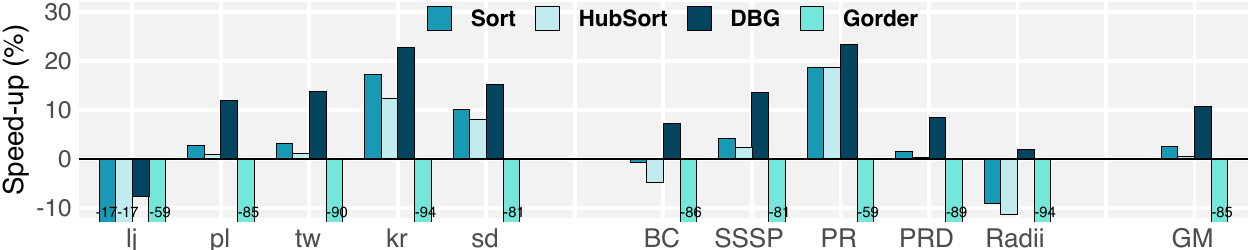}}}
    
    \subfloat[\small Application speed-up of \sam over the RRIP baseline on top of different reordering techniques.]{\label{fig:grasp:sam-compatibility}{\includegraphics{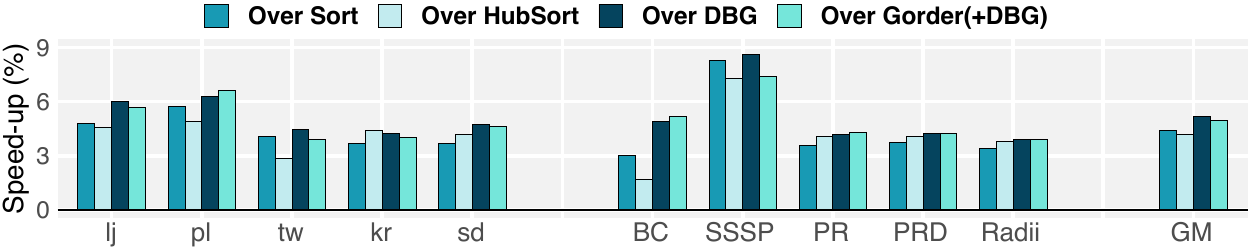}}}
    
    \caption{\label{fig:grasp:graphin}Reordering Techniques + GRASP: the left group shows speed-up for a dataset across all applications while the right group shows speed-up for an application across all datasets.}
\end{figure}

In this section, we first summarize the performance of skew-aware techniques -- Sort, HubSort~\cite{fc} and DBG -- for graph applications processing high-skew datasets.  We also evaluate \gorder{}~\cite{gorder}, a complex vertex reordering approach. Note that the software techniques are evaluated on a real machine with 40 hardware threads as mentioned in Sec.~\ref{sec:dbg:soft-eval} of Chapter~\ref{ch:dbg}. 

Fig.~\stitchref{fig:grasp:graphin}{fig:grasp:sw-perf} shows the speed-up for these software techniques after accounting for their reordering cost over the baseline with no reordering.
Among skew-aware techniques, all techniques are effective on largest of the datasets (\eg \kr{} and \sd) and long iterative applications (\eg \pr). As these techniques rely on a low cost approach for reordering, the reordering cost is amortized quickly when the application runtime is high, making these solutions practically attractive. Averaged across all application and dataset pairs, skew-aware techniques yield a net speed-up of 2.6\% for Sort, 0.6\% for HubSort and 10.8\% for DBG.

Unsurprisingly, \gorder{} causes a significant slowdown on all datapoints due to its large reordering cost, yielding an \mbox{average} speed-up of -85.4\%. Thus, \gorder{} is less practical when \mbox{compared} to simple yet effective skew-aware techniques.%

\subsubsection{Generality of \sam \label{sec:grasp:sam-compatibility}}
As software vertex reordering techniques offer different trade-offs in preserving graph structure and reducing reordering cost, it is important for \sam to not be coupled to any one software technique. In this section, we evaluate \sam with different reordering techniques, both skew-aware and complex ones. While skew-aware techniques are readily compatible with \samnospace, \gorder{} requires a simple tweak as follows.

After applying \gorder{} on an original dataset, we apply DBG to further reorder vertices, which results in a vertex order that retains most of the \gorder{} ordering while also segregating hot vertices in a contiguous region, making \gorder{} compatible with \samnospace. 

Fig.~\stitchref{fig:grasp:graphin}{fig:grasp:sam-compatibility} shows the speed-up for \sam over RRIP on top of the same reordering technique as the baseline.
As with DBG, \sam consistently provides a speed-up across datasets and applications on top of other reordering techniques as well. On average, \sam yields a speed-up of 4.4\%, 4.2\%, 5.2\% and 5.0\% on top of Sort, HubSort, DBG and \gorder, respectively. The result confirms that \sam complements a broad class of existing software reordering techniques.

\subsection{GRASP vs Optimal Replacement (OPT) \label{sec:grasp:oracle}}

In this section, we compare GRASP with Belady's optimal replacement policy (OPT)~\cite{opt}. As OPT requires the perfect knowledge of the future, we generate the traces of LLC accesses (up to 2 billion for each trace) for the applications processing graph datasets reordered using DBG on the simulation baseline configuration specified in Sec.~\ref{sec:grasp:hw-eval-method}.
We apply OPT on each trace for five different LLC sizes -- 1MB, 4MB, 8MB, 16MB and 32MB -- to obtain the minimum number of misses for a given cache size and report the percentage of misses eliminated over {\em LRU} on the same LLC size.

\begin{figure}
    \centering
    \includegraphics{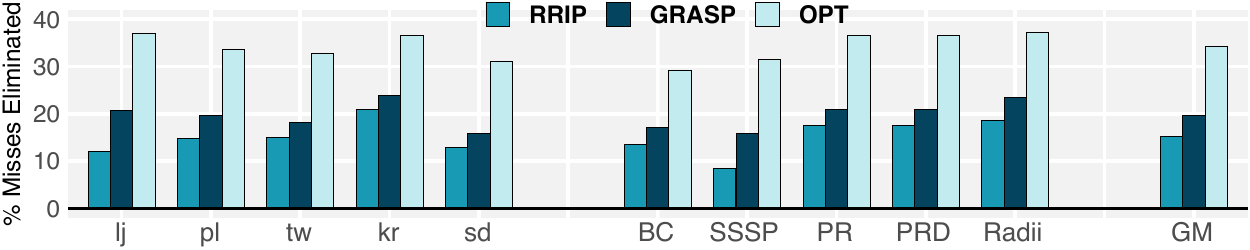}
    \caption{Percentage of misses eliminated over LRU.}
    \label{fig:grasp:oracle}
\end{figure}

\noindenttitle{Miss reduction on 16MB LLC:}
Fig.~\ref{fig:grasp:oracle} shows the results for OPT along with RRIP and GRASP for 16MB LLC size.
OPT eliminates 34.3\% of total misses over LRU. In comparison, GRASP eliminates 19.7\% of misses (vs 15.2\% for RRIP). Overall, GRASP is 57.5\% effective in eliminating misses when compared to OPT, an offline technique with perfect knowledge of the future. While GRASP is the most effective among the online techniques, the results also show that the remaining opportunity (difference between OPT and GRASP) is still significant, which warrants further research in this~direction.

\noindenttitle{Sensitivity of GRASP to LLC size: }
Table~\ref{tab:grasp:size-sensitivity-oracle} shows the average percentage of misses eliminated by RRIP, \sam and OPT for different LLC sizes over LRU. With the increase in LLC size, \sam becomes more effective at eliminating misses over LRU (average miss reduction of 15.4\% for 1MB vs 21.2\% for 32MB). This is expected, as the larger LLC size allows \sam to provide preferential treatment to more hot vertices. In general, yet larger LLC sizes are expected to benefit even more from \sam until the LLC size becomes large enough to accommodate all hot vertices.

\begin{table}[t]
    \small
    \centering
    \begin{tabular}{|l||c|c|c|c|c|c|}
        \hline
        \centering{Technique} & 1MB	&  4MB & 8MB & 16MB & 32MB \\ \hline \hline
        RRIP & 15.9\% &	16.4\% &	15.7\% &	15.2\% &	16.2\% \\ \hline
        GRASP & 15.4\% &	17.0\% &	18.1\% &	19.7\% &	21.2\% \\ \hline
        OPT	& 27.5\% &	32.2\% &	33.3\% &	34.3\% &	34.5\% \\ \hline
    \end{tabular}
    \caption{Percentage of misses eliminated over LRU for different LLC size.}
    \label{tab:grasp:size-sensitivity-oracle}
\end{table}

\section{Related Work}
\label{sec:grasp:related}

\noindenttitle{Shared-memory graph frameworks:}
A significant amount of research has focused on designing high performance shared-memory frameworks for graph applications. Majority of these frameworks are vertex-centric~\cite{ligra, galois, gap, graphmat, graphlab, graphchi} and use CSR or its variants to encode a graph, making GRASP readily compatible with these frameworks. More generally, GRASP requires classification of only the Property Array(s), making it independent of the specific data structure used to represent the graph, which further increases compatibility across the spectrum of frameworks. Thus, we expect GRASP to reduce misses across frameworks, though absolute speed-ups will likely vary.

\noindenttitle{Distributed-memory graph frameworks:}
Distributed graph processing frameworks can also benefit from GRASP. For example, 
PGX~\cite{pgx.d} and PowerGraph~\cite{powergraph} proposed duplicating high degree vertices in the graph partitions to reduce high communication overhead across computing nodes. These optimizations are largely orthogonal to \sam cache management. As such, \sam can be applied to distributed graph processing by caching high-degree vertices within each node's LLC to improve node-level cache behavior.

\noindenttitle{Streaming graph frameworks:}
In this work, we have assumed that graphs are static.
In practice, graphs may evolve over time and a stream of graph updates (\ie addition or removal of vertices or edges) are interleaved with graph-analytic queries (e.g., computing PageRank of vertices or computing shortest path from different root vertices). For such deployment settings, a CSR-based structure is infeasible. Instead, researchers have proposed various data structures for graph encoding that can accommodate fast graph updates and allow space-efficient versioning~\cite{stinger, llama, aspen}. Meanwhile, each graph query is performed on a consistent view (\ie static snapshot) of a graph. For example, Aspen~\cite{aspen}, a recent graph-streaming framework, uses Ligra (a static graph-processing framework) in the back-end to run graph-analytic queries. Thus, the observations made in this paper regarding cache thrashing due to the irregular access patterns of the Property Array, as well as skew-aware reordering and GRASP being complementary in combating cache thrashing, are also relevant for dynamic graphs.

For static graphs, vertex reordering cost is amortized over multiple graph traversals for a single graph query (as shown in Fig.~\stitchref{fig:grasp:graphin}{fig:grasp:sw-perf}). However, for dynamic graphs, reordering cost can be further amortized over multiple graph queries.
Intuitively, addition or deletion of some vertices or edges in a large graph would not lead to a drastic change in the degree distribution, and thus unlikely to change which vertices are classified hot in a short time window. Therefore, skew-aware reordering can be applied at periodic intervals to improve cache behavior after a series of updates has been made to a graph, amortizing reordering cost over multiple graph queries.

\noindenttitle{Hardware prefetchers:} 
Modern processors typically employ prefetchers that target stride-based access patterns and thus are not amenable to graph analytics.  
Researchers have proposed custom prefetchers at L1-D that specifically target indirect memory access patterns of graph analytics~\cite{imp,prefetch-data-structure}. 
Nevertheless, prefetching can only {\em hide} memory access latency. Unlike cache replacement, prefetching cannot reduce memory bandwidth pressure or DRAM energy expenditure. 
Indeed, prior work observes that even the ideal, 100\% accurate, prefetcher for graph analytics is bottlenecked by memory bandwidth~\cite{imp}. In contrast, GRASP reduces bandwidth pressure by reducing LLC misses, and thus is complementary to prefetching.

\section{Conclusion}
\label{sec:grasp:conclusion}
{
In this chapter, we explored how to design hardware cache management to tackle cache thrashing at LLC for the domain of graph analytics.
We showed that state-of-the-art history-based predictive cache management techniques are deficient in the presence of cache thrashing stemming from irregular access patterns of graph applications processing large graphs. In response, we introduced GRASP, specialized cache management for LLC for graph analytics on natural graphs. GRASP's specialized cache policies exploit the high reuse inherent in hot vertices while maintaining the flexibility to capture reuse in other cache blocks. GRASP leverages software reordering optimizations such as DBG to enable a lightweight interface that allows hardware to reliably pinpoint hot vertices amidst irregular access patterns. In doing so, GRASP avoids the need for a storage-intensive prediction mechanism or additional metadata storage in the LLC. GRASP requires minimal hardware support, making it attractive for integration into commodity server processors to enable acceleration for the domain of graph analytics. %
Finally, GRASP delivers consistent performance gains on high-skew datasets, while preventing slowdowns on low-skew datasets.
}

\chapter{Conclusions and Future Work\label{ch:conclusion}}
\section{Contributions}

In this section, we summarize the main contributions made in the preceding chapters.

\subsection{Leeway -- Domain-Agnostic Cache Management}
In Chapter~\ref{ch:leeway}, we highlighted the limitations of state-of-the-art history-based predictive techniques in achieving high performance 
in the face of variability. 
To address those limitations, we argued for variability-tolerant mechanisms and policies for cache management.
As a step in that direction, we proposed Leeway, a history-based predictive technique employing two variability-tolerant features.
First, Leeway introduces a new metric, Live Distance, that captures the largest interval of temporal reuse for a cache block, providing a conservative estimate of a cache block's useful lifetime.
Second, Leeway implements a robust prediction mechanism that identifies dead blocks based on their past Live Distance values. To maximize cache efficiency in the face of variability, Leeway monitors the change in Live Distance values at runtime using its reuse-aware policies to adapt to the observed access patterns.
Meanwhile, Leeway embeds prediction metadata with cache blocks in order to avoid critical path history table look-ups on cache hits and reduce the on-chip network traffic, in contrast to the state-of-the-art techniques that access history table on every cache access (including cache hits).
On a variety of applications and deployment scenarios, Leeway consistently provides good performance that generally matches or exceeds that of state-of-the-art techniques.

\subsection{DBG -- Lightweight Vertex Reordering}
In Chapter~\ref{ch:dbg}, we studied existing skew-aware reordering techniques that seek to improve cache efficiency for graph analytics by reducing the cache footprint of hot vertices. 
We demonstrated the inherent tension between reducing the cache footprint of hot vertices and preserving original graph structure, which limits the effectiveness of existing skew-aware reordering techniques. In response, we proposed Degree-Based Grouping (DBG), a lightweight vertex reordering software technique that employs
coarse-grain reordering to preserve graph structure while reducing the cache footprint of hot vertices. 
On a variety of graph applications and datasets, DBG achieves higher average performance than all existing skew-aware techniques and nearly matches the average performance of the state-of-the-art complex reordering technique.

\subsection{GRASP -- Domain-Specialized Cache Management}
In Chapter~\ref{ch:grasp}, we explored how to design hardware cache management to tackle cache thrashing at LLC for the domain of graph analytics.
We showed that state-of-the-art history-based predictive cache management techniques are deficient in the presence of cache thrashing stemming from irregular access patterns of graph applications processing large graphs. In response, we introduced GRASP, specialized cache management for LLC for graph analytics on natural graphs. GRASP's specialized cache policies exploit the high reuse inherent in hot vertices while maintaining the flexibility to capture reuse in other cache blocks. GRASP leverages software reordering optimizations such as DBG to enable a lightweight interface that allows hardware to reliably pinpoint hot vertices amidst irregular access patterns. In doing so, GRASP avoids the need for a storage-intensive prediction mechanism or additional metadata storage in the LLC. GRASP requires minimal hardware support, making it attractive for integration into commodity server processors to enable acceleration for the domain of graph analytics. %
Finally, GRASP delivers consistent performance gains on high-skew datasets, while preventing slowdowns on low-skew datasets.

\section{Critical Analysis}
In this section, we perform a critical analysis of the proposals presented in the prior chapters.

\subsection{Hardware Overheads}
Hardware overhead of a cache management technique may hinder its commercial adoption. 
Leeway, like the state-of-the-art Hawkeye and most other history-based techniques, requires a PC signature to be propagated through the core pipeline all the way to the LLC. 
Leeway also requires slightly higher storage than the prior techniques (e.g., 44KB for Leeway vs 31KB for Hawkeye)
to store recency state and other prediction metadata. However, it is noteworthy that the total storage requirement for Leeway is only 1.4\% of LLC capacity. More importantly, Leeway accesses the history table completely off the critical path, unlike Hawkeye, and requires significantly fewer number of look-ups than prior techniques. 

GRASP altogether removes the requirement of history table, and in turn, propagation of a PC signature too. Instead, reuse predictions rely on a new interface, which software uses to pass semantic information of the application to the hardware. While the interface is lightweight, it does require a new LLC component that is physically placed near the core. While such distributed design of LLC components may not pose a technical challenge, it may incur extra organizational cost by requiring additional communication between core, cache design and verification teams.

Overall, the hardware overheads of our proposals are generally at or below par with the state-of-the-art techniques. Meanwhile, they generally provide higher performance improvements compared to the state-of-the-art techniques across a variety of applications and deployment scenarios, making them promising candidates among the high-performance prior techniques for commercial adoption.

\subsection{Evaluation Methodology}
In this thesis, we use a simulation-based methodology to evaluate various cache management techniques. Our decision to restrict ourselves to simulation infrastructures, and therefore trading off accuracy and cost for speed and ease of evaluation, is influenced by the prohibitive cost to evaluate the architectural modifications in real chips. We follow a well accepted practice for architecture research in both academia and industry to evaluate performance impact of microarchitecture features by simulations. Having said that, we do note that our proposals presented in this thesis are backed by intuitive reasoning and sound modeling of cache statistics (e.g., modeling of miss rate or MPKI) to ensure reproducibility of results on real chips.

\subsection{Evaluation of Other Emerging Domains}
In this thesis, we proposed domain-specialized cache management only for the domain of graph analytics. In practice, there are numerous other emerging domains such as data analytics, machine learning and other big data applications (e.g., popular data center applications such as web search and data serving) that could potentially benefit from the domain-specialized cache management. We do not characterize those applications as studying the fundamental cache access patterns of all (or a subset of) applications from a given domain requires significant time, resources and domain expertise. However, doing so may not be a barrier for a commercial entity, which wishes to accelerate a particular domain of interest that is considered of a high value for their business. Therefore, we envision that in the future systems, for selected high-value domains, LLC will be managed via domain-specialized cache management (such as GRASP) and for the rest of the applications, LLC will be managed via a robust domain-agnostic technique such as Leeway. It is noteworthy that each domain-specialized cache management technique may not necessarily require a unique software-hardware interface as the interface can be made abstract (as done for GRASP), and can be generalized to meet the requirements of a set of domains.

\section{Future Work}
In this section, we highlight limitations of our proposals presented in the preceding chapters and highlight potential future directions for the research in cache management.

\subsection{Inclusive/Exclusive Cache Hierarchy}
As explained in Chapter~\ref{ch:background}, a cache hierarchy can be maintained as {\em fully-inclusive}, as {\em fully-exclusive} or as {\em non-inclusive non-exclusive (NINE)}. In this thesis, we simulated Leeway and GRASP under NINE LLC. %
Leeway and GRASP (as well as state-of-the-art history-based predictive techniques) employ aggressive prediction mechanisms to reduce cache thrashing. For example, Leeway bypasses the insertion for cache blocks that are predicted dead on arrival by forwarding data directly to the higher-level caches and GRASP inserts cache blocks that are expected to have no reuse with the least priority, immediately making them eviction candidates. 
While such mechanisms are useful in reducing cache pollution, and in turn, improving application performance for NINE LLC, they cannot be readily ported to fully-inclusive and fully-exclusive LLC as discussed below.

\noindenttitle{Fully-inclusive LLC:} For the fully-inclusive LLC, a cache block eviction at LLC requires a back invalidation to evict the same cache block from all the higher-level caches to maintain inclusion. 
Under such an inclusion policy, bypassing, by definition, is not possible as LLC must contain the cache blocks present in any higher-level caches.
Similarly, other aggressive mechanisms may not always be beneficial for fully-inclusive LLC as cache blocks that do not exhibit any reuse at LLC may exhibit high reuse at the higher-level caches. Evicting such cache blocks from fully-inclusive LLC triggers back invalidation, leading to premature evictions of these cache blocks from the higher-level caches. Therefore, accommodating such aggressive thrash-resistant mechanisms for fully-inclusive LLC may require coordination across different levels of the cache hierarchy such as {\em Query Based Selection (QBS)}~\cite{qbs}. 
While QBS has been shown to work for recency-friendly techniques like LRU or NRU, integrating QBS for aggressive thrash-resistant techniques such as Leeway (or prior history-based techniques) remains an open question as discussed below.

QBS selects a provisional victim (e.g., LRU cache block) and queries the higher-level caches (e.g., L1, L2 or both) to check if they contain a provisionally selected victim cache block.
If they do, QBS infers that the provisional victim has long temporal reuse in the higher-level caches, and thus gives it a {\em second chance} by increasing the priority of the provisional victim (e.g., by moving the victim to the MRU position). 
Subsequently, QBS attempts to find another victim, such as the second least recently cache block and so on. 
Meanwhile, if the provisional victim is not present in the higher-level caches, QBS evicts the block from LLC. 
Intuitively, the time window for a block to move from the MRU position to the LRU position at LLC under recency-friendly techniques is reasonably big, which allows the higher-level caches to completely exploit the reuse for the cache blocks having short temporal reuse. 
Thus, QBS policy is effective for recency-friendly techniques as it can differentiate cache blocks with long temporal reuse from the blocks with short temporal reuse in the higher-level caches.
However, combining QBS with aggressive thrash-resistant techniques at LLC pose a challenge. Consider an example of SHiP, which
inserts a significant fraction of cache blocks at the LRU position, leaving little time for the higher-level caches to fully exploit the reuse of many cache blocks. Therefore, a significant fraction of victim cache blocks are likely to be present in the higher-level caches, forcing QBS to provide them second chance. However, doing so defeats the purpose of their insertion at the LRU position as these blocks are unlikely to exhibit any reuse.

\noindenttitle{Fully-exclusive LLC:} For the fully-exclusive LLC, on LLC hit, a cache block is moved from LLC to L2, which involves an eviction at LLC and an insertion at L2. 
Thus, by design, in a single generation of a cache block, the block can incur at most one hit. 
Under such an inclusion policy, a cache block is evicted from LLC on a cache hit, and thus looses reuse information (e.g., Live Distance for Leeway) that, otherwise, can be accumulated over the block's on-chip residency. 
One potential way to mitigate this is by utilizing the cache directory. The directory keeps track of the coherence state for each cache block. The directory is usually inclusive of all on-chip cache blocks even when the LLC is not. Thus, directory can be augmented to accumulate reuse information per cache block during the block's on-chip residency.

\subsection{Removing PC-Dependency for Reuse Predictions at LLC}
Like Leeway, most of the prior history-based predictive techniques rely on a PC-based reuse correlation for reuse prediction~\citepcandcrc{}. Thus, they require 
propagating a PC signature through the core pipeline all the way to the LLC.
While a PC signature requires far fewer bits than a full PC address (\eg 14-bits for a PC signature vs 48-bits for a full PC address), number of bits needed to be added in a cache request to accommodate a PC signature is still non-trivial, which so far has hindered the commercial adoption of PC-based predictive techniques for LLC management. This calls for new mechanisms to predict reuse of cache blocks that do not rely on PC signatures, but provide performance that is on par, if not above, with the PC-based predicting techniques. 

GRASP employs one such mechanism that leverages a lightweight software support.
GRASP not just eliminates the need for propagating a PC signature but also eliminates the need for storage-intensive history tables altogether. GRASP requires propagating only a 2-bit Reuse Hint to the LLC on each cache access to guide cache policy decisions.

\subsection{Overhead of Software Vertex Reordering Techniques}
Software vertex reordering techniques are effective when the time required for the reordering is less than the reduction in the execution time of an application due to improved cache efficiency. For applications that have small execution time, reordering cost of a vertex reordering technique may not be amortized, resulting in a net slowdown (e.g., SSSP from one root traversal in Fig.~\ref{fig:dbg:net-cost-sssp} of Chapter~\ref{ch:dbg}). However, we believe, there are two future research directions that have potential to amortize reordering cost even for such applications.%

\noindenttitle{Integrating reordering techniques with graph generation: }
In this thesis, we assumed that the graph datasets are readily available, and thus also assumed that the spatio-temporal locality in real-world datasets (specifically for the structured datasets) exists without any overhead. In practice, such ordering may be a positive side effect of dataset generation algorithm (e.g., crawling webpages in certain order) or it may have been achieved by post-processing a dataset (e.g., graph datasets available from The Laboratory for Web Algorithmics have been ordered with the Layered Label Propagation technique~\cite{llp}). Thus, there exist an opportunity to integrate skew-aware reordering techniques with the dataset generation process; by doing so, we can eliminate the need to regenerate CSR-like structure post vertex reordering, which dominates the reordering cost. At the very least, the cost of a reordering technique should be compared to the cost of a post-processing technique used over the raw dataset to understand the cost-benefit trade-offs of techniques from different domains.

\noindenttitle{Amortizing reordering costs on dynamic graphs: }
In this thesis, we assumed that graphs are static, and thus have evaluated a net speed-up conservatively assuming only one graph application (or query) over the reordered dataset (refer to Fig.~\ref{fig:dbg:all-cost-perf} in Chapter~\ref{ch:dbg}). In practice, a graph may evolve over time and a stream of graph updates (i.e., addition or removal of vertices or edges) are interleaved with graph-analytic queries. For such a deployment, graph reordering may provide an even greater benefit as the reordering cost can be amortized not only over multiple graph traversals of a single query, but also over multiple graph queries. Intuitively, addition or removal of some vertices or edges in a large graph would not lead to a drastic change in the degree distribution, and thus unlikely to change which vertices are classified hot in a short time window. Therefore, reordering techniques may need to be re-applied at large periodic intervals (\ie after a series of updates has been made to a graph) to improve cache behavior,
amortizing the cost of reordering over multiple graph queries performed in a given interval.

\section{Concluding Remarks}
In this thesis, we emphasized the need for robust cache management mechanisms and policies for LLC to minimize cache misses in the face of variability in the reuse behavior of cache blocks. To that end, we proposed two cache management techniques, employing new variability-tolerant features such as a new metric (Live Distance) and adaptive reuse-aware policies by Leeway, and software-guided cache management for graph analytics by GRASP. While these features are used by our proposed techniques in a specific way, we believe, they can potentially be integrated with other cache management techniques to make them robust in addressing variability in reuse prediction for LLC.

\nocite{c-dead}
\nocite{c-crc2-5reuse}
\nocite{c-leeway}
\nocite{c-dbg}
\nocite{c-grasp-patent}
\nocite{c-grasp-poster}
\nocite{c-grasp}

\else

\fi

\singlespace

\printbibliography

@manual{championship,
    author = {A. R. Alameldeen and A. Jaleel and M. K. Qureshi and J. Emer},
    title = "{JILP Workshop on Computer Architecture Competitions: Cache Replacement Championship}",
    year = {2010},
    series = {JWAC-1:CRC},
    month = jun,
    note = "\url{http://www.jilp.org/jwac-1}",
}

@inproceedings{burst,
    author = {Liu, Haiming and Ferdman, Michael and Huh, Jaehyuk and Burger, Doug},
    title = "{Cache Bursts: A New Approach for Eliminating Dead Blocks and Increasing Cache Efficiency}",
    booktitle = {IEEE/ACM International Symposium on Microarchitecture},
    year = {2008},
    series = {MICRO-41},
    month = nov,
    doi = {10.1109/MICRO.2008.4771793},
    publisher = {IEEE},
}

@article{counter,
    author = {Kharbutli, Mazen and Solihin, Yan},
    title = "{Counter-Based Cache {Replacement} and Bypassing Algorithms}",
    journal = {IEEE Transactions on Computers},
    year = {2008},
    volume={57},
    number={4},
    doi = {10.1109/TC.2007.70816},
    publisher = {IEEE},
}

@INPROCEEDINGS{cmpsim,
    author = {Aamer Jaleel and Robert S. Cohn and Chi-Keung Luk and Bruce Jacob},
    title = "{CMP\$im: A Pin-Based On-The-Fly Multi-Core Cache Simulator}",
    year = {2008},
    booktitle = {International Workshop on Modeling, Benchmarking and Simulation (MoBS)},
}

@manual{crc2,
    author = {J. Kim and P. V. Gratz},
    title = "{The 2nd Cache Replacement Championship, co-located with ISCA}",
    year = {2017},
    month = jun,
    series = {CRC2},
    note = "\url{http://crc2.ece.tamu.edu}",
}

@manual{champsim,
    author = {},
    key = {champsim},
    title = "{ChampSim: A Trace-based Cycle-accurate Simulator}",
    year = {2017},
    month = jun,
    note = "\url{https://github.com/ChampSim/ChampSim}",
}

@inproceedings{simpoints,
    author = {Perelman, Erez and Hamerly, Greg and Van Biesbrouck, Michael and Sherwood, Timothy and Calder, Brad},
    title = "{Using SimPoint for Accurate and Efficient Simulation}",
    booktitle = {Proceedings of the ACM SIGMETRICS International Conference on Measurement and Modeling of Computer Systems},
    doi = {10.1145/781027.781076},
    publisher = {Association for Computing Machinery},
    year = {2003},
    month = jun,
    series = {SIGMETRICS’03}
}

@manual{cloudsuite,
    author={},
    key={cloudsuite},
    title = "{CloudSuite: The Benchmark Suite of Cloud Services}",
    year = {2012},
    note = "\url{http://cloudsuite.ch}"
}

@article{reusedist,
    author = {Das, Subhasis and Aamodt, Tor M. and Dally, William J.},
    title = "{Reuse Distance-Based Probabilistic Cache Replacement}",
    journal = {ACM Transactions on Architecture and Code Optimization},
    publisher = {Association for Computing Machinery},
    year = {2015},
    month = oct,
    doi = {10.1145/2818374},
    volume = {12},
    number = {4},
}

@misc{snapnets,
    author       = {Jure Leskovec and Andrej Krevl},
    title        = "{{SNAP Datasets}: {Stanford} Large Network Dataset Collection}",
    note          = "\url{http://snap.stanford.edu/data}",
    year         = 2014,
}

@inproceedings{twitter,
    author = {Kwak, Haewoon and Lee, Changhyun and Park, Hosung and Moon, Sue},
    title = "{W}hat is {T}witter, a social network or a news media?",
    month = apr,
    publisher = {Association for Computing Machinery},
    doi = {10.1145/1772690.1772751},
    series = {WWW’10},
    booktitle = {International Conference on World Wide Web},
    year = {2010},
}

@MISC{twitter_mpi,
    title = "{Twitter (MPI) network dataset -- {KONECT}}",
    organization = {The Koblenz Network Collection},
    year = {2017},
    note = "\url{http://konect.uni-koblenz.de/networks/twitter\_mpi}"
}

@MISC{konect-friendster,
    organization = {The Koblenz Network Collection},
    title = {Friendster network dataset -- KONECT},
    year = {2016},
    note = "\url{http://konect.uni-koblenz.de/networks/friendster}"
}

@inproceedings{PaRMAT,
    author = {Khorasani, Farzad and Gupta, Rajiv and Bhuyan, Laxmi N.},
    title = "{Scalable SIMD-Efficient Graph Processing on GPUs}",
    publisher = {IEEE},
    month = oct,
    series = {PACT'15},
    doi = {10.1109/PACT.2015.15},
    booktitle = {International Conference on Parallel Architectures and Compilation Techniques},
    year = {2015}
}

@inproceedings{gorder,
    author = {Wei, Hao and Yu, Jeffrey Xu and Lu, Can and Lin, Xuemin},
    title = {Speedup Graph {Processing} by Graph Ordering},
    month = jun,
    publisher = {Association for Computing Machinery},
    doi = {10.1145/2882903.2915220},
    booktitle = {International Conference on Management of Data},
    series = {SIGMOD’16},
    year = {2016},
}

@ARTICLE{sniper,
    author = {Trevor E. Carlson and Wim Heirman and Stijn Eyerman and Ibrahim Hur
        and Lieven Eeckhout},
    title = "{An Evaluation of High-Level Mechanistic Core Models}",
    month = aug,
    publisher = {Association for Computing Machinery},
    volume = {11},
    number = {3},
    doi = {10.1145/2629677},
    journal = {ACM Transactions on Architecture and Code Optimization},
    year = {2014},
}

@inproceedings{galois,
    author = {Nguyen, Donald and Lenharth, Andrew and Pingali, Keshav},
    title = "{A Lightweight {Infrastructure} for Graph Analytics}",
    month = nov,
    publisher = {Association for Computing Machinery},
    doi = {10.1145/2517349.2522739},
    booktitle = {Proceedings of the ACM Symposium on Operating Systems Principles},
    series = {SOSP’13},
    year = {2013},
}

@techreport{pagerank,
    number = {1999-66},
    author = {Lawrence Page and Sergey Brin and Rajeev Motwani and Terry Winograd},
    title = "{The {PageRank} Citation Ranking: Bringing Order to the Web}",
    type = {Technical Report},
    publisher = {Stanford InfoLab},
    year = {1999},
    institution = {Stanford InfoLab},
    note = "\url{http://ilpubs.stanford.edu:8090/422}",
}

@article{power-law,
	author = {Barab{\'a}si, Albert-L{\'a}szl{\'o} and Albert, R{\'e}ka},
	title = "{Emergence of Scaling in {Random} Networks}",
	volume = {286},
	number = {5439},
	doi = {10.1126/science.286.5439.509},
	year = {1999},
	publisher = {American Association for the Advancement of Science},
	journal = {Science}
}

@inproceedings{community2,
    author = {Leskovec, Jure and Lang, Kevin J. and Dasgupta, Anirban and Mahoney, Michael W.},
    title = "{Statistical Properties of Community Structure in Large Social and Information Networks}",
    month = apr,
    publisher = {Association for Computing Machinery},
    doi = {10.1145/1367497.1367591},
    series = {WWW’08},
    booktitle = {International Conference on World Wide Web},
    year = {2008},
}

@inproceedings{powergraph,
    author = {Gonzalez, Joseph E. and Low, Yucheng and Gu, Haijie and Bickson, Danny and Guestrin, Carlos},
    title = "{PowerGraph: Distributed Graph-parallel Computation on {Natural} Graphs}",
    booktitle = {{USENIX} Symposium on Operating Systems Design and Implementation},
    year = {2012},
    note = "\url{https://www.usenix.org/conference/osdi12/technical-sessions/presentation/gonzalez}",
    publisher = {{USENIX}},
    month = oct,
    series = {OSDI'12}
}

@inproceedings{ligra,
    author = {Shun, Julian and Blelloch, Guy E.},
    title = "{Ligra: A Lightweight Graph Processing Framework for Shared Memory}",
    booktitle = {Proceedings of the ACM SIGPLAN Symposium on Principles and Practice of Parallel Programming},
    year = {2013},
    publisher = {Association for Computing Machinery},
    month = feb,
    doi = {10.1145/2442516.2442530},
    series = {PPoPP ’13}
}

@INPROCEEDINGS{pgx.d, 
    author={S. Hong and S. Depner and T. Manhardt and J. Van Der Lugt and M. Verstraaten and H. Chafi}, 
    booktitle={International Conference for High Performance Computing, Networking, Storage and Analysis}, 
    title="{PGX.D: a fast distributed graph processing engine}", 
    month = nov,
    publisher = {Association for Computing Machinery},
    doi = {10.1145/2807591.2807620},
    series = {SC’15},
    year={2015}, 
}

@INPROCEEDINGS{rmat,
    author = {Deepayan Chakrabarti and Yiping Zhan and Christos Faloutsos},
    title = {R-MAT: A recursive model for graph mining},
    doi = {10.1137/1.9781611972740.43},
    booktitle = {SIAM International {Conference} on Data Mining},
    year = {2004},
    month = apr,
}

@inproceedings{power-law-internet,
    author = {Faloutsos, Michalis and Faloutsos, Petros and Faloutsos, Christos},
    title = "{On Power-law Relationships of the Internet Topology}",
    month = aug,
    publisher = {Association for Computing Machinery},
    doi = {10.1145/316188.316229},
    series = {SIGCOMM ’99},
    booktitle = {The Conference on Applications, Technologies, Architectures, and Protocols for Computer Communication},
    year = {1999},
}

@inproceedings{prefetch-data-structure,
    author = {Ainsworth, Sam and Jones, Timothy M.},
    title = "{Graph Prefetching Using Data Structure Knowledge}",
    publisher = {Association for Computing Machinery},
    doi = {10.1145/2925426.2926254},
    booktitle = {International Conference on Supercomputing},
    series = {ICS’16},
    month = jun,
    year = {2016},
}

@INPROCEEDINGS{graphicionado, 
    author={T. J. Ham and L. Wu and N. Sundaram and N. Satish and M. Martonosi}, 
    booktitle={Proceedings of the ACM/IEEE International Symposium on Microarchitecture}, 
    title="{Graphicionado: A high-performance and energy-efficient {accelerator} for graph analytics}", 
    publisher = {IEEE Press},
    doi = {10.1109/MICRO.2016.7783759},
    series = {MICRO-49},
    month = oct,
    year={2016}, 
}

@inproceedings{imp,
    author = {Yu, Xiangyao and Hughes, Christopher J. and Satish, Nadathur and Devadas, Srinivas},
    title = "{IMP: {Indirect} Memory Prefetcher}",
    publisher = {Association for Computing Machinery},
    doi = {10.1145/2830772.2830807},
    series = {MICRO-48},
    month = dec,
    booktitle = {Proceedings of the ACM/IEEE International Symposium on Microarchitecture},
    year = {2015},
}

@inproceedings{graphchi,
    author = {Kyrola, Aapo and Blelloch, Guy and Guestrin, Carlos},
    title = "{GraphChi: Large-scale Graph Computation on Just a PC}",
    booktitle = {USENIX Symposium on Operating Systems Design and Implementation},
    publisher = {{USENIX}},
    month = oct,
    note = "\url{https://www.usenix.org/conference/osdi12/technical-sessions/presentation/kyrola}",
    year = {2012},
    series = {OSDI'12},
}

@misc{xeon,
    author={},
    key={xeon},
    title="{Intel Xeon Processor E5-2630 v4}",
    note =  "\url{https://ark.intel.com/products/92981/Intel-Xeon-Processor-E5-2630-v4-25M-Cache-2_20-GHz}",
    organization={Intel Corporation},
    year = {2016}
}

@MISC{konect-wl,
    organization = {The Koblenz Network Collection},
    title = "{Wikipedia, English network dataset -- KONECT}",
    year = {2017},
    note = "\url{http://konect.uni-koblenz.de/networks/dbpedia-link}"
}

@inproceedings{graphlab,
  author    = {Yucheng Low and
              Joseph Gonzalez and
              Aapo Kyrola and
              Danny Bickson and
              Carlos Guestrin and
              Joseph M. Hellerstein},
  title     = "{GraphLab: {A} New Framework For Parallel Machine Learning}",
  publisher = {AUAI Press},
  booktitle = {The Conference on Uncertainty
              in Artificial Intelligence},
  year      = {2010},
  month = jul,
  note = "\url{https://dl.acm.org/doi/10.5555/3023549.3023589}",
  series = {UAI’10}
}

@article{graphmat,
    author = {Sundaram, Narayanan and Satish, Nadathur and Patwary, Md Mostofa Ali and Dulloor, Subramanya R. and Anderson, Michael J. and Vadlamudi, Satya Gautam and Das, Dipankar and Dubey, Pradeep},
    title = "{GraphMat: High Performance Graph Analytics Made Productive}",
    journal = {Proceedings of the VLDB Endowment},
    publisher = {VLDB Endowment},
    month = jul,
    volume = {8},
    number = {11},
    doi = {10.14778/2809974.2809983},
    year = {2015},
}

@inproceedings{rcm,
    author = {Cuthill, E. and McKee, J.},
    title = "{Reducing the Bandwidth of Sparse Symmetric Matrices}",
    booktitle = {Proceedings of the National Conference},
    month = aug,
    publisher = {Association for Computing Machinery},
    doi = {10.1145/800195.805928},
    year = {1969},
    series = {ACM ’69}
}

@INPROCEEDINGS{recall,
    author={K. Lakhotia and S. Singapura and R. Kannan and V. Prasanna},
    booktitle={IEEE International Conference on High Performance Computing},
    title="{ReCALL: Reordered Cache Aware Locality Based Graph Processing}",
    doi = {10.1109/HiPC.2017.00039},
    publisher = {IEEE},
    month = dec,
    year={2017},
    series = {HiPC'17},
}

@INPROCEEDINGS{barren,
    author={X. Tong and A. Moshovos},
    booktitle={International {Conference} on Computer Design},
    title="{BarTLB: Barren page resistant TLB for managed runtime languages}",
    publisher = {IEEE},
    month = oct,
    doi = {10.1109/ICCD.2014.6974692},
    year={2014},
    series = {ICCD'14}
}

@ARTICLE{CHDFS, 
    author={J. Banerjee and W. Kim and S. -. Kim and J. F. Garza}, 
    journal={IEEE Transactions on Software Engineering}, 
    title="{Clustering a DAG for CAD databases}", 
    publisher = {IEEE Press},
    volume = {14},
    number = {11},
    month = nov,
    doi = {10.1109/32.9055},
    year={1988}, 
}

@INPROCEEDINGS{SlashBurn, 
    author={U. Kang and C. Faloutsos}, 
    booktitle={IEEE International Conference on Data Mining}, 
    title="{Beyond `Caveman Communities': Hubs and Spokes for Graph Compression and Mining}", 
    publisher = {IEEE},
    doi = {10.1109/ICDM.2011.26},
    series = {ICDM'11},
    month = dec,
    year={2011}, 
}

@inproceedings{LDG,
    author = {Stanton, Isabelle and Kliot, Gabriel},
    title = "{Streaming Graph Partitioning for Large Distributed Graphs}",
    booktitle = {ACM SIGKDD International Conference on Knowledge Discovery and Data Mining},
    year = {2012},
    month = aug,
    doi = {10.1145/2339530.2339722},
    series = {KDD'12},
}

@article{METIS,
    author = {Karypis, George and Kumar, Vipin},
    title = "{Multilevel k-way Partitioning Scheme for Irregular Graphs}",
    journal = {Journal of Parallel and Distributed Computing},
    month = jan,
    doi = {10.1006/jpdc.1997.1404},
    publisher = {Elsevier BV},
    year = {1998},
    volume = {48},
    number = {1},
}

@INPROCEEDINGS{rabbit, 
    author={J. Arai and H. Shiokawa and T. Yamamuro and M. Onizuka and S. Iwamura}, 
    booktitle={IEEE International Parallel and Distributed Processing Symposium}, 
    title="{Rabbit Order: Just-in-Time Parallel Reordering for Fast Graph Analysis}", 
    doi = {10.1109/IPDPS.2016.110},
    publisher = {IEEE},
    month = may,
    series = {IPDPS'16},
    year={2016}, 
}

@INPROCEEDINGS{xmem,
    author={Nandita Vijaykumar and Abhilasha Jain and Diptesh Majumdar and Kevin Hsieh and Gennady Pekhimenko and Eiman Ebrahimi and Nastaran Hajinazar and Phillip B. Gibbons and Onur Mutlu},
    booktitle={International Symposium on Computer Architecture},
    title="{A Case for Richer Cross-Layer Abstractions: Bridging the Semantic Gap with Expressive Memory}",
    year={2018},
    month = jun,
    doi = {10.1109/ISCA.2018.00027},
    publisher = {IEEE Press},
    series = {ISCA'18},
}

@inproceedings{whirlpool,
    author = {Mukkara, Anurag and Beckmann, Nathan and Sanchez, Daniel},
    title = "{Whirlpool: {Improving} Dynamic Cache Management with Static Data Classification}",
    month = mar,
    publisher = {Association for Computing Machinery},
    doi = {10.1145/2872362.2872363},
    series = {ASPLOS ’16},
    booktitle = {International Conference on Architectural Support for Programming Languages and Operating Systems},
    year = {2016},
}

@article{radii,
    author = {Magnien, Cl{\'e}mence and Latapy, Matthieu and Habib, Michel},
    title = "{Fast Computation of Empirically Tight Bounds for the Diameter of Massive Graphs}",
    journal = {Journal of Experimental Algorithmics},
    month = feb,
    publisher = {Association for Computing Machinery},
    volume = {13},
    doi = {10.1145/1412228.1455266},
    articleno = {10},
    year = {2009},
}

@article{gap,
  author    = {Scott Beamer and
               Krste Asanovic and
               David A. Patterson},
  title     = "{The {GAP} Benchmark Suite}",
  journal   = {CoRR},
  year      = {2015},
  note       = "\url{http://arxiv.org/abs/1508.03619}",
}

@misc{pld,
    author       = {},
    key = {pld},
    organization = {Web Data Commons},
    title        = "{Hyperlink Graphs}",
    note          = "\url{http://webdatacommons.org/hyperlinkgraph}",
    year = {2018}
}

@inproceedings{hats,
    author = { Anurag Mukkara and 
    Nathan Beckmann and 
    Maleen Abeydeera and 
    Xiaosong Ma and 
    Daniel Sanchez},
    title = "{Exploiting Locality in Graph Analytics through Hardware-Accelerated Traversal Scheduling}",
    month = oct,
    doi = {10.1109/MICRO.2018.00010},
    publisher = {IEEE Press},
    series = {MICRO-51},
    booktitle = {Proceedings of the ACM/IEEE International Symposium on Microarchitecture},
    year = {2018},
}

@article {community1,
	author = {Girvan, M. and Newman, M. E. J.},
	title = "{Community structure in social and biological networks}",
	year = {2002},
	volume = {99},
	number = {12},
	month = jun,
	doi = {10.1073/pnas.122653799},
	publisher = {National Academy of Sciences},
	journal = {The National Academy of Sciences}
}

@misc{inteldie,
    author={},
    key={inteldie},
    title="{Intel Broadwell Microarchitectures}",
    note="\url{https://en.wikichip.org/wiki/intel/microarchitectures/broadwell\_(client)}",
    year={2016}
}

@misc{amddie,
    author={},
    key={amddie},
    title="{AMD Zen Microarchitecutres}",
    note="\url{https://en.wikichip.org/wiki/amd/microarchitectures/zen}",
    year={2017}
}

@book{PseudoLRU,
    author = {Handy, Jim},
    title = "{The Cache Memory Book}",
    year = {1993},
    isbn = {0123229855},
    note = "\url{https://dl.acm.org/doi/10.5555/157953}",
    publisher = {Academic Press Professional, Inc.},
}

@book{hp,
    author={Hennessy, John L. and Patterson, David A.},
    title = "{Computer Architecture, Sixth Edition: A Quantitative Approach}",
    year = {2017},
    isbn = {0128119055},
    publisher = {Morgan Kaufmann Publishers Inc.},
    edition = {6th},
    note = "\url{https://dl.acm.org/doi/10.5555/3207796}"
}

@INPROCEEDINGS{gippr,
    author={D. A. {Jiménez}},
    booktitle={Proceedings of the ACM/IEEE International Symposium on Microarchitecture},
    title="{Insertion and promotion for tree-based {PseudoLRU} last-level caches}",
    publisher = {Association for Computing Machinery},
    doi = {10.1145/2540708.2540733},
    series = {MICRO-46},
    month = dec,
    year={2013},
}

@inproceedings{rrip,
    author = {Jaleel, Aamer and Theobald, Kevin B. and Steely,Jr., Simon C. and Emer, Joel},
    title = "{High Performance Cache Replacement Using Re-reference Interval Prediction (RRIP)}",
    booktitle = {International {Symposium} on Computer Architecture},
    year = {2010},
    publisher = {Association for Computing Machinery},
    doi = {10.1145/1815961.1815971},
    series = {ISCA'10},
    month = jun,
}

@inproceedings{ship,
    author = {Wu, Carole-Jean and Jaleel, Aamer and Hasenplaugh, Will and Martonosi, Margaret and Steely,Jr., Simon C. and Emer, Joel},
    title = "{SHiP: Signature-based Hit Predictor for High Performance Caching}",
    booktitle = {Proceedings of the ACM/IEEE International Symposium on Microarchitecture},
    year = {2011},
    publisher = {Association for Computing Machinery},
    doi = {10.1145/2155620.2155671},
    series = {MICRO-44},
    month = dec,
}

@inproceedings{sampler,
    author = {Khan, Samira Manabi and Tian, Yingying and {Jiménez}, Daniel A.},
    title = "{Sampling Dead Block Prediction for Last-Level Caches}",
    booktitle = {Proceedings of the ACM/IEEE International Symposium on Microarchitecture},
    publisher = {IEEE Computer Society},
    year = {2010},
    doi = {10.1109/MICRO.2010.24},
    series = {MICRO-43},
    month = dec,
}

@inproceedings{lasttouch,
    author = {Lai, An-Chow and Falsafi, Babak},
    title = "{Selective, Accurate, and Timely Self-invalidation Using Last-touch Prediction}",
    booktitle = {International Symposium on Computer Architecture},
    year = {2000},
    doi = {10.1109/ISCA.2000.854385},
    month = jun,
    publisher = {IEEE},
    series = {ISCA'00}
}

@inproceedings{mdpp, 
    author={E. Teran and Y. Tian and Z. Wang and D. A. {Jiménez}}, 
    booktitle={IEEE International Symposium on High-Performance Computer Architecture}, 
    title="{Minimal disturbance placement and promotion}", 
    doi = {10.1109/HPCA.2016.7446065},
    month = mar,
    publisher = {IEEE},
    year={2016}, 
    series = {HPCA'16}
}

@inproceedings{tadip,
    author = {Jaleel, Aamer and Hasenplaugh, William and Qureshi, Moinuddin K. and Sebot, Julien and Steely,Jr., Simon and Emer, Joel},
    title = "{Adaptive Insertion Policies for Managing Shared Caches}",
    booktitle = {International Conference on Parallel Architectures and Compilation Techniques},
    year = {2008},
    month = oct,
    publisher = {Association for Computing Machinery},
    series = {PACT’08},
    doi = {10.1145/1454115.1454145},
}

@inproceedings{dip,
    author = {Qureshi, Moinuddin K. and Jaleel, Aamer and Patt, Yale N. and Steely, Simon C. and Emer, Joel},
    title = {Adaptive Insertion Policies for High Performance Caching},
    booktitle = {International Symposium on Computer Architecture},
    year = {2007},
    series = {ISCA'07},
    month = jun,
    publisher = {Association for Computing Machinery},
    doi = {10.1145/1250662.1250709},
}

@inproceedings{reftrace,
    author = {Lai, An-Chow and Fide, Cem and Falsafi, Babak},
    title = "{Dead-block Prediction \& Dead-block Correlating Prefetchers}",
    booktitle = {International Symposium on Computer Architecture},
    year = {2001},
    series = {ISCA'01},
    month = may,
    doi = {10.1145/379240.379259},
    publisher = {Association for Computing Machinery},
}

@INPROCEEDINGS{cache-hints1-soft-assisted,
    author={P. {Jain} and S. {Devadas} and D. {Engels} and L. {Rudolph}},
    booktitle={IEEE/ACM International Conference on Computer Aided Design},
    title="{Software-assisted cache replacement mechanisms for embedded systems}",
    year={2001},
    publisher = {IEEE},
    month = nov,
    series = {ICCAD'01},
    doi = {10.1109/ICCAD.2001.968607}
}

@INPROCEEDINGS{cache-hints2,
    author={Zhenlin Wang and K. S. {McKinley} and A. L. {Rosenberg} and C. C. {Weems}},
    booktitle={International Conference on Parallel Architectures and Compilation Techniques},
    title="{Using the compiler to improve cache replacement decisions}",
    month = sep,
    doi = {10.1109/PACT.2002.1106018},
    publisher = {IEEE Computer Society},
    series = {PACT’02},
    year={2002},
}

@article{cache-hints3,
    author = {Beyls, Kristof and D'Hollander, Erik H.},
    title = "{Generating Cache Hints for Improved Program Efficiency}",
    doi = {10.1016/j.sysarc.2004.09.004},
    journal = {Journal of Systems Architecture},
    year = {2005},
    month = apr,
    volume = {51},
    number = {4},
    publisher = {Elsevier BV},
}

@inproceedings{cache-hints1,
    author = {Lebeck, Alvin R. and Raymond, David R. and Yang, Chia-Lin and Thottethodi, Mithuna},
    title = "{Annotated Memory References: A Mechanism for Informed Cache Management}",
    publisher= {Springer Berlin Heidelberg},
    booktitle = {Euro-Par Conference on Parallel Processing},
    doi = {10.1007/3-540-48311-X_177},
    month = aug,
    year = {1999},
    series = {Euro-Par'99}
}

@inproceedings{cache-hints5-pacman,
    author = {Brock, Jacob and Gu, Xiaoming and Bao, Bin and Ding, Chen},
    title = "{Pacman: Program-assisted Cache Management}",
    booktitle = {International Symposium on Memory Management},
    doi = {10.1145/2491894.2466482},
    year = {2013},
    month = jun,
    publisher = {Association for Computing Machinery},
    series = {ISMM’13}
    
}

@inproceedings{EAF,
    author = {Seshadri, Vivek and Mutlu, Onur and Kozuch, Michael A. and Mowry, Todd C.},
    title = "{The Evicted-address Filter: A Unified Mechanism to Address Both Cache Pollution and Thrashing}",
    publisher = {Association for Computing Machinery},
    doi = {10.1145/2370816.2370868},
    booktitle = {International Conference on Parallel Architectures and Compilation Techniques},
    year = {2012},
    series = {PACT’12},
    month = sep,
}

@article{opt,
    author = {Belady, L. A.},
    title = "{A Study of Replacement Algorithms for a Virtual-storage Computer}",
    journal = {IBM Systems Journal},
    year = {1966},
    volume={5},
    number={2},
    publisher = {IBM Corp.},
    doi = {10.1147/sj.52.0078},
}

@inproceedings{pseudo,
    author = {Chaudhuri, Mainak},
    title = "{Pseudo-LIFO: The Foundation of a New Family of Replacement Policies for Last-level Caches}",
    doi = {10.1145/1669112.1669164},
    booktitle = {Proceedings of the ACM/IEEE International Symposium on Microarchitecture},
    year={2009},
    month = dec,
    series = {MICRO-42}
}

@inproceedings{pd,
    author = {Duong, Nam and Zhao, Dali and Kim, Taesu and Cammarota, Rosario and Valero, Mateo and Veidenbaum, Alexander V.},
    title = "{Improving Cache Management Policies Using Dynamic Reuse Distances}",
    month = dec,
    publisher = {IEEE Computer Society},
    doi = {10.1109/MICRO.2012.43},
    series = {MICRO-45},
    booktitle = {Proceedings of the ACM/IEEE International Symposium on Microarchitecture},
    year = {2012},
}

@INPROCEEDINGS{shepherd,
    author={K. {Rajan} and G. {Ramaswamy}},
    booktitle={Proceedings of the ACM/IEEE International Symposium on Microarchitecture},
    title="{Emulating Optimal Replacement with a Shepherd Cache}",
    doi = {10.1109/MICRO.2007.25},
    year={2007},
    series = {MICRO-40},
    month = dec,
    publisher = {IEEE},
}

@inproceedings{timekeeping,
    author = {Hu, Zhigang and Kaxiras, Stefanos and Martonosi, Margaret},
    title = "{Timekeeping in the Memory System: Predicting and Optimizing Memory Behavior}",
    doi = {10.1109/ISCA.2002.1003579},
    month = may,
    publisher = {IEEE},
    booktitle = {International Symposium on Computer Architecture},
    year = {2002},
    series = {ISCA'02}
}

@INPROCEEDINGS{timekeeping2,
    author={G. {Keramidas} and P. {Petoumenos} and S. {Kaxiras}},
    booktitle={International Conference on Computer Design},
    title={Cache replacement based on reuse-distance prediction},
    publisher = {IEEE},
    doi = {10.1109/ICCD.2007.4601909},
    series = {ICCD'07},
    month = oct,
    year={2007},
}

@INPROCEEDINGS{segmentedLRU,
    author={Hongliang Gao and Chris Wilkerson},
    booktitle={In JILP Workshop on Computer Architecture Competitions: Cache Replacement Championship},
    title="{A dueling segmented LRU replacement algorithm with adaptive bypassing}",
    year={2010},
    series = {JWAC-1:CRC},
    month = jun,
    note = "\url{http://www.jilp.org/jwac-1}",
}

@article{turnoffdead,
    author = {Abella, Jaume and Gonz\'{a}lez, Antonio and Vera, Xavier and O'Boyle, Michael F. P.},
    title = "{IATAC: A Smart Predictor to Turn-off L2 Cache Lines}",
    journal = {ACM Transactions on Architecture and Code Optimization (TACO)},
    year = {2005},
    month = mar,
    publisher = {Association for Computing Machinery},
    volume = {2},
    number = {1},
    doi = {10.1145/1061267.1061271},
}

@INPROCEEDINGS{cachedecay,
    author={S. {Kaxiras} and {Zhigang Hu} and M. {Martonosi}},
    booktitle={International Symposium on Computer Architecture},
    title="{Cache decay: exploiting generational behavior to reduce cache leakage power}",
    year={2001},
    publisher = {IEEE},
    doi = {10.1109/ISCA.2001.937453},
    series = {ISCA'01},
    month = jun,
}

@article{letcachesdecay,
    author = {Hu, Zhigang and Kaxiras, Stefanos and Martonosi, Margaret},
    year = {2003},
    title = "{Let Caches Decay: Reducing Leakage Energy via Exploitation of Cache Generational Behavior}",
    journal = {ACM Transactions on Computer Systems},
    publisher = {Association for Computing Machinery},
    volume = {20},
    number = {2},
    doi = {10.1145/507052.507055},
    month = may,
}

@inproceedings{perceptron,
    author={Teran, Elvira and Wang, Zhe and Jim{\'e}nez, Daniel A.},
    publisher = {IEEE Press},
    booktitle={Proceedings of the IEEE/ACM International Symposium on Microarchitecture},
    title="{Perceptron Learning for Reuse Prediction}",
    year={2016},
    doi = {10.1109/MICRO.2016.7783705},
    month = oct,
    series = {MICRO-49}
}

@inproceedings{multi,
    author = {Jim{\'e}nez, Daniel A. and Teran, Elvira},
    title = "{Multiperspective Reuse Prediction}",
    booktitle = {Proceedings of the IEEE/ACM International Symposium on Microarchitecture},
    year = {2017},
    doi = {10.1145/3123939.3123942},
    publisher = {Association for Computing Machinery},
    month = oct,
    series = {MICRO-50}
}

@inproceedings{c-dead,
    author={Faldu, Priyank and Boris Grot}, 
    booktitle={International Workshop on Duplicating, Deconstructing and Debunking, co-located with ISCA}, 
    title="{LLC Dead Block Prediction Considered Not Useful}", 
    year={2016}, 
    month = jun,
    series = {WDDD-13}
}

@inproceedings{hawkeye,
    author={Jain, Akanksha and Lin, Calvin},
    booktitle={International Symposium on Computer Architecture},
    title="{Back to the Future: Leveraging Belady's Algorithm for Improved Cache Replacement}",
    month = jun,
    publisher = {IEEE Press},
    doi = {10.1109/ISCA.2016.17},
    series = {ISCA’16},
    year={2016},
}

@inproceedings{c-leeway,
    author={Priyank Faldu and Boris Grot}, 
    booktitle={International Conference on Parallel Architectures and Compilation Techniques},
    title="{Leeway: Addressing Variability in Dead-Block Prediction for Last-Level Caches}",
    year={2017}, 
    doi = {10.1109/PACT.2017.32},
    publisher = {IEEE},
    month = sep,
    series = {PACT'17},
}

@patent{c-grasp-patent,
    author={Priyank Faldu and Jeffrey Diamond and Avadh Patel}, 
    number={10417134},
    title="{Cache Memory Architecture and Policies for Accelerating Graph Algorithms}",
    year={2019},
    month=sep,
    type={patentus},
    holder={{Oracle International Corporation}},
}

@inproceedings{c-grasp-poster,
    author={Priyank Faldu and Jeff Diamond and Boris Grot}, 
    booktitle={International Conference on Parallel Architectures and Compilation Techniques},
    title="{POSTER: Domain-Specialized Cache Management for Graph Analytics}",
    doi = {10.1109/PACT.2019.00051},
    month = sep,
    publisher = {IEEE},
    year={2019}, 
    series = {PACT'19}
}

@inproceedings{c-grasp,
    author={Priyank Faldu and Jeff Diamond and Boris Grot}, 
    booktitle={IEEE International Symposium on High-Performance Computer Architecture},
    title="{Domain-Specialized Cache Management for Graph Analytics}",
    year={2020}, 
    doi = {10.1109/HPCA47549.2020.00028},
    series = {HPCA'20},
    publisher = {IEEE},
    month = feb,
}

@inproceedings{c-dbg,
    author={Priyank Faldu and Jeff Diamond and Boris Grot}, 
    booktitle={IEEE International Symposium on Workload Characterization},
    title="{A Closer Look at Lightweight Graph Reordering}",
    year={2019},
    doi = {10.1109/IISWC47752.2019.9041948},
    series = {IISWC'19},
    publisher = {IEEE},
    month = nov,
}

@INPROCEEDINGS{fc, 
    author={Y. Zhang and V. Kiriansky and C. Mendis and S. Amarasinghe and M. Zaharia}, 
    booktitle={IEEE International Conference on Big Data}, 
    title="{Making caches work for graph analytics}",
    doi = {10.1109/BigData.2017.8257937},
    series = {Big Data'17},
    publisher = {IEEE},
    month = dec,
    year={2017}, 
}

@INPROCEEDINGS{harmony,
    author={Jain, Akanksha and Lin, Calvin},
    title="{Rethinking Belady's Algorithm to Accommodate Prefetching}",
    booktitle={International Symposium on Computer Architecture},
    doi = {10.1109/ISCA.2018.00020},
    month = jun,
    publisher = {IEEE Press},
    series = {ISCA’18},
    year={2018}
}

@INPROCEEDINGS{hubcluster,
    author={Balaji, Vignesh and Lucia, Brandon},
    title="{When is Graph {Reordering} an {Optimization}? Studying the Effect of Lightweight Graph Reordering Across Applications and Input Graphs}",
    booktitle = {IEEE International Symposium on Workload Characterization},
    doi = {10.1109/IISWC.2018.8573478},
    month = sep,
    year={2018},
    series = {IISWC'18}
}

@INPROCEEDINGS{llama,
    author={P. {Macko} and V. J. {Marathe} and D. W. {Margo} and M. I. {Seltzer}},
    booktitle={IEEE International Conference on Data Engineering},
    title="{LLAMA: Efficient graph analytics using Large Multiversioned Arrays}",
    year={2015},
    doi = {10.1109/ICDE.2015.7113298},
    month = apr,
    series = {ICDE'15}
}

@inproceedings{aspen,
    author = {Dhulipala, Laxman and Blelloch, Guy E. and Shun, Julian},
    title = "{Low-latency Graph Streaming Using Compressed Purely-functional Trees}",
    booktitle = {International Conference on Programming Language Design and Implementation},
    year = {2019},
    month = jun,
    publisher = {Association for Computing Machinery},
    doi = {10.1145/3314221.3314598},
    series = {PLDI 2019}
}

@inproceedings{stinger,
    author = {Ediger, David and McColl, Robert and Riedy, Jason and Bader, David A.},
    title = "{Stinger: High performance data structure for streaming graphs}",
    month = sep,
    doi = {10.1109/HPEC.2012.6408680},
    booktitle = {IEEE International Conference on High Performance Extreme Computing},
    year = {2012},
    series = {HPEC'12},
}

@inproceedings{crc2-1shippp,
    author = {V. Young and C. Chou and A. Jaleel and M. K. Qureshi},
    title = "{SHiP++: Enhancing Signature-Based Hit Predictor for Improved Cache Performance}",
    booktitle = {International Workshop on Cache Replacement Championship, co-located with ISCA},
    year = {2017},
    month = jun,
    series = {CRC2},
    note = "\url{http://crc2.ece.tamu.edu}",
}

@inproceedings{crc2-2lime,
    author = {J. Wang and L. Zhang and R. Panda and L. John},
    title = "{Less is More: Leveraging Belady’s Algorithm with Demand-based Learning}",
    booktitle = {International Workshop on Cache Replacement Championship, co-located with ISCA},
    year = {2017},
    month = jun,
    series = {CRC2},
    note = "\url{http://crc2.ece.tamu.edu}",
}

@inproceedings{crc2-3multi,
    author = {D. A. Jiménez},
    title = "{Multiperspective Reuse Prediction}",
    booktitle = {{International} Workshop on Cache Replacement Championship, co-located with ISCA},
    year = {2017},
    month = jun,
    series = {CRC2},
    note = "\url{http://crc2.ece.tamu.edu}",
}

@inproceedings{crc2-4expected,
    author = {A. Vakil-Ghahani and S. Mahdizadeh-Shahri and M. Lotfi-Namin and M. Bakhshalipour and P. Lotfi-Kamran and H. Sarbazi-Azad },
    title = "{Cache Replacement Policy Based on Expected Hit Count}",
    booktitle = {International Workshop on Cache Replacement Championship, co-located with ISCA},
    year = {2017},
    month = jun,
    series = {CRC2},
    note = "\url{http://crc2.ece.tamu.edu}",
}

@inproceedings{c-crc2-5reuse,
    author = {P. Faldu and B. Grot},
    title = "{Reuse-Aware Management for Last-Level Caches}",
    booktitle = {International Workshop on Cache Replacement Championship, co-located with ISCA},
    year = {2017},
    month = jun,
    series = {CRC2},
    note = "\url{http://crc2.ece.tamu.edu}",
}

@inproceedings{crc2-6hawkeye,
    author = {A. Jain and C. Lin},
    title = "{Hawkeye Cache Replacement: Leveraging Belady’s Algorithm for Improved Cache Replacement}",
    booktitle = {International Workshop on Cache Replacement Championship, co-located with ISCA},
    year = {2017},
    month = jun,
    series = {CRC2},
    note = "\url{http://crc2.ece.tamu.edu}",
}

@inproceedings{crc2-7red,
    author = {J. Díaz and P. Ibáñez and T. Monreal and V. Viñals and J. Llabería},
    title = "{ReD: A Policy Based on Reuse Detection for a Demanding Block Selection in Last-Level Caches}",
    booktitle = {International Workshop on Cache Replacement Championship, co-located with ISCA},
    year = {2017},
    month = jun,
    series = {CRC2},
    note = "\url{http://crc2.ece.tamu.edu}",
}

@inproceedings{pipp,
    author = {Xie, Yuejian and Loh, Gabriel H.},
    title = "{PIPP: Promotion/Insertion Pseudo-partitioning of Multi-core Shared Caches}",
    booktitle = {International Symposium on Computer Architecture},
    year = {2009},
    month = jun,
    publisher = {Association for Computing Machinery},
    doi = {10.1145/1555754.1555778},
    series = {ISCA’09}
}

@ARTICLE{lrfu,
    author={ Donghee Lee and Jongmoo Choi and Jong-Hun Kim and S. H. Noh and Sang Lyul Min and Yookun Cho and Chong Sang Kim},
    journal={IEEE Transactions on Computers},
    title="{LRFU: a spectrum of policies that subsumes the least recently used and least frequently used policies}",
    months = dec,
    publisher = {IEEE Computer Society},
    volume = {50},
    number = {12},
    doi = {10.1109/TC.2001.970573},
    year={2001},
}

@inproceedings{mlp,
    title={A case for MLP-aware cache replacement},
    publisher = {IEEE Computer Society},
    doi = {10.1109/ISCA.2006.5},
    series = {ISCA’06},
    month = may,
    author={Qureshi, Moinuddin K. and Lynch, Daniel N and Mutlu, Onur and Patt, Yale N},
    booktitle={International Symposium on Computer Architecture},
    year={2006},
}

@inproceedings{dbpvictimcache,
    author = {Khan, Samira M. and Jim\'{e}nez, Daniel A. and Burger, Doug and Falsafi, Babak},
    title = "{Using Dead Blocks as a Virtual Victim Cache}",
    year = {2010},
    month = sep,
    booktitle = {Proceedings of the International Conference on Parallel Architectures and Compilation Techniques},
    series = {PACT ’10},
    publisher = {Association for Computing Machinery},
    doi = {10.1145/1854273.1854333},
}

@article{stack-distance,
    author = {Mattson, Richard L. and Gecsei, Jan and Slutz, D. R. and Traiger, I. L.},
    title = "{Evaluation Techniques for Storage Hierarchies}",
    publisher = {IBM Corp.},
    volume = {9},
    number = {2},
    doi = {10.1147/sj.92.0078},
    month = jun,
    journal = {IBM Systems Journal},
    year = {1970},
}

@inproceedings{rd2,
    author = {Ding, Chen and Zhong, Yutao},
    title = "{Predicting Whole-program Locality Through Reuse Distance Analysis}",
    booktitle = {Proceedings of the ACM SIGPLAN Conference on Programming Language Design and Implementation},
    series = {PLDI ’03},
    year = {2003},
    month = may,
    publisher = {Association for Computing Machinery},
    doi = {10.1145/781131.781159},
}

@inproceedings{rd1,
    author = {Sen, Rathijit and Wood, David A.},
    title = "{Reuse-based Online Models for Caches}",
    booktitle = {Proceedings of the ACM SIGMETRICS International Conference on Measurement and Modeling of Computer Systems},
    publisher = {Association for Computing Machinery},
    doi = {10.1145/2465529.2465756},
    series = {SIGMETRICS'13},
    month = jun,
    year = {2013},
}

@INPROCEEDINGS{rd-daniel, 
    author={Nathan Beckmann and Daniel Sanchez}, 
    booktitle={IEEE International Symposium on High-Performance Computer Architecture}, 
    title="{Modeling Cache Performance Beyond LRU}", 
    doi = {10.1109/HPCA.2016.7446067},
    year={2016}, 
    month = mar,
    series = {HPCA'16}
}

@INPROCEEDINGS{nuca,
    author = {Changkyu Kim and Doug Burger and Stephen W. Keckler},
    title = "{An adaptive, non-uniform cache structure for wire-delay dominated on-chip caches}",
    booktitle = {Proceedings of the International Conference on Architectural Support for Programming Languages and Operating Systems},
    year = {2002},
    month = oct,
    series = {ASPLOS X},
    publisher = {Association for Computing Machinery},
    doi = {10.1145/605397.605420},
}

@inproceedings{llp,
    author = {Boldi, Paolo and Rosa, Marco and Santini, Massimo and Vigna, Sebastiano},
    title = "{Layered Label Propagation: A Multiresolution Coordinate-free Ordering for Compressing Social Networks}",
    booktitle = {Proceedings of the International Conference on World Wide Web},
    year = {2011},
    month = mar,
    publisher = {Association for Computing Machinery},
    doi = {10.1145/1963405.1963488},
    series = {WWW'11}
}

@INPROCEEDINGS{qbs,
    author={A. {Jaleel} and E. {Borch} and M. {Bhandaru} and S. C. {Steely Jr.} and J. {Emer}},
    booktitle={Proceedings of the IEEE/ACM International Symposium on Microarchitecture},
    title="{Achieving Non-Inclusive Cache Performance with Inclusive Caches: Temporal Locality Aware (TLA) Cache Management Policies}",
    month = dec,
    publisher = {IEEE Computer Society},
    doi = {10.1109/MICRO.2010.52},
    year={2010},
    series = {MICRO-43}
}

@inproceedings{road,
    author = {Rossi, Ryan A. and Ahmed, Nesreen K.},
    title = "{The Network Data Repository with Interactive Graph Analytics and Visualization}",
    publisher = {AAAI Press},
    booktitle = {Proceedings of the AAAI Conference on Artificial Intelligence},
    year = {2015},
    series = {AAAI’15},
    month = jan,
    note="\url{http://networkrepository.com/road-road-usa.php}",
}

@inproceedings{adaptivecaches,
    author = {Subramanian, Ranjith and Smaragdakis, Yannis and Loh, Gabriel H.},
    title = "{Adaptive Caches: Effective Shaping of Cache Behavior to Workloads}",
    year = {2006},
    series = {MICRO-39},
    doi = {10.1109/MICRO.2006.7},
    booktitle = {Proceedings of the IEEE/ACM International Symposium on Microarchitecture},
    publisher = {IEEE Computer Society},
    month = dec,
}

@inproceedings{doubledip,
    title="{Double-DIP: Augmenting DIP with adaptive promotion policies to manage shared L2 caches}",
    author={Kron, Jonathan D and Prumo, Brooks and Loh, Gabriel H},
    booktitle={The Workshop on Chip Multiprocessor Memory Systems and Interconnects},
    year={2008}
}

@inproceedings{kpc,
    author = {Kim, Jinchun and Teran, Elvira and Gratz, Paul V. and Jim\'{e}nez, Daniel A. and Pugsley, Seth H. and Wilkerson, Chris},
    title = "{Kill the Program Counter: Reconstructing Program Behavior in the Processor Cache Hierarchy}",
    year = {2017},
    month = apr,
    doi = {10.1145/3037697.3037701},
    publisher = {Association for Computing Machinery},
    booktitle = {Proceedings of the International Conference on Architectural Support for Programming Languages and Operating Systems},
    series = {ASPLOS'17}
}

\end{document}